\documentclass[prd,nofootinbib,aps,floats,floatfix,amsmath,amssymb,secnumarabic,twocolumn]{revtex4} %
\usepackage{bm,amsmath,mbboard}
\usepackage{graphicx}
\usepackage[utf8]{inputenc}
\usepackage{color}
\usepackage{hyperref}
\usepackage{slashed}
\usepackage{cancel}
\usepackage{verbatim}
\usepackage{amsmath}
\numberwithin{equation}{section}
\let\oldtheequation\theequation
\renewcommand\theequation{\arabic{section}.\arabic{equation}}
\newcommand{\be}{\begin{equation}}
\newcommand{\ee}{\end{equation}}
\newcommand{\bea}{\begin{eqnarray}}
\newcommand{\eea}{\end{eqnarray}}
\newcommand{\sss}{\scriptscriptstyle}

\newcommand{\nn}{\nonumber}
\newcommand{\ua}{\uparrow}
\newcommand{\da}{\downarrow}
\newcommand{\phm}{\phantom{-}}
\newcommand{\tr}{{\rm tr}}
\newcommand{\Det}{{\rm Det}}
\def\ul#1{\underline{#1}}
\def\sfrac#1#2{{\textstyle{#1\over #2}}}
\makeatletter
\newcommand*\bigcdot{\mathpalette\bigcdot@{.5}}
\newcommand*\bigcdot@[2]{\mathbin{\vcenter{\hbox{\scalebox{#2}{$\m@th#1\bullet$}}}}}
\makeatother
\newcommand{\spm}{\raisebox{0.03cm}{$\sss\pm$}}
\newcommand{\smin}{\raisebox{0.03cm}{$\sss -$}}
\newcommand{\lcdot}{\hbox{\raisebox{-0.05cm}{$\cdot$}}\,}

\begin{document}
\title{Finemensch Lectures on the Strong Interactions}
\author{Robert P.\ Finemensch}
\affiliation{Loresitzen Laboratory, Pacific Institute of Technology,
Primavera, California}
\author{revised by Gerald X.\ Gilbert-Thorple}
\affiliation{University of Southern North Dakota, Hoople, North Dakota}

\begin{abstract}  These twenty-two lectures, with exercises, comprise the extent of what  was
meant to be a  full-year graduate-level course on the strong interactions and QCD, given at
Pactech in 19$xx$-$xy$.  The course was cut short by the illness that led to Finemensch's death.  
Several of the lectures were finalized in collaboration with Finemensch for an anticipated
monograph  based on the course.   The others, while retaining Finemensch's idiosyncrasies, are
revised similarly to those he was able to check. His distinctive approach and manner of
presentation are manifest throughout.  Near the end he suggests a  novel, nonperturbative
formulation of quantum field theory in $D$ dimensions. Supplementary material is  provided
in appendices and ancillary files, including verbatim transcriptions of  three lectures  and
the corresponding audiotaped recordings.

\end{abstract}
\maketitle

\tableofcontents
\makeatletter
\let\toc@pre\relax
\let\toc@post\relax
\makeatother 
%\listoffigures
%\listoftables

\section*{Preface}

During the last year of my Ph.D.\ at Pactech in 19$xx$-$xy$, I was looking
for a course to TA that would not take too much time from finishing my
dissertation.  I had heard that Finemensch did not assign homework in his
courses, and in my naivet\'e asked him if I could be his teaching
assistant for a new course that had been announced, on quantum
chromodynamics.  After checking my credentials with my supervisor Joe
Preeskool, he agreed.  Only afterwards did I realize that the TA in
Finemensch's courses was generally the person who did the transcription
of the notes to create the monograph that would follow.  This was not
the easy job I had bargained for, and I persuaded Stephen R\"osti to
assign several other TAs to the course to divide the labor.  We took
turns rewriting the lectures into publishable form, which Finemensch
would revise before considering final.  Little did I suspect that I
was only postponing my task by $\sim$30 years.

Unfortunately most of those corrected drafts became dispersed with the other TAs,
who have left physics.  In my possession are seven lectures that I prepared 
for publication, at least some of which were  revised by Finemensch. (These are denoted by an asterisk $^*$ in the
section headings.)   As for the rest, I  report what is in my class notes, trying to
convey their intent as best I can.   Based upon the rather extensive revisions he
made to some of my first drafts, the sections he did not check are unlikely to do
justice to all of his intended meanings.  Certain parts call for elaboration, but I abstain
from  restoring longer explanations where I have no record of what Finemensch actually
said.  These fully revised lectures can be found in sections \ref{sect6},
\ref{sect7}, \ref{sect8}, \ref{sect9},  \ref{sect12}.   I was able to supplement my
notes in some places with his own (mostly very sketchy) lecture notes, that are
available from the Pactech Archives, Folder 41.7 of the Finemensch Papers.

For the lectures of Jan.\ 5, 12  and  14, 19$xy$, 
I was able to refer to tape recordings that were kindly
provided by Nura K.\ Atpug, one of the former TAs.   I have placed verbatim 
transcriptions of these lectures in the appendix, as a supplement to the 
more conventional versions in the main document.  
The  quality of the recordings makes it
impossible to reproduce every word, and ellipses indicate words or passages
that I could not make out.  This is especially the case toward the end of long
explanations, where Finemensch's voice would tend to diminish greatly, whereas at
the beginning he might almost be shouting.  These recordings are available
alongside the lectures as supplementary material.    I have preserved as much as possible his original words to convey the
style of delivery, which was considerably more colorful and colloquial than the
tone he  adopted in the drafts to be published.  The reader who compares these
``raw'' versions with the revised ones will understand why it was sometimes challenging
to correctly capture Finemensch's intended meanings.  

One thing you may notice, and
that struck me as an educator now myself, is
that Finemensch was never in a rush to explain anything (although at times he would
speak very fast), nor did he eschew repeating himself, perhaps in several
different ways, to try to get his point across.   And of course there was his
bent for telling stories, which I had forgotten about in the context of this
course,  since I had omitted them from my written notes.  The ``interlude,''
section \ref{interlude}, which were Finemensch's remarks at the start of the new
term, is kept in the main body of the text; it has a few interesting
stories, and shows that he would make time to help a high school student
with his geometry.

I have the impression that in some places Finemensch
had not prepared carefully and was working things out on the spot, sometimes
getting them not quite right, and at times  seemingly meandering through the
material.  This was apparent for example in the early lectures on QCD, where in
subsequent class sessions he came back and revised previous equations to correct
the  details.  It is interesting that no notes corresponding to the
QCD lectures appear in the Pactech Archives folder, suggesting he was speaking
extemporaneously.  
There is also repetition of already introduced material. Perhaps
this was a deliberate pedagogical strategy, since it gave the students time to
digest the concepts and to see it being derived from scratch.  It is also
possible that his terminal illness was interfering with his ability to prepare
as well as he might have liked to.   These detours would have been
smoothed over in the version destined for publication, had there been time for
him to revise the notes.

Although there were no homework assignments, there were some
recommended  problems that are included in the lectures.  
Moreover about a month before the end of the first
term, when students were starting to think about the upcoming final
exams, Finemensch decided that each of them should do an original
research project relating to QCD.  I recall that many  were dumbfounded when
this announcement was made.  Such an unexpected demand made by a
lesser instructor would have created some outcry,  but to a decree
from the great man nobody objected, and everyone somehow managed to
carry out the task: it was a privilege.  Finemensch of course graded the
projects himself, and he comments on them in the interlude section.

One may wonder what the specific content of the unfinished part of the
course might have been. Finemensch announces at the beginning of the second
term that it will be half on perturbative methods followed by
nonperturbative.  At that time he was interested in QCD in $1+1$
dimensions, as an exactly solvable model that might shed light on the
real theory.  He started working with a few graduate students on this
subject, including Sandeep Triviadi.  

His private course notes reveal a different
direction; around 20 of the 60 pages are devoted to reformulating vector
spaces and calculus in arbitrary noninteger dimensions, which he discusses 
in lectures 20-21.  His intent was to combine this with Schwinger's
functional formulation of field theory, presented in lectures 21-22, to overcome
the difficulty of defining the path integral in noninteger
dimensions.  Also in those notes is some material on chiral symmetry breaking by the
axial anomaly and theta vacua in QCD, that he did not have time to present.
No doubt the students would
have been exposed to his ideas for deepening our understanding of the
strong interactions, had he lived until the end of the course.

Finemensch was an inspiring teacher, presenting everything in an incisive and
fascinating way, that obviously had his own mark on it.  He reinvented the subject
as was his wont, even if he was not the first to discover, for example, the
Faddeev-Popov procedure for gauge fixing the path integral.  In the final meetings,
he was too weak to stand at the board, and he delivered the lectures while seated.  He
died less than three weeks following the last lecture.  His  passion for
transmitting the excitement of physics to a new generation never waned.

Sorry this took so long, professor.

\begin{flushright}
Gerald X.\ Gilbert-Thorple \\
Hoople, 2020$\qquad\qquad\quad\ \ \,$\\
\end{flushright}

\begin{figure*}[t]
\vspace{0.5cm}
\centerline{
\includegraphics[width=0.4\textwidth]{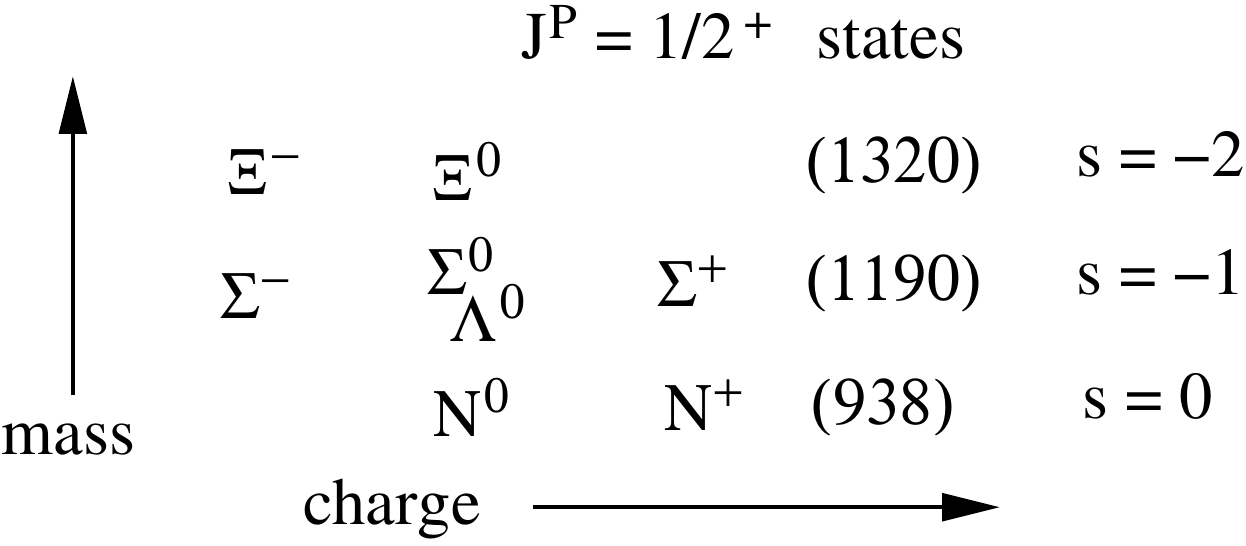}
\hfil\includegraphics[width=0.45\textwidth]{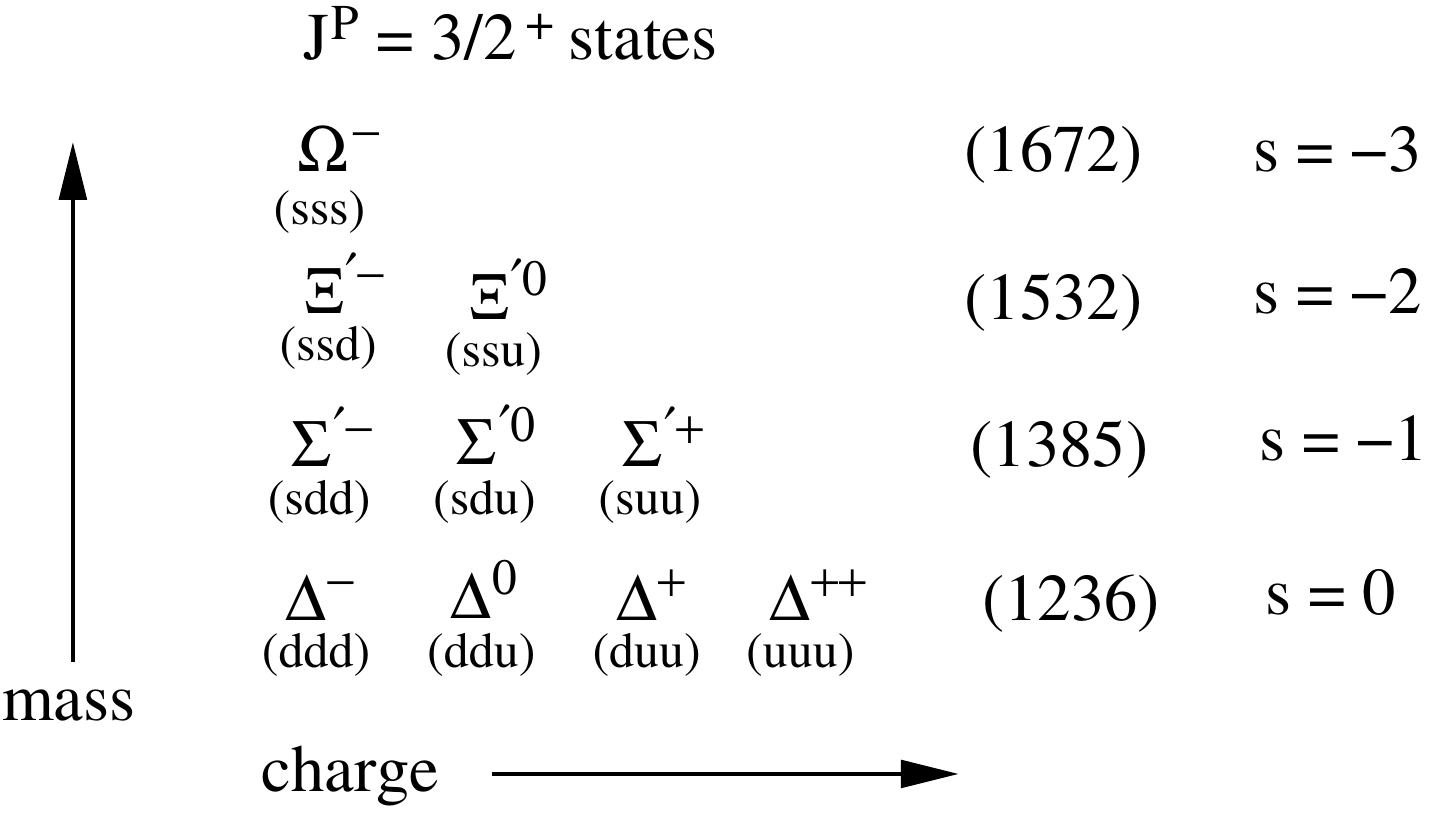}}
\caption{The baryon octet (left) and decuplet (right).
Masses indicated in MeV/$c^2$.
}
\label{baryons}
\end{figure*}

\section{The quark model (10-15-$xx$)}
We begin our exploration of the strong interactions with a survey of 
the hadronic particles, interpreted from the quark model perspective.
The spin-$1/2$ baryons are arranged in an octet in the plane of
mass versus charge, and likewise the spin-$3/2$ baryons form a
decuplet, as shown in fig.\ \ref{baryons}.  The quark content is
indicated for the decuplet states, where the quarks $u,d,s$ have
charge $+2/3,-1/3,-1/3$ respectively, and we take the opposite
convention for the sign of strangeness than is usual.

Detailed properties of the baryons can be understood within the quark
model by constructing the flavor/spin wave functions for the states.
Consider the $\Delta^0$ state ($J=3/2$), whose upper two spin states are given by
\be
\Delta^0:\left\{\begin{array}{ll} ddu\, \ua\ua\ua,& J_z = 3/2\\
	ddu\, \sfrac{1}{\sqrt{3}}(\ua\ua\da + \ua\da\ua + \da\ua\ua),&
	J_z = 1/2\end{array}\right.
\ee
Compare this to the neutron and $\Sigma^0$ ($J=1/2)$,
\bea
	N^0&:&\   ddu\, \sfrac{1}{\sqrt{6}}
	(-2\ua\ua\da + \ua\da\ua + \da\ua\ua), \quad
	J_z = 1/2\nonumber\\
	{\Sigma}^0&:&\   uds\, \sfrac{1}{\sqrt{6}}
	(-2\ua\ua\da + \ua\da\ua + \da\ua\ua)
\label{neutron}
\eea
For $\Delta^0$ and $N^0$, the coefficients of the 
$\ua\da\ua + \da\ua\ua$ spin terms had to be equal, since
they are symmetric under interchange of the first two quarks,
which have identical flavors ($dd$).  However this is not a constraint
for the $uds$ baryons, so there must exist an additional state
$\Lambda^0$
\be
	\Lambda^0:\ uds\ \sfrac{1}{\sqrt{2}}(\ua\da\ua - \da\ua\ua)
\ee
that has isospin 0.  The fact that the mass eigenstates are also
eigenstates of isospin indicates that $u$ and $d$ are approximately
degenerate, compared to the scale of the hadron masses.

\begin{figure}[b]
\vspace{0.5cm}
\centerline{
\includegraphics[width=0.3\textwidth]{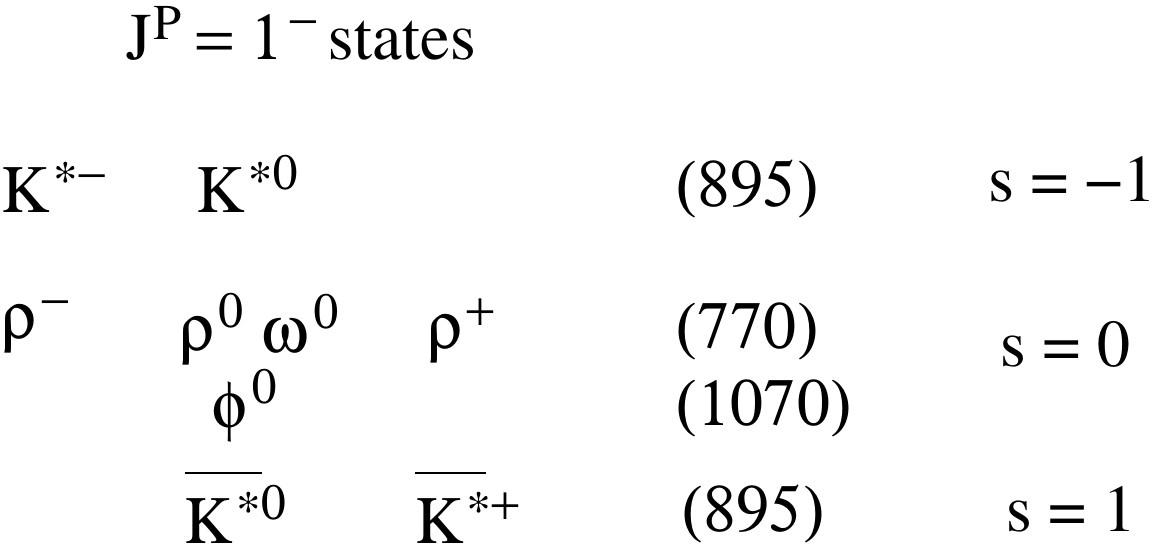}}
\caption{The vector meson nonet.}
\label{vector-meson}
\end{figure}

Similarly the mesons can be arranged into multiplets, as we
illustrate for the $J=1^-$ vector mesons in fig.\
\ref{vector-meson}.
The wave functions are given by
\bea
	\omega&:&\ \sfrac{1}{\sqrt{2}}
	(u\bar u+d\bar d)\ \ua\ua\nn\\
	\rho^+&:&\ u\bar d\ \ua\ua; \quad \rho^0:\ \sfrac{1}{\sqrt{2}}
	(u\bar u-d\bar d)\ \ua\ua;\quad
	\rho^-:\ d\bar u\ \ua\ua\nn\\
	{K^{*}}^-&:&\ s\bar u\  \ua\ua;\quad {K^{*}}^0:\ s\bar d\ \ua\ua; 
	\nn\\
	\overline{K^{*}}^{\,+}&:&\ u\bar s\  \ua\ua;\quad 
	\overline{ K^{*}}^{\,0}:\ d\bar s\ \ua\ua;\nn\\
	\phi&:&\ s\bar s\ \ua\ua 
\eea
It is interesting to notice that the $\omega$ and $\rho^0$ are very
close to each other in mass.  What do we learn about the strong
interactions from this near-degeneracy?  Apparently, the strong
interactions conserve isospin.

It is also interesting to observe that $\phi$ decays much faster into
$KK$ than into pions.  This is an example of Zweig's rule (OZI
suppression), that can be pictured diagrammatically by the statement
that\\

\vskip-0.2cm
\centerline{\includegraphics[width=0.45\textwidth]{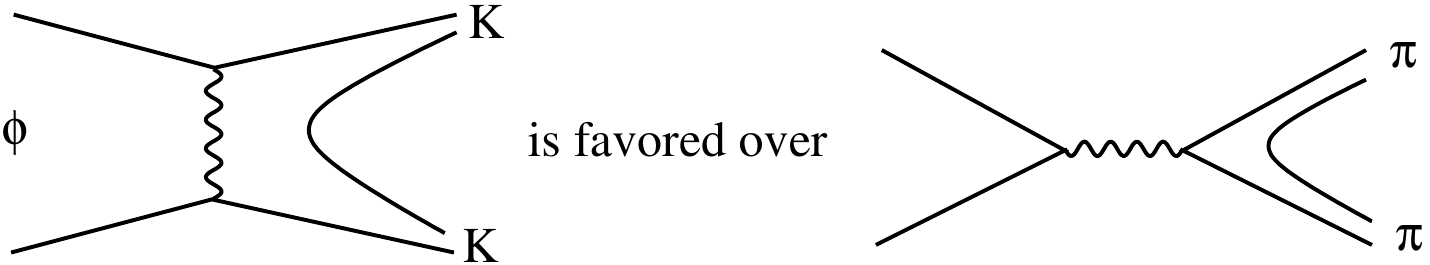}}
One might wonder whether OZI suppression in this example is somehow
related to the degeneracy of the $\phi$-$\omega$ system.  In fact
there is a connection: if $\omega$ had some $s\bar s$ content rather
than being purely made from $u\bar u$ and $d\bar d$, which would spoil
the degeneracy, then by the same mixing $\phi$ would also have light
quark content, allowing for decays into pions without going through
the annihilation diagram. 

The pseudoscalar mesons ($J^P=0^-$) have a different flavor structure from the vector
mesons, apart from the similarities between the two isotriplets
$\rho$ and $\pi$,
\be
	[\pi^+,\ \pi^-,\ \pi^0]:\ \sfrac{1}{\sqrt{2}}(\ua\da-\da\ua)
	\left[u\bar d,\ d\bar u, \sfrac{1}{\sqrt{2}}(u\bar u-d\bar d)
	\right]
\ee
In this case there is mixing between the isosinglets,
\bea
	\eta\,(546)&\sim& (u\bar u + d\bar d) - 1.4\, s\bar s\nn\\
	\eta'(960)&\sim& (u\bar u + d\bar d) + 0.7 \, s\bar s\
\eea
Why isn't $\eta$ purely $(u\bar u + d\bar d)$, in analogy to $\omega$,
which would have made it approximately degenerate with the pions?
This has to do with chiral symmetry breaking, which is specific to
QCD and not accounted for by the quark model.

An interesting prediction of the quark model is electromagnetic
matrix elements, that determine the baryon magnetic moments.
We consider those of the proton and the neutron, where the proton
wave function is
\be
	|p\rangle = (uud)\,\sfrac{1}{\sqrt{6}}(-2\ua\ua\da + \ua\da\ua + \da\ua\ua) 
\label{proton}
\ee
The magnetic moment is given by
\be
	\left\langle p\left| {q\hbar\over 2 m}\,\sigma_z\right|p\right\rangle
\ee
where $q$ is the charge operator acting on the quarks, and $m=m_p/3$
is the constituent quark mass.  
Using (\ref{proton}),
\bea
	q\sigma_z|p\rangle = {uud\over\sqrt{6}}\Big( \!\!\!\!\!\!\!
	\!\!\!
	&&{2e\over3}(-2\ua\ua\da + \ua\da\ua - \da\ua\ua)\nn\\
	&+&{2e\over3}(-2\ua\ua\da - \ua\da\ua + \da\ua\ua)\nn\\
	&-&{e\over 3}(+2\ua\ua\da + \ua\da\ua + \da\ua\ua)\Big)
\eea
we find
\be
	\left\langle p\left| 
{q\hbar\over 2 m}\,\sigma_z\right|p\right\rangle = 3\, \mu_N
\ee
where $\mu_N = e\hbar/2m_p$ is the nuclear magneton.  The analogous
calculation for the neutron (see eq.\ (\ref{neutron})) gives $-2\,\mu_N$.  These predictions are
compared to the measured values in the table \ref{tab1}.\footnote{In class, RPF only presented the $n$ and
$p$ values, and omitted the ``corrected'' predictions.  I have restored these and
some of the related discussion from his private notes.}  

\begin{table}[b]
\begin{center}
\setlength\tabcolsep{1.5pt} % default value: 6pt
\begin{tabular}{|c|c|c|c|c|c|c|}\hline
         & p        & n       & $\Lambda$      & $\Sigma^+$    & $\Sigma^-$     & $\Xi^-$ \\ 
\hline
$\mu_{\rm th.}$ & $\phm 3$ & $-2$    &                & $3$           &   $-1$         &   $-1$       \\
$\mu_{\rm exp.}$   & $2.79$   & $-1.93$ & $\smin0.61\spm.03$ & $2.83\spm.25$ & $\smin1.48\spm.37$&  $\smin1.85\spm.75$  \\
$\mu_{\rm corr.}$  & $2.59$   & $-1.73$ & $-0.58$        & $2.50$        & $-0.96$        &  $-0.49$\\
\hline
\end{tabular}
\setlength\tabcolsep{6pt} % default value: 6pt
\label{tab1}
\caption{Predicted and observed baryon magnetic moments, in the quark model.}
\end{center}
\end{table}

These predictions can be corrected, as shown in the 
third row of the table, by taking a more realistic value of the constituent $u$ and $d$ quark masses, $m_q=1085/3\cong 362$\,MeV
instead of $m_p/3$.\footnote{It is not explained in his notes where the number 1085 
comes from; probably it is a consequence of taking the spin-spin interactions into account
in the baryon mass calculation.}  Further improvement
might arise from taking into account isospin breaking; the $u$ and $d$ masses are not exactly the same.  We must
certainly take SU(3) flavor breaking into account for the $s$ quark, whose constituent mass is $m_s = 1617/3=539\,$MeV.  We can then
predict the other magnetic moments as
\bea
	{q\sigma_z\over m}|\Lambda^0\rangle &=& {sud\over\sqrt{2}}{q\sigma_z\over m}\left(\ua\ua\da-\ua\da\ua\right)
	\to -{1\over 3 m_s}\nn\\
	{q\sigma_z\over m}|\Sigma^+\rangle &=& {sud\over\sqrt{6}}{q\sigma_z\over m}
	\left(2\da\ua\ua-\ua\da\ua -\ua\ua\da\right)\nn\\
	&\to&\sfrac46\left[+\sfrac{1}{3m_s}+\sfrac{2}{3m_q}+\sfrac{2}{3m_q}\right]+\sfrac13\left[-\sfrac{1}{3m_s}\right]\nn\\
        &=& \sfrac{1}{9m_s} + \sfrac{8}{9m_q}
\eea
\bea
	{q\sigma_z\over m}|\Sigma^-\rangle &=& {sdd\over\sqrt{6}}{q\sigma_z\over m}
	\left(2\da\ua\ua-\ua\da\ua -\ua\ua\da\right)\nn\\
	&\to&\sfrac46\left[+\sfrac{1}{3m_s}-\sfrac{1}{3m_q}-\sfrac{1}{3m_q}\right]+\sfrac13\left[-\sfrac{1}{3m_s}\right]\nn\\
        &=& \sfrac{1}{9m_s} - \sfrac{4}{9m_q}\nn\\
	{q\sigma_z\over m}|\Xi^-\rangle &=& {dss\over\sqrt{6}}{q\sigma_z\over m}
	\left(2\da\ua\ua-\ua\da\ua -\ua\ua\da\right)\nn\\
        &\to& \sfrac{1}{9m_q} - \sfrac{4}{9m_s}
\eea
These are in rather good agreement with the data, except for the $\Xi^-$.

The nonrelativistic quark model can also be used to predict the axial vector current
matrix elements, $\sigma_z\gamma^5$. In the quark model we find that $\sigma_z\gamma^5$
gives $+1$ for $u$ and $-1$ for $d$, leading to the prediction
\be
g_A = \sfrac46[1+1+1] + \sfrac13[-1] = 5/3 
\ee
for the proton,\footnote{This calculation, also taken from his written notes, seems to be based on 
unstated insights from the extended quark model analysis that takes into account the small components
of the Dirac spinors.}   which is high compared to the experimental value $1.253\pm0.007$.

{\bf Exercise.}  What kind of baryon states do you expect when there
is one unit of internal angular momentum?

\begin{figure}[b]
\centerline{
\includegraphics[width=0.25\textwidth]{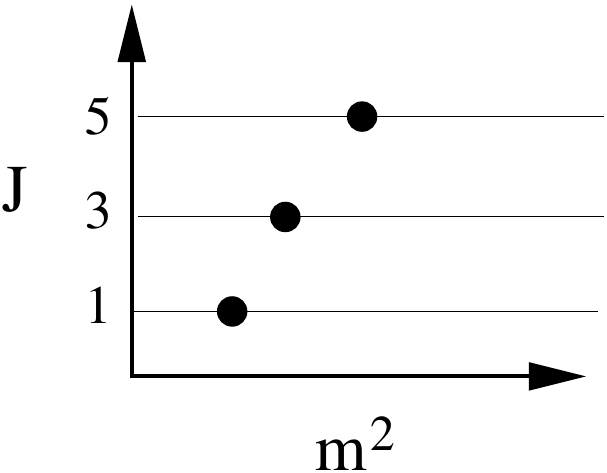}}
\caption{A Regge trajectory.}
\label{regge}
\end{figure}
\begin{figure}[t]
\centerline{
\includegraphics[width=0.35\textwidth]{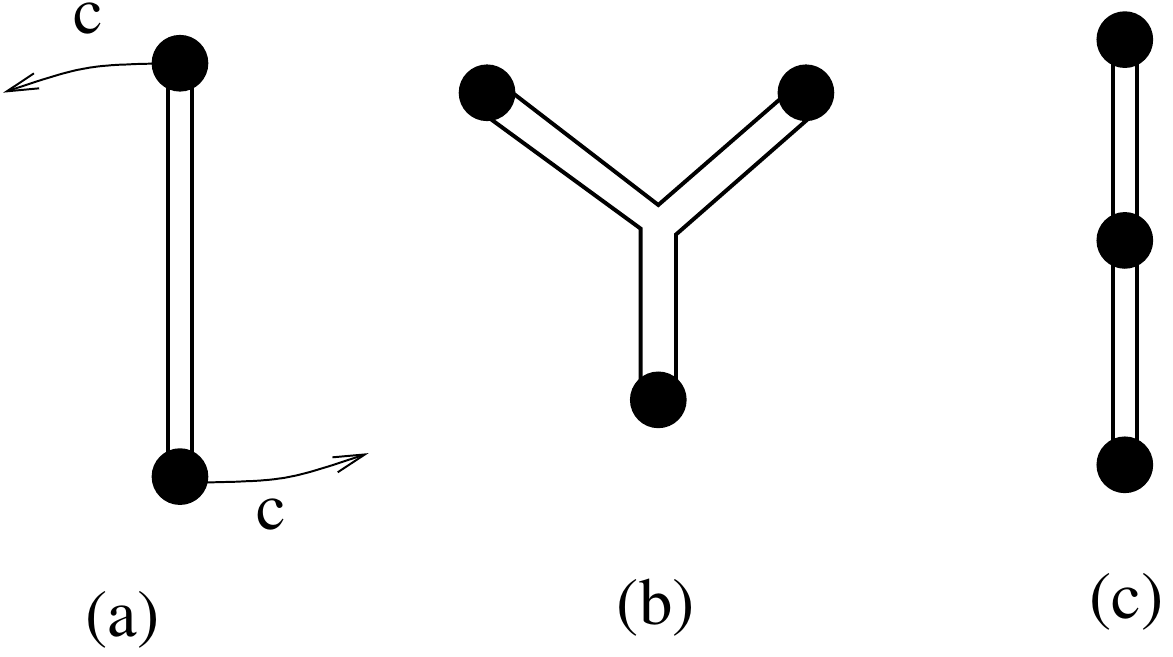}}
\caption{String model for mesons (a)  and  baryons (b,c).}
\label{string}
\end{figure}

\section{Other phenomenological models (10-20-$xx$)}
There are complementary phenomenological models for describing the
strong interactions, which we briefly review here.  The first is the
relativistic string, which was inspired by the observed Regge
trajectories.  These are plots of the spin versus mass squared of
hadronic resonances, as in fig.\ \ref{regge}.  One considers only
families of resonances having the same parity, requiring $J$ to jump
by $\Delta J = 2$.  Empirically the trajectory is linear, which was not predicted
by the quark model.  However the relativistic string, illustrated
in fig.\ \ref{string}, gets the correct relation.\footnote{This is worked out
in the next lecture.}\   In the simplest
version of the model, the masses of quarks or antiquarks on the 
ends of the string are neglected, and one cares only about the
constant tension $T = $ energy/length of the strings, which represent
flux tubes of the strong interaction field.  One finds
that $m^2$ (the energy squared) is proportional to the angular
momentum, with the endpoints of the string moving at the speed of
light.  To explain the linear Regge trajectories of baryons in this 
picture, one could imagine flux tube configurations as in 
fig.\ \ref{string}(b,c).  Configuration (c) would obviously lead to
the same prediction of linear Regge trajectories as for mesons.

\begin{figure}[b]
\centerline{
\includegraphics[width=0.45\textwidth]{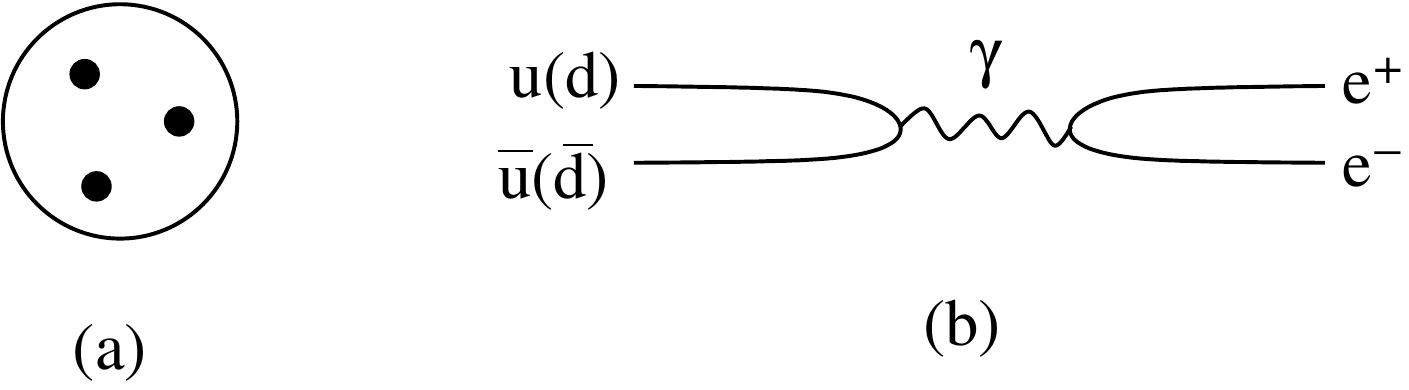}}
\caption{(a) Bag model for baryons. (b) Electromagnetic decay of
$\rho$ meson.}
\label{bag}
\end{figure}

In the bag model (fig.\ \ref{bag}(a)), quarks in a hadron ``push away the
vacuum'' and move around in this evacuated region with nearly zero
mass.  It takes energy to make the hole, which is interpreted as the
hadron mass.  An application is to the decay $\rho\to e^+ e^-$
(fig.\ \ref{bag}(b)), where we recall that $\rho = (u\bar u - d\bar
d)/\sqrt{2}$.  One must know the wave function of the $q\bar q$ bound
state at the origin, $\psi(0)$, to estimate the amplitude.  Although
the bag model gives a reasonable estimate for $\psi(0)$, the bag
nevertheless turns out to be too stable to get the rate right.
One needs to make it more dynamical than in the bag model
picture, so that it shrinks more easily when the $q$ and $\bar q$
are close to each other.  And the bag should turn into a flux tube
when the pair is well-separated.

\begin{figure}[t]
\centerline{
\includegraphics[width=0.25\textwidth]{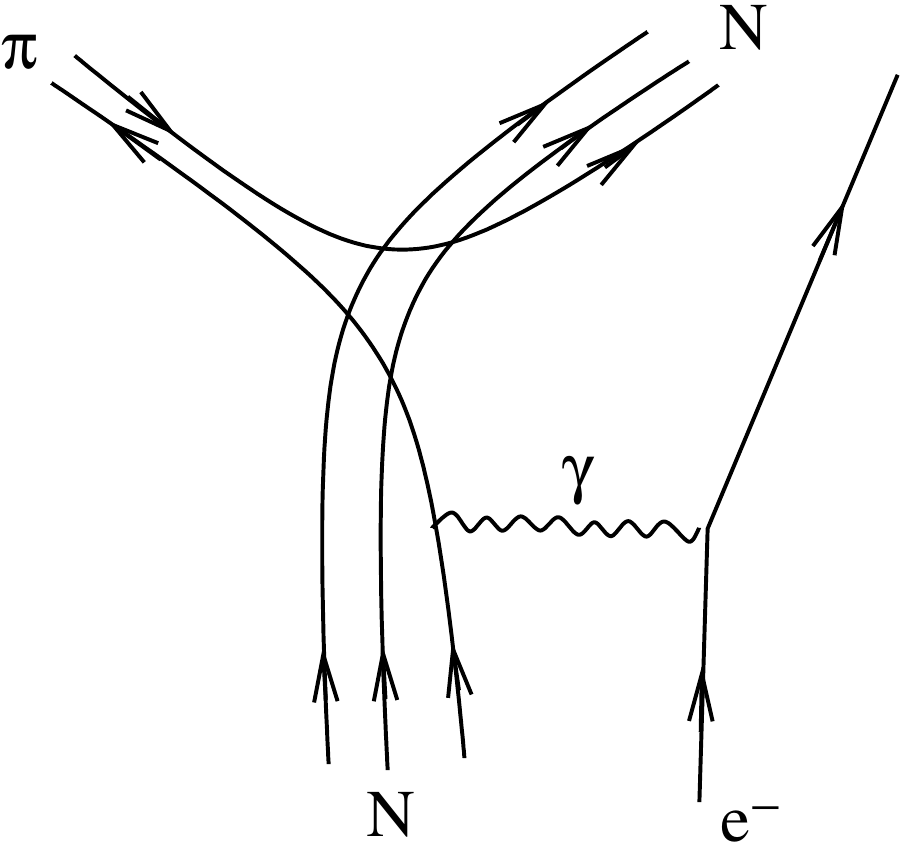}}
\caption{High-energy electron-nucleon scattering}
\label{parton}
\end{figure}

The parton model is useful for describing high-energy processes,
including inelastic scattering of electrons on nucleons.  An example is shown
in fig.\ \ref{parton}, for the case of $e N\to e N\pi$.
Let us think about the partons in the initial state nucleon, in a
reference frame where it is moving to the right with 4-momentum
$p^\mu = (E,p,0,0)$ and $E\sim 10\,$GeV for example, and the
virtual photon 4-momentum is $q^\mu = (0,\vec Q)$.  A parton
in the nucleus will have momentum components 
\be
	p_\parallel = x p,\quad p_\perp \sim 300\,{\rm MeV}
\ee
parallel and perpendicular to the beam, respectively, where $x$ is
the momentum fraction of the parton that interacts with the
virtual photon.  In this frame, its momentum just gets reversed after
scattering, $px \to -px = Q/2$, since its energy
changes by $q^0 = 0$.  In an arbitrary frame
we can write the momentum fraction as
\be
	x = {Q^2\over 2 p Q} = {-q^2\over -2p\cdot q}\, .
\ee

The momentum distribution of partons in the nucleon can be thought of
as coming from their respective wave functions, written in momentum
space.  Naively, we would expect the probability to find a quark
with momentum in the interval $[5,\,5.5]\,$GeV in a 10 GeV proton
to be the same as for the interval $[10,\,11]\,$GeV in a 20 GeV
proton.  Each parton has its own
probability distribution
\bea
	u(x),\, d(x),\,s(x): &\quad {\rm quarks}\nn\\
	\bar u(x),\, \bar d(x),\, \bar s(x): &\quad {\rm antiquarks}\nn\\
	g(x):&\quad {\rm  gluon}
\eea
For example $u(x)$ is the probability density for finding a $u$ quark
with momentum fraction $x$.  Of course these definitions depend upon
which hadron the parton belongs to.  If we define the above functions
as belonging to the proton, then the amplitude for the photoproduction
process is proportional to 
\be
	\sfrac49 u(x) + \sfrac19 d(x) + \sfrac19 s(x) + \sfrac49 \bar
u(x) + \cdots,\quad (e p \to e N\pi)
\ee
for scattering on protons, whereas it is 
\be
	\sfrac49 d(x) + \sfrac19 u(x) + \sfrac19 s(x) + \sfrac49 \bar
u(x) + \cdots,\quad (e n \to e N\pi)
\ee
for scattering on neutrons since $u(x)$ in a proton must be equal
to $d(x)$ in a neutron.  The amplitudes will of course also depend
upon $Q$.

{\bf Challenge.} Compute the width for $\phi\to e^+ e^-$.

\section{Deep inelastic scattering; electron-positron annihilation (10-22-$xx$)}
In the last lecture we saw that the electron-proton scattering cross
section is proportional to a function
\be
	F^{ep}(x) = \sfrac49(u + \bar u) + \sfrac19(d+\bar d)
	+\sfrac19(s+\bar s)
\ee
where $u(x)$ is the probability density for finding a $u$ quark with 
momentum fraction $x$ in the proton.  The momentum distribution
functions are subject to constraints
\bea
	1 &=& \left({\rm proton\atop charge}\right) = \int\left[\sfrac23(u\!-\!\bar u)
	\!-\!\sfrac13(d\!-\!\bar d)\! -\!\sfrac13(s\!-\!\bar s)\right]dx\nn\\
	0 &=& \left({\rm neutron\atop charge}\right) = 
	\int\left[\sfrac23(d\!-\!\bar d)
	\!-\!\sfrac13(u\!-\!\bar u) \!-\! \sfrac13(s\!-\!\bar s)\right]dx\nn\\
	0 &=& {\rm nucleon\ strangeness} = \int(s-\bar s)\,dx
\eea
where again we assumed the neutron is related to the proton by
interchange $u\leftrightarrow d$.  From these it follows that
\bea
	\int(u-\bar u)\,dx &=& 2\nn\\
	\int(d-\bar d)\,dx &=& 1\nn\\
	\int(s-\bar s)\,dx &=& = 0
\eea
It has been shown that as $x\to 0$, the distribution functions have
the behavior
\be
	u=\bar u = d = \bar d \sim {0.24\over x}
\ee
Likewise, $s=\bar s$ scales as $1/x$.  This behavior can be understood
as coming from brehmsstrahlung of soft gluons, which have a
distribution of $1/x$.  These (virtual) gluons decay into soft
quark-antiquark pairs,\\
\centerline{
\includegraphics[width=0.25\textwidth]{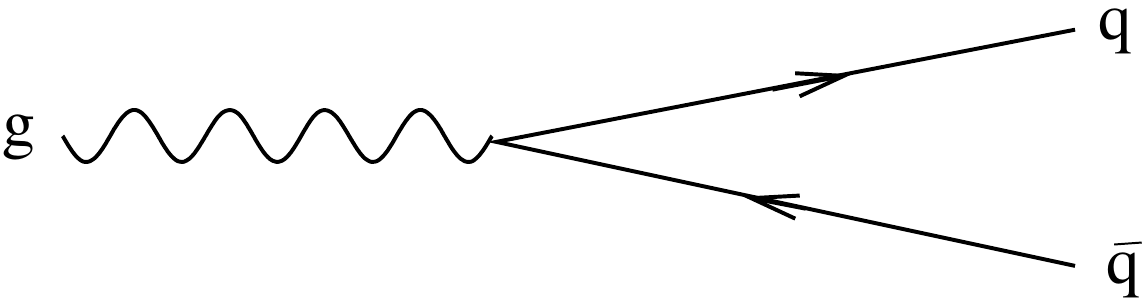}}
explaining why all flavors have the same
$1/x$ dependence at low $x$, regardless of whether they are particles
or antiparticles: gluons can decay into all flavors equally.  The
gluons are known to comprise a significant fraction of the total partons,
\be
	\int_0^1dx\,g(x) \approx 0.44
\ee

\begin{figure}[t]
\centerline{
\includegraphics[width=0.3\textwidth]{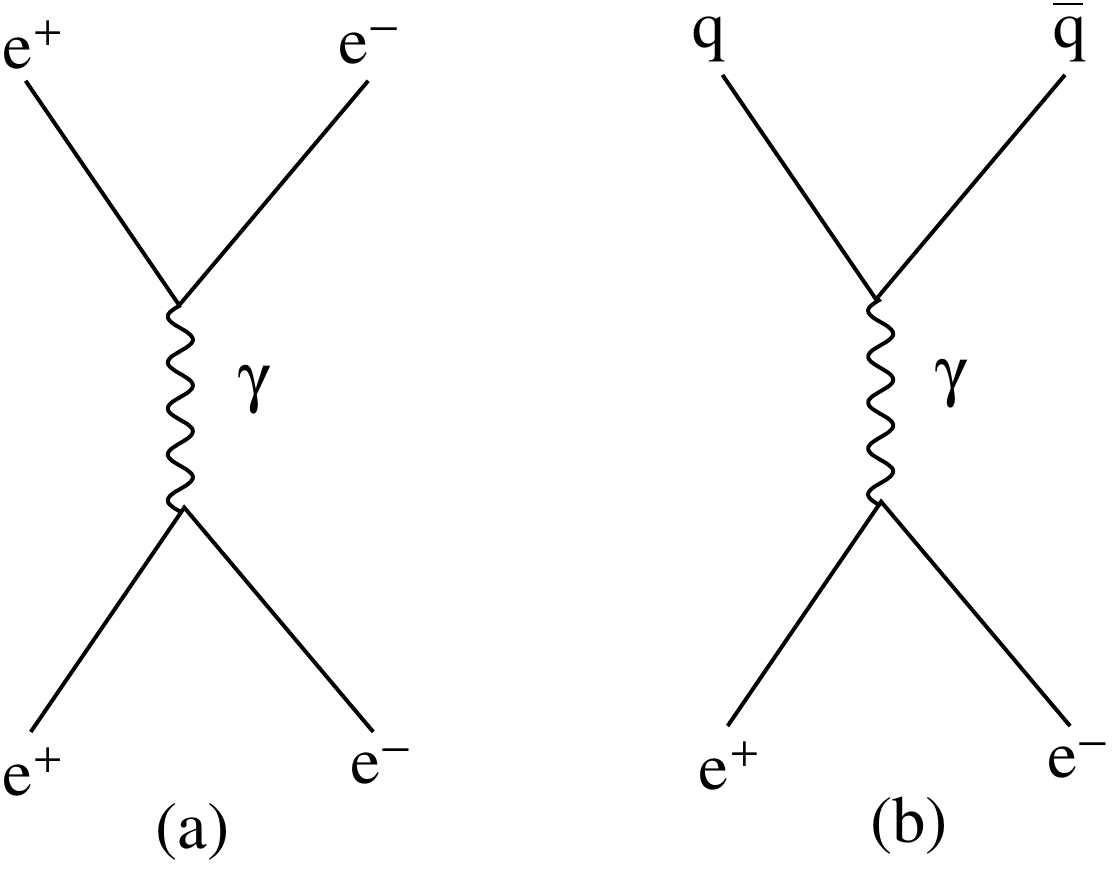}}
\caption{Electron-positron annihilation into $e^+e^-$ (a)
and $q\bar q$ (b).}
\label{ee-ann}
\end{figure}

So far, we have taken for granted that we know the charges of the
quarks.  One experiment that constrains the charges is annihilation of
electrons and positrons, fig.~\ref{ee-ann}.  Denoting the cross
section for annihilation into $e^+e^-$ by $\sigma_{\rm el}$, we can
express that for annihilation into $u\bar u$ as\footnote{RPF has apparently
ignored the $t$-channel
contribution to $\sigma_{\rm el}$ here.}
\be
	\sigma_{u\bar u} \sim \frac49\,\sigma_{\rm el}\cdot 3
\ee
where the final factor of 3 is for the number of colors, which we must
determine by some independent means.  We have assumed here that final
state interactions can be neglected. Similarly for 
$\sigma_{d\bar d}$ or $\sigma_{s\bar s}$ we get 
$\sfrac19 \,\sigma_{\rm el}\cdot 3$.  For the inclusive cross
section to produce hadrons, we add the three flavors together to 
obtain
\be
	{\sigma_{e^+e^-\to{\rm hadrons}}\over \sigma_{\rm el}}
	= \frac43 +\frac13 + \frac13 = 2
\ee
assuming the energy is below the $c$ quark threshold.  Above
this threshold, $2\to 10/3$, and above the $b$ quark threshold
$10/3\to 11/3$.

\begin{figure}[t]
\centerline{
\includegraphics[width=0.3\textwidth]{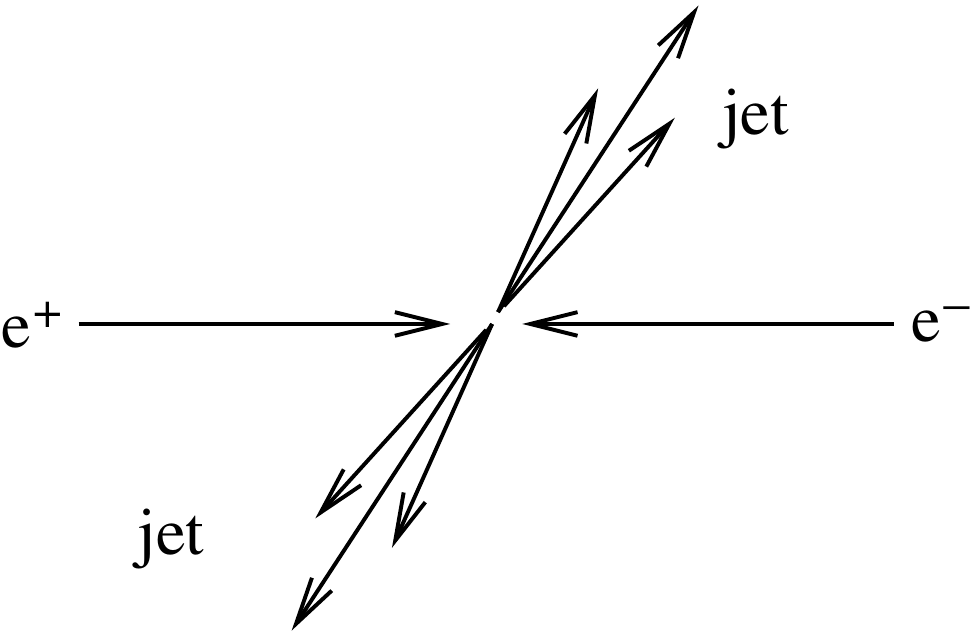}}
\caption{Electron-positron annihilation into hadronic jets.}
\label{ee-jet}
\end{figure}

\begin{figure}[b]
\centerline{
\includegraphics[width=0.4\textwidth]{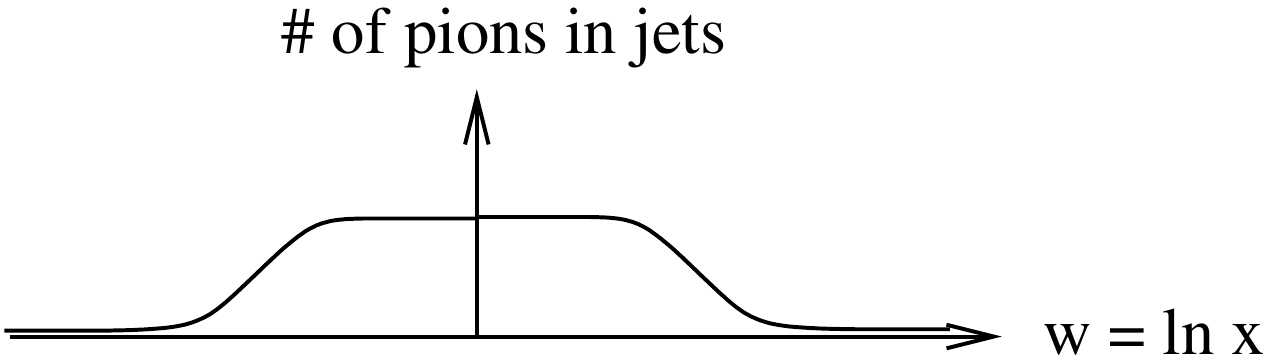}}
\caption{Distribution of pions with momentum fraction $x$
in jets.}
\label{dist}
\end{figure}

Or course what we really see is not quarks in the final state, but
rather jets (fig.\ \ref{ee-jet}), primarily $K$'s and $\pi$'s, with smaller admixtures
of nucleons and antinucleons.  One can define probability distribution
functions for hadrons in the jets in analogy to those of the partons,
for example $\pi(x)$, where now the momentum fraction is defined as
\be
	x = {\hbox{energy of $\pi$}\over \hbox{energy of $e^+e^-$}}
\ee
Like for the quarks, these distributions go like $1/x$ at small $x$.
Denote the components of the $\pi$ momentum parallel to and transverse
to the average jet momentum as $p_{\parallel}$ and $p_\perp$.  Then
\be
	{dx\over x} = {dp_{\parallel}\over E}
	= {dp_{\parallel}\over \sqrt{m_\pi^2 + p_\perp^2 + 
	p_\parallel^2}}
\ee
This shows that at small $x$ the distribution of particles with
momentum fraction $x$ is flat as a function of $w=\ln x$, 
fig.\ \ref{dist}.
Lorentz transforming to a frame where the average momentum of the
two jets does not add to zero causes the distribution to be translated
to the right or left in $w$.
More generally, we can define {\it fragmentation functions}
$D^q_h(x)$ that denote the probability distribution for producing a
hadron  of type $h$ and momentum fraction $x$ from a jet originating
from a quark of flavor $q$.  These can be measured in deep inelastic
scattering experiments.

The formation of two jets from the breaking of a string of strong
interaction flux is in some ways analogous to a simpler problem, 
the spontaneous emission of an electron-positron pair from a constant
electric field.  Solving the Dirac equation in this background, I find
that the probability of pair production is $\exp(- {e^2\over
8\pi}\,{m^2+\vec p^{\,2}\over \vec E^2})$.  Now imagine that the electric
field is created by two charged plates that are moving together with
velocity $v$.  The pairs should be produced with net total momentum.
But locally they are created from a uniform field, so how do they know
they should have net momentum?

A related problem concerns the wave function of quarks in a stationary
proton versus a moving proton.  The wave function is not a
relativistic invariant, nor even something that transforms nicely.\\
{\bf Exercise.} From the Schr\"odinger equation, how does the solution
for the wave function $\psi$ transform when the potential changes by
the Galilean transformation $V(\vec r) \to V(\vec r-\vec v t)$?
(Answer: $\psi\to e^{im(\vec v\cdot\vec r-\frac12\vec v^{\,2} t)}\psi(\vec
 r-\vec v t,t)$.)

\begin{figure}[t]
\centerline{
\includegraphics[width=0.4\textwidth]{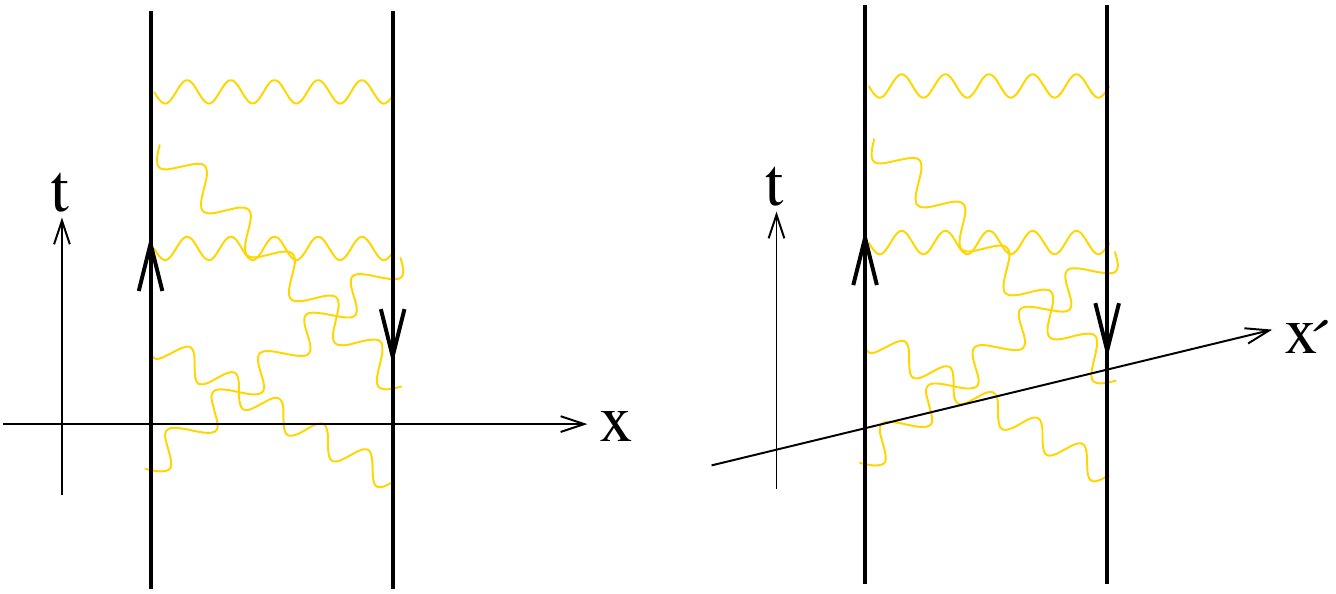}}
\caption{Relativity of simultaneity for positronium consituents
in the rest frame versus a boosted frame.}
\label{positronium}
\end{figure}

Similarly, the wave function for positronium, $\psi(\vec x_1,\vec
x_2,t)$, depending on the positions of its constituents, changes in
a complicated way, as can be understood by Lorentz transforming and
noting that in the new frame, the events that were $(t,\vec x_1)$ 
and $(t,\vec x_2)$ in the original frame are no longer simultaneous;
see fig.\ \ref{positronium}.
We have to evolve one particle forward and the other backward in time
to find the new wave function.  Hence $\psi$ in the new frame,
call it $\psi'$, is not just a function of the original $\psi$, but
rather $\psi' = f(\psi,H)$, depending also on the Hamiltonian $H$ of
the system.

\begin{figure}[t]
\centerline{
\includegraphics[width=0.25\textwidth]{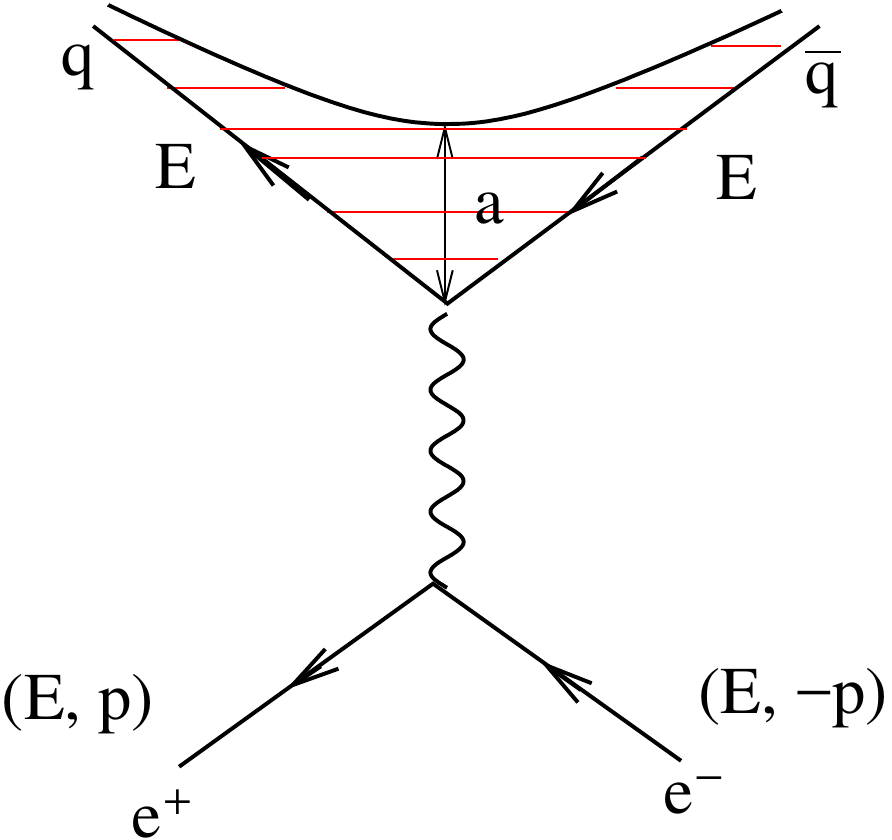}}
\caption{Hadronization after $e^+e^-\to q\bar q$ production, showing
the breaking of the QCD string.}
\label{ee-qq}
\end{figure}

We can say something more quantitative about the distributions of
transverse and longitudinal momenta however.  The center of mass
energy of the $e^+e$ is $2E \cong 2|\vec p|$.  The produced quarks
carry less momentum because of the mass produced in the QCD string
during hadronization, $E = \sqrt{p_\parallel^2 + p_\perp^2 + m^2}
= xp\sqrt{1 + (p_\perp^2+m^2)/(xp)^2}$ 
(recalling that $p_\parallel = x p$).  Hence\footnote{This equation
which holds at large $xp$ does not seem to be needed for what
follows.}
\be
	E - p_\parallel \sim {\#\over x p}\,.
\ee
A measure of momentum loss is given by the sum over the different
kinds of hadronic particles $h$ produced,
\bea
	|\Delta\vec p| &=& \sum_h \int (E-p_\parallel){d p_\parallel\over E}
	C_h\, d^2 p_\perp\nn\\ &=& \sum_h \int d^2 p_\perp\, C_h\,
	(p_\parallel - E)\Big|^p_0\nn\\
	&=& \sum_h \int d^2 p_\perp\, C_h\,\sqrt{p_\perp^2 +
m_h^2}
\eea
where $C_h$ depends on the particle.  As shown, one can do the integral over
$p_\parallel$ exactly.  This gives us a way of measuring the string
tension $T$, since the loss of momentum is given by Newton's law,
\be
	\sum_{L+R} |\Delta\vec p| = T \times a
\ee
where the sum is over both left and right sides of the diagram (the two
jets) and $a$ is the time it takes to break the string, as shown in
fig.\ \ref{ee-qq}.

The theory of the string tension is not very quantitative.  It comes from
Regge trajectories, but these are not known for highly stretched
strings, and moreover in the real situation there are quarks at the
ends of the strings, that have not been taken into account.

\begin{figure}[t]
\centerline{
\includegraphics[width=0.25\textwidth]{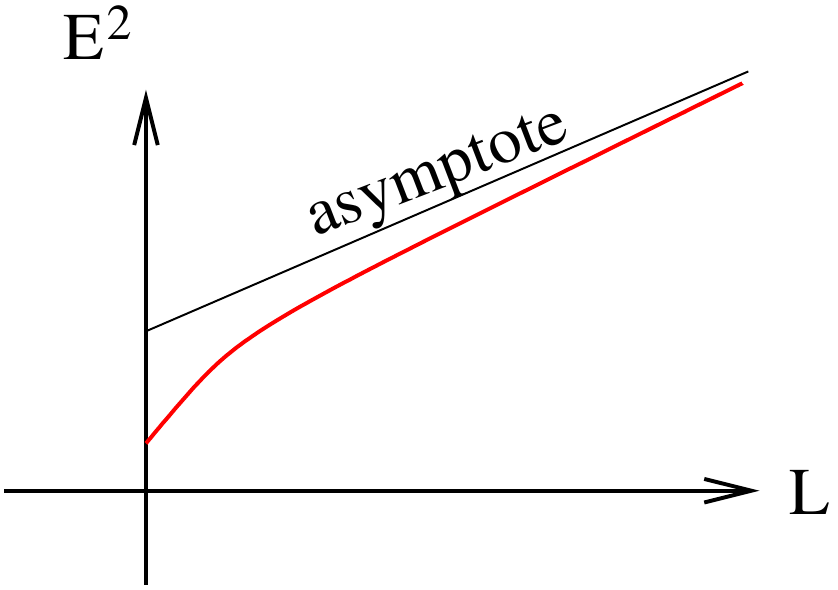}}
\caption{Regge trajectory for the flexible bag model.}
\label{flex-bag}
\end{figure}

{\bf Exercise.} The ``flexible bag'' model takes the quark mass to
depend on the quark separation in a hadron, with Hamiltonian
$H = \sfrac12 m(r) {d\over dt}{\vec r}\cdot {d\over dt}{\vec r} + 
V(r)$, $V(r) = kr$, $m(r) = \mu r$.  The quark mass represents the
inertia of the gluon field (bag).  Prove that $E^2 \sim aL +b + c/L
+\dots$ for large angular momentum $L$.  Find $E$
for $L = 0$ and for large $L$.\footnote{RPF in his private notes devotes four pages to
working this out, first classically for circular orbits, then quantum mechanically using
exponential and Gaussian variational anz\"atze for the wave function.}  [RPF shows graphically his result in 
fig.\ \ref{flex-bag}.]\\
{\bf Problem.}  What happens for the relativistic treatment of
the string?  We must formulate a relativistic equation.  String
theory!\footnote{This was around the time of the first string
revolution.  I reproduce the following answer from RPF's private notes.}\\
{\bf Solution:} The proper tension $T$ is the energy per unit length.  The radial variable
goes from $0$ to $a$, so the velocity varies as $v(r) = r/a$ along the string, which rotates
at angular frequency $\omega = 1/a$.  The differential force acting on an element of the
string is $dF = \omega\mu v dr$ where $\mu = T/\sqrt{1-v^2}$.  Therefore the total energy
and angular momenta are
\bea
	E &=& 2\int_0^a {T\,dr\over \sqrt{1-v^2}} = 2 T a\int_0^1 {dv\over\sqrt{1-v^2}} = 
\pi T a\nn\\
	J &=& 2\int_0^a \mu v r\, dr = 2 Ta^2\int_0^1{v^2\,dv\over \sqrt{1-v^2}}=
	\sfrac{\pi}{2} T a^2\nn
\eea
and we understand the Regge trajectory behavior, $E^2 = 2\pi T J$.  Comparing to data,
$2\pi T = 1.05\,$GeV$^2$, giving $T = 0.167$\,GeV$^2$.

\bigskip
{\bf References}.\footnote{These were originally given at the end of lecture 5,
but logically they belong here since they pertain to quark models.} 

Quark model references:
\begin{trivlist}
\item
  O.~W.~Greenberg,
  ``Spin and Unitary Spin Independence in a Paraquark Model of Baryons and Mesons,''
  Phys.\ Rev.\ Lett.\  {\bf 13}, 598 (1964).
  doi:10.1103/PhysRevLett.13.598
\item
 O.~W.~Greenberg and M.~Resnikoff,
  ``Symmetric Quark Model of Baryon Resonances,''
  Phys.\ Rev.\  {\bf 163}, 1844 (1967).
  doi:10.1103/PhysRev.163.1844
\item
R.~P.~Finemensch, M.~Kissingher and F.~Ravendahl,
  ``Past and future matrix elements from a relativistic quark model,''
  Unphys.\ Rev.\ D {\bf 3}, 2706 (1971).
  doi:10.1103/UnPhysRevD.3.2706
\item Isgur\footnote{No specific references are given, but probably RPF had in
mind Isgur's papers from 1978-1979 on the quark model.}
\end{trivlist} 

MIT bag model references:
\begin{trivlist}
\item
  A.~Chodos, R.~L.~Jaffe, K.~Johnson, C.~B.~Thorn and V.~F.~Weisskopf,
  ``A New Extended Model of Hadrons,''
  Phys.\ Rev.\ D {\bf 9}, 3471 (1974).
  doi:10.1103/PhysRevD.9.3471
\item
 T.~A.~DeGrand, R.~L.~Jaffe, K.~Johnson and J.~E.~Kiskis,
  ``Masses and Other Parameters of the Light Hadrons,''
  Phys.\ Rev.\ D {\bf 12}, 2060 (1975).
  doi:10.1103/PhysRevD.12.2060
\end{trivlist} 

\section{Quantum Chromodynamics (10-27-$xx$)}
We will denote color indices by $a,b,\dots = r,b,g$ and flavor by $f$
so that the noninteracting part of the quark Lagrangian is\footnote{RPF
omits 
coupling constants and numerical factors in eqs.\
(\ref{qcd-int}-\ref{gauge-trans}), but restores them in (\ref{Lqcd}).}
\be
	\sum_f\bar\psi^f_{\bar a}(i\slashed{\partial} - \mu_f)\psi^f_a
\ee
while the interaction term (for a single flavor) is
\be
	\bar\psi_{\bar a}\, A^\mu_{\bar a b}\gamma_\mu \psi_b
\label{qcd-int}
\ee
and the gluon kinetic term is\footnote{RPF uses $E_{\mu\nu}$ and $F_{\mu\nu}$
interchangeably for the field strength.}
\be
	\tr(\partial_\mu A_\nu^{b\bar a}- \partial_\nu A_\mu^{b\bar
a})^2 \equiv \tr (E_{\mu\nu}^{b\bar a})^2
\label{gluon-kin}
\ee
(Here only the noninteracting part of the field strength is used.)
The gauge transformations are given by
\bea
\label{gauge-trans}
	\psi' &=& \Lambda\psi\nn\\
	A'_\mu &=& \Lambda^\dagger A_\mu \Lambda 
	+\Lambda^\dagger i\partial_\mu\Lambda\nn\\
	E_{\mu\nu} &=&  \partial_\mu A_\nu- \partial_\nu A_\mu
	-[A_\mu,A_\nu]\nn\\ &\to& \Lambda^\dagger E_{\mu\nu}\Lambda =
E'_{\mu\nu}
\eea
The fully gauge invariant QCD Lagrangian is
\bea
\label{Lqcd}
	{\cal L}_{QCD} &=&
\sum_f\bar\psi_f[i\slashed{\partial}-\slashed{A}-\mu_f]	\psi_f\nn\\
	&+& {1\over 2 g^2}\tr\, E_{\mu\nu} E_{\mu\nu}
\eea
The quark-gluon interaction can also be written as
\be
	{\cal L}_{q- \rm int} = \tr\, A_\mu J_\mu
\ee
with the current
\be
	J_\mu = \sum_f J_{\mu,f};\qquad J_{\mu,f}^{\bar b a} = 
	\bar\psi^{\bar b}_f\gamma_\mu\psi_f^a
\ee
Notice that there are $6\times 4\times 3$ (quark) and $4\times 8$
(gluon) field degrees of freedom at each point in spacetime.

{\bf Exercises.}  1. If $B(x)$ is a $3\times 3$ matrix transforming
as $B' = \Lambda^\dagger B \Lambda$, show that $\partial_\mu B$
does not transform homogeneously in this way, but $\partial_\mu B
-i[A_\mu,B]$ does.  {\it I.e.,} $\partial_\mu -i[A_\mu,\quad] \equiv
D_\mu$ is the covariant derivative for fields transforming in the
octet representation.    \\

2. Prove that $[D_\mu,D_\nu]B = i[E_{\mu\nu},B]$.\\

3. If $A_\mu\to A_\mu + \delta A_\mu$ then $\delta E_{\mu\nu} = 
D_\mu\delta A_\nu - D_\nu \delta A_\mu$.\\

4. By varying $A_\mu$ in ${\cal L}$, show that $D_\mu E_{\mu\nu}=
g^2 J_\nu$, where $J$ is the quark current.\\

5. Derive the equation of motion of the quarks,
$(i\slashed{\partial}-\mu_f-\slashed{A})\psi = 0.$\\

6.  From this, show that $D_\mu J_{\mu,f}=0$; also $\partial_\mu
\tr\,J_{\mu,f} = 0$.\\

7.  Show that \#4 is meaningless unless $D_\mu J_\mu = \sum_f D_\mu
J_{\mu,f} = $.  Hint: \#2.\\

8. Show (identically, not as a consequence of the equations of motion
but due to the form of $E_{\mu\nu}$)
that $D_\mu E_{\nu\sigma} + D_\nu E_{\sigma\mu} + D_\sigma 
E_{\mu\nu} = 0$.  Note that if $\tilde E_{\mu\nu} \equiv\sfrac12
\epsilon_{\mu\nu\sigma\tau}E_{\sigma\tau}$ then this can be written
as $D_\mu \tilde E_{\mu\nu} = 0$.\\

9. Define color electric and magnetic fields $E_x = E_{xt}$ {\it etc.}
and $B_z= E_{xy}$ {\it etc.} (check the signs).  Then rewrite the 
field equations in terms of these quantities.  Let $\mathbb{E} =
(E_x,E_y,E_z)$ {\it etc}.  Show that
\bea
	\mathbb{D}\cdot\mathbb{B} &=& 0\nn\\
	\mathbb{D}\times\mathbb{E} +D_t \mathbb{B} &=& 0\nn\\
	\mathbb{D}\cdot\mathbb{E} &=& g^2\rho \hbox{\ where\ }
	\rho = J_t\nn\\
	\mathbb{D}\times\mathbb{B} -D_t \mathbb{E} &=& g^2 \mathbb{J}
\eea
where $\mathbb{D} = \bbnabla-i[\mathbb{A},\ \ ]$ and $D_t =
\partial_t-i [A_t,\ \ ]$.

Define the matrices 
%\footnote{RPF does not call them ``Gell-Mann''
%matrices.  In fact he rarely identified anything in terms of 
%historical names.  Below I have introduced ``Gauss's Law'' but
%RPF referred to it as ``Maxwell's equation.'' In a later lecture when he 
%describes Pauli-Villars regularization, he refrains from naming it
%as such.  No mention of `t Hooft or Veltman is made in connection with
%dimensional regularization.  I don't think this was meant to take
%credit away from the inventors, but rather reflected his emphasis on
%the physics content over the historical origins.}
\bea 
\lambda_1 &=& \!\!\!\!\left(\begin{array}{rrr} 0 & 1 & 0 \\1 & 0 & 0\\ 0 & 0 &
0\end{array}\right)\,  
\lambda_2 = \left(\begin{array}{rrr} 0 & -i & 0 \\i & 0 & 0\\ 0 & 0 &
0\end{array}\right)\, 
\lambda_3 = \left(\begin{array}{rrr} 1 & 0 & 0 \\0 & -1 & 0\\ 0 & 0 &
0\end{array}\right)\nn\\
\lambda_4 &=& \left(\begin{array}{rrr} 0 & 0 & 1 \\0 & 0 & 0\\ 1 & 0 &
0\end{array}\right)\, 
\lambda_5 = \left(\begin{array}{rrr} 0 & 0 & -i \\0 & 0 & 0\\ i & 0 &
0\end{array}\right)\,
\lambda_6 = \left(\begin{array}{rrr} 0 & 0 & 0 \\0 & 0 & 1\\ 0 & 1 &
0\end{array}\right)\nn\\
\lambda_7 &=& \left(\begin{array}{rrr} 0 & 0 & 0 \\0 & 0 & -i\\ 0 & -i &
0\end{array}\right)\,\  
\lambda_8 = \sqrt{\sfrac13}\left(\begin{array}{rrr} 1 & 0 & 0 \\0 & 1 & 0\\ 0
& 0 &-2\end{array}\right)\nn\\ 
\lambda_0 &=& \sqrt{\sfrac23}\left(\begin{array}{rrr} 1 & 0 & 0 \\0 & 1 & 0\\
0 & 0 &
1\end{array}\right)
\eea
Our convention is that $\tr(\lambda_i\lambda_j) = 2\delta_{ij}$.  We define
$A_\mu = \sfrac12 A^a_\mu\lambda_a$.  There is no $A^0_\mu$ term because
this would correspond to an extra U(1) force.  Let $\vec A_\mu =
(A^1_\mu,\dots, A^8_\mu)$.

{\bf Exercises, continued.}  10.  Show that $\Lambda^\dagger \partial_\mu\Lambda$ is traceless, {\it
i.e.,} it has no $\lambda_0$ component.

The structure constants $f_{ijk}$ are defined through
\be
	\left[(\sfrac12\lambda_i), (\sfrac12\lambda_j)\right] = i f_{ijk}\,
(\sfrac12\lambda_k)
\ee
One can show that $f_{ijk}$ is totally antisymmetric.
We also define the anticommutator
\be
	\{\lambda_i, \lambda_j\} = 2\,d_{ijk}\,\lambda_k
\ee
Unlike $f_{ijk}$, we can find the extra generator $k=0$ amongst those
on the right-hand side.\footnote{This seems to be a notational innovation
 of RPF.}

11.  Show that $\partial_\mu A_\nu^i - \partial_\nu A_\mu^i - f_{ijk}A_\mu^j
A_\nu^k = E^i_{\mu\nu}$.  We can rewrite this as 
$\partial_\mu\vec A_\nu - \partial_\nu\vec A_\mu - \vec A_\mu\,^\times
	\vec A_\nu = \vec E_{\mu\nu}$
by defining a cross product in color space as
\be
	(\vec C\,^\times \vec D)_i = f_{ijk}\,C_j\, D_k
\ee
Similarly define the dot product
\be
	\vec C\, ^{\bigcdot} \vec D = \sum_i C_i D_i
\ee

12. Prove that \[\vec C\,^{\bigcdot}(\vec C\,^\times \vec D) = 0\]
and
\[\vec A\,^\times (\vec B\,^\times \vec C) + 
\vec B\,^\times (\vec C\,^\times \vec A) + 
\vec C\,^\times (\vec A\,^\times \vec B) = 0\]
However (you don't need to prove this), the familiar identity
$\vec A\,^\times (\vec B\,^\times \vec C) = \vec B(\vec A\,^{\bigcdot}
\vec C) - \vec C(\vec A\,^{\bigcdot}
\vec B)$ is only true for SU(2) and not for general SU($N$).\\

13. Show that 
\be
	C = D_\mu B\implies \vec C = \partial_\mu\vec B - \vec
A_\mu\,^{\bigcdot}\vec B
\ee

14. Rewrite ${\cal L}$ using component notation.

\subsection{Geometry of color space}

Consider successive transformations $\psi' = \Lambda\Psi$, $\psi'' = M\psi'$.
Then $\psi'' = N\psi \equiv M\Lambda\psi$ obviously. This is an example of the
group multiplication law for the color rotations.  For many purposes we 
may be interested in infinitesimal rotations,
$\Lambda = 1 + i\mathbf{a}$.  Under this, the gauge field transforms as
\bea
	A'_\mu &=& (1-i\mathbf{a})A_\mu(1 + i\mathbf{a}) + 
(1-i\mathbf{a})\partial_\mu(1 + i\mathbf{a})\nn\\
	&=& A_\mu -i[\mathbf{a},A_\mu] +i\partial_\mu \mathbf{a}\nn\\
	&=&  A_\mu  +i D_\mu \mathbf{a}
\label{infgt}
\eea
So it is always possible to impose temporal gauge, $A_0=0$, since this
only requires solving a first order differential equation.  In the
following however we will discuss a difficulty that arises when
charges are present.

What happens to a quark's color as it is transported through a gluon 
field?  The transformation between two sets of color axes separated by 
a distance $\Delta x_\mu$ can be written as $1+iA_\mu\, \Delta x_\mu$.
Now suppose that every set of axes is changed locally by a rotation
$\Lambda(x)$.  Then the new transformation relating the two sets of axes
is
\be
	\Lambda^\dagger(x+\Delta x)(1+iA_\mu\, \Delta x_\mu)\Lambda(x)
	\equiv 1 + iA'_\mu\, \Delta x_\mu
\ee
Therefore 
\be
	A'_\mu = \Lambda^\dagger A_\mu\Lambda + \Lambda^\dagger
i\partial_\mu\Lambda
\label{gtrans}
\ee
which is the finite version of (\ref{infgt}).

\subsection{Quark-antiquark potential}
It would be very satisfying if we could justify some of the phenomenological
approaches I considered earlier, using QCD as a starting point. Heavy
quarkonium systems, being approximately nonrelativistic, are the simplest
systems to consider, and can be described by a potential of the form
\be
	V \sim {\alpha\over r} + b r + \dots
\ee
where I have omitted the spin-spin and spin-orbit interactions. (In the
complete Hamiltonian there is also an annihilation term $H_A$ that can cause transitions
like $u\bar u\leftrightarrow s\bar s$, that give rise to $\eta$-$\eta'$ mixing.)  The terms written describe the linearly confining 
potential representing the mass of the string connecting the quark to the antiquark, and
the Coulomb-like interaction, which might rather be something like $e^{-\mu r}/r$.  

One can also
make predictions for relativistic systems like the vector mesons; see 
S.\ Godfrey, N.\ Isgur, Phys.\ Rev.\ D32, 189 (1985).  Then it is advantageous to use
harmonic oscillator wave functions as a basis for computing matrix elements of the
Hamiltonian to get a good approximate solution and compare to the data.  Not only can one
compute the mass spectrum, but also strong interaction decay amplitudes, such as for
$\phi\to K\bar K$.
But all of this still
relies on making a reasonable guess for the form of the potential, and 
it would be preferable to derive these interactions directly from QCD.

Let us recall how the analogous calculation works in QED, for the potential between
a proton and an electron.  We start with the fundamental interactions,
\be
	{\cal L} = \sfrac14 F_{\mu\nu}F_{\mu\nu} + \bar\psi_e{\slashed D}\psi_e
	+ \bar\psi_p{\slashed D}\psi_p
\ee
and from this we would like to derive the nonrelativistic effective Hamiltonian
\be
	H_{NR} = {p^2\over 2\mu} + V(r)
\ee
The potential can be computed perturbatively, by Fourier transforming the amplitude,
\be
	V(r) \sim\ \  \raisebox{-0.75cm}{\includegraphics[width=0.25\textwidth]{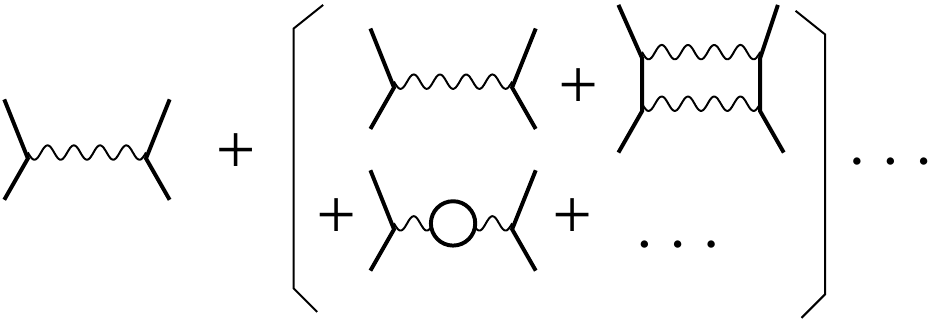}}
\ee
But for QCD we know that the linear term is a nonperturbative effect, so a different
approach is needed.

A better way might be to solve the classical equations of motion for
the gauge field, in the presence of a source term, where the Lagrangian is
\bea
	{\cal L} &=& \sfrac14 F_{\mu\nu}F_{\mu\nu}  + A_\mu J_\mu\nn\\
	&=&\sfrac12\left(\partial_\mu A_\nu\partial_\mu A_\nu 
	-\partial_\mu A_\nu\partial_\nu A_\mu\right) + A_\mu J_\mu\nn\\
	&\to& -\sfrac12\left(A_\nu\square A_\nu + A_\nu \partial_\mu\partial_\nu A_\mu\right)
	+ A_\mu J_\mu
\eea
This gives the equation of motion
\bea
	-\square A_\mu + \partial_\mu\partial_\nu A_\nu + J_\mu &=& 0% \implies\nn\\
%	\underbrace{-\square\partial_\mu A_\mu +\square\partial_\nu A_\nu} 
%+ \partial_\mu J_\mu &=& 0\nn\\
%0 \qquad\qquad\qquad\quad  &&
\label{eq4-28}
\eea
which by taking the divergence implies the current is conserved, $\partial_\mu J_\mu = 0$.
Since $\partial\cdot F = \partial_\mu F_{\mu\nu} = \partial_\mu(\partial_\mu A_\nu - 
\partial_\nu A_\mu)$, eq.\ (\ref{eq4-28}) can be written in the gauge-invariant form
\be
	\partial_\nu F_{\mu\nu} = J_\mu
\label{eq4-29}
\ee

Now we must solve (\ref{eq4-29}) when the source is $J_0 = e\left(\delta(\vec r) - 
\delta(\vec r-\vec a)\right)$, supposing that the two charges are located at the origin and
at $\vec r= \vec a$ respectively.
In electrodynamics this is easy, thanks to the linearity of the theory.  We just superpose
the solutions from the two sources, call them
\be
	\vec E_1 = {q_1\over r^2}\,\hat r,\quad \vec E_2 = {q_2\over r'^2}\,\hat r'
\ee
where $\vec r^{\,\prime} = \vec r - \vec a$.  Then we can compute the interaction energy
by integrating the energy density in the fields, ${\cal E}\sim |\vec E_1+\vec E_2|^2$:
\be
	V(a) = \int d^{\,3}x\, {\cal E}_{\rm int} =  \int d^{\,3}x\, 2\vec E_1\cdot \vec E_2
\ee
This shows how one might be able to compute the quark-antiquark potential without relying
on perturbation theory; it would require knowing the classical solution for the gluons
fields in the presence of a static source. 

\subsection{Classical solutions}

Therefore we would like to solve for the chromoelectric field in the presence of
a source.  However it is no longer possible write this only in terms of the
field strength, as we could for QED.    In the temporal gauge $A_0=0$,
the Gauss's law equation is
\be
\mathbb{D}\cdot\mathbb{E} = \bbnabla\cdot\mathbb{E} - i\left(\mathbb{A}\cdot\mathbb{E}
	-\mathbb{E}\cdot\mathbb{A}\right) = g^2\,\rho 
\label{eq4-27}
\ee
where
\be
	\rho = J_0 = \sum_f\bar\psi^b_f\gamma_0\psi_f^a
\label{eq4-28a}
\ee
which is a matrix in color space.  Before fixing the gauge, \be
	\mathbb{D}\cdot\mathbb{E} = \mathbb{D}\cdot \left(\bbnabla A_0
	-\partial_0 \mathbb{A} - [\mathbb{A},A_0]\right) \to
	- \mathbb{D}\cdot  \partial_0 \mathbb{A}
\ee
so another way of writing (\ref{eq4-27}) in $A_0=0$ gauge is 
\be
 	- \mathbb{D}\cdot  \partial_0 \mathbb{A} = g^2\rho
\ee

However we should first verify that there is no obstacle to 
transforming to the
$A_0=0$ gauge when external charges are present.  An issue, as I will
show, is whether one can consider the source to be static. 
Starting from some configuration with $A_0\neq 0$, we would like
to construct the gauge transformation that makes $A_0'=0$, by solving
eq.\ (\ref{gtrans}) with $\mu=0$.  One can guess that it is 
a time-ordered exponential, 
\be
	\Lambda = P \exp\left(i\int_{t_0}^t dt' A_0(t')\right)
\ee
and verify that this is a solution, since
\bea
	\partial_0\Lambda &=& \partial_0\left(1 +
	 i \int_{t_0}^tdt' A_0(t')\right.\nn\\
	&-& \left. \int_{t_0}^t dt_1 A_0(t_1) \int_{t_0}^{t_1} dt_2 A_0(t_2) 
	+ \dots \right)\nn\\
	&=& A_0(t) - A_0(t)\int_{t_0}^tdt' A_0(t') + \dots
\eea
So there is no difficulty in transforming to the temporal gauge.  But 
we must also consider how the source (\ref{eq4-28}) transforms:
\be
	\rho\to \Lambda^\dagger\rho\Lambda
\ee
Recall that the quark changes its color when it emits a gluon;
that's
why the charge is a matrix.  Does the gauge dependence mean that
it makes no sense to ask what is the potential between two spatially
separated charge matrices?

Before getting too ambitious and trying to solve with the source having both a quark and an
antiquark,  let's first imagine the seemingly easier case of a single quark, even though the
solution is not expected to fall off at large distances.  
Suppose that a quark starts out being red.  At a later time,
after emitting or absorbing a gluon, it is some linear
combination of red, green, blue:
\be	
	q(t) = \left(\begin{array}{c} r(t)\\ b(t)\\ g(t)\end{array}
	\right)
\ee
where $|r|^2 + |b|^2 + |g|^2 = 1$, say.  It gives a color charge
matrix of the form $\rho_{\bar a b} = \bar q^{\bar a}\gamma_0 q^b$.
Clearly a nontrivial solution will have time dependence, associated with the
fact that the source is not a color singlet.  To avoid this, we would have
to include the antiquark contribution, so as to form a
gauge-invariant source, 
\be
	\rho = \bar q^a(x)\,\left[P e^{i\int_x^{x+r}A_\mu dx^\mu}\right]_{ab}\, q^b(x+r)
\label{eq4-35}
\ee
which is no longer a matrix, since we have traced over the color indices.
But this trace is not actually present in the equation of motion 
(\ref{eq4-27}), which has the explicit form
\bea
\mathbb{D}\cdot\mathbb{E} &=& \bbnabla\cdot\mathbb{E} - i[\mathbb{A}\cdot,\mathbb{E}]\nn\\
&=& \bbnabla\cdot\dot{\mathbb{A}} - i[\mathbb{A}\cdot,\dot{\mathbb{A}}] = g^2\rho
\eea
in $A_0=0$ gauge.  In this form it is clear that $\tr\, \mathbb{D}\cdot\mathbb{E}=0$ since
every term is proportional to an SU(3) generator.  Therefore (as we already knew)
only the traceless part of
$\rho$ can act as a source for the gluons.

 Let's rewrite (\ref{eq4-27}) in the SU(3) vector notation I introduced previously,
\be
	\nabla_i \dot{\vec A}_i - i A_i^a\dot A_i^b [T^a,T^b] =
	g^2\vec\rho\,.
\ee
Using $[T^a,T^b] = i f_{abc}T^c$ and rescaling $A\to g^2 A$, 
this becomes
\be
	\nabla_i \dot{\vec A}_i + g^2{\vec A}_i^{\ \times}\dot{\vec A}_i = 
	\vec\rho\,.
\ee
Now let $\vec A_i = t\,\vec E_i$; then since $\vec E_i^{\ \times}\vec
E_i = 0$, we get Gauss's law $\nabla_i\vec E_i = \vec\rho$.  So it looks like
we have succeeded in finding a class of solutions, that looks like 
just eight copies of the Abelian problem.  
Not so fast!  In electrodynamics, there is no difficulty in setting the magnetic
field to zero for a static charge configuration.  But in QCD the color field
sources itself, and now it is no longer obvious that we can set $B=0$.
Since $B$ contains the term $[A,A]$, 
it would vanish for special charge distributions where $\rho^3$ and
$\rho^8$ (whose generators are diagonal) are the only nonzero components.  
But such solutions are
not helpful for understanding the distinctive properties of QCD,
in particular the confining potential.

More generally there could be an integration constant, $\vec A_i = \vec
a_i(\vec x) + t\,\vec E_i$, giving the extra term
\be
	\nabla_i\vec E_i + g^2 \vec a_i^{\ \times}\vec E_i = \vec\rho
\ee
in Gauss's law.  
What is the physical significance of $\vec a_i$?  Recall that
\be
	B_z = E_{xy} = \partial_x A_y - \partial_y A_x - [A_x,A_y]
\ee
so that $\vec a_i(x)$ gives a time-independent contribution to the
chromomagnetic field,
\bea
	B_z &=& \partial_x a_y - \partial_y a_x + t[\partial_x E_y
	-\partial_y E_x]\\
         &-& [a_x + tE_x,a_y+tE_y]\nn\\
	&=& \partial_x a_y - \partial_y a_x - [a_x,a_y]\nn\\
	&+& t\left(\partial_x E_y - \partial_y E_x - [E_x,a_y] - 
	[a_x,E_y]\right)  - t^2[E_x,E_y]\nn
\eea
The time-dependent terms still vanish for the special charge distributions
$\rho^3$, $\rho^8$, while the time-independent one  vanishes if in addition
$a^3_i$ and $a^8_i$ are curl-free.

In electrodynamics, a static electric and magnetic field in temporal
gauge are described by $A_i = a_i + t E_i$ with $\nabla\times E = 0$
and $\partial_t a_i = \partial_t E_i = 0$.  We can't seem to do that
here:
\bea
	E_i &=& -\partial_i A_0 + \partial_0 A_i + [A_i,A_0]\nn\\
	    &=& \partial_0 A_i \hbox{\ \ in temporal gauge};\nn\\
	B_x &=& \partial_yA_z - \partial_z A_y - [A_y,A_z]
\eea
because the commutator in $B$ generically gives rise to time dependence.
This seems to imply that we cannot impose $A_0=0$ gauge when charges
are present.  In electrodynamics it is more common to express a static
solution as $\vec E = \vec\nabla A_0$ in Coulomb gauge where;
$\vec\nabla\cdot\vec A = 0$.  Then
\be
	\vec\nabla\cdot\vec E \sim \vec\nabla\cdot {\hat r\over r^2}
	= \partial_i{x_i\over r^3} = {3\over r^3} -
\left(\frac32\right) {x_i\,2x_i\over
r^5} = 0
\ee
But we previously showed that it is always possible to go to temporal
gauge; why should it matter what gauge we choose?

One reason it could matter is the gauge-covariance of the source.
Suppose that $\rho$ was initially static
in a gauge where $A_0\ne 0$.  When we transform to temporal gauge,
it is no longer static!  Instead
\be
	\Lambda^\dagger \rho\Lambda = \left(P\, e^{i\int^t dt'
A_0(t')}\right)^\dagger \rho \left(P\, e^{i\int^t dt'
A_0(t')}\right)
\ee
which is $t$-dependent, unlike in the Abelian case.  One might
try to fix the problem by rewriting 
Gauss's law in terms of gauge invariant quantities on the
left-hand side of the equation.  Using $\mathbb{D}\cdot\mathbb{E} = 
\bbnabla\cdot\mathbb{E} - i[\mathbb{A},\mathbb{E}]$, it would read
\be
	\bbnabla\cdot\mathbb{E} =  g^2\rho +  i[\mathbb{A},\mathbb{E}]
\ee
But that  doesn't work, since $\mathbb{E}$ itself is
not gauge invariant!

Eq.\ (\ref{eq4-35}) suggests that it might be possible to find 
 a solution where the charge remains static if
we work instead in an axial gauge with $\mathbb{n}\cdot\mathbb{A}=0$
for some spacelike vector $\mathbb{n}$, for example $A_z = 0$, in 
the case where the quark and antiquark are separated along the $z$
direction.  Then the gauge transformation needed to transform from  
$A_z\neq 0$ to $A'_z = 0$ is
\be
	\Lambda = P e^{i\int^z dz' A_z(z')}
\ee
If the initial gauge field was static, $\dot A_\mu=0$, then
$\Lambda^\dagger\rho\Lambda$ remains static.  In fact, the same
argument would have worked in temporal gauge since then
$Pe^{i\int^t dt' A_0(t')} = Pe^{i t A_0} = e^{itA_0}$, which would be
consistent, but then $\vec E = 0$.

In summary, it seems to be difficult to find the classical  gauge configurations that would
explain  the origin of the quark-antiquark potential.\footnote{This statement was not
in my notes; it conveys my impression that
RPF was explaining from memory the sequence of difficulties he encountered when looking
for classical solutions, some time prior to the course.  
There is no record of these attempts in his personal notes.}

\section{QCD Conventions (10-29-$xx$)}  

In the previous lectures we may have been a bit careless with
numerical factors and signs.  Let's now try to get all of these
right and establish a consistent set of conventions.
First, we can verify that the quark Lagrangian should read
$\bar\psi(i\slashed{\partial}-\slashed{A})\psi$ to be invariant
under the gauge transformations 
\bea
	\psi&\to&\Lambda\psi,\qquad 
	\bar\psi\to\bar\psi\Lambda^\dagger,\nn\\
	A_\mu&\to& \Lambda\, A_\mu\,\Lambda^\dagger + i(\partial_\mu \Lambda)
	\,\Lambda^\dagger
\label{gaugetrans}
\eea

Second, we carry out the gauge transformations
\bea
	\partial_\mu A_\nu - \partial_\nu A_\mu &\to& 
	\Lambda(\partial_\mu A_\nu - \partial_\nu
A_\mu)\Lambda^\dagger\\
	&+& (\partial_\mu \Lambda)A_\nu\Lambda^\dagger - 
	(\partial_\nu \Lambda)A_\mu\Lambda^\dagger\nn\\
       &+& \Lambda A_\nu \partial_\mu\Lambda^\dagger - 
	\Lambda A_\mu \partial_\nu\Lambda^\dagger\nn\\
	&+& i\partial_\mu((\partial_\nu\Lambda)\Lambda^\dagger)
	- i\partial_\nu((\partial_\mu\Lambda)\Lambda^\dagger)\nn
\eea
and 
\bea
	[A_\mu,A_\nu] &\to & [\Lambda A_\mu\Lambda^\dagger +
i(\partial_\mu\Lambda)\Lambda^\dagger,
\Lambda A_\nu\Lambda^\dagger +
i(\partial_\nu\Lambda)\Lambda^\dagger]\nn\\
	&=& \Lambda [A_\mu,A_\nu]\Lambda^\dagger + i[\Lambda
A_\mu\Lambda^\dagger,(\partial_\nu\Lambda)\Lambda^\dagger]\\
	&+& i[(\partial_\mu\Lambda)\Lambda^\dagger,\Lambda A_\nu\Lambda^\dagger]
	-
[(\partial_\mu\Lambda)\Lambda^\dagger,(\partial_\nu\Lambda)\Lambda^\dagger]\nn
\eea 
Using $(\partial_\nu\Lambda)\Lambda^\dagger =
-\Lambda\partial_\nu\Lambda^\dagger$, we can show that the terms involving
derivatives of $\Lambda$ cancel in the linear combination
\be
	F_{\mu\nu} = \partial_\mu A_\nu - \partial_\nu A_\mu + i [A_\mu,A_\nu]
\ee 
which is therefore the covariant field strength.  The chromoelectric and
magnetic fields are
\bea	
	E_i &=& F_{0i} = \partial_0 A_i - \partial_i A_0 +i[A_0,A_i]\nn\\
	B_x &=& F_{yz} = \partial_y A_z - \partial_z A_y +i[A_y,A_z]
\eea 

The full Lagrangian is
\be
	{\cal L} = {1\over 2 g^2}\tr\,F_{\mu\nu}F_{\mu\nu} 
	+ \bar\psi(i\slashed{\partial} - \slashed{A} -m)\psi
\ee
Let's vary it with respect to $A$ to find the equation of motion. 
The variation of the first term is
\bea 
	\delta{\cal L} &=&  {1\over g^2}\tr\, F_{\mu\nu}\,\delta F_{\mu\nu} +\dots
	\\
	&=& {1\over g^2}\tr\,\Big(2\delta A_\mu \partial_\nu F_{\mu\nu}
	+ i([\delta A_\mu,A_\nu] + [A_\mu,\delta A_\nu])F_{\mu\nu}\Big)\nn\\
&=& {2\over g^2}\tr\,\Big(\delta A_\mu \partial_\nu F_{\mu\nu}
	+ i[\delta A_\mu,A_\nu]F_{\mu\nu}\Big)
\eea
Then\footnote{At this point RPF writes $\delta{\cal L}/\delta A^a_\mu$ on the
left side, but on the right side 
gives the variation of  ${\cal L}$ with
respect to  $A_\mu^{a\bar b}$ labeled by the
$(3,\bar 3)$ indices, rather than varying with respect to $A^a_\mu$ labeled
by the adjoint index $a$.  This gives a result twice as large as it should
be (due to the normalization of the generators), which RPF recognizes as being wrong and therefore concludes that the gluon
kinetic term should really be normalized as $(1/4g^2)\tr\,F^2$.  This may be another case of him extemporizing.  I have 
corrected the derivation here.} writing $A_\mu = A^a_\mu T_a =  A^a_\mu (\sfrac12\lambda_a)$,
\bea
g^2{\delta{\cal L}\over \delta A^a_\mu} &=& 
{2}\tr\,\Big(T_a\, \partial_\nu F_{\mu\nu}
	+ i[T_a,A_\nu]F_{\mu\nu}\Big)\nn\\
	&=& \partial_\nu F^a_{\mu\nu} + {2i}\,A_\nu^b\, F_{\mu\nu}^c
	\tr\Big([T^a,T^b]\,T^c\Big)\nn\\
	&=& \partial_\nu F^a_{\mu\nu} + i[A_\nu,F_{\mu\nu}]^a
\eea
where we used 
\be
\tr([T^a,T^b]\,T^c) = \frac{i}{2}f_{abc}= \tr([T^b,T^c]\,T^a)
\ee
Therefore
\be
	\partial_\nu F^a_{\mu\nu} + i[A_\nu,F_{\mu\nu}]^a = g^2
	\bar\psi T^a\gamma_\mu \psi
\ee

{\bf Exercise.}  Show that in Coulomb gauge $\partial_i A_i=0$, the 
Gauss's law constraint becomes\footnote{In the lectures, RPF derives this but I leave it as an
exercise.  Part of the derivation involves assuming that $\partial_0 A_i=0$
in the commutator $[A_i, \partial_0 A_i]$, which seems not generally true.}
\be
	\nabla^2 A_0 + 2 i g[A_i,\partial_i A_0] - g^2\left[A_i,[A_i,A_0]\right] = g\rho
\ee
after rescaling $A_\mu \to g A_\mu$.  $\rho$ is the matrix charge defined in 
eq.\ (\ref{eq4-28a}).

\section{Geometry of color space$^*$  (11-3,5-$xx$)}
\label{sect6}
\footnote{This section, which was revised
by RPF, combines lectures 6 and 7, given on Nov.\ 3 and 5, 19$xx$.  It repeats 
some material that was presented
earlier.  I retained the redundancies in the interest of historical accuracy.}
Our discussion of the QCD Lagrangian has been of a largely algebraic nature to
this point, but much intuition can be gained by considering the local color
symmetry in geometric terms.  At each point in spacetime we imagine there exists
a set of axes in the color space, which may vary in its relative orientation
from place to place.  This freedom to rotate color frames independently at each
point is embodied in the SU(3) transformation matrices $\Lambda(x)$, under which
a quark transforms as
\be
	\psi'(x) = \Lambda(x)\psi(x)\, .
\label{eq6-1}
\ee
Since one rotation may be followed by another,
\bea
	\psi''(x) &=& \Lambda'(x)\psi'(x) = \Lambda'(x)\Lambda(x)\psi(x)\nn\\
	&\equiv& \Lambda''(x)\psi(x)\,;\nn
\eea 
they form a group, with $\Lambda'' = \Lambda'\Lambda$ being the group
multiplication law.  Requiring that $\Lambda$ not change the length of a color
vector is equivalent to demanding that $\Lambda^\dagger\Lambda = 1$.  Thus the
$\Lambda$'s would represent the group U(3) of unitary $3\times 3$ matrices.
However U(3) contains a U(1) subgroup, matrices of the form
$e^{i\theta}\mathbb{1}$, which would give rise to an additional long-range
interaction like the electromagnetic force.  To eliminate this we note that
\be
	\det\Lambda'' = \det\Lambda'\det\Lambda\nn
\ee
represents the U(1) transformations (it is Abelian), so we should make the
restriction
\be
	\det\Lambda = 1,\nn
\ee
{\it i.e.,} $\Lambda$ is a {\it special} unitary matrix, hence the group is
SU(3). 

The transformation law for the gluon field has a less immediately obvious
interpretation than that for the quarks, eq.\ (\ref{eq6-1}).  For infinitesimal
rotations $\Lambda = 1 + i\mathbf{a},$\footnote{Neither RPF nor
I noticed the inconsistency with eq.\ (\ref{gaugetrans}), which is the correct
version having $i\partial_\mu$ instead of $\partial_\mu$ in the first line.}
\bea
	A'_\mu &=& (1-i\mathbf{a})A_\mu(1 + i\mathbf{a}) + 
(1-i\mathbf{a})\partial_\mu(1 + i\mathbf{a})\nn\\
	&=& A_\mu -i[\mathbf{a},A_\mu] +i\partial_\mu \mathbf{a}\nn\\
	&=&  A_\mu  +i D_\mu \mathbf{a}
\label{eq6-2}
\eea
where $D_\mu$ is the covariant derivative.  How can this be understood
geometrically?

To answer this, it must first be realized that there is, {\it a priori}, no way
of telling whether a color frame at point $x$ is parallel to one at $x+\Delta
x$, because the color space is completely unrelated to spacetime.  An analogy is
trying to choose local tangent frames on a curved space, such as the surface of
a two-sphere, that are ``parallel'' to each other.  It is not possible to do
without defining a law of parallel transport for vectors, so that we know what
it means for two vectors at different locations to be parallel.  Similarly in
QCD one needs a rule for comparing orientations of nearby color frames.  This is
the function of the gauge field $A_\mu(x)$, in much the same way as the metric
tensor (to be more precise, the Christoffel symbol) defines parallel transport
in the geometry of curved space.  {\it Define} the relative orientation between
two nearby color frames, at $x$ and $x+\Delta x$, to be given by the rotation
matrix
\be
	U(x,\Delta x) = 1 + i A_\mu(x)\Delta x^\mu\, .
\label{eq6-3}
\ee
Now suppose that every set of axes is rotated by $\Lambda(x)$, depending on the
position $x$.  Then the new transformation relating the frames at $x$ and
$x+\Delta x$ is
\bea
	U'(x,\Delta x) &=& \Lambda^\dagger(x+\Delta x)U(x,\Delta x)
	\Lambda(x)\nn\\
	&=& 1 + i A'_\mu(x)\Delta x^\mu\,
\eea
It follows that 
\be
	A'_\mu(x) = \Lambda^\dagger(x) A_\mu(x) \Lambda(x) + i
	\Lambda^\dagger(x)\partial_\mu \Lambda(x)
\label{eq6-4}
\ee
which shows that our geometric interpretation of $A_\mu(x)$ agrees with its
previously determined transformation law.

\begin{figure}[t]
\centerline{
\includegraphics[width=0.25\textwidth]{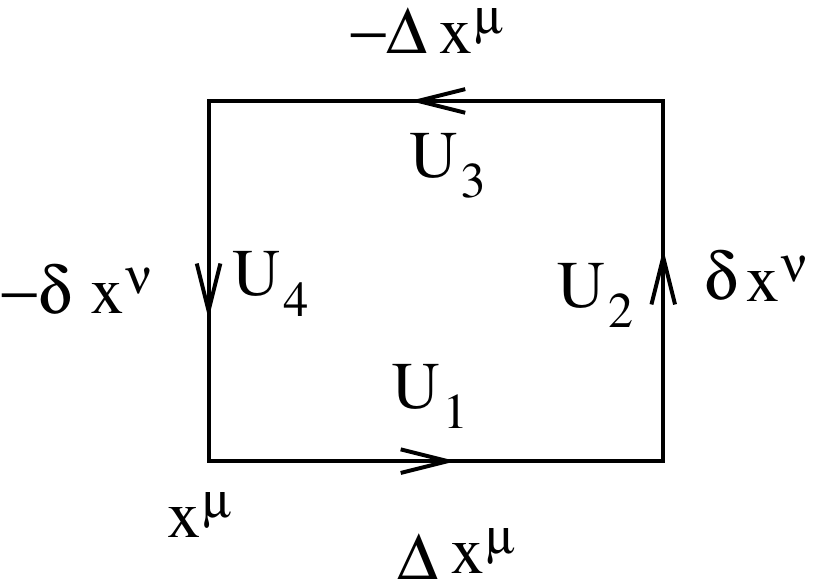}}
\caption{Parallel transport of a quark around a closed loop.}
\label{plaquette}
\end{figure}

If one was to take a quark at $x$ with color vector $\vec q$ and parallel-transport it to $x+\Delta x$, its color would change to $U(x,\Delta x)\vec q$.
Of particular interest is the change in $\vec q$ when transported around a
closed loop, such as the one shown in fig.\ \ref{plaquette}.  Let
$U_1=U(x,\Delta x)$, $U_2 = U(x+\Delta x,\delta x)$, $U_3 = U(x+\Delta x+\delta
x,-\Delta x)$, $U_4 = U(x+\delta x, -\delta x)$.  The transformation of $\vec q$ in
going around the loop is
\be
	\vec q^{\,\prime} = U_4 U_3 U_2 U_1\vec q = U_{\rm tot}\vec q
\label{eq6-5}
\ee
One notices that $U_{\rm tot}(x)$ has a simple transformation under local
SU(3) rotations,
\be
	U_{\rm tot}(x) \to \Lambda^\dagger(x) U_{\rm tot}(x)\Lambda(x)\,,\nn
\ee
which is just how the field strength $F_{\mu\nu}(x)$ transforms.  This is
not an accident: if you expand $U_{\rm tot}$ in terms of the gauge field as
in (\ref{eq6-3}), you will find that
\be
	U_{\rm tot} = 1 + iF_{\mu\nu}(x)\Delta x^\mu\delta x^\nu
\label{eq6-6}
\ee 
plus terms of order $(\Delta x)^2$ and $(\delta x)^2$.  Notice that 
$\Delta x^\mu\delta x^\nu$ is the area of the loop.  So $F_{\mu\nu}(x)$ tells us
how much color rotation a quark suffers under transformations around
infinitesimal loops.  It is analogous to the Riemann tensor, which does the same
thing for vectors in curved space.

With the concept of parallel transport in hand, covariant differentiation
becomes quite transparent.  If $\psi(x)$ is a quark field, it is not
$\psi(x+\Delta x) - \psi(x)$ that is of physical interest, because this includes
the difference due to arbitrary orientations of the local color axes.  We should
rather compare $\psi(x+\Delta x)$ with $\psi(x)$ transported to $x+\Delta x$.
Therefore define the covariant derivative as
\be
	\Delta x^\mu D_\mu \psi(x) = \psi(x + \Delta x) -
U(x,\Delta x)\psi(x)\nn
\ee
or equivalently
\be
	D_\mu\psi(x) = (\partial_\mu - i A_\mu)\psi(x)\, .
\label{eq6-7}
\ee
Similarly for the field strength,
\be
	\Delta x^\alpha D_\alpha F_{\mu\nu}(x) = F_{\mu\nu}(x+\Delta x) 
	-U^\dagger(x,\Delta x) F_{\mu\nu}(x)U(x,\Delta x)\nn
\ee
which implies
\be
	D_\alpha F_{\mu\nu}(x) = \partial_\alpha F_{\mu\nu}(x) 
	-i[A_\alpha,F_{\mu\nu}]
\label{eq6-8}
\ee

\begin{figure}[b]
\centerline{
\includegraphics[width=0.25\textwidth]{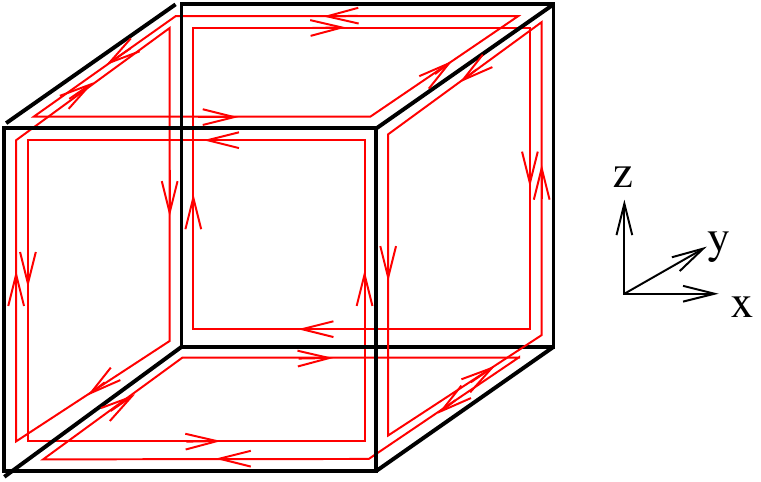}}
\caption{Geometric interpretation of the Bianchi identity.}
\label{bianchi}
\end{figure}

Even as seemingly abstract an equation as the Bianchi identity can be understood
geometrically.  This is one of the statements you were asked to prove
previously, 
\be
	D_\alpha F_{\beta\gamma} + D_{\gamma} F_{\alpha\beta} + D_\beta
F_{\gamma\alpha} = 0\, .
\label{eq6-9}
\ee
For concreteness, let $(\alpha,\beta,\gamma) = (x,y,z)$.  Then the first
term is $D_x F_{yz}$.  In terms of fig.\ \ref{bianchi}, this is the change in color
axis orientation around the top loop minus that of the bottom loop.  The  use of
$D_z$ rather than $\partial_z$ means that the bottom loop was parallel
transported to the position of the top loop before making the subtraction.
The contribution to $D_z F_{xy}$ from each link of the cube is denoted by a
line with an arrow that shows the relative sign of the contribution.  From the 
figure, it is easy to see that when the remaining terms in (\ref{eq6-9}) are
included, each link will contribute twice, once in each direction.  Therefore
the sum is zero.

\begin{figure}[t]
\centerline{
\includegraphics[width=0.25\textwidth]{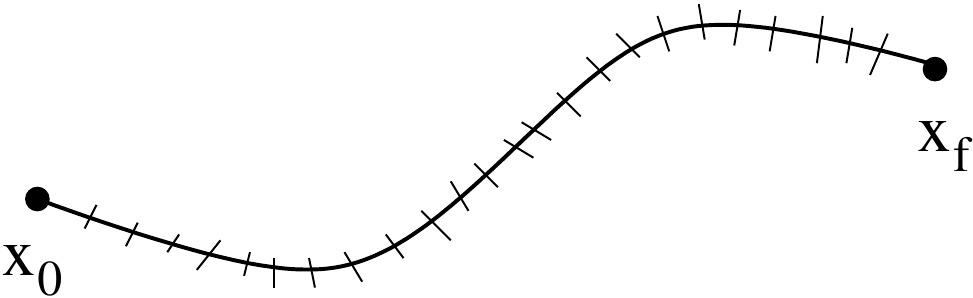}}
\caption{Path connecting two points in spacetime.}
\label{path}
\end{figure}

The matrix $U(x,\Delta x)$ that connects nearby color axes can be used to
construct the rotation connecting color frames that are separated by a finite
distance.  Choose a path connecting the two points and divide it into
small increments, labeled by $x_i^\mu$, such that $\Delta x_i^\mu = 
x_{i+1}^\mu - x_i^\mu$ as shown in fig.\ \ref{path}.  Then the rotation matrix
between $x_0^\mu$ and $x_f^\mu$ is
\be
	S(x_0,x_f) = \prod_i\left(1 + i A_\mu(x_i)\Delta x_i^\mu\right)\, .
\label{eq6-10}
\ee 
As $\Delta x_i^\mu \to 0$, this becomes equivalent to $\prod_i e^{i
A_\mu(x_i)\Delta x_i^\mu}$.  We would like to write it as $\exp(i\int A_\mu\,
dx^\mu)$, which would be true if the $A_\mu$ were numbers, but since the $A_\mu$
don't commute at different positions, we cannot add the exponents.  Instead one
defines the path ordering operator
\be
	P e^{b_2 + b_1} = e^{b_2}e^{b_1},\nn
\label{eq6-11}
\ee
where it is understood that $b_2$ is farther along the path than $b_1$.
Therefore
\bea 
	S(x_0, x_f) &=& P\exp\left(i\int_{x_0}^{x_f} dx^\mu\, A_\mu \right)\nn\\
\label{eq6-12}
	 &=& 1 + i\int_{x_0}^{x_f} dx^\mu\, A_\mu\\
	 & - &
	\int_{x_0}^{x_f} dx^\mu\, \int_{x}^{x_f} dx'^\nu\, A_\nu(x') A_\mu(x)
	+\cdots\nn
\eea
Notice that $S$ is by no means unique; it depends upon the path chosen.
Under a gauge transformation however,
\be
	S(x_0,x_f) = \Lambda^\dagger(x_f) S \Lambda(x_0)
\label{eq6-13}
\ee	
regardless of the path.

{\bf Exercise.}  Using the definition (\ref{eq6-12}), show that 
\be
	P\exp\left(-\int_{x_0}^{x_f} dx^\mu\, \Lambda^\dagger(x)\partial_\mu
	\Lambda(x)\right) = \Lambda^\dagger(x_f)\Lambda(x_0) \nn
\ee
Hint: consider the differential equation satisfied by (\ref{eq6-12}) for
$\partial S/\partial x^\mu$.

The connection $S$ is useful for making bilocal operators gauge invariant.
For example, in the full interacting theory, the two-point function for the
field strength vanishes because of gauge invariance, since
\bea
	\langle 0| F_{\alpha\beta}(x_0) F_{\mu\nu}(x_f)|0\rangle =
\qquad\qquad\qquad\qquad\qquad\qquad\nn\\
\langle 0| \Lambda^\dagger (x_0) F_{\alpha\beta}(x_0)\Lambda(x_0) 
	\Lambda^\dagger (x_f) F_{\mu\nu}(x_f) \Lambda(x_f)|0\rangle\nn
\eea 
for an arbitrary $\Lambda(x)$.  This can only be satisfied if $\langle FF\rangle
= 0$.  However the function
\be
	\tr\left(S^\dagger(x_0,x_f) F_{\alpha\beta}(x_0)S(x_0,x_f) F_{\mu\nu}(x_f)
	\right)\nn
\ee
is gauge invariant, and has a meaningful nonvanishing expectation value.

The statement that $\langle FF\rangle
= 0$ for $F$'s at distinct points implies some way of defining expectation
values without tampering with the gauge symmetry of the path integral over
$A_\mu$, to be discussed later on.  In practice, it is necessary to choose a
condition that fixes the gauge, by associating $F_{\mu\nu}(x)$ with a unique
vector potential $A_\mu(x)$.  For example, it is always possible to demand that
$A_0=0$ by transforming $A_\mu\to \Lambda^\dagger A_\mu \Lambda +
i\Lambda^\dagger \partial_\mu\Lambda$, where
\be
	\Lambda^\dagger A_0\Lambda + i\Lambda^\dagger \partial_\mu\Lambda = 
0\nn
\ee
One can show that the solution to this equation is
\be
	\Lambda = T\,\exp\left(i\int^t A_0(t')\right) dt' \nn
\ee
where $T$ is the same as $P$ but for a purely timelike path.

Another gauge condition that is conceptually useful is to minimize the quantity
\be
	\int d^{\,4}x\, \tr(A_\mu(x))^2
\label{eq6-14}
\ee
In this gauge, $P\,\exp(i\int dx\, A)$ does not change much if the path is varied
only slightly.  Therefore it is possible to define a global orientation for
color axes on short enough distance scales: a quark that looks red at point $x$
will still look red after parallel transport if it is not carried too far. 
Because of this it makes some sense to say that quarks of the same color repel
each other, whereas quarks that are antisymmetric in their colors attract each
other, as will be shown in the next lecture.  In an arbitrary gauge it would not
be meaningful to say that two blue quarks repel each other, unless they were at
the same position, since what is blue at one place may not be blue at another.

The above choice of gauge is closely related to another more familiar one.
Under an infinitesimal gauge transformation $A_\mu\to A_\mu+ D_\mu\alpha$, 
the change in (\ref{eq6-14}) is 
\be
	2 \int d^{\,4}x\, \tr  A_\mu D_\mu\alpha = -2\int d^{\,4}x\, \tr (\alpha\,D_\mu
A_\mu)
\label{eq6-15}
\ee
If $\int\tr(A_\mu^2)$ is at a minimum, then (\ref{eq6-15}) must vanish for all
$\alpha(x)$.  This implies
\be
	D_\mu A_\mu = \partial_\mu A_\mu = 0
\label{eq6-16}
\ee
However the two gauges are not equivalent, because the first one asks for the 
absolute minimum of $\int\tr(A_\mu^2)$, whereas $\partial_\mu A_\mu =0$
only requires that $\int \tr(A_\mu)^2$	be at a local minimum.  Therefore
$\partial_\mu A_\mu =0$ may have many solutions, and it does not uniquely fix
the gauge.  This problem was first discussed by Gribov in the context of the
path integral.

\subsection{Omitted material}
\footnote{This material appears in my original notes but was omitted from the 
revised version above.}
A synopsis of the popular gauge choices is
\bea
	A_t=0 && \hbox{Weyl}\nn\\
	\partial_\mu A_\mu = 0 && \hbox{Lorentz}\nn\\
	\bbnabla\cdot\mathbb{A}=0 && \hbox{Coulomb}
\eea
In addition, there is an analog to (\ref{eq6-14}) due to Mandula, which is to 
minimize $\int d^{\,3}x\,\mathbb{A}:\mathbb{A}$.

{\bf Exercise.}  Show that $E_{\mu\nu}\,^{\bigcdot}\widetilde E_{\mu\nu} \equiv
\epsilon_{\mu\nu\alpha\beta} E_{\mu\nu}\,^{\bigcdot}E_{\alpha\beta}$ could be
added to the Lagrangian (usually written as the action
$\theta\int d^{\,4}x\, E_{\mu\nu}\widetilde E_{\mu\nu}$), but it makes no
contribution to the equations of motion: it is a total derivative.

\section{Semiclassical QCD$^*$ (11-10-$xx$)}
\label{sect7}

I know you are eager to move on to the {\it quantum} theory of chromodynamics,
now that we have studied it at the classical level, but there always has to be
some professor deterring you by saying ``before we do that, let's look at
such-and-such!''  Accordingly, before we quantize QCD I want to discuss a
somewhat tangential but very important issue: can we explain the properties of
the hadrons, even {\it qualitatively}, with the theory of QCD?  That is, we
would like to see that we are at least going in roughly the right direction
before we invest all our effort in it.  For example, it would be quite
discouraging if at the lowest level of analysis QCD predicted that the three
quarks in a baryon will want to fly apart.  

\begin{figure}[t]
\centerline{
\includegraphics[width=0.175\textwidth]{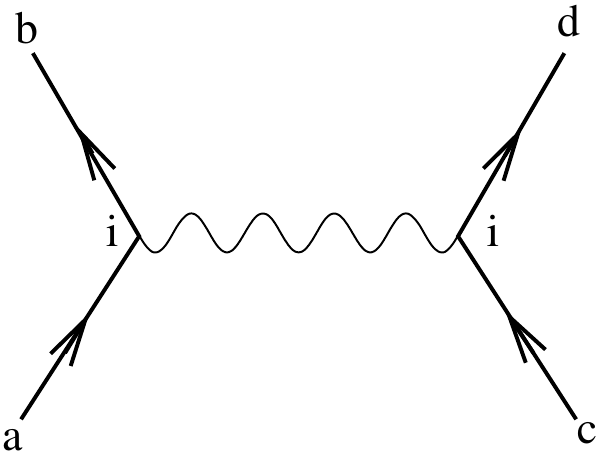}\hfil
\includegraphics[width=0.175\textwidth]{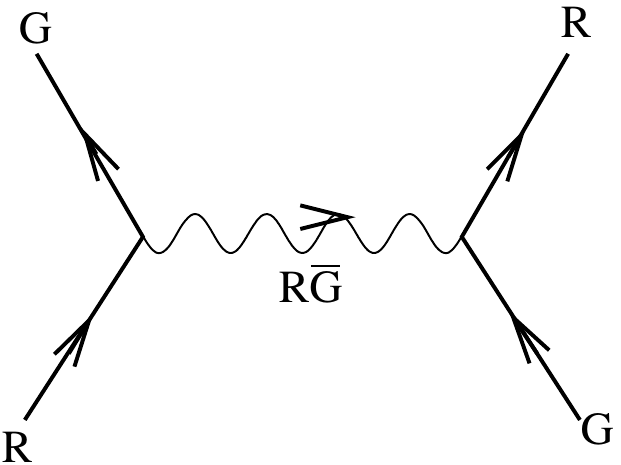}}
\caption{(a) Left: quark-quark scattering by gluon exchange.
(b) Right: same, with particular choice of colors.}
\label{qqscatt}
\end{figure}

But we shall see that it {\it does}
work, and we won't even have to do that much work ourselves to see it, if we
just remember a few things from quantum {\it electro}dynamics.  This is because
at lowest order in the coupling constant $g$, the interaction between two quarks
is given by essentially the same Finemensch diagram as that for electron-electron
scattering, fig.\ \ref{qqscatt}.  The only difference is that in the case of
QCD, each vertex comes with a group theory factor $\lambda^i_{ab}$ and
$\lambda^i_{cd}$ to account for the fact that the quarks are changing color
when they exchange a gluon of color $i$.  
The sum over intermediate gluon colors then gives a factor 
$\vec\lambda_{ab}\,^{\bigcdot}\vec\lambda_{cd}$ in the amplitude.\footnote{
These should be accompanied by extra factors of $1/2$ from $T^a = \lambda^a/2$.}

Of course we know that fig.\ \ref{qqscatt}(a) is not a good approximation for the
quark-quark scattering in a hadron, because the coupling is large.  But we just
want to see that it is going in the right {\it direction}, when we do make
this approximation.  Knowing that a single photon exchange gives rise to the
Coulomb potential in electrodynamics, we can immediately write the quark-quark
potential from fig.\ \ref{qqscatt}(a) as
\be
	V(r) = {g^2\over r}\, \vec\lambda_{ab}\,^{\bigcdot}\vec\lambda_{cd}
\label{eq7-2}
\ee
Now all that remains is to evaluate the $\vec\lambda\,^{\bigcdot}\vec\lambda$
factor in the color channels appropriate for baryons and mesons.  This could be
done by using fancy group theory techniques, but I find it useful to take a
more simple-minded approach at first.  It is not necessary to know group
theory---all one needs to know is that there are three colors!

Let us suppose that quark 1 (on the left) starts out being red, and converts to
green by emitting a red-antigreen gluon.  In order to conserve color, quark 2
must have started out being green and turn to red when it absorbs the gluon. 
Using the convention
\be
	(\hbox{R,\ G\, B})^T = (\hbox{red,\ green,\, blue})^T
\label{eq7-3}
\ee
for the components of a color vector, there are two $\lambda$ matrices
contributing to the process shown in fig.\ \ref{qqscatt}(b), namely
$\lambda^1$ and $\lambda^2$,
\be
\lambda^1 = \left(\begin{array}{rrr} 0 & 1 & 0 \\1 & 0 & 0\\ 0 & 0 &
0\end{array}\right),\   
\lambda^2 = \left(\begin{array}{rrr} 0 & -i & 0 \\i & 0 & 0\\ 0 & 0 &
0\end{array}\right)\,\ 
\label{eq7-5}
\ee
due to the fact that a $R\bar G$ gluon corresponds to a particular linear
combination, $\lambda^1+i\lambda^2$.  Therefore the contribution to 
$\vec\lambda_{ab}\,^{\bigcdot}\vec\lambda_{cd}$ from fig.\ \ref{qqscatt}(b)
is 
\be
	1\cdot 1 + i(-i) = 2
\label{eq7-6}
\ee

\begin{figure}[b]
\centerline{
\includegraphics[width=0.2\textwidth]{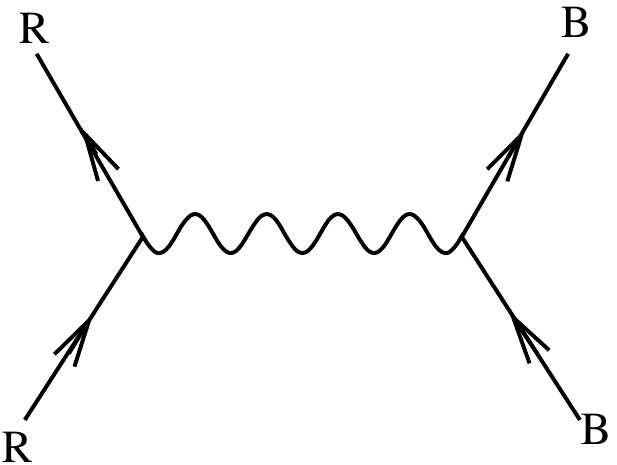}}
\caption{A color-conserving $q$-$q$ scattering process.}
\label{qqscatt3}
\end{figure}

Now suppose that the colors of the initial state quark did not change.  This
could happen if gluons corresponding to the color-diagonal generators
$\lambda^3$ and $\lambda^8$ were exchanged,
\be
\lambda_3 = \left(\begin{array}{rrr} 1 & 0 & 0 \\0 & -1 & 0\\ 0 & 0 &
0\end{array}\right),\quad
\lambda_8 = \sqrt{\sfrac13}\left(\begin{array}{rrr} 1 & 0 & 0 \\0 & 1 & 0\\ 0
& 0 &-2\end{array}\right)\,.
\label{eq7-7}
\ee
For example, the contribution to $\vec\lambda\,^{\bigcdot}\vec\lambda$ from fig.\
\ref{qqscatt3} is
\be
	1\cdot 0 + {1\over\sqrt{3}}\left(-{2\over\sqrt{3}}\right) = -\frac23
\label{eq7-9}
\ee
From (\ref{eq7-6}) and (\ref{eq7-9}) we could guess that the general expression
for $\vec\lambda\,^{\bigcdot}\vec\lambda$ is
\be
	\vec\lambda_{ab}\,^{\bigcdot}\vec\lambda_{cd} = 2\, P^{c.e.}_{ab,cd} - 
\frac23 \delta_{ab}\delta_{cd}
\label{eq7-10}
\ee
where $P^{c.e.}$ is the color-exchange operator,
\be
	P^{c.e.}_{ab,cd} = \left\{\begin{array}{rl} 1 & \hbox{\ if\ } a=d 
	\hbox{\ and\ } b=c\\
	0 & \hbox{\ otherwise}\end{array}\right.
\label{eq7-11}
\ee
As a check, look at the graph
\be
\raisebox{-0.75cm}{\includegraphics[width=0.1\textwidth]{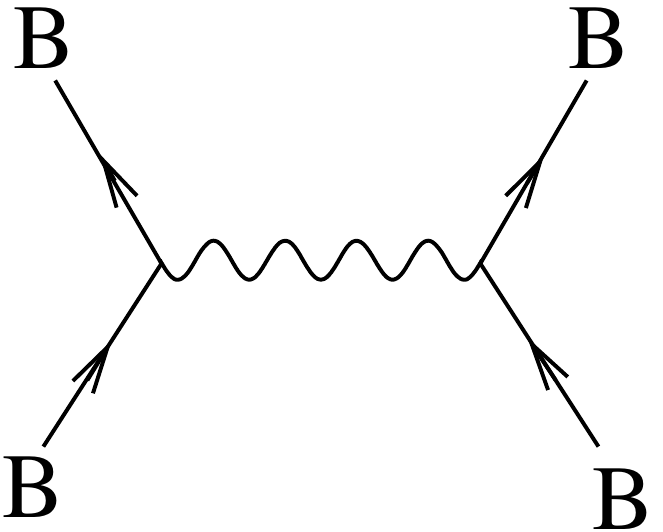}} : \left(-{2\over \sqrt{3}}\right)
	\left(-{2\over\sqrt{3}}\right) = \frac43
\label{eq7-12}
\ee	
which has only a $\lambda^8$-type gluon.  This agrees with (\ref{eq7-10}) since
$2P^{c.e.}_{BB,BB} -\sfrac23\delta_{BB}\delta_{BB} = 4/3$.

\begin{figure}[b]
\centerline{
\includegraphics[width=0.4\textwidth]{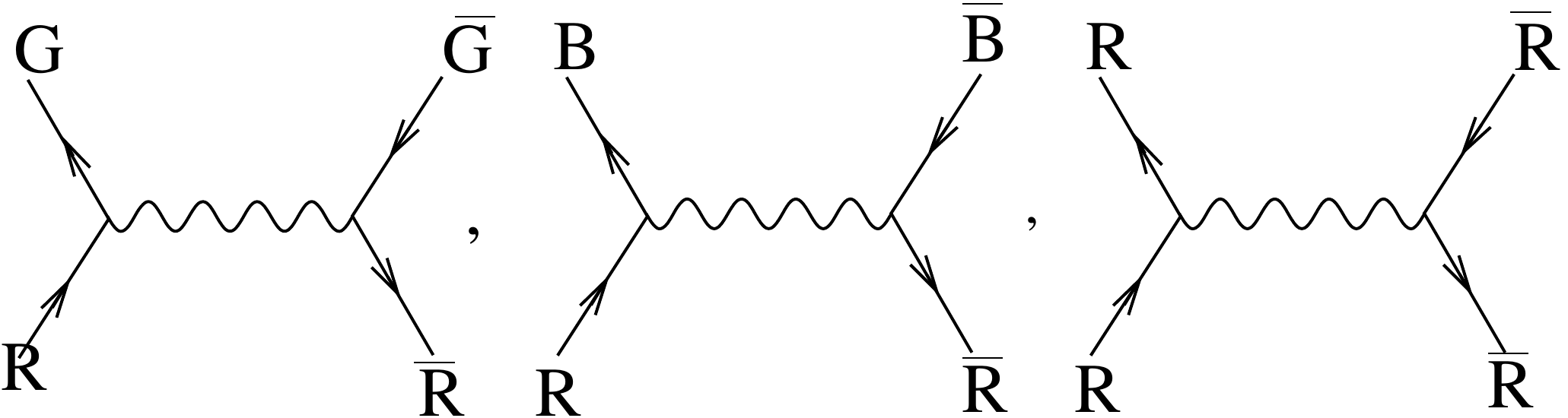}}
\caption{Scattering of $q$-$\bar q$ within a meson.}
\label{qqbscatt}
\end{figure}

So far we have not concerned ourselves about whether the initial state was
symmetric or antisymmetric in color.  If it was antisymmetric, it would be an
eigenstate of the color exchange operator with eigenvalue $-1$.  (Of course, it
would also be an eigenstate of the identity operator $\delta_{ab}\delta_{cd}$
with eigenvalue $+1$.)  On such a state,
\be
	\vec\lambda\,^{\bigcdot}\vec\lambda\, |\psi\rangle = \left(2 P^{c.e.} -
\sfrac23\right)|\psi\rangle = -\sfrac83 |\psi\rangle\,.
\label{eq7-13}
\ee
It means that in the antisymmetric channel, the quark-quark potential is
\be
	V(r) = -\frac83\,{g^2\over r}\,.
\label{eq7-14}
\ee
The sign is important: since electron-electron scattering is repulsive, the
relative sign here tells us that quarks that are antisymmetric in their colors
{\it attract}.  This is precisely what we want: in a baryon the quarks are in a
completely antisymmetric state, which is the only way to make a color-neutral
object out of three color triplets.  So we understand, in a rough way, why
protons, neutrons, $\Delta$'s, {\it etc.} exist.

The mesons can be understood similarly.  In this case the initial state is
color symmetric, 
\be
	|\psi\rangle = {1\over\sqrt{3}}\left(R\bar R + B\bar B + G\bar
G\right)\,.
\label{eq7-15}
\ee
If we focus on the $R\bar R$ part, there are three graphs to consider, fig.\
\ref{qqbscatt}.  They contribute $-\vec\lambda\,^{\bigcdot}\vec\lambda$
factors (the extra minus sign coming from the coupling of vectors to
antiparticles) of 
\be
	-\left( 1\cdot 1 + i(-i)\right) = -2,\quad -2, \quad -\left(1\cdot 1
	+\sfrac{1}{\sqrt{3}}\cdot \sfrac{1}{\sqrt{3}}\right) = -\sfrac43
\label{eq7-17}
\ee
respectively.  Obviously the $B\bar B$ and $G\bar G$ parts of $|\psi\rangle$ do
the analogous thing, so that the quark-antiquark potential for color-symmetric
states is 
\be
	V(r) = -\frac{16}{3}\, {g^2\over r}\,.
\label{eq7-18}
\ee
Graphs like fig.\ \ref{qqbscatt} give an {\it attractive} potential for
electrons and positrons in QED, and we see that quarks and antiquarks in a meson
must also attract each other.

One notices that the $q\bar q$ force in mesons is twice as strong as the $qq$
force in baryons.  But it is interesting to note that the total force {\it per
quark} is the same in each system, since
\be
	{1\over 2\hbox{\ quarks}}\left({16\over 3}\right) = \frac83
	\hbox{\ for mesons}
\ee
and
\be
	{1\over 3\hbox{\ quarks}}\times\left(3\ \hbox{\ pairwise forces\ }
	\right)
	\times \left({8\over 3}\right) = \frac83
	\hbox{\ for baryons}
\ee

{\bf Problem.}  Show that the general $q\bar q$ interaction due to one-gluon
exchange (summed over gluon colors) can be written as
\be
	\sfrac23\,\mathbb{1} - 6|s\rangle\langle s|\nn
\ee
where $\mathbb{1}$ is the identity operator in color space, and
$|s\rangle\langle s|$ projects onto the color singlet state,
\be
	|s\rangle = \sfrac{1}{\sqrt{3}}\left(R\bar R + B\bar B + G\bar G\right)
\nn
\ee

\subsection{Spin-spin interactions}

\begin{figure}[t]
\centerline{
\includegraphics[width=0.45\textwidth]{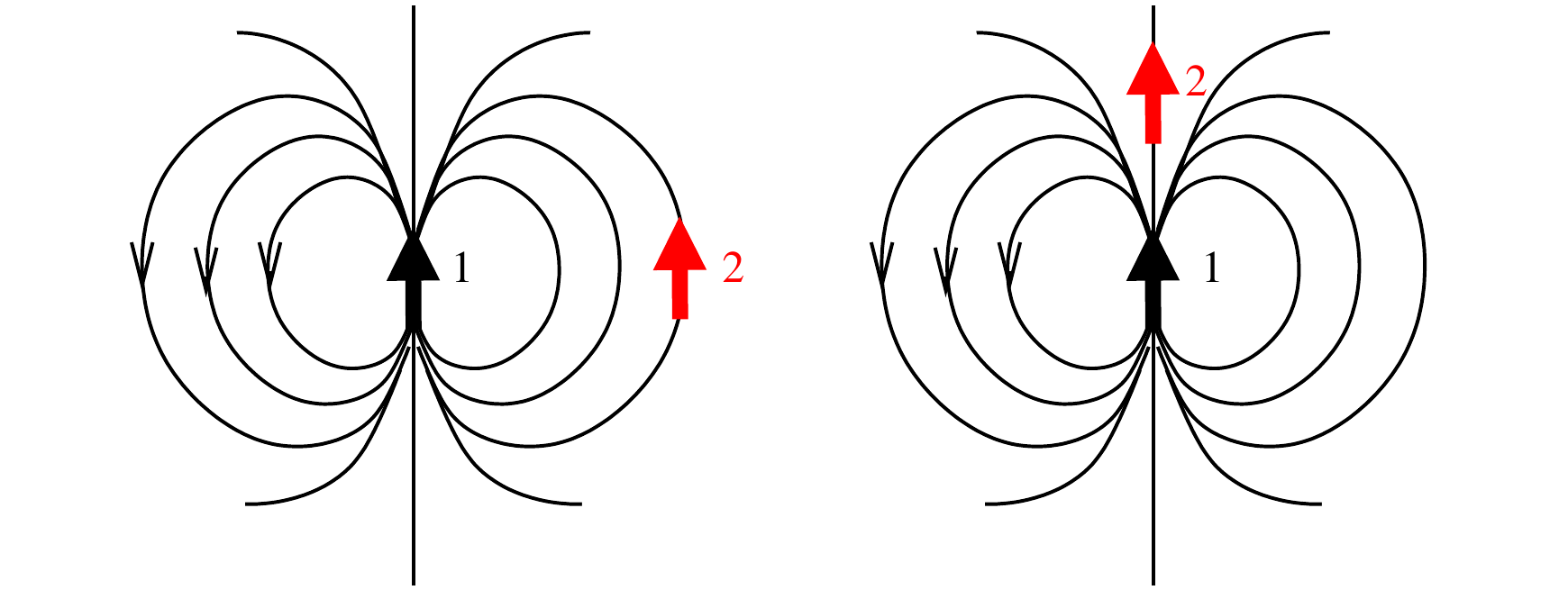}}
\caption{An electron spin (2) in the dipole field
produced by another electron (1).}
\label{dipole}
\end{figure}

So far, so good: QCD explains why the hadrons exist, even at this crude level of
approximation where $g$ was taken to be small.  Now we would like to see if it
explains some more detailed observations, like the nondegeneracy of the
$\Delta^0$ and the neutron:
\bea
	\label{eq7-19}
	\Delta^0&:& \raisebox{0.1cm}{\rule{0.5cm}{0.4pt}}\ 1232\,{\rm MeV};\ 
	|\psi\rangle = (udd)(\uparrow\uparrow\uparrow)\\
	N &:& \raisebox{0.1cm}{\rule{0.5cm}{0.4pt}}\ \ 935\,{\rm MeV};\ 
	|\psi\rangle = (udd)\sfrac{1}{\sqrt{6}}(2\downarrow\uparrow\uparrow
	-\uparrow\downarrow\uparrow - \uparrow\uparrow\downarrow)\nn
\eea
If the masses were coming solely from the constituent masses of the quarks,
these states would be degenerate due to their identical quark content.
The only difference between them seems to be their spin wave functions.
Therefore their mass splitting must be due to spin-dependent forces.  This comes
as no surprise since the photon-exchange graph gives a spin-spin interaction
as well as the Coulomb interaction in QED.  Let us recall what the sign of the
force is in electromagnetism.  Specifically, if we could make an $s$-wave from
two electrons, would their spins tend to align or anti-align?  We draw the
second spin in the magnetic field of the first, with both of them pointing up;
see fig.\ \ref{dipole}.  When the second spin is beside the first, the
interaction energy is positive, since spin 2 would prefer to flip so as 
to align with the $\vec B$ field.  When spin 2 is above spin 1,
the interaction energy is negative.  To make an $s$-wave we must average over
the positions of spin 2 relative to spin 1 in a spherically symmetric way.  One
can show that at any nonzero radius, the interaction energy is zero when this is
done.

\begin{figure}[t]
\centerline{
\includegraphics[width=0.2\textwidth]{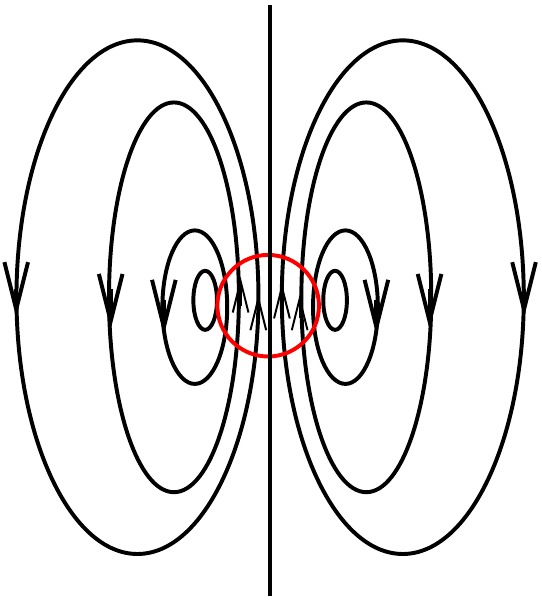}}
\caption{Effect of a nonpointlike spin 2 (red) overlapping the central region 
of the magnetic field produced by spin 1.}
\label{dipole2}
\end{figure}

However, zero is not the correct answer, as we already know.  The problem is
that the spins have been idealized as pointlike objects.  If they actually have
a small but finite spatial extent, the magnetic field of spin 1 will look like
fig.\ \ref{dipole2}.  Because of the interior region, the integration over
positions of spin 2 give a net positive interaction energy, and two
electrons in an $s$-wave tend to align.  This is known as the Fermi
interaction; it is the first term in the expression
\bea
	V_{\rm spin}(\vec r) &\propto& -\frac{8\pi}{3} \vec\sigma_1\!\cdot\!
\vec\sigma_2 \, \delta^{(3)}(\vec r)\nn \\
&+& {1\over r^3}\big(\vec\sigma_1\!\cdot\!
\vec\sigma_2 - 3(\vec\sigma_1\!\cdot\! \hat r)(\vec\sigma_2\!\cdot\! \hat r)
	\big)\,.
\label{eq7-20}
\eea
The second term comes from the exterior region that we discussed previously;
its angular average is zero, as we noted.  

From this we can deduce that the color magnetic moment interaction energy for
two quarks, in a relative $s$-state, is {\it positive} if the spins are aligned
and the colors are antisymmetric.  This is because the sign of the spin force
relative to that of the Coulomb force is determined by the Lorentz indices of
the diagram in fig.\ \ref{qqscatt}, so this relative sign must be the same for
QCD and QED.  Hence the $\Delta$ is heavier than the $N$---it takes more
energy to line up all the spins.

What about the $\Sigma^0$ and the $\Lambda$?  They have identical quark
content, and equal numbers of aligned spins, yet the $\Sigma^0$ is more massive.
However, the spin wave functions are not the same,
\bea
	|\Sigma^0\rangle &=& (sud)\,\sfrac{1}{\sqrt{6}}
	(2\downarrow\uparrow\uparrow - \uparrow\downarrow\uparrow
	-\uparrow\uparrow\downarrow)\nn\\
	|\Lambda\rangle &=& (sud)\,\sfrac{1}{\sqrt{2}}(\uparrow\uparrow\downarrow
	-\uparrow\downarrow\uparrow) 
\label{eq7-21}
\eea
We must remember that the magnetic moment of a quark is inversely proportional
to its mass, so the $s$ quark has smaller spin interactions.  In the limit that
its magnetic moment is neglected, the $\Sigma^0$ is $4/6$ aligned and $2/6$
anti-aligned in its spins, whereas the $\Lambda$ is completely anti-aligned.
Hence the $\Lambda$ is expected to be lighter, as observed.

To be more quantitative, suppose the spin-spin coupling has strength $a_{qq}$,
$a_{qs}$, $a_{ss}$ between two light quarks, one light and one strange, and two
strange quarks, respectively.  Then the spin-spin coupling for $\Sigma^0$ and
$\Lambda$ particles is
\be
	\hat{\cal O} = a_{qq}\,(\sigma_u\!\cdot\!\sigma_d) +
a_{qs}\,(\sigma_s\!\cdot\!\sigma_u + \sigma_s\!\cdot\!\sigma_d)
\label{eq7-22}
\ee
where $a_{qs} < a_{qq}$, or equivalently, defining a spin-exchange operator
$P^{s.e.}_{qq'}$, 
\be
	\hat{\cal O} =   a_{qq}\left(2 P^{s.e.}_{ud}-1\right) + 
	a_{qs}\left(2 P^{s.e.}_{su}-1\right)
	+ a_{qs}\left(2 P^{s.e.}_{sd}-1\right) 
\ee
Acting on $\Sigma^0$,
\bea
	P^{s.e}_{ud}(2\downarrow\uparrow\uparrow - \uparrow\downarrow\uparrow
	-\uparrow\uparrow\downarrow) &=& 
2\downarrow\uparrow\uparrow - \uparrow\uparrow\downarrow
	-\uparrow\downarrow\uparrow,\nn\\
	P^{s.e}_{su}(2\downarrow\uparrow\uparrow - \uparrow\downarrow\uparrow
	-\uparrow\uparrow\downarrow) &=&  
2\uparrow\downarrow\uparrow - \downarrow\uparrow\uparrow
	-\uparrow\uparrow\downarrow,\nn\\
	P^{s.e}_{sd}(2\downarrow\uparrow\uparrow - \uparrow\downarrow\uparrow
	-\uparrow\uparrow\downarrow) &=&
	2\uparrow\uparrow\downarrow - \uparrow\downarrow\uparrow
	-\downarrow\uparrow\uparrow\nn\\
\label{eq7-24}
\eea
Therefore
\bea
\hat{\cal O}|\Sigma^0\rangle  &=& \left[(2-1)a_{qq} + (-2-2)a_{qs}\right]
	|\Sigma^0\rangle\nn\\
	&=& \left(a_{qq} - 4\,a_{sq}\right)|\Sigma^0\rangle
\label{eq7-25}
\eea
Similarly one finds that 
\be
	\hat{\cal O}|\Lambda\rangle = \left[(-2-1)a_{qq} + (2-2)a_{qs}\right]
	|\Lambda\rangle = -3|\Lambda\rangle
\label{eq7-26}
\ee
Therefore the mass difference is $M_\Sigma - M_\Lambda = 4(a_{qq}-a_{qs})$,
which is positive since $a_{qq} > a_{qs}$.

{\bf Problem.}  Find the mass splittings of the rest of the baryon $\sfrac12^+$
octet and $\sfrac32^+$ decuplet states.  Assuming that
\be
	a_{qq}-a_{qs} = a_{qs}-a_{ss}\nn
\ee
prove the Gell-Mann--Okubo formula,
\be
	2(M_{\Xi} + M_N) = 3 M_\Lambda + M_\Sigma \nn
\ee

\begin{figure}[b]
\centerline{
\includegraphics[width=0.07\textwidth]{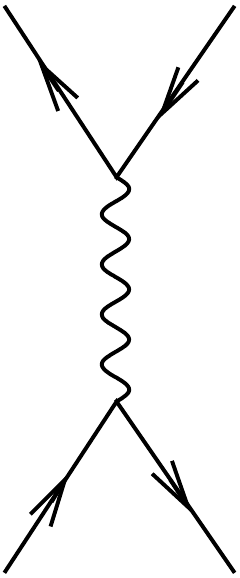}}
\caption{Annihilation diagram for positronium, not present for quarkonium.}
\label{ann}
\end{figure}

A similar analysis can be done for the mesons, and it is observed that the level
splittings of heavy quarkonium excitations, such as the $\psi$ and $\Upsilon$
systems, are similar to those of positronium.  However there is an interesting
distinction between the spin forces in quarkonium and those of positronium.  In
the latter an extra contribution to the spin-spin interaction arises from the 
annihilation diagram, fig.\ \ref{ann}.  However, at lowest order in $g$, no such
process can occur for quarkonium.  This is because the $q\bar q$ pair in a meson
forms a color singlet, which cannot annihilate into a colored object like a
gluon.  It cannot even annihilate into a pair of gluons, for it is spin $1^-$,
a state not available to two gluons.  It requires at least three gluons, which
means a high power of the coupling constant, $g^6$, which is rather small at
the scale of the separation between a heavy $q$ and $\bar q$ in the $\psi$ or
$\Upsilon$.  Also the numerical coefficient of the annihilation amplitude 
is small, making the width for disintegration of $\psi$ or
$\Upsilon$ into hadrons quite narrow.  Hence the OZI rule is understood for
these particles, at least.

\bigskip
{\bf References}\footnote{The following references (here modernized to include
DOIs) and discussion of Isgur's work are
present in my original notes, but somehow got omitted from the
revised versions submitted to RPF.}
\begin{trivlist}
\item
 A.~De Rujula, H.~Georgi and S.~L.~Glashow,
  ``Hadron Masses in a Gauge Theory,''
  Phys.\ Rev.\ D {\bf 12}, 147 (1975).
  doi:10.1103/PhysRevD.12.147
\item
T.~Appelquist, R.~M.~Barnett and K.~D.~Lane,
  ``Charm and Beyond,''
  Ann.\ Rev.\ Nucl.\ Part.\ Sci.\  {\bf 28}, 387 (1978).
  doi:10.1146/annurev.ns.28.120178.002131
\item
 N.~Isgur,
  ``Hadronic structure with QCD: From $\alpha$ to $\omega$ (via $\psi$ and $\Upsilon$),''
  AIP Conf.\ Proc.\  {\bf 81}, 1 (1982).
  doi:10.1063/1.33447
\end{trivlist}

In the last reference, Isgur supposed that the potential between
quarks in a hadron is $V(r_{ij}) = \sfrac12k r_{ij}^2 + U(r_{ij})$
where $U$ is a perturbation.  One needs to evaluate $\langle U\rangle$
and $\langle U|x_i-x_j|^2\rangle$ expectation values.  This gives two
parameters to fit the spectrum.  He discusses the fact that even heavy quark systems have 
sufficiently large wave functions at large distances such that
the linear term in $V(r) = -\alpha/r + b r$ is important.
But the whole potential approach is
approximate for various reasons, including relativistic effects
and annihilation/creation processes.\footnote{This paragraph,
present in my notes, was omitted from the original revised version.}

\section{Quantization of QCD$^*$ (11-12-$xx$)}
\label{sect8}

We now turn to the quantization of QCD.  Recall that the action for the gluon
fields, interacting with a non\-dynamical source, is
\be
	S[A] = \int\left({1\over 4 g^2}\, \vec F_{\mu\nu}\,\!^{\bigcdot}\vec F_{\mu\nu} 
	+\vec J_\mu\,\!^{\bigcdot}\vec A_\mu\right) d^{\,4}x
\label{eq8-1}
\ee
I will assume that you are familiar with the path integral formulation of
quantum mechanical amplitudes.  If $A_i$ and $A_f$ are the initial and final
configurations of the gluon field, the transition amplitude for going from 
$A_i$ to $A_f$ is
\be
	K[A_f,A_i] \equiv \int_i^f {\cal D}A\, e^{i S[A]}
\label{eq8-2}
\ee
where the integral is supposed to be over all field configurations $A_\mu^a(\vec
x,t)$ such that $A^a_\mu(\vec x,t_i) = A_{i,\mu}^a(\vec x)$ and 
$A^a_\mu(\vec x,t_f) = A_{f,\mu}^a(\vec x)$.  Formally, the measure is defined
as an infinite product over all points in spacetime between $t=t_i$ and $t=t_f$,
\be
	{\cal D}A = \prod_{\vec x,t,\mu} d^{\,8}\!A_\mu(\vec x,t)
\label{eq8-3}
\ee
The eight-dimensional measure $d^{\,8}\!A_\mu$ simply means the product over the eight
components of color, $\prod_i dA^i_\mu$.\footnote{This sentence may seem
extraneous, but RPF was correcting my misconception that the Haar measure for
the group manifold was somehow incorporated.}\ \  The action (\ref{eq8-1})
is invariant under the local color gauge transformations of $A_\mu$,
\be 
	A'_\mu = \Lambda^\dagger A_\mu \Lambda + i\Lambda^\dagger\partial_\mu
	\Lambda
\label{eq8-4}
\ee
This implies that the measure ${\cal D}A$ is also gauge invariant, since it
transforms as 
$dA'_\mu = \Lambda^\dagger dA_\mu\Lambda$ at each point, and the Jacobian
of this transformation is trivial when we consider $d^{\,8}\!A_\mu$.
{\bf Problem.}  Prove this.\footnote{The Jacobian matrix is  $\sfrac12\tr(\lambda^a\Lambda^\dagger
\lambda^b\Lambda)$.  One can show its determinant is trivial by considering
an infinitesimal transformation to leading order, and using 
$\det = \exp\tr\ln$.}

Now as you know, this path integral is plagued with infinities.  One
rather trivial kind is the infinite volume of spacetime.  Another sort, the 
ultraviolet divergences, comes from the uncountably infinite dimensional nature
of the measure, an integral for each point of spacetime.  This kind I want to
ignore for the moment---it can be cured by approximating spacetime as a discrete
lattice, in some gauge invariant way.  This has been discussed by Wilson, and we
shall describe it later.  It is the basis for a numerical method to evaluate the
path integral.

But in QCD we are still left with another infinity, due to the gauge
symmetry itself.  If we represent the space of functions $A_\mu(x)$ in two
dimensions, we have trajectories of gauge fields that are related to one another
by local color rotations:
\centerline{
\includegraphics[width=0.3\textwidth]{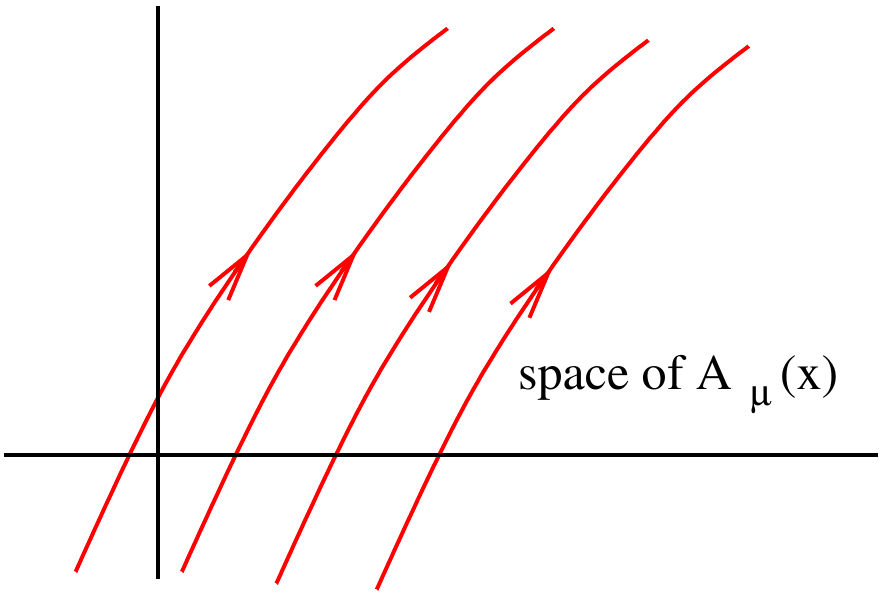}}
This means that the expectation value of a physically relevant operator,
that is gauge invariant, will diverge like the volume of local gauge
transformations in function space,
\be
	\int{\cal D}A\, e^{iS[A]} \sim \int{\cal D}\Lambda(\vec x,t) = \infty
\label{eq8-6}
\ee
Put another way, there are directions in which $A$ can change ({\it i.e.,} by
gauge transformations) for which the integrand is invariant and the region of
integration is infinite.

However we can get a finite and meaningful result by defining a kind of
expectation value of any gauge invariant functional $F[A]$ of the field, and
limiting ourselves to computing only such quantities,
\be
	\langle F\rangle = \lim_{{\cal R}\to\infty} {\int{\cal D}A\, F\,
e^{iS[A]} \over \int{\cal D}A\, e^{iS[A]}}
\label{eq8-7}
\ee
where ${\cal R}$ is a relatively finite region in the space of gauge fields;
for example we might limit the range of each $A_\mu$ to be $[-M,+M]$ for some
large value of $M$, the same in the numerator and denominator, and then take the
limit as $M\to\infty$.  (I say ``relatively finite'' because there are still
infinitely many integration variables, one for each point in spacetime, unless we
go to a lattice.)  Gauge invariance is broken temporarily by this procedure
because $|A'_\mu|$ need not be less than $M$ even if $|A_\mu|$ is, but this
should be no problem in the limit ${\cal R}\to\infty$.

Having found a finite and gauge-invariant definition of amplitudes, we are free
to choose a gauge.  A convenient choice is $A_0=0$.  Recall that it
is always possible to reach this gauge from a configuration where $A_0^{\rm old}
\neq 0$, using the transformation matrix
\be
	\Lambda(\vec x,t) = P \,\exp\left(i\int^t A_0^{\rm old}(\vec
x,t')\right)dt'\,.
\label{eq8-9}
\ee
In general, the path integral measure ${\cal D}A_\mu^{\rm old}$ changes by
a Jacobian factor when we transform the variables $A_\mu^{\rm old}\to
A_\mu^{\rm new}$, where $A_0^{\rm new}=0$.  How can this be?  The measure was
supposed to be gauge invariant!  But this is true only for gauge transformations
that are {\it independent of $A_\mu$}.  Nevertheless, we will show that for the
special case of (\ref{eq8-9}), the Jacobian is still almost trivial, even though
$\Lambda(\vec x,t)$ depends on $A_\mu$.  This is because $\Lambda$ depends only
on $A_0$, and thus for the three space directions at least, $d^{\,8}A'_m = 
d^{\,8}A_m$ (we use $n,m,$ {\it etc.} for the spatial components of $\nu,\mu$,
while the time component is 0.)  The expectation value of an
operator can now be written as
\bea
	&&{\int_{{\cal R}^{\rm old}} e^{iS[\mathbb{A}^{\rm old},A_0^{\rm old}]}
	F[\mathbb{A}^{\rm old},A_0^{\rm old}] \,
	\prod_{n=1}^3 {\cal D}A_n^{\rm old}\,{\cal D}A_0^{\rm old}\over
	\int_{{\cal R}^{\rm old}} e^{iS[\mathbb{A}^{\rm old},A_0^{\rm old}]}
	\prod_{n=1}^3 {\cal D}A_n^{\rm old}\,{\cal D}A_0^{\rm old}}\nn\\
	&=& {\int_{{\cal R}^{\rm new}} e^{iS[\mathbb{A}^{\rm new},0]}
	F[\mathbb{A}^{\rm new},0] \,
	\prod_{n=1}^3 {\cal D}A_n^{\rm new}\,{\cal D}A_0^{\rm old}\over
	\int_{{\cal R}^{\rm new}} e^{iS[\mathbb{A}^{\rm new},0]}
	\prod_{n=1}^3 {\cal D}A_n^{\rm new}\,{\cal D}A_0^{\rm old}}\nn\\
\label{eq8-11}
\eea
by making the gauge transformation (\ref{eq8-9}).
In the second expression, the integrands are independent of $A_0$, and the
factors of $\int{\cal D}A_0^{\rm old}$ cancel between numerator and denominator,
using the definition (\ref{eq8-7}).

If $F$ were not a gauge invariant operator, we could always replace it by its
gauge-averaged expression,
\be
	\hat F[A] = {\int d\Lambda \, F[A']\over \int d\Lambda}
\label{eq8-12}
\ee
where $A'$ is as in (\ref{eq8-4}), and $d\Lambda$ is the invariant group
measure,
satisfying
\be
	\int d\Lambda\, f(\Lambda) =  \int d\Lambda\, f(\Lambda\Lambda_0)
\label{eq8-13}
\ee
for any SU(3) matrix $\Lambda_0$ and any function $f$.

{\bf Problem:} If $F$ is not gauge invariant, show that $\langle F\rangle$ as
defined in (\ref{eq8-7}) is the same as $\langle\hat F\rangle$.

Now the path integral is reduced to the simpler expression,
\be
	Z = \int e^{iS[\mathbb{A},A_0]}{\cal D}\mathbb{A}
\label{eq.8-14}
\ee
where
\be
	S = {1\over 2 g^2}\int d^{\,4}x\left(\vec{\mathbb{E}}
	\ ^{\bigcdot}\!\!\!\cdot
	\vec{\mathbb{E}} - \vec{\mathbb{B}}
	\ ^{\bigcdot}\!\!\!\cdot\vec{\mathbb{B}}\right)\,.
\label{eq8-15}
\ee
Since $A_0=0$,
\be
	\vec{\mathbb{E}} = -\partial_0\vec{\mathbb{A}} = -\dot{\vec{\mathbb{A}}}
\label{eq8-16}
\ee
and $\mathbb{B}$ is just as it was before,\footnote{RPF had written
$\vec{\mathbb{B}} = \bbnabla_\times\vec{\mathbb{A}} - \vec{\mathbb{A}}\,^\times
	\vec{\mathbb{A}}$.  I have corrected it here and in some subsequent
equations to indicate the operation needed for the spatial indices of
the interaction term.
The spatial cross product introduces a factor of 2 that must be compensated.
\label{Bcorr}}
\be
	\vec{\mathbb{B}} = \bbnabla_\times\vec{\mathbb{A}} -\sfrac12\,
\vec{\mathbb{A}}\,^\times_\times 
	\vec{\mathbb{A}}
\label{eq8-17}
\ee
(for example, $B_z = \partial_x A_y - \partial_y A_x - [A_x,A_y]$).
In terms of the gauge field, the Lagrangian is
\be
	L = {1\over 2 g^2}\int d^{\,3}x\,\left(\dot{\vec{\mathbb{A}}}
\ ^{\bigcdot}\!\!\!\cdot\dot{\vec{\mathbb{A}}} - (\bbnabla_\times
\vec{\mathbb{A}} - 
\sfrac12\,\vec{\mathbb{A}}\,^\times_\times
	\vec{\mathbb{A}})^2\right)\,.
\label{eq8-18}
\ee
This is analogous to the Lagrangian for a particle moving in a potential
\be
	L = \sfrac12 m\,\dot x^2(t) - V(x(t))
\label{eq8-19}
\ee
where $\vec{\mathbb{A}}$ is like the position of the particle, 
$\dot{\vec{\mathbb{A}}}\!:\!{\vec{\mathbb{A}}}$ plays the role of the kinetic term, and
$\vec{\mathbb{B}}\!:\!\vec{\mathbb{B}}$ is the potential.  This
situation is unique to the $A_0=0$ gauge; in other gauges we would have terms
like $\dot{\vec{\mathbb{A}}}\!:\!\vec{\mathbb{A}}$, and the
separation between kinetic and potential energy would no longer be so clean.

For simplicity we suppressed the source term.  In $A_0=0$ gauge, it is
\be
	L_{\rm source} = \int d^{\,3}x\,
	 \vec{\mathbb{A}}\,^{\bigcdot}\,\!\!\!\!\cdot\vec{\mathbb{J}}\,,
\label{eq8-20}
\ee
and one sees that the charge density $\rho$ does not enter.  Then the equation
of motion for $\vec{\mathbb{A}}$ implied by the full Lagrangian is
\be
	\ddot{\vec{\mathbb{A}}} - \mathbb{D}_\times\mathbb{B} = g^2 \mathbb{J}
\label{eq8-21}
\ee
We also have the nondynamical equations
\be
	\mathbb{D}\cdot\mathbb{B} = 0,\quad
	\dot{\mathbb{B}} + \mathbb{D}_\times\mathbb{E} = 0
\label{eq8-22}
\ee
which are identities, due to the way $\mathbb{B}$ and $\mathbb{E}$ are defined 
in terms of $\mathbb{A}$.  Recall that we found one further equation
by varying the covariant form of the Lagrangian, namely Gauss's law,
\be
	\mathbb{D}\cdot\mathbb{E} = g^2\rho
\label{eq8-23}
\ee

{\bf Problems.} Gauss's law does not seem to arise from the path integral
formulation in $A_0=0$ gauge.  What happened to it?

What condition must the source $J_\mu$ obey in order for
$\int d^{\,4}x\, \vec J_\mu \,^{\bigcdot}\vec{A_\mu}$ to be gauge invariant?  What is the physical
significance of this condition when $J_\mu$ is the quark current?

Show that the equations of motion of $\psi$ imply that $D_\mu
J_\mu = 0$. 

\section{Hamiltonian formulation of QCD$^*$ (11-17-$xx$)}
\label{sect9}

Just as in ordinary quantum mechanics the state of a system is specified by a
wavefunction $\psi(x,t)$, in the purely gluonic version of QCD we can
characterize physical states by a wave functional $\Psi[\mathbb{A}(\vec x,t)]$
that depends on the gauge field $\mathbb{A}$.  (We continue to work in $A_0=0$
gauge.)  In this lecture we shall explore the analogy somewhat, and discuss the
Hamiltonian formalism for evolving $\Psi$ in time.  

Perhaps the closest analogy to the situation in field theory, where we have
infinitely many dynamical variables, would be a lattice of atoms, say,
interacting with each other through some potential $V$ that depends on the
positions $\vec q(\vec n,t)$, where $\vec n$ is a lattice vector telling us
which atom is being referred to.  The action is
\be
	S = \int dt\, L = \int dt\,\left(\sfrac12 m\sum_{\vec n}\left|
	\dot{\vec{q}}(\vec n,t)\right|^2 - V\left(\vec q(\vec n,t)\right)\right)\,.
\label{eq9-1}
\ee
Now if $\psi_i(\vec q(\vec n))$ is the wave function at some initial time $t_i$,
then the amplitude for reaching a state $\psi_f(\vec q(\vec n))$ at a later time
$t_f$ is given by an ordinary integral
\bea
	&&\int\prod_{\vec n} d^{\,3}q_i(\vec n)\, d^{\,3}q_f(\vec n)\\
	&&\qquad\psi_f(\vec q_f(\vec n))\, 
	K\Big(\vec q_f(\vec n),t_f; \vec q_i(\vec n),t_i\Big)\, \psi_i(\vec
q_i(\vec n))\nn
\label{eq9-2}
\eea
and the function $K$ that propagates the initial state is given by a path
integral
\be
	K\Big(\vec q_f(\vec n),t_f; \vec q_i(\vec n),t_i\Big) =
	\prod_{\vec n}{\cal D}\vec q(\vec n,t)\, e^{iS[\vec q(\vec n,t)]}
\label{eq9-3}
\ee
where the integral is over functions $\vec q(\vec n,t)$ satisfying
\bea
	\vec q(\vec n,t_i) &=& \vec q_i(\vec n)\,,\nn\\
	\vec q(\vec n,t_f) &=& \vec q_f(\vec n)\,.
\label{eq9-4}
\eea

In the same way, we can assign to each state in QCD a wave functional
$\Psi[\mathbb{A}(\vec x,t)]$ such that $|\Psi|^2$ is the probability density
for the gauge field to have the value $\mathbb{A}(\vec x,t)$, for each point in
space, at a given time.  Here $\mathbb{A}$ corresponds to $\vec q$, and the
position $\vec x$ corresponds to the lattice vector $\vec n$ in the atomic
crystal analog.  The kernel for time evolution of $\Psi[\mathbb{A}]$ was given
in (\ref{eq8-2}), which is the analog of (\ref{eq9-3}).

Alternatively, the time evolution of $\Psi$ can be described in differential
rather than integral form---Schr\"odinger's equation!  To do this, we must
first find the Hamiltonian.  In the finite system (\ref{eq9-1}), the canonically
conjugate momenta are
\be
	\vec p(\vec n) = {\partial L\over \partial \dot{\vec q}(\vec n)}
	= m\, \dot{\vec q}(\vec n)
\label{eq9-5}
\ee
They can be represented by
\be
	\vec p(\vec n) = {1\over i}\,{\partial \over \partial \vec q(\vec n)}
\label{eq9-6}
\ee
(taking $\hbar = 1$) since $-i\partial/\partial\vec q$ has the same commutation
relation 
with $\vec q$ as $\vec p$ has canonically.  Similarly in QCD the momentum
conjugate to $A^a_n(\vec x)$ is
\be
	p^a_n(\vec x) = {\delta L\over \delta \dot A^a_n(\vec x)}
	= \dot A^a_N(\vec x) = - E^a_n(\vec x)
\label{eq9-7}
\ee
where we used the Lagrangian (\ref{eq8-18}) after rescaling $\mathbb{A}\to
g\mathbb{A}$, and the operation that appears is a {\it functional} derivative,
\be
	{\delta  A^a_n(\vec x)\over \delta A^b_m(\vec x')}
	= \delta_{ab}\,\delta_{nm}\,\delta^{(3)}(\vec x-\vec x')
\label{eq9-8}
\ee
This is the natural generalization of the partial derivative to the case of
infinitely many variables, labeled by a continuous index $\vec x$.  Notice that
$\vec p$ is just minus the color electric field.  It can also be written as
\be
	p^a_n(\vec x) = {1\over i}\, {\delta\over \delta A^a_n(\vec x)}
\label{eq9-9}
\ee
similarly to (\ref{eq9-6}).  Now the Hamiltonian can be constructed.  For a discrete
system like the lattice, it is
\be
	H = \left(\sum_{\vec n} \vec p(\vec n) \cdot \dot{\vec q}(\vec n)
	-L\right)_{\dot{\vec q} = \vec p/m}\, .
\label{eq9-10}
\ee
For QCD, one simply replaces the sum with an integral, so that 
\bea
%\label{eq9-11a}
	H &=& \sfrac12 \int d^{\,3}x\left(\vec{\mathbb{E}}^2 + \vec{\mathbb{B}}^2
	\right)\nn\\
	&=& \sfrac12\int d^{\,3}x\left(-\left(\delta\over \delta
\vec{\mathbb{A}}\right)^{\!\!2} + \vec{\mathbb{B}}^2
	\right)\,.
\label{eq9-11b}
\eea
Then the Schr\"odinger equation for $\Psi[\mathbb{A}]$ is
\be
	{1\over i}\, {\partial\over\partial t}\Psi = H\Psi\,.
\label{eq9-12}
\ee

Previously we noted that three of the four Maxwell equations of QCD emerged from
the gauge-fixed Lagrangian (\ref{eq8-18}), and the definitions of $\mathbb{E}$
and $\mathbb{B}$, but Gauss's law,
\be
	{\mathbb{D}}\hbox{\raisebox{-0.05cm}{$\cdot$}}\vec{\mathbb{E}} = g^2\vec\rho\nn\,,
\ee
did not appear.  However, the {\it time derivative} of Gauss's law can be
deduced as follows:
\bea
	{\partial\over\partial t}\left(\mathbb{D}\hbox{\raisebox{-0.05cm}{$\cdot$}}
	\vec{\mathbb{E}}\right) &=& - {\partial\over\partial t}\left(
	\bbnabla \hbox{\raisebox{-0.05cm}{$\cdot$}}\,\dot{\vec{\mathbb{A}}}
	- i \vec{\mathbb{A}}\ \hbox{\raisebox{-0.05cm}{$\cdot$}}\!\!\!^\times
	\dot{\vec{\mathbb{A}}}\right)\nn\\
	&=& -\bbnabla \hbox{\raisebox{-0.05cm}{$\cdot$}}\,\ddot{\vec{\mathbb{A}}}
	+ i \vec{\mathbb{A}}\ \hbox{\raisebox{-0.05cm}{$\cdot$}}\!\!\!^\times
	\ddot{\vec{\mathbb{A}}}\nn\\
	&=& \mathbb{D}\hbox{\raisebox{-0.05cm}{$\cdot$}}\, \ddot{\vec{\mathbb{A}}}
\label{eq9-13}
\eea
Using the equation of motion for $\mathbb{A}$, eq.\ (\ref{eq8-21}), this becomes
\be
	{\partial\over\partial t}\left(\mathbb{D}\hbox{\raisebox{-0.05cm}{$\cdot$}}
	\vec{\mathbb{E}}\right) = 
	-\mathbb{D} \hbox{\raisebox{-0.05cm}{$\cdot$}}\,(\mathbb{D}
\hbox{\raisebox{-0.05cm}{$\times$}}\vec{\mathbb{B}}) - g^2\mathbb{D}\hbox{\raisebox{-0.05cm}{$\cdot$}}
	\vec{\mathbb{J}}
\label{eq9-14}
\ee
The middle term would vanish trivially if $\mathbb{D}$ was the ordinary
gradient, but since the components of $\mathbb{D}$ do not commute, more care is
required.  One finds that $\mathbb{D} \hbox{\raisebox{-0.05cm}{$\cdot$}}\,(\mathbb{D}
\hbox{\raisebox{-0.05cm}{$\times$}}\vec{\mathbb{B}})$ is
\be
	[D_x,D_y]B_z \hbox{ plus cyclic permutations.}\nn
\ee
But this is just $i[F_{xy},B_z]$, as you showed in a previous exercise, which 
vanishes because $F_{xy} = B_z$.  Furthermore the source is covariantly 
conserved, 
\be
	D_\mu J_\mu = 0,\nn
\ee
so that 
\be
	\mathbb{D} \hbox{\raisebox{-0.05cm}{$\cdot$}}\,\vec{\mathbb{J}} 
	= -{\partial\over\partial t}\vec\rho\nn
\ee
in $A_0=0$ gauge.  (In our conventions, $A_\mu B_\mu = -A_0 B_0 + A_i B_i$,
and $\rho = -J_0$.\footnote{This choice of the metric signature is not
consistent throughout the lectures.})  Therefore the condition
\be
	{\partial\over\partial t}(\mathbb{D}
\hbox{\raisebox{-0.05cm}{$\cdot$}}\,\vec{\mathbb{E}} )
	= g^2{\partial\over\partial t}\vec\rho
\label{eq9-15}
\ee
is a consequence of the equations of motion. Consequently,
if the wave functional satisfied
\be
	\left[ \mathbb{D}\hbox{\raisebox{-0.05cm}{$\cdot$}}\,\vec{\mathbb{E}} -
	g^2\vec\rho\right]\Psi[\mathbb{A}(x)] = 0
\label{eq9-16}
\ee
at some initial time, it would continue to do so forever.  Therefore Gauss's law
can be implemented by imposing it as a constraint on the state of the system,
$\Psi$.

Notice that (\ref{eq9-16}) is a functional differential equation, since
$\mathbb{E}$ is to be interpreted as $-i\delta/\delta\mathbb{A}$.  Moreover
(\ref{eq9-16}) is an infinite set of constraints, one at each point
in space.  One might wonder whether solutions exist, since the
constraint operator
\be
	\vec C(\vec x) \equiv \mathbb{D}\hbox{\raisebox{-0.05cm}{$\cdot$}}\,\vec{\mathbb{E}}(\vec x)
	-g^2\vec\rho(\vec x)
\label{eq9-17}
\ee
does not commute at different positions,
\be
	[\vec C(\vec x),\vec C(\vec x')]\neq 0\,.\nn
\ee
To be consistent, we require that this new operator also
annihilates the wave functional.  If the commutator is a linear
combination of $\vec C$'s there is no problem, but if not we might
generate more and more constraints, to the point that no solution
existed.  It turns out to be nicer to investigate this not with the
$\vec C$'s directly, but rather with their weighted averages, defined
by
\be
	\Gamma(\mu) \equiv \int \mu^i(\vec x)\, C^i(\vec x)\, d^{\,3}x\, .
\label{eq9-18}
\ee
One can show that
\be
	[\Gamma(\mu),\Gamma(\nu)] = \Gamma(\lambda)
\label{eq9-19}
\ee
where
\be
	\vec\lambda = \vec\mu\,^\times\vec\nu
\label{eq9-20}
\ee
Thus the commutators produce no new constraints; instead they form a
closed algebra.  It is the algebra of the color group SU(3),
for we could define generators of local color transformations
\be
	\hat\Gamma(\mu) = \mu^i(x) T^i\nn
\ee
in the three-dimensional ({\it i.e.} {\it fundamental})
representation of SU(3), and they would satisfy the same relations
(\ref{eq9-19},\ref{eq9-20}) as the $\Gamma(\mu)$.

{\bf Problem.}  Prove eqs.\ (\ref{eq9-19},\ref{eq9-20}).

It is therefore not surprising that the $\Gamma$ operators generate
gauge transformation on the state $\Psi$.  That is,
\be
	e^{i\Gamma(\mu)}\Psi[\mathbb{A}] = \Psi[\mathbb{A}']
\label{eq9-21}
\ee
where $\mathbb{A}'$ is the gauge field obtained from $\mathbb{A}$
by transforming with the matrix
\be
	\Lambda(\vec x) = e^{i\vec\mu(\vec x)\,^{\bigcdot}\,\vec T}
\label{eq9-22}
\ee

The alert reader may wonder how it is possible to do gauge
transformations, since we have already fixed the gauge to $A_0=0$.
However the transformation (\ref{eq9-22}) is time-independent, so
$\Lambda^\dagger\partial_0\Lambda=0$ and any such $\Lambda$ will keep
$A_0=0$.  Hence the Gauss's law constraint on $\Psi$ means
that $\Psi$ must be {\it invariant} under the residual gauge
transformations that preserve the $A_0=0$ condition.  We can prove
this directly from the equation itself: let
\be
	\left[\int \vec\alpha\,^{\bigcdot}\left(
\mathbb{D}\lcdot\vec{\mathbb{E}} - g^2\vec\rho(\vec x)\right) d^{\,3}x\right]
	\,\Psi[\mathbb{A}] = 0
\label{eq9-23}
\ee
for some $\vec\alpha(\vec x)$.  For simplicity suppose that there are
no quarks, so that $\rho=0$.  Then (\ref{eq9-23}) can be rewritten as
\be
	\left[\int \mathbb{D}\vec\alpha(\vec x):\vec{\mathbb{E}}(\vec x)\,
	d^{\,3}x\right]\, \Psi[\mathbb{A}] = 0\nn
\ee
where we have integrated by parts.  (The reader should satisfy himself
that partial integration works for covariant derivatives.)  Using the
operator form of $\mathbb{E} = -i\delta/\delta\mathbb{A}$, this becomes
\be
	\int \mathbb{D}\vec\alpha(\vec
x):{\delta\Psi\over\delta\vec{\mathbb{A}}}\,d^{\,3}x\ = 0
\label{eq9-24}
\ee
   Now the 
functional version of Taylor's theorem says that
\be
	\Psi[\vec{\mathbb{A}}(x) + \vec{\mathbb{a}}(x)]
	= \Psi[\vec{\mathbb{A}}(x)] + 
	\int \vec{\mathbb{a}}:{\delta\Psi[\vec{\mathbb{A}}(x)]\over
	\delta\vec{\mathbb{A}}(x)}\, d^{\,3}x
\label{eq9-25}
\ee
to first order in $\vec{\mathbb{a}}$.  So for infinitesimal
$\vec\alpha(\vec x)$,  Gauss's law is equivalent to
\be
	\Psi[\vec{\mathbb{A}}(x) + \mathbb{D}\vec\alpha(x)] = 
	\Psi[\vec{\mathbb{A}}(x)]
\label{eq9-26}
\ee
which is just the result claimed, since $\mathbb{D}\vec\alpha(x)$
is the effect of an infinitesimal gauge transformation.  

However not all gauge transformations can be built up from
infinitesimal ones.  This is most easily demonstrated for the 
SU(2) subgroup of SU(3) generated by 
\be
	\vec\sigma = (\lambda_1,\lambda_2,\lambda_3)\nn
\ee
Imagine that we have cut off the infinite volume of space by
introducing a large radius $R$.  It is easy to check that the matrix
\be
	\Lambda = \sqrt{1-{x^2\over R^2}} + i{\vec x\cdot\vec\sigma\over R}
\label{eq9-27}
\ee
is unitary and has determinant $1$, yet cannot be written in the form
of $\exp(i\vec\alpha\,^{\bigcdot}\vec\lambda)$.  We should therefore
require that $\Psi[\mathbb{A}]$ be invariant under all such ``large''
gauge transformations, as well as the ordinary ones.

\section{Perturbation Theory  (11-19-$xx$)}
\begin{figure}[t]
\centerline{
\includegraphics[width=0.45\textwidth]{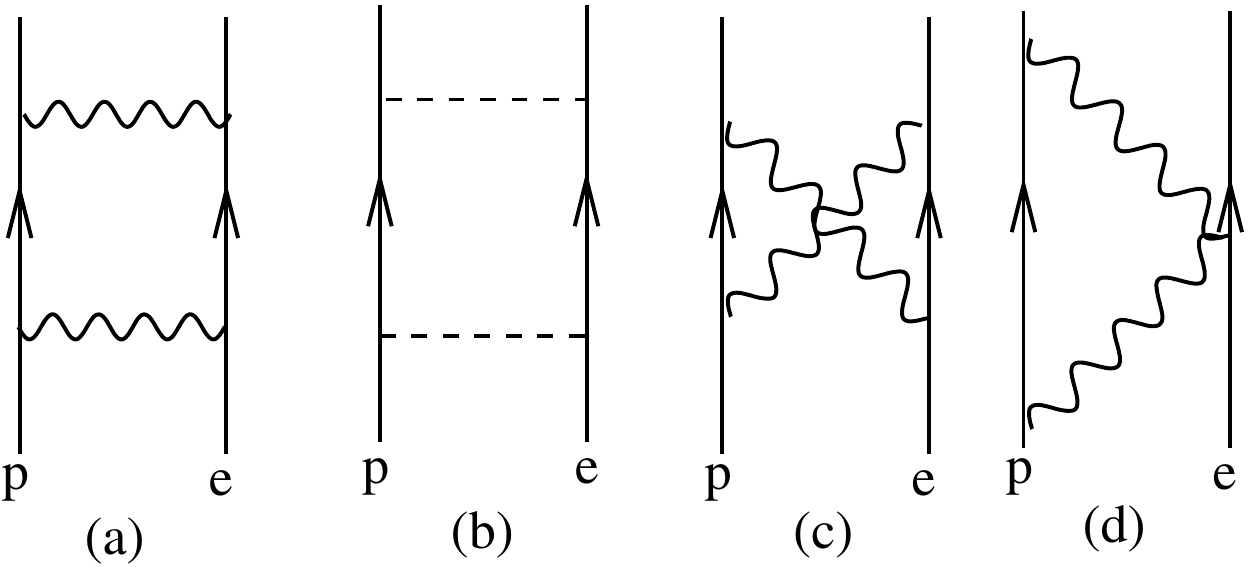}}
\caption{Diagrams for the hydrogen atom.}
\label{hatom}
\end{figure}

Now we must confront the question, how do we calculate anything
quantitatively in this theory?  Perturbation theory (P.T.) is not very useful
for bound state properties since the coupling $g$ is large.  On the
other hand, we have some experience with nonperturbative processes 
even in nonrelativistic quantum mechanics; the hydrogen atom is not
a perturbative problem---it is an exact nonperturbative solution to 
Schr\"odinger's equation.  But it is nevertheless made more accurate by the
smallness of the coupling $e$.  Consider fig.\ \ref{hatom}(a).  Since
photon exchange is relatively infrequent, we can replace the photon
exchanges by instantaneous effective interactions depicted in 
fig.\ \ref{hatom}(b).  The Schr\"odinger equation takes an initial
state of the proton and the electron and propagates them freely via
$-(\partial/\partial x)^2$ plus interactions $V(x)$.  This only works
if the coupling constant is small; otherwise diagrams like fig.\
\ref{hatom}(c,d) become too important and we would have to find some
other kind of effective interaction potential to represent their 
effect.

\begin{figure}[b]
\centerline{
\includegraphics[width=0.3\textwidth]{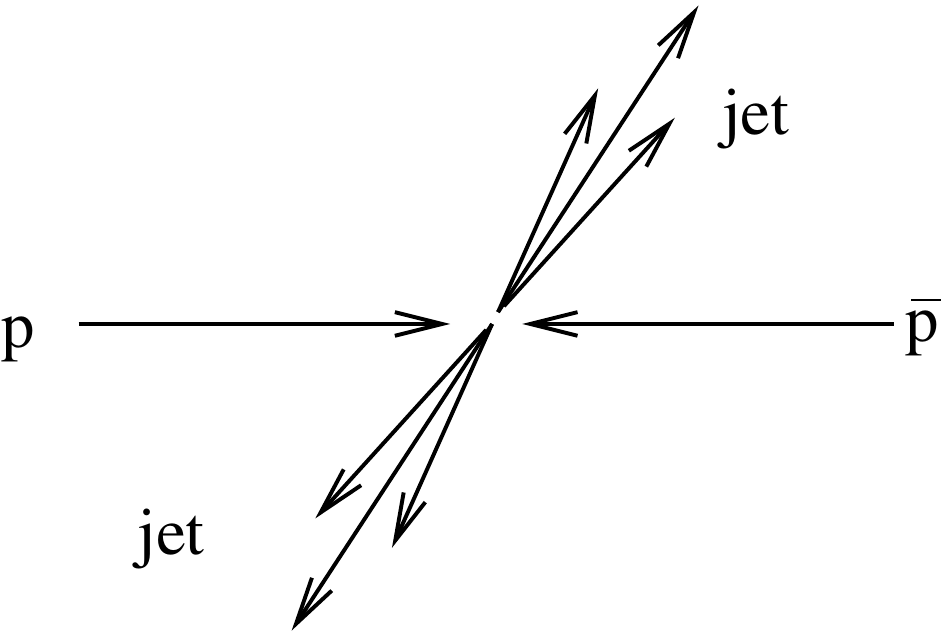}}
\caption{Electron-positron annihilation into hadronic jets.}
\label{pp-jet}
\end{figure}

But there are some processes for which perturbation theory in QCD
works relatively well, for example $p\bar p$ collisions (fig.\
\ref{pp-jet}).  At sufficiently high energies, only one gluon might
be exchanged, similarly to fig.\ \ref{ann}.  One can then predict
the dynamics of the jets rather precisely using perturbation theory.

\subsection{Review of P.T.\  from the path integral}
Recall the massless scalar field theory with Lagrangian
\be
	{\cal L} = \frac12\left(\dot\phi^2 -(\nabla\phi)^2\right)
	+ \frac{g}{3!}\,\phi^3\nn - S(x)\,\phi 
\ee
The equation of motion is
\be
	\ddot\phi - \nabla^2\phi \equiv \square\phi = {g\over 2}\phi^2
- S
\ee
Let us ignore interactions for the moment and focus on the source
term.  Going to Fourier space, the equation of motion becomes
\be
	(\omega^2 - \vec k^2)\phi_k =  k^2\phi_k = S_k
\ee
To solve for $\phi_k$, a prescription must be given for treating the
pole of the propagator, {\it i.e.,} we must add $i\epsilon$,\footnote{Throughout these lectures, 
RPF avoids writing factors
of $i$ for the vertices and the propagators.  Later, he will claim
that the only factors of $i$ that are necessary to keep track of
can be associated with loops.  I have not checked whether his rules
for the cubic and quartic gluon interactions give rise to the
correct sign of interference for the diagrams contributing to
four-gluon scattering, but his later claim implies that he did so.
\label{fn24}}
\be
	\phi_k = {S_k\over k^2 + i\epsilon}
\ee
Putting back the interaction, we have the rule that the amplitude
for three $\phi$ particles to interact is $g$.  Then for example
the $s$-channel scattering diagram is
\be
\raisebox{-0.5cm}{\includegraphics[width=0.15\textwidth]{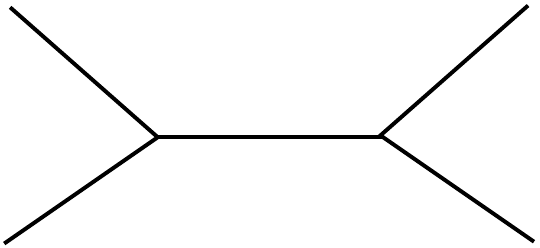}} =
{g^2\over
k^2+i\epsilon}
\ee

Let's now compare this to the case of QCD in $A_0=0$ gauge, where the
propagator comes from the free Lagrangian
\be
{\cal L} = {1\over 2}\,\left(\dot{\vec{\mathbb{A}}}
\ ^{\bigcdot}\!\!\!\cdot\dot{\vec{\mathbb{A}}} - 
(\bbnabla_\times\vec{\mathbb{A}})^2\right)\, .		
\ee
Here we have rescaled $\mathbb{A}\to g\mathbb{A}$ to get the 
coupling out of the propagator and back into the interactions.
Including a classical source $\vec{\mathbb{S}}$, the equation of motion is
\be
	\ddot{\vec{\mathbb{A}}} -
\bbnabla\times(\bbnabla\times\vec{\mathbb{A}})
	= -\vec{\mathbb{S}}
\ee
or equivalently
\be
	\ddot{\mathbb{A}}^i - \nabla^2\mathbb{A}^i + \bbnabla
	(\bbnabla\cdot\mathbb{A}^i) = -\mathbb{S}^i
\ee
Hence, going to Fourier space,
\be
	(\omega^2-\mathbb{k}^2)\,\mathbb{A} + \mathbb{k}(\mathbb{k}\cdot
	\mathbb{A}) = \mathbb{S}
\label{eq10-8}
\ee
(omitting the gauge index $i$ and subscript $k$ for brevity).  Next
dot $\mathbb{k}$ into this equation to get
\be
	(\omega^2-\mathbb{k}^2)\,\mathbb{k}\cdot\mathbb{A} + 
	\mathbb{k}^2\, \mathbb{k}\cdot
	\mathbb{A} = \omega^2\, \mathbb{k}\cdot\mathbb{A} = \mathbb{k}\cdot\mathbb{S}\,.
\ee
We can therefore eliminate $\mathbb{k}\cdot\mathbb{A}$ from eq.\
(\ref{eq10-8}) and find
\be
	k^2\, \mathbb{A} = {1\over \omega^2}\left(\mathbb{S} - \mathbb{k}
	(\mathbb{k}\cdot\mathbb{S})\right)\, .
\ee
In analogy to the scalar field example, one can read off the propagator
\be
	\mathbb{P}(k) =  {\mathbb{1} - \mathbb{k}\mathbb{k}/\omega^2\over k^2 +i\epsilon}
\ee
in the form of a $3\times 3$ matrix for the spatial components of
the gauge field, with the understanding that $\mathbb{k}\mathbb{k}$
denotes the outer product of the spatial momenta components.

\subsection{Perturbation theory for QCD}

Now we would like to perturbatively compute path integrals involving gauge
invariant functionals $F[\mathbb{A},0]$ (showing explicitly that
$A_0$ is set to zero),
\be
\langle F\rangle =    {\int e^{iS[\mathbb{A},0]}\,F[\mathbb{A},0]\,{\cal D}\mathbb{A}
\over 
\int e^{iS[\mathbb{A},0]}\,{\cal D}\mathbb{A}}
\ee
where the action is
\be
	S[\mathbb{A},0] = \sfrac12\int\left[
	\dot{\vec{\mathbb{A}}}:\dot{\vec{\mathbb{A}}}
	- \vec{\mathbb{B}}:\vec{\mathbb{B}}\right]\,d^{\,4}x
	+ g\int \vec{\mathbb{A}}:\vec{\mathbb{J}} \,d^{\,4}x
\ee
with $\vec{\mathbb{B}} = \bbnabla_\times\vec{\mathbb{A}} - 
(g/2)\, \vec{\mathbb{A}}^\times_\times\vec{\mathbb{A}}$ and
the current
\be
	\mathbb{J}_n^i = \bar\psi(\vec x,t)\,\gamma_n\,{\lambda^i\over 2}
	\,\psi(\vec x,t) \qquad (n=1,2,3)\,.
\ee

The amplitude for a quark-gluon interaction can be written	as
\be
\raisebox{-1.3cm}{\includegraphics[width=0.15\textwidth]{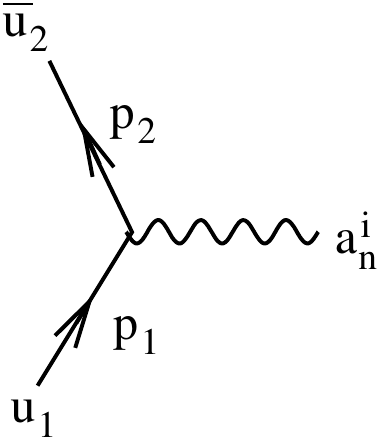}} =
\sqrt{4\pi g^2}\sum_{n,i}
\left(\bar u_2 \gamma_n\,{\lambda^i\over 2}\, u_1\right) a_n^i
\ee
where $a_n^i$ is the combined spin-color polarization of the gluon
and the coupling is rescaled so that the Coulomb interaction is
$g^2/r$ instead of $g^2/(4\pi r)$.  Therefore the gluon-exchange
diagram giving rise to the potential is 
\vskip-0.3cm
\be
\raisebox{-1.3cm}{\includegraphics[width=0.13\textwidth]{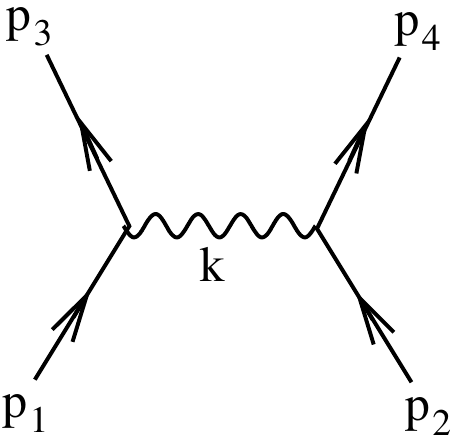}} =
\pi g^2 
\left(\bar u_3 \gamma_n{\lambda^i}u_1\right) \left({\delta_{nm} - 
	{k_m k_n\over\omega^2}\over k^2}\right)\delta_{ij}
\label{eq10-16}
\ee
\vskip-1.5cm
\centerline{$\times \left(\bar u_4
\gamma_m{\lambda^j}u_2\right)$} 
\vskip 1cm
\noindent This can be put into a simpler form using the conservation of the
quark currents, {\it e.g.,}
\bea
	k_\mu(\bar u_3\gamma_\mu\lambda^i u_1) &=& 
	\bar u_3(\slashed{p}_1-\slashed{p}_3)\lambda^i u_1\nn\\
	&=& \bar u_3(m-m)\lambda^i u_1 = 0
\eea
since the gluon vertex conserves flavors.  Therefore $k_n\bar
u_3\gamma_n\lambda^i u_1$ = $\omega\, \bar
u_3\gamma_0\lambda^i u_1$ and we can rewrite the right-hand side of
(\ref{eq10-16}) as\footnote{RPF writes $\delta_{\mu\nu}$ for
the Minkowski metric tensor, even though he is not working in Euclidean space. 
 Moreover he normally does not distinguish 
between covariant
and contravariant Lorentz indices.}
\be
	\pi g^2 
\left(\bar u_3 \gamma_\mu{\lambda^i}u_1\right)
\left({-\delta_{\mu\nu}\over k^2}\right)\delta_{ij}
\left(\bar u_4
\gamma_\nu{\lambda^j}u_2\right)
\ee
in which the gluon propagator takes a Lorentz covariant form.  This
is an illustration of the fact that physical amplitudes are
independent of the choice of gauge.  Nevertheless we must make
{\it some} choice.  Consider the gluon Lagrangian with no choice
of gauge imposed,
\be
	{\cal L} \sim \left[\partial_\mu A_\nu - \partial_\nu A_\mu
	- A_\mu^{\,\times}\! A_\nu\right]^2
\ee
The noninteracting part has the structure $k^2\delta_{\mu\nu} -
k_\mu k_\nu$ in momentum space, which is noninvertible.  Hence
we need to fix the gauge to define the gluon propagator.  

Another
interesting observation is that when varying the full action, the
equation of motion takes the form
\bea
	\square A_\nu - \partial_\nu\partial_\mu A_\mu &=& S_\nu\nn\\
	\implies k^2 A_\nu - k_\nu(k\cdot A) &=& S_\nu(k)\nn\\
	\implies 0 &=& k\cdot S
\eea
where the source $S_\nu$ now includes contributions that are nonlinear
in $A$ for nonvanishing background gauge fields, in addition to the
quark current contribution.  In this case, current conservation is
more complicated than for QED, where the current comes only from the
charged fermions. 

\subsection{Unitarity}

Consider the lowest order processes contributing to gluon 
propagation (postponing for the moment the issue of gluon loops):\footnote{The factor of $i$ associated with the loop will be explained
at the end of the lecture.}
\be
\raisebox{-1.05cm}{\includegraphics[width=0.1\textwidth]{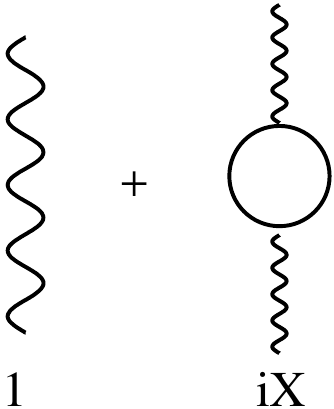}}
 = 1 + iX
\ee	
The probability associated with this amplitude is $1 + i(X-X^*)
+\dots$.  This means that $i(X-X^*)$ is the probability of 
${\it not}$ producing a gluon from a gluon, which is the probability
of instead producing a $q\bar q$ pair:
\be
\includegraphics[width=0.2\textwidth]{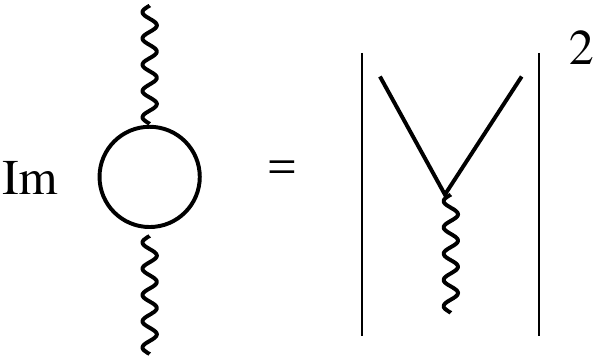}
\ee	
However if we try to do the same thing with gluons in the loop, and
using a covariant gauge for the gluon propagator, the analogous
relation breaks down.  To fix it, we need to add ghost fields,
\be
\includegraphics[width=0.35\textwidth]{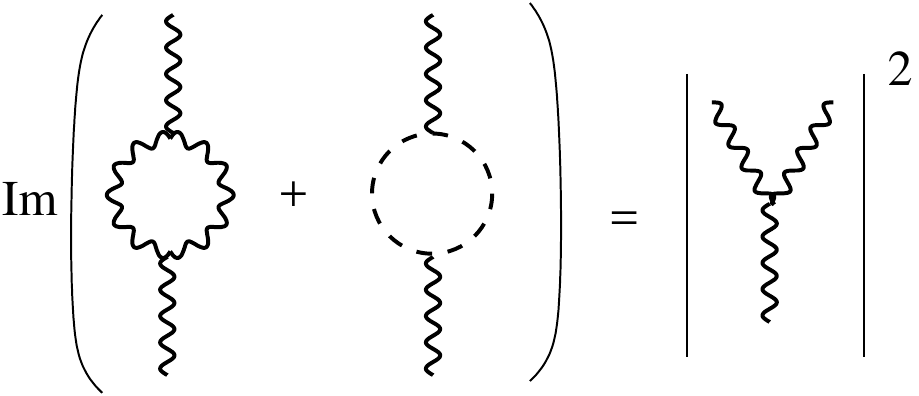}
\label{eq10-23}
\ee
The ghost fields (dashed line) do not appear on the right-hand side of
eq.\ (\ref{eq10-23}) because ghosts never enter into a final state.
Alternatively, we can work in $A_0=0$ gauge instead of covariant gauge,
and dispense with the ghosts.  

Before Faddeev and Popov solved the problem of consistently
incorporating ghosts at any order of perturbation theory, I was 
trying to figure out how to make it work beyond one loop, without
success.  At that time it did not occur to me to use $A_0=0$ gauge, 
because of a prejudice based on experience with QED.  People
originally tried to formulate QED in a nonrelativistic gauge,
$\bbnabla\cdot\mathbb{A}=0$, but nobody understood how to renormalize
this theory in a relativistic way.  

Another way to try to get around the gauge fixing problem is to
temporarily give the gluons a mass.  Then the propagator exists, and
one can try to take $m_g\to 0$ at the end of the calculation.  However
for processes beyond tree level, this limit turns out not to exist.

\subsection{Gluon self-interactions}

For now we will avoid the ghosts by continuing in $A_0=0$ gauge.
The next step is to write the rules for the gluon self-interaction
vertices.  We continue to write combined spin/color polarization
vectors $a,b,c,\dots$ corresponding to a plane wave solution
$A^i_\nu = (a_\nu^i e^{i q_a^\mu x_\mu} +b_\nu^i e^{i q_b^\mu x_\mu}
+c_\nu^i e^{i q_c^\mu x_\mu}$, and taking all $q_i$ to 
point inwards toward the vertex, so that $\sum_i q_i^\mu = 0$.
Recall that the cubic interaction Lagrangian is
\be
	g(\partial_\mu A_\nu - \partial_\nu A_\mu)A_\mu\,\!^\times A_\nu
\ee
By substituting the plane wave solution for the fields, we can read
off the rule for the 3-gluon vertex,
\be
	g(q_{a}^\mu a_\nu - q_{b}^\nu a_\mu)\,^{\bigcdot}(b_\mu\,\!^\times
c_\nu) + \hbox{\ cyclic permutations}
\ee
The permutations can be reorganized into the form
\be
\raisebox{-1.75cm}{\includegraphics[width=0.15\textwidth]{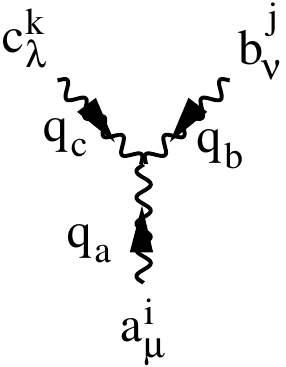}} =
\sqrt{4\pi g^2}\left(
\begin{aligned}
\phantom{+}(q^a-q^c)_\nu \left(b_\nu \,^{\bigcdot}(a_\mu\,\!^\times c_\mu)\right)\\
          +(q^b-q^a)_\nu \left(c_\nu \,^{\bigcdot}(b_\mu\,\!^\times a_\mu)\right)\\
          +(q^c-q^b)_\nu \left(a_\nu \,^{\bigcdot}(c_\mu\,\!^\times b_\mu)\right)
\end{aligned}
\right)
\ee
after rescaling the coupling as before.  Similarly, for the four-gluon amplitude, we
obtain
\be
	\raisebox{-1.4cm}{\includegraphics[width=0.15\textwidth]{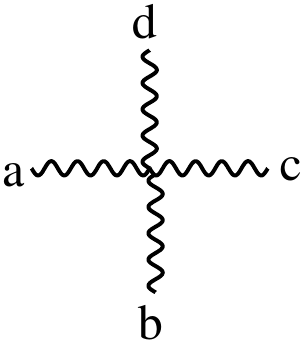}} 
\begin{aligned}	={4\pi g^2\over 4}
(a_\mu\,\!^{\times}b_\nu)\,^{\bigcdot}
	(c_\mu\,\!^{\times}d_\nu)\\ + \hbox{\ symmetric permutations}
\end{aligned}
\ee
For completeness, the gluon propagator is again
\be
	\includegraphics[width=0.1\textwidth]{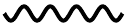} = 
	{\delta_{mn} - k_m k_n/\omega^2\over k^2}\delta_{ij}
\ee
and the rules for quarks are
\be
	\includegraphics[width=0.1\textwidth]{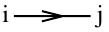} = 
	{\delta_{ij}\over \slashed{p} - m_q}	
\ee
and\footnote{RPF aligns $\vec p_2$ against the flow
of fermion number since the spinor is $\bar u_2$, not $\bar v_2$.  We will see this
again in section \ref{scattsect}.  The point of defining final state momenta
as being negative is to make the Mandelstam variables all look the same, all
involving plus signs rather than minus signs.\label{psigns}} 
\be
	\raisebox{-1.2cm}{\includegraphics[width=0.15\textwidth]{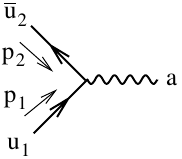}} =
	\sqrt{4\pi g^2}\,\bar u_2(\,\vec a_\mu\,^{\bigcdot}\vec\lambda\,\gamma_\mu) u_1 
\ee

We can rewrite the gluon propagator in a more covariant-looking form by
introducing the 4-vector $\eta_\mu = (1,0,0,0)$.  Taking $q_\mu = (\omega,\vec k)$,
we have $\omega = \eta\cdot q$ and $(0,\vec k) = q_\mu - \eta_\mu(\eta\cdot q)$,
so
\be
	{k_n k_m\over\omega^2} = {\left(q_\mu - \eta_\mu(\eta\cdot q)\right)
	\left(q_\nu - \eta_\nu(\eta\cdot q)\right)\over (\eta\cdot q)^2}
\ee
Moreover $-\delta_{mn} = \delta_{\mu\nu} - \eta_\mu\eta_\nu$.  The gluon propagator
(ignoring color indices) becomes 
\be
	P_{\mu\nu} = {-\delta_{\mu\nu} + {q_\mu\eta_\nu + q_\nu\eta_\mu\over
q\cdot\eta} - {q_\mu q_\nu \eta^2\over (\eta\cdot q)^2}\over q^2 + i\epsilon}
\label{eq10-32}
\ee
which has the property $P_{\mu\nu}\eta_\nu = 0.$  All of the $\eta$-dependence
must drop out of physical amplitudes for them to be Lorentz invariant.  We
will see in section \ref{sect12} that the $q_\mu q_\nu$ term can be eliminated by an
appropriate gauge fixing procedure.

\subsection{Loops}
Thus far all the rules have been unencumbered by any factors of $i$.  We can
consistently push them into the rules for loops, integrating over the internal
momentum of the loop,
\bea
	i\int {d^{\,4}p\over (2\pi)^4} && \hbox{\ for gluon loops}\nn\\
	-i\int {d^{\,4}p\over (2\pi)^4} && \hbox{\ for quark loops}
\eea
These are the only factors of $i$ that one ever needs.\footnote{See note \ref{fn24}}

\footnote{The following material appears in my notes as an interruption of the
new subject that has just been introduced, section \ref{scattsect}, as though
it suddenly occurred to RPF that he had meant to discuss it earlier.  In characteristic
showman fashion, he removed his own belt to do the demonstration.}
The minus sign for fermion loops can be understood as the result of doing a
360$^\circ$ rotation, illustrated by taking the two ends of a belt and exchanging
their positions while keeping the orientations of ends of the belts fixed.
Although each fermion by itself undergoes only a 180$^\circ$ rotation, relative to
each other it is 360$^\circ$, which as we know for fermions introduces a relative sign,
symbolized by the kink in the belt. 
\be
\includegraphics[width=0.35\textwidth]{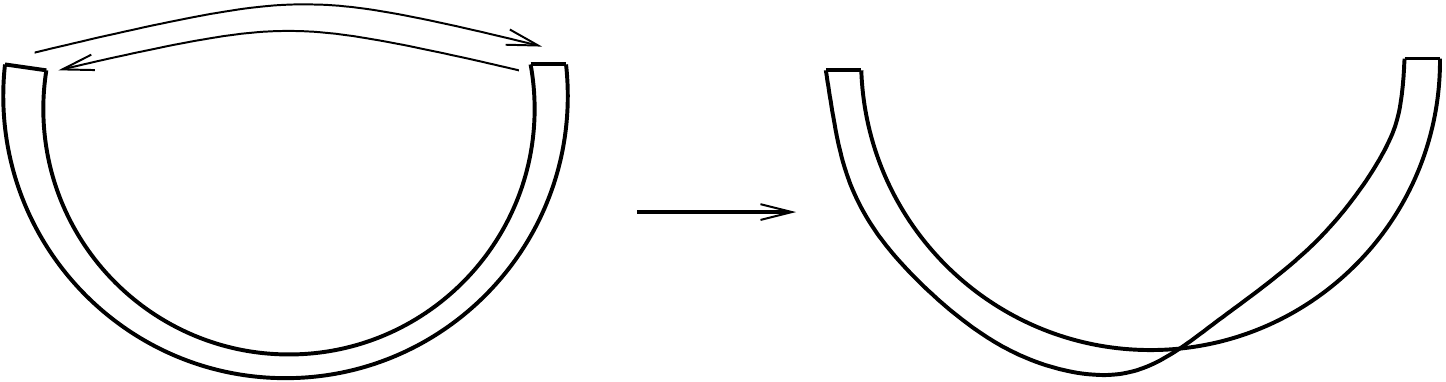}\nn
\ee
Similarly when we exchange the positions of two fermions in diagrams such as
\be
\includegraphics[width=0.2\textwidth]{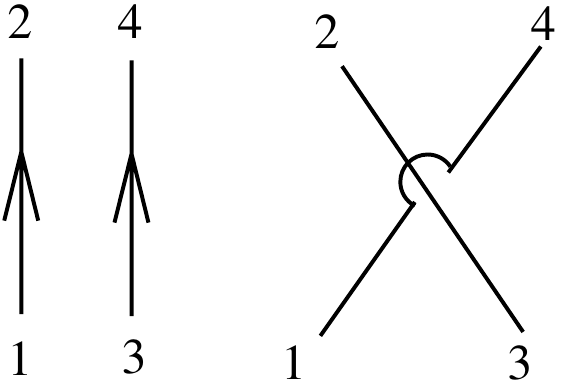}
\raisebox{1cm}{,}\nn
\ee
they differ from each other by a minus sign because of Fermi statistics.  There is
inherently a 360$^\circ$ rotation of their relative orientations.
 A similar exchange occurs when there
is a fermion loop, leading to the minus sign in the diagrammatic rule.

\begin{figure}[t!]
\centerline{
\includegraphics[width=0.4\textwidth]{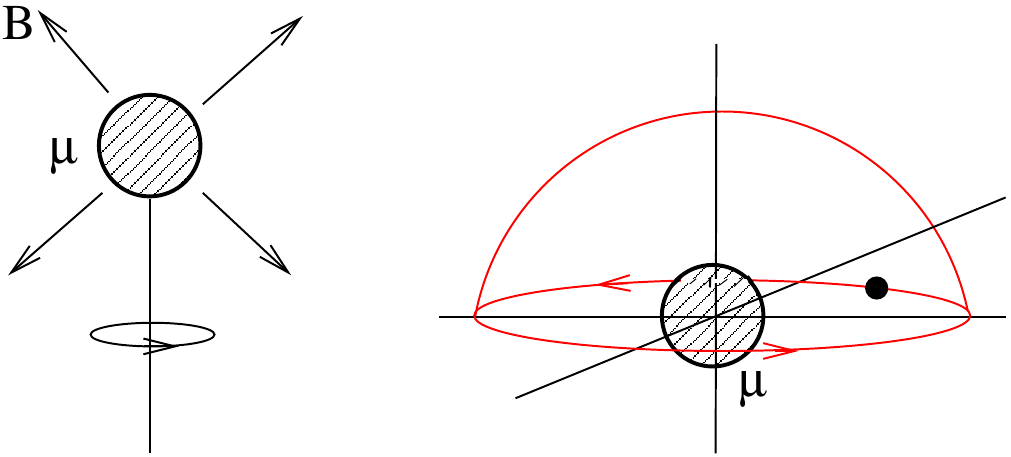}}
\caption{Left: magnetic monopole with a charge transported around its
Dirac string.  Right: the charge is transported around the monopole within the
plane. }
\label{monopole}
\end{figure}

An interesting aside illustrates the origin of this sign for a composite
system that behaves like a fermion.  This is the combination of a magnetic
monopole of strength $\mu$ and an electrically charged scalar, 
with charge $q$, separated by some distance, 
$\vec r$\ \footnote{Characteristically, RPF does not call this by its 
common name, dyon.  Several pages of his personal notes are devoted to this problem. }
\be
\includegraphics[width=0.15\textwidth]{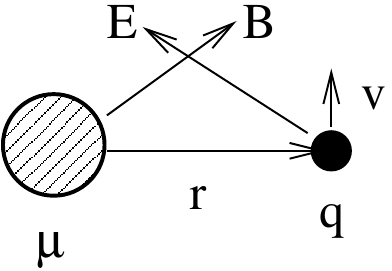}\nn
\ee
Even in the absence of any relative motion between the two constituents, 
this system has angular momentum in the $\vec r$ direction, which can be deduced
by imagining that we try to move the charge with velocity $\vec v$ as shown.
Since $q$ is moving in a magnetic field, it experiences a force that gives a
torque on the system, as if it were a gyroscope, 
showing that it has angular momentum.

Since $\mathbb{B} = \bbnabla\times\mathbb{A}$, $\mathbb{B}$ can never have
a divergence unless there is a Dirac string.  If $\mu$ is quantized such that
$\mu q = \hbar/2$, then the change in phase of an electron as it moves around the string is
$e^{ie\int \vec A\cdot\vec dx}$, which is $-1$ if $\mu$ is quantized properly.
This is easiest to see in the right-hand part of  fig.\ \ref{monopole}  by considering the phase change of a 
charge moving in the plane of the monopole using Green's theorem,
\bea
	e^{iq\int A\cdot dx} &=& e^{iq\Phi}\nn\\
	&=& e^{iq(\sfrac12 4\pi\mu)} = e^{iq 2\pi\mu}\nn\\
	&=& e^{i\pi}
\eea 
where $\Phi$ is the flux of $B$ through the upper hemisphere.  But we have merely 
rotated the system by 360$^\circ$, so this phase change shows that it behaves
like a particle of spin $1/2$. 

\begin{figure}[t]
\centerline{
\includegraphics[width=0.3\textwidth]{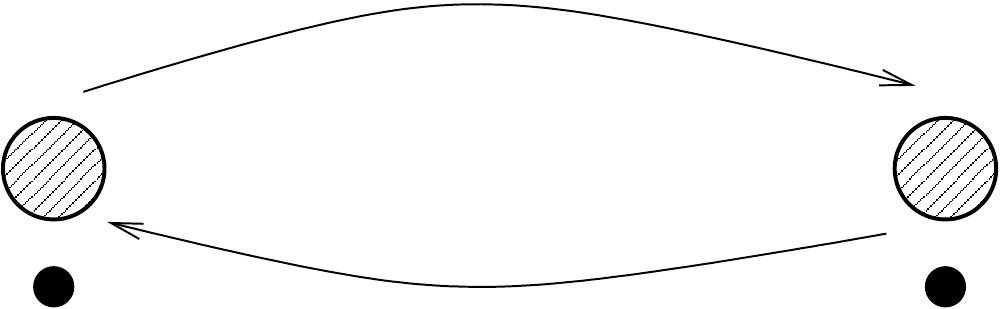}}
\caption{Interchange of two monopole-charge systems.}
\label{2dyon}
\end{figure}

Moreover if we consider two such systems such as in fig.\ \ref{2dyon}, and
interchange them, their combined wave function acquires a phase of $-1$.  It comes
from the combined phase changes of the charges as they move around the opposite
monopole by 180$^\circ$.
	
\section{Scattering processes (11-24-$xx$)}
\label{scattsect}
\begin{figure}[t]
\centerline{
\includegraphics[width=0.25\textwidth]{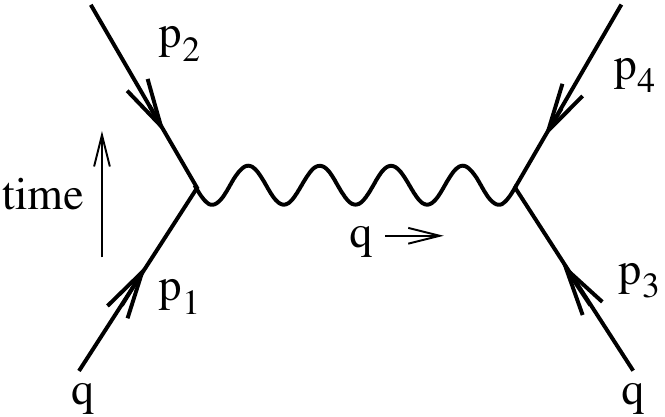}}
\caption{Quark-quark scattering.}
\label{qqscatt4}
\end{figure}

Let us now consider the scattering of two quarks as shown in 
fig.\ \ref{qqscatt4}.\footnote{The unconventional choice of momentum 
labels is deliberate; see note \ref{psigns}.}\ \   This can 
be measured by doing $p$-$p$ scattering, since the parton model allows us to
relate the two processes.  The parton distribution functions are measured by
deep inelastic scattering experiments.

\begin{figure}[t]
\centerline{
\includegraphics[width=0.15\textwidth]{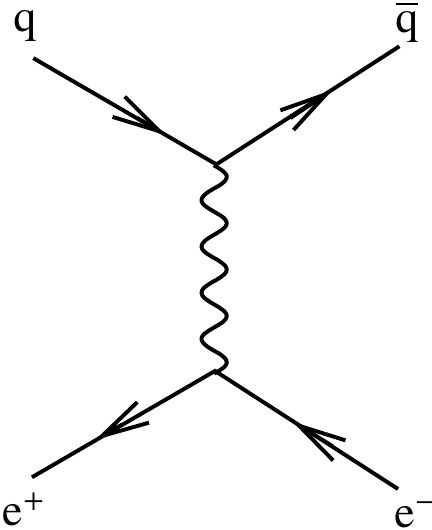}}
\caption{Electromagnetic production of jets.}
\label{epemqq}
\end{figure}

Most of the time the scattering does not produce jets, but these are the
observables we are interested in.  The scattered quarks determine the directions
of the jets.  This part of the problem---how quarks hadronize into jets---can be understood from the observations of
$e^+ e^-\to $ [2 hadron jets] through the electromagnetic process shown in fig.\
\ref{epemqq}.  

Of course it is also possible to get jets originating from gluons produced in the 
QCD scattering process.  This has to be taken into account, but for simplicity
we will start with the quark production process.  Our goal is to measure $\alpha_s$
through scattering.  Historically this analysis helped to design the experiments
observing these processes, and QCD helped to tune the phenomenological models
needed to make the predictions.

Recall the gluon propagator
(\ref{eq10-32}) in $\eta\cdot A=0$ gauge. The terms involving $\eta$ drop out of
the amplitude because of the conservation of the external quark currents.  Then
the amplitude becomes
\be
  T = 	[4\pi]\,g^2\,\left( \bar u_2\gamma_\mu{\lambda^i\over 2} u_1\right)\, 
	{1\over q^2}\,\left(
	\bar u_4\gamma_\mu{\lambda^i\over 2} u_3\right)
\label{eq11-1}
\ee
Conventionally we express it in terms of the Mandelstam variables. Neglect
quark masses, since we want high enough energies to get clean jets.  Then
\bea
	s &=& (p_1+p_3)^2 = 4E^2 \hbox{\ in c.m. frame}\nn\\
	t &=& (p_1+p_2)^2 = -(\hbox{momentum transfer})^2\nn\\
	u &=& (p_1+p_4)^2
\eea
which have the property $s+t+u = \sum_{i=1}^4 m_i^2$ (left as an exercise for the 
reader to prove).  In the center-of-mass frame,
\be
	\includegraphics[width=0.25\textwidth]{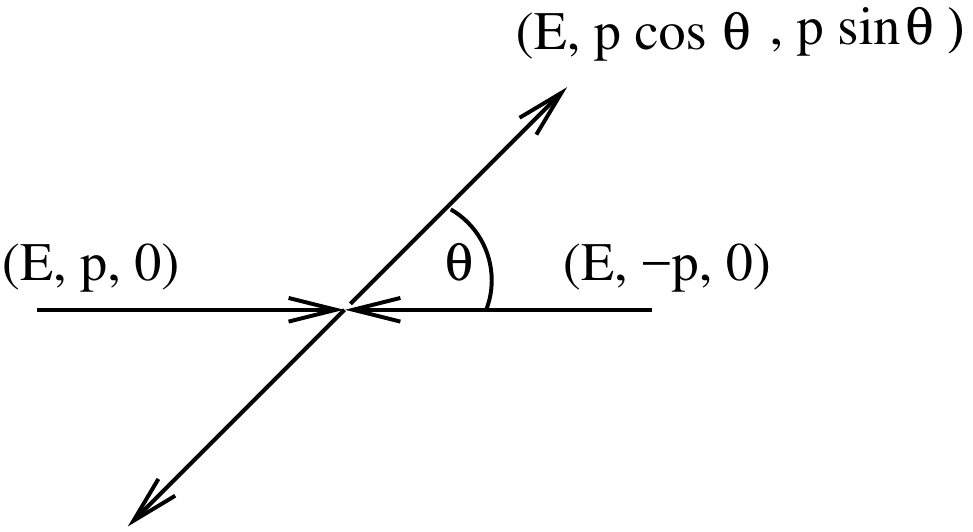}\nn
\ee
the momentum transfer is given by 
\bea
	t &=& \left(0,\, p(1-\cos\theta),\, p\sin\theta\right)^2\nn\\
	&=& -2 p^2(1-\cos\theta) = q^2
\eea

Next we need $|T|^2$, which depends upon the polarizations of the quarks.  If these
are not measured, then we are only interested in 
\be
	\overline{|T|^2} = \sum_{\rm init.\ spins\atop \&\  colors}
\sum_{\rm final\ spins\atop \&\  colors}{|T|^2\over \left({\rm no.\ of\ spins\ \&\
\atop colors\ in\ init.\ state}\right)}
\ee
If we measure the final state polarizations but the incoming beams are unpolarized,
then we should omit the sum over final state spins.

In eq.\ (\ref{eq11-1}) we have implicitly assumed that the spinors $u_i$ are
products of spin and color factors, which we can write as
\be
	u_1 = U_1\, \alpha_1, {\it\ etc.}
\ee
Then
\bea
	|T|^2 &=& \left(4\pi g^2\over q^2\right)^2\left[\bar
U_2\bar\alpha_2\gamma_\mu{\lambda^i\over 2}U_1\alpha_1\right]
\left[\bar
U_4\bar\alpha_4\gamma_\mu{\lambda^i\over 2}U_3\alpha_3\right]\nn\\
&\times& \left(
\left[\bar
U_2\bar\alpha_2\gamma_\nu{\lambda^j\over 2}U_1\alpha_1\right]
\left[\bar
U_4\bar\alpha_4\gamma_\nu{\lambda^j\over 2}U_3\alpha_3\right]
\right)^*\nn\\
&\to& \sum_{\rm spins}\left(4\pi g^2\over q^2\right)^2\\
&\times&
\left(\bar U_2\gamma_\mu U_1\right)\left(\bar U_4\gamma_\mu U_3\right)
\Big(\left(\bar U_2\gamma_\nu U_1\right)\left(\bar U_4\gamma_\nu
U_3\right)\Big)^*\nn\\
&\times&\left(\bar\alpha_2{\lambda^i\over 2}\alpha_1\right)
\left(\bar\alpha_4{\lambda^i\over 2}\alpha_3\right)
\left(\bar\alpha_2{\lambda^j\over 2}\alpha_1\right)^*
\left(\bar\alpha_4{\lambda^j\over 2}\alpha_3\right)^*\nn
\eea
The sum on spins gives, for example,
\be
	u_2\bar u_2 = \slashed{p}_2 + m
\ee
leading to the traces
\be
	\tr\Big(\gamma_\nu(\slashed{p}_2 + m)\gamma_\mu(\slashed{p}_1+m)\Big)
\tr\Big((\slashed{p}_4 + m)\gamma_\mu(\slashed{p}_3+m)\gamma_\nu\Big)\nn
\ee
Similarly, the sum on colors gives
\be
	\sum_{\rm colors}\left(\bar\alpha_2{\lambda^i\over 2}\alpha_1 \bar\alpha_1
	{\lambda^j\over 2}\alpha_2\right) = \tr{\lambda^i\over 2}{\lambda^j\over 2}
	= \frac12 \delta_{ij}
\ee
Recall that
\bea
	\tr\,\gamma_\nu\slashed{p}_2\gamma_\mu\slashed{p}_1 &=& 
4\Big(p_{2\nu} p_{1\mu} + p_{2\mu} p_{1\nu} - (p_1\!\cdot\! p_2)\delta_{\mu\nu}\Big)\nn\\
	\tr\,\slashed{p}_4\gamma_\mu\slashed{p}_3\gamma_\nu &=& 
4\Big(p_{4\nu} p_{3\mu} + p_{4\mu} p_{3\nu} - (p_3\!\cdot\! p_4)\delta_{\mu\nu}\Big)\nn
\eea
neglecting masses.  The total color factor is
\be
	{\delta_{ij}\delta_{ij}\over 4} = \frac84 = {N^2-1\over 4}
\ee
where we indicated the more general result for SU(N) at the end, and we have not
yet included the $1/N^2$ from averaging over the quark colors.  Putting everything
together,
\bea
	\overline{|T|^2} &=& \left((4\pi g^2)^2\over 2^2 N^2
q^4\right)16\!\cdot\!\sfrac14 (N^2-1)\nn\\
	&\times& \left[2(p_2\!\cdot\! p_4)(p_1\!\cdot\! p_3) + 2(p_1\!\cdot\!
p_4)(p_1\!\cdot\! p_3)\right]\nn\\
&=& {N^2-1\over 4 N^2}\left(16\over 4\right){(4\pi g^2)^2\over t^2}\sfrac12(s^2+u^2)
\eea
and the differential cross section is\footnote{RPF had written 
$|T|^2/(8\pi^2 s^2)$; I have restored the missing factor of $1/2$. For same-flavor quarks,
there should be yet another factor of $1/2$.}
\be
	{d\sigma\over dt} = {\overline{|T|^2}\over 16\pi^2 s^2}
\ee
where $dt = 2E^2\, d\cos\theta$.

Similar results can be found for quark-gluon ($QG$) and gluon-gluon ($GG$) scattering.
At low momentum transfer, and apart from an overall proportionality constant,
their relative squared matrix elements go as
\bea	
	QQ\to QQ &=& \frac89\, \frac{s^2}{t^2}\nn\\
	QG\to QG &=& 2\, \frac{s^2}{t^2}\nn\\
	GG\to GG &=& \frac92\, \frac{s^2}{t^2}
\eea
We see that the formulas simplify at small $t$.

\section{Gauge fixing the path integral$^*$ (12-1-$xx$)}
\label{sect12}

We now have a complete set of rules for calculating amplitudes perturbatively
in the gauge $A_t=0$.  (It is not perfectly complete because we have not yet
specified how to deal with the infinities arising from loop diagrams; this will be
the subject of some later chapters.)  Ordinarily I would not complicate matters by
introducing an additional formalism that gives the same answers in the end, but Faddeev
and Popov (Phys.\ Lett.\ B25 (1967) pp.\ 29-30) have invented another way of fixing the gauge that is so elegant and
useful that it deserves mention.  It allows one to evaluate the path integral with
an arbitrary gauge condition, of which $A_t=0$ is just a special case.

Before deriving the method, it will be useful to know a technique that allows the
path integral in $A_t=0$ gauge,
\be
	\int e^{iS[\mathbb{A},0]}\, {\cal D}^{\,3}\mathbb{A}(\mathbb{x},t)\,,
\label{eq12-1}
\ee
to be rewritten as an integral over all four of the $A_\mu$, with an extra
$A_t$-dependent term in the Lagrangian.  Notice that (\ref{eq12-1}) is the same as
\be
	\int e^{iS[\mathbb{A},\phi]}\, {\cal D}^{\,3}\mathbb{A}\,,
\label{eq12-2}
\ee
for any function $\phi$.  One way to see this is to gauge transform $\mathbb{A}$
in (\ref{eq12-1}) to
\bea
	\mathbb{A}' &=& \Lambda^\dagger \mathbb{A} \Lambda +
	\Lambda^\dagger\bbnabla\Lambda\nn\\
	A'_t &=& \phi = \Lambda^\dagger\dot\Lambda\,,
\label{eq12-3}
\eea
where $\Lambda$, which is determined by solving $\dot\Lambda = \phi\Lambda$, 
does not depend upon $\mathbb{A}$, hence ${\cal D}^{\,3}\mathbb{A}$ is invariant.
Eq.\ (\ref{eq12-2}) can also be written as
\be
		\int e^{iS[\mathbb{A},A_t]}\, {\cal D}^{\,3}\mathbb{A}
		\,\delta[A_t-\phi]
\label{eq12-4}	
\ee
Since it does not depend on $\phi$, we can functionally integrate over $\phi$ with
some weight, say
\be
	\exp\left({i\mu^2\over 2 g^2}\int d^{\,4}x\, \phi^2\right)
\label{eq12-5}
\ee	
and change the path integral by only an overall multiplicative factor.  This factor
has no effect on an expectation value of a gauge-invariant functional,
\be
	\langle F\rangle = {\int {\cal D}\phi\, e^{ic\!\int\!\phi^2}
	\!\!\!\int e^{iS[\mathbb{A},A_t]}\, F(\mathbb{A},A_t)\, 
	{\cal D}^{\,3}\mathbb{A}\,
	{\cal D}A_t\, \delta[A_t-\phi]\over
	\int {\cal D}\phi\, e^{ic\!\int\!\phi^2}
	\!\!\!\int e^{iS[\mathbb{A},A_t]}\,
	{\cal D}^{\,3}\mathbb{A}\,
	{\cal D}A_t\, \delta[A_t-\phi]
}
\label{eq12-6}
\ee
since it cancels between numerator and denominator.  Now the $\phi$ integral is
trivial because of the delta functional, and the path integral is
\be
	\int e^{iS[\mathbb{A},A_t] + {i\mu^2\over 2 g^2}\int\! A_t^2\, d^{\,4}x}\,
	{\cal D}^{\,4}\!A
\label{eq12-7}
\ee
Because of the new term in the action, the gluon propagator now exists, even though
$A_t$ is no longer fixed to be zero.

{\bf Exercise.}  Show that the propagator for (\ref{eq12-7}) is
\bea
	P_{\mu\nu}(k) &=& {1\over k^2}\left(-\delta_{\mu\nu} + {k_\mu\eta_\nu\over
	k\cdot\eta} + {k_\nu\eta_\mu\over k\cdot\eta}
	- {k_\mu k_\nu\eta^2\over (k\cdot\eta)^2}\right) \nn\\
	&-&{  k_\mu k_\nu\eta^2\over \mu^2(k\cdot\eta)^2}
\label{eq12-8}
\eea
where $\eta_\mu = (1,0,0,0)$.

It would be nice to simplify the propagator by choosing $\mu^2=-k^2$, which is 
impossible because $\mu^2$ is just a constant, not a Fourier transform variable.
But the same thing can be accomplished by using
\be
	e^{i{\mu^2\over 2g^2}\int d^{\,4}x\,(\partial_\mu\phi)^2}
\label{eq12-9}
\ee
instead of (\ref{eq12-5}) as the weight factor.  Then we get (\ref{eq12-8}), 
but with $\mu^2\to -k^2\mu^2$, and $\mu^2$ can  be chosen so that the last two
terms in (\ref{eq12-8}) cancel.  This is the justification for saying, in a
previous lecture, that the $k_\mu k_\nu$ term in the propagator was irrelevant. 

\begin{figure}[t]
\centerline{
\includegraphics[width=0.4\textwidth]{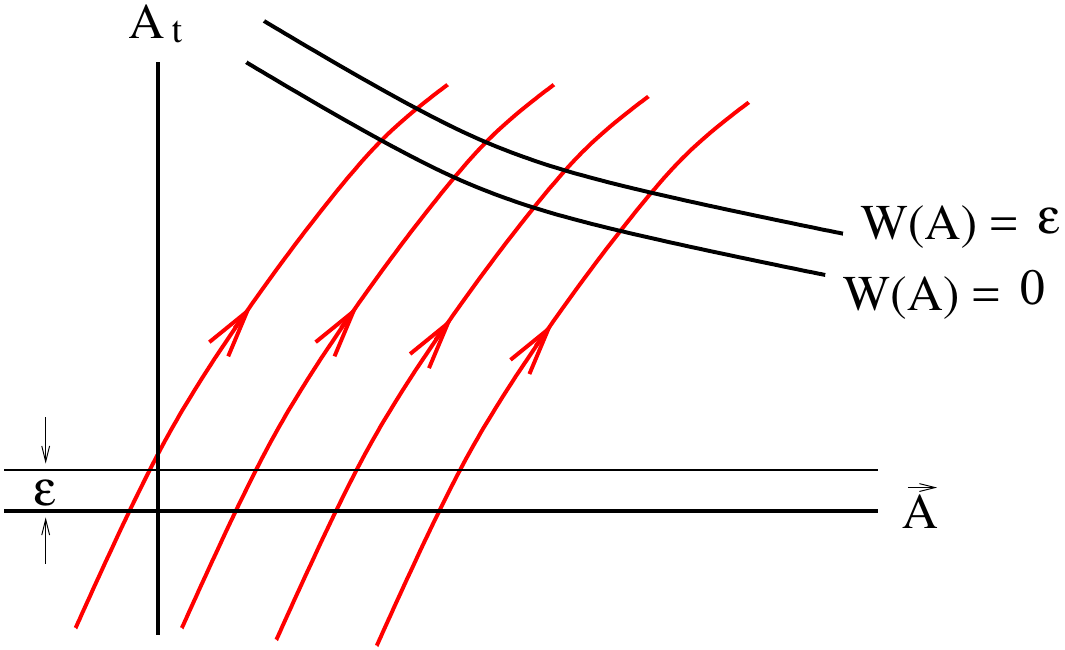}}
\caption{Gauge orbits sliced by a gauge condition $W$.}
\label{orbit2}
\end{figure}

We are aiming for an expression similar to (\ref{eq12-7}), but for some arbitrary
gauge condition, not necessarily $A_t = 0$.  For this the Faddeev-Popov procedure
will be needed, which since I didn't invent it myself, I claim is extremely subtle!
Suppose we wanted the gauge condition to be $\partial_\mu A_\mu = 0$.  Then the
path integral must look something like
\be
	\int e^{iS[A]}\, \delta[\partial_\mu A_\mu]\, {\cal D}^{\,4}\!A\,.
\label{eq12-10}
\ee
But this is not quite right, even though it would be right for the $A_t=0$ case.
Consider the space of all gauge field configurations, represented schematically
by plotting $A_t$ along one axis and $\mathbb{A}$ along the other.  Let $W(A)=0$
be the desired gauge condition, represented by a surface in the function space,
that cuts across the trajectories of gauge-equivalent $A_\mu(x)$'s, called
``gauge orbits'' (see fig.\ \ref{orbit2}).\ \ Previously we integrated over the
line $A_t=0$.  The delta functional $\delta[A_t]$ can be interpreted as the limit 
of a less singular constraint, which is to integrate over the strip between $A_t=0$
and $A_t=\epsilon$, divide by $\epsilon$, and take $\epsilon\to 0$.  However if we
try to do the same thing for the surface $W(A)=0$, the gauge orbits will not
necessarily cross the strip at the same angle everywhere, and the simple constraint
$\delta[W(A)]$ will weight some orbits too much, some too little, as one moves
along the surface.  An extra factor is needed to compensate for the varying length
of the orbits crossing the strip.

Call this compensating factor $\Delta(A)$.  It will be shown that
\be
	\Delta(A)^{-1} = \int{\cal D}g\,\delta[W(A^g)]
\label{eq12-11}
\ee
where
\be
	A^g_\mu = \Lambda^\dagger(g)\,A_\mu\,\Lambda(g) + \Lambda^\dagger(g)\,
	i\partial_\mu\Lambda(g)\,,
\label{eq12-12}
\ee
is the gauge-transformed $A_\mu$
and $\Lambda(g)$ is the matrix representation of the abstract group element
$g$.  ${\cal D}g$ stands for the invariant group measure at each point in
spacetime.  It has the property
\bea
	\int {\cal D}g\, \delta[W(A^g)] &=& \int {\cal D}g\, \delta[W(A^{hg})]\nn\\
	&=& \Delta(A^h)^{-1}
\label{eq12-13}
\eea
for any $h$ in SU(3), so we see that $\Delta(A)$ is gauge invariant.

To define the path integral, insert a factor of $1 = \int\Delta(A)\int{\cal D}g\,
\delta[W(A^g)]$ into $\int e^{iS[A]}\,{\cal D}^{\,4}\!A$.  The expectation value of 
$F[A]$ is then
\bea
	\langle F\rangle &=& {\int e^{iS[A]}\,F\,{\cal D}^{\,4}\!A\over
	\int e^{iS[A]}\,{\cal D}^{\,4}\!A}\\
	&=& {\int e^{iS[A]}\,F\,\Delta(A)\,\delta[W(A^g)]\,{\cal D}^{\,4}\!A\,{\cal D}g\over
	\int e^{iS[A]}\,\Delta(A)\,\delta[W(A^g)]\,{\cal D}^{\,4}\!A\,{\cal D}g}\,.\nn
\label{eq12-14}
\eea
The nice thing about these integrals is that they don't depend on $g$ (assuming, as
usual, that $F$ is gauge invariant).  Make the change of variables $A\to
A^{g^{-1}}$.  Because each factor in the integrals is gauge invariant except for
the delta functional, the $g$-dependence disappears, and the (infinite) integrals
$\int{\cal D}g$ cancel between numerator and denominator.  We are therefore left
with the path integral
\be
	Z=\int e^{iS[A]}\,\Delta(A)\, \delta[W(A)]\, {\cal D}^{\,4}\!A\,,
\label{eq12-15}
\ee
as claimed.

Eq.\ (\ref{eq12-15}) is the desired generalization of (\ref{eq12-1}) for $A_t=0$
gauge, but it is not in a very useful form for explicit computations.  $\Delta(A)$
is some horribly complicated functional which in general nobody knows how to
compute.  Fortunately, it is not necessary to know $\Delta(A)$ for all values of
$A$, but only where $W(A) = 0$, and there it {\it can} be determined.  Take
$W(A)=\partial_\mu A_\mu$, for example.  We must evaluate  
\be
	\int \delta[\partial_\mu A_\mu^g]\,{\cal D}g\Big|_{\partial_\mu A_\mu =
0}\, .
\label{eq12-16}
\ee
We first assume that there is a unique solution to $\partial_\mu A_\mu^g=0$,
such that $g$ is the identity when $\partial_\mu A_\mu$ is already zero.  Therefore
we can focus on infinitesimal gauge transformations, 
\be
	\vec A^g_\mu = \vec A_\mu + D_\mu\vec\alpha = \vec A_\mu + 
	(\partial_\mu - A_\mu^\times)\vec\alpha\,.
\ee
Then (\ref{eq12-16}) becomes
\be
	\int\delta[\partial_\mu A_\mu + \partial_\mu D_\mu\alpha]\,{\cal D}\alpha\Big|_{\partial_\mu A_\mu =
0}\,.
\label{eq12-18}
\ee
Recall that for finite-dimensional integrals,
\be
	\int d^{\,n}x\, \delta^{(n)}(M_{ab}x^b) = {1\over|\det M|}\,.\nn
\ee
In the present case, eq.\ (\ref{eq12-18}), we get a functional determinant,
\be
	\Det^{-1}(\partial_\mu D_\mu)\,,\nn
\ee
which depends on $A_\mu$ through $D_\mu$.  Therefore the path integral
(\ref{eq12-5}) is
\be
	\int e^{iS[A]}\,\delta[\partial_\mu A_\mu]\,\Det(\partial_\mu D_\mu)\,
	{\cal D}A
\label{eq12-19}
\ee
in the gauge $\partial_\mu A_\mu = 0$.  

The next step is to reexpress the determinant so that it looks like a new term in
the action.  If instead of $\Det(\partial_\mu D_\mu)$ we had 
$\Det^{-1/2}(\partial_\mu D_\mu)$, we could use the functional generalization of
the formula 
\be
	\int d^{\,n}x\, e^{-\frac12 x^a M_{ab} x^b} = {(2\pi)^{n/2}\over\sqrt{\det
M}}\,.\nn 
\ee
However, there is an analogous formula for anticommuting variables that does what
is needed,
\be
	\int{\cal D}P\,{\cal D}\bar P\, e^{\bar P M P} = \det M\, .
\label{eq12-20}
\ee
Here $P$ is an anticommuting function, $\{P(x),P(x')\} = 0$, $\bar P$ is its
complex conjugate, and $M$ is a differential operator.  (The reader who is
unfamiliar with this type of integral should work through the following exercise.)

{\bf Exercise.}  Complex anticommuting variables are defined to satisfy $\theta^2 =
\bar\theta^2 = \{\theta,\bar\theta\}=0$.  The complete table of integrals for such
variables is, by definition, 
\be
	\int d\theta = \int d\theta \, \bar\theta = 0;\quad \int d\theta\,\theta = 1
\nn
\ee
(and similarly for the complex conjugates).  If there are $2N$ variables
$\theta_i$, $\bar\theta_i$, $i=1,\dots,N$, then they all anticommute with each
other.  By the above rule, the only nonvanishing integrals over all the $\theta_i$,
$\bar\theta_i$ are
\be
	\int d^{\,n}\theta\, d^{\,n}\bar\theta\, \left[ \bar\theta_1\theta_1
	\dots \bar\theta_N\theta_N\right] = 1\,,\nn
\ee
and those integrals that differ from it by a permutation of the variables in the
integrand.  (If the permutation is odd, the integral will be $-1$ instead of $+1$.)
Convince yourself that 
\be
	\int d^{\,n}\theta\, d^{\,n}\bar\theta\, e^{\bar\theta_a M_{ab}\theta_b}
	= \det M\,.\nn
\ee

Applying this technique to the path integral gives
\be
	\int e^{iS[A]}\, \delta[\partial_\mu A_\mu]\,
	e^{i\int \bar P(x)\partial_\mu D_\mu P(x)\, d^{\,4}x}
	\,{\cal D}P\,{\cal D}\bar P\,, 
\label{eq12-21}
\ee
where the $P$'s transform in the octet representation of SU(3), since that is the
representation on which $D_\mu$ acted in (\ref{eq12-18}).

The new fields $P$, $\bar P$ are called Faddeev-Popov {\it ghosts} to underscore
the fact that they are not physical fields like quarks or gluons, but only a
mathematical convenience.  Because they are anticommuting, they behave like
fermions in the sense that ghost loops contribute a factor of $-1$ in a Finemensch
diagram; but they are not fermions, for their spin is zero.  The ghost action can
be put into a more conventional form by integrating by parts,
\be
	\partial_\mu \bar P \,D_\mu P = \partial_\mu \bar P\, \partial_\mu P
	- \partial_\mu \bar P\cdot (A_\mu\,^{\!\times} P)\,.
\label{eq12-22}
\ee

The Finemensch rules for ghosts are seen to be 
\bea
	-1 \hbox{\ for loops;} &&\nn\\
	\hbox{propagator:}\ \  \includegraphics[width=0.1\textwidth]{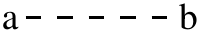}
	&=& {1\over k^2}\,\delta_{ab}\,;\nn\\
\hbox{coupling}:\ \
\raisebox{-0.75cm}{\includegraphics[width=0.1\textwidth]{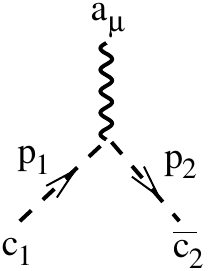}} &=&
	p_2^\mu\, \vec{\bar c}_2^{\ \bigcdot}(\vec a_\mu\,\!\!^\times \vec c_1) \nn
\eea
where $\vec c_1$ and $\vec{\bar c}_2$ are octet color vectors, just like
$\vec a_0,\dots,\vec a_3$.  In addition, the gauge field propagator is simply
\be
	\includegraphics[width=0.15\textwidth]{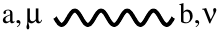} = {\delta_{\mu\nu}\,
	\delta_{ab}\over k^2} \nn
\ee
in this gauge, since
\be
	\sfrac14\int d^{\,4}x\,\left(\partial_\mu A_\nu - \partial_\nu
A_\mu\right)^2 = \sfrac12\int d^{\,4}x\,\left( (\partial_\mu A_\nu)^2 - (\partial_\mu
A_\mu)^2\right)\nn
\ee
after integrating by parts, and $\partial_\mu A_\mu = 0$.

Now we are almost done, but the expression (\ref{eq12-21}) is still not easy to
use since we don't know how to do Gaussian integrals with a constraint.  This is
where the trick introduced at the beginning of the chapter comes in.  If instead of
$\partial_\mu A_\mu = 0$ one used $\partial_\mu A_\mu = f$, $\Delta(A)$ would be
the same as before, and (\ref{eq12-21}) would be
\be
	\int e^{iS[A]}\, \Delta(A)\, \delta[\partial_\mu A_\mu -f]\,
	{\cal D}^{\,4}\!A\, .
\label{eq12-23}
\ee
This expression does not really depend on $f$ because it was obtained from the same
starting point, $\int e^{iS[A]}\,{\cal D}^{\,4}\!A$, for any $f$.  So again it is
permissible to integrate over $f$ with a weight factor 
$\exp(i\int f^2\, d^{\,4}x)$.
The final result is
\be
	\int e^{iS[A] + {i\over 2 g^2}\int(\partial_\mu A_\mu)^2\,d^{\,4}x
	+ i\int \partial_\mu \bar P D_\mu P\, d^{\,4}x}\,
	{\cal D}^{\,4}\!A\,{\cal D}\bar P\, {\cal D}P\, .
\label{eq12-24}
\ee
The second term in the action cancels the similar term in $S[A]$ so that the
gluon propagator is still $\delta_{\mu\nu}/k^2$, but there is no longer any 
restriction on $A_\mu$ in the integral.

\footnote{The following paragraph was added to my revision of the lecture by RPF.}  In deriving this we assumed that for a given vector field $A_\mu$
the gauge transformation $g$ needed to arrange that the divergence of the new
field is zero, $\partial_\mu A_\mu^g=0$, is unique.  This was found to be false
by Gribov.  Thus $\partial_\mu A_\mu=0$ does not completely specify the gauge.
Thus Faddeev's argument looks imperfect---first there are several places in $g$
where there are contributions to (\ref{eq12-14}).  In addition our ghost gives
$\Det(\partial_\mu D_\mu)$ but our analysis from (\ref{eq12-18}) wants the 
absolute value $|\Det(\partial_\mu D_\mu)|$.  There is much confusion, but I think
(from some studies I made some time ago) the final integral is 
really correct.\footnote{In my course notes, I have some elaboration of this point:
``RPF conjectures that the sign of the $\Det$ changes only if the 
gauge condition does not uniquely determine $A_\mu$.  In this case,
he believes that integrating over the different solutions gives a
compensating error that makes up for using $\Det$ instead of
$|\Det|$.''}\ \ 
At any rate no error would be expected in perturbation theory because for
configurations with small $A_\mu$ the gauge that makes $\partial_\mu A_\mu=0$
is unique.  Gribov's ambiguity appears only for sufficiently large $A$.  We shall
see examples of it later.

{\bf Exercise.}  Rewrite (\ref{eq12-24}) for a general gauge condition, $W(A)$.

\section{Quark confinement$^*$ (12-3-$xx$)}
\label{sect13}
The utility of being able to quantize QCD in a variety of different gauges  is that
some gauges are particularly convenient for certain applications.  In
electrodynamics, the Coulomb gauge has the virtue that fields satisfying the gauge
condition
\be
	\bbnabla\cdot\mathbb{A} = 0
\label{eq13-1}
\ee
represent truly physical, transverse degrees of freedom.  It also has some
peculiarities. The Hamiltonian has a nonlocal, instantaneous interaction (action
at a distance), which must combine with the interactions of the transverse photons
so that the net force propagates at the speed of light.  

Interactions that look
instantaneous are well suited to Schr\"odinger's equation, which requires the
potential between particles at equal times.  It would be quite awkward to
explicitly describe finite-velocity forces in the Schr\"odinger equation because
the potential for one particle at a time $t$ would depend on the positions of the
others at the retarded times, and one would need the past histories of all the
particles to propagate the system forward in time.
In what follows we will derive the Coulomb gauge path integral for QCD in the
Hamiltonian form (which is closely related to the Schr\"odinger picture) and see
some indications of quark confinement.

Write the gauge field as
\be
	A^i_\mu = (\phi^i,\, \mathbb{A}^i)\,.
\label{eq13-2}
\ee
Then in the gauge (\ref{eq13-1}), the vacuum amplitude is
\be
	X = \int
e^{iS[A]}\,\Det(\bbnabla\cdot\mathbb{D})\,\delta[\bbnabla\cdot\mathbb{A}]\,
	{\cal D}^{\,3}\!\mathbb{A}\,{\cal D}\phi\,,
\label{eq13-3}
\ee
where $\mathbb{D} = (\bbnabla - \mathbb{A}^\times)$ is the covariant derivative in
the adjoint representation.  The constraint in (\ref{eq13-3}) implies that if
$\mathbb{A}$ is expanded in plane waves, $\mathbb{C}(\mathbb{k})
e^{i\mathbb{k}\cdot\mathbb{x}}$, then $\mathbb{k}\cdot\mathbb{C} = 0$. The color
electric and magnetic fields are (omitting the gauge index for 
brevity)\footnote{See note [\ref{Bcorr}].}
\bea
	\mathbb{E} &=& (-\bbnabla\phi + \dot{\mathbb{A}} + \mathbb{A}\,^\times\phi)
	= \dot{\mathbb{A}} - \mathbb{D}\phi\\
\label{eq13-4}
	\mathbb{B} &=& \bbnabla_\times\mathbb{A} -
\sfrac12\,\mathbb{A}^\times_\times\mathbb{A}
\label{eq13-5}
\eea
and the action is 
\be
	S[A] = {1\over 2 g^2}\int(\mathbb{E}^2 - \mathbb{B}^2)\,d^{\,4}x + 
	\int(\rho\phi - \mathbb{J}\cdot\mathbb{A})\,d^{\,4}x 
\label{eq13-6}
\ee
in the presence of an external source $J^\mu = (\rho,\mathbb{J})$.  Notice
that $\exp({i\over 2 g^2}\int\mathbb{E}^2)$ can be rewritten as
\be
	e^{{i\over 2 g^2}\int\mathbb{E}^2\, d^{\,4}x} = c\int e^{-{i g^2\over 2}
	\int\bbPi^2\, d^{\,4}x + i\int \bbPi\cdot\mathbb{E}\, d^{\,4}x
}\,{\cal D}^3\bbPi
\label{eq13-7}
\ee
by completing the square.  Then the amplitude becomes
\bea
	X &=& \int e^{-{i g^2\over 2}
	\int\bbPi^2\, d^{\,4}x + i\int \bbPi\cdot(\dot{\mathbb{A}}-\mathbb{D}\phi)
	\, d^{\,4}x + {i\over 2 g^2}\int{\mathbb{B}^2\, d^{\,4}x }}\,\nn\\
&\times& e^{i\int(\rho\phi-\mathbb{J}\cdot\mathbb{A})\,d^{\,4}x }
\,\Det(\bbnabla\cdot\mathbb{D})\, {\cal D}^{\,3}\mathbb{A}_T\,{\cal D}\phi\,
{\cal D}^3\bbPi\nn\\
\label{eq13-8}
\eea
where $\delta[\bbnabla\cdot\mathbb{A}]$ has been eliminated by integrating only
over the transverse part of $\mathbb{A}$, denoted by $\mathbb{A}_T$, and the
integrand is evaluated at $\bbnabla\cdot\mathbb{A}=0$.

Now any vector field can be split into a longitudinal and a transverse part,
\bea
	\bbPi &=& \bbPi_{\rm transverse} + \bbPi_{\rm longitudinal}\nn\\
	&\equiv& \mathbb{P} + \bbnabla f
\label{eq13-9}
\eea
where $\bbnabla\cdot\mathbb{P} = 0$, by definition.  Then 
\bea X &=& \int\exp\Bigg[-{ig^2\over 2}\int\left(\mathbb{P}\!\cdot\!\mathbb{P} +
\underline{2\mathbb{P}\!\cdot\!\bbnabla f} + (\bbnabla f)^2\right)\,d^{\,4}x \nn\\
&+& i\int\mathbb{P}\!\cdot\!\left(\dot{\mathbb{A}}\!-\!\underline{\bbnabla\phi} + \mathbb{A}^\times\phi
\right)\,d^{\,4}x + i\int(\rho\phi\!-\!\mathbb{J}\!\cdot\!\mathbb{A})\,d^{\,4}x\nn\\
&+& i\int\bbnabla f\!\cdot\!(\underline{\dot{\mathbb{A}}}-\mathbb{D}\phi)\,d^{\,4}x
+ {i\over 2 g^2}\int \mathbb{B}\!\cdot\!\mathbb{B}\, \,d^{\,4}x\Bigg]\nn\\
&\times& \Det(\bbnabla\!\cdot\!\mathbb{D})\, {\cal D}\mathbb{A}_T\,{\cal D}\phi\,
{\cal D}\mathbb{P}\,{\cal D}f\, .
\label{eq13-10}
\eea
The underlined terms can be eliminated by integrating by parts and using the
fact that $\bbnabla\cdot\mathbb{P} = \bbnabla\cdot\mathbb{A} = 0$.  An overall
constant, $\Det(\bbnabla)$, has been omitted from the functional measure
${\cal D}f$.  Grouping the remaining $\phi$-dependent factors together and
integrating over $\phi$ gives a delta functional,
\be
	\delta[\mathbb{P}\,^\times\mathbb{A} + \mathbb{D}\cdot\bbnabla f + \rho]
\label{eq13-11}
\ee
which says that $f$ must satisfy
\be
	f = -(\mathbb{D}\cdot\bbnabla)^{-1}(\rho +
\mathbb{P}\,^\times\mathbb{A})\,.
\label{eq13-12}
\ee
This is reminiscent of the analogous equation in electrodynamics,
\be
	\tilde f = -{1\over\nabla^2}\rho\,,\nn
\ee
whose solution is
\be
	\tilde f = {1\over 4\pi}\int {\rho(\mathbb{R}')\over
	|\mathbb{R}-\mathbb{R}'|} \, d^{\,3}\mathbb{R}'\,,
\label{eq13-13}
\ee
but the complexity of the QCD version (\ref{eq13-12}) prevents us from obtaining
such a nice closed-form solution for $f$.

There is a fortunate simplification from the constraint (\ref{eq13-11}) however.
The delta functional produces a factor of 
\be
	\Det^{-1}(\mathbb{D}\cdot\bbnabla)
\label{eq13-14}
\ee
when the integral over $f$ is performed (recall that $\delta(ax) = \delta(x)/a$).  If it was $\Det^{-1}(\bbnabla\cdot\mathbb{D})$ instead, it would
cancel the Faddeev-Popov determinant in (\ref{eq13-10}).  However,
${\bbnabla\cdot\mathbb{D}}$ is the same as $\mathbb{D}\cdot\bbnabla$ in the present
case, since 
\be
	\mathbb{D}\cdot\bbnabla g = (\bbnabla - \mathbb{A}^\times)\cdot\bbnabla g
	= \nabla^2 g - \mathbb{A}^\times_{\ \bigcdot}\bbnabla g\nn
\ee
and
\be
	\bbnabla\cdot\mathbb{D} = \bbnabla\cdot(\bbnabla - \mathbb{A}^\times)g
	= \nabla^2 g - (\bbnabla\cdot\mathbb{A})^\times g - \mathbb{A}^\times_{\
\bigcdot}\bbnabla g\,;\nn
\ee
the $\bbnabla\cdot\mathbb{A}$ term vanishes in the path integral (\ref{eq13-10})
which can therefore be written as 
\bea
	X &=& \int e^{i\int\mathbb{P}\cdot\dot{\mathbb{A}} - {\cal H}(\mathbb{P},
	 \mathbb{A})]\,d^{\,4}x}\,{\cal D}\mathbb{A}_T\,{\cal D}\mathbb{P}\,;\\
\label{eq13-15}
	{\cal H} &=& {1\over 2 g^2}\left((\nabla f)^2 
	+ \mathbb{P}\cdot\mathbb{P}
	+ \mathbb{B}\cdot\mathbb{B} + \mathbb{J}\cdot\mathbb{A}\right)
\label{eq13-16}
\eea

This is the Hamiltonian form of the path integral.  Let us recall from ordinary 
quantum mechanics the connection between it and the Lagrangian form,
\be
	X_L = \int e^{iS}{\cal D}Q(t)
\label{eq13-17}
\ee
where $Q$ is the particle coordinate, $S= \int L dt$, and the Lagrangian is
\be
	L = {m\over 2}\dot Q^2 - V(Q)
\label{eq13-18}
\ee
for a particle of mass $m$.  From $L$ one derives the canonical momentum
\be
	P = {\partial L\over \partial\dot Q} = m\dot Q
\label{eq13-19}
\ee
and Hamiltonian
\be
	H = (P\dot Q - L)\Big|_{\dot Q = P/m} = {P^2\over 2m} + V(Q)\,.
\label{eq13-20}
\ee
The Hamiltonian path integral is given by
\bea
	X_H &=& \int e^{i\int[P(t)\dot Q(t) - H(P,Q)]dt}\,{\cal D}P(t)\,
	{\cal D}Q(t)\nn\\
	&=&\int e^{i\int[P(t)\dot Q(t) - {P^2\over 2m} - V(Q)]dt}\,{\cal D}P\,
	{\cal D}Q\nn\\
\label{eq13-21}
\eea
This is seen to be the same as $X_L$ after completing the square and integrating
over $P$. However, it would do no good to carry out the integral over $\mathbb{P}$
in the case of QCD, (\ref{eq13-15}), because it would introduce a complicated
functional determinant depending on $\mathbb{A}_T$, due to the way $\mathbb{P}$
enters $\bbnabla f$, (\ref{eq12-12}).

We now concentrate on the interaction Hamiltonian from (\ref{eq13-16}),
\bea
	{\cal H}_I &=& {1\over 2 g^2}\int (\bbnabla f)^2 d^{\,3}x = -{1\over 2 g^2}
	\int f\nabla^2 f\, d^{\,3}x \nn\\
	&=& -{1\over 2 g^2}\!\!\int\! \left[\rho + \mathbb{P}^\times_{\ \!\bigcdot}\mathbb{A}
	\right]{1\over \mathbb{D}\cdot\bbnabla}\nabla^2\!{1\over \mathbb{D}\cdot\bbnabla}
	\left[\rho +  \mathbb{P}^\times_{\ \!\bigcdot}\mathbb{A}\right]d^{\,3}x \nn\\
\label{eq13-22}
\eea
You will recall that $\rho$ is the charge density of the external quark field. 
This suggests interpreting $\mathbb{P}^\times_{\ \!\bigcdot}\mathbb{A}$ as the
color charge density of the transverse gluons.  To see that this interpretation
makes sense, recall that a complex scalar field $\phi = (\phi_1+i\phi_2)$ has the
charge density 
\bea
	\rho &=& {1\over 2i}\left(\phi^*\dot\phi - \dot\phi^*\phi)\right)
	= {1\over 2i}\left(\phi^* \pi^* - \pi\phi\right)\nn\\
	&=& (\phi_1\pi_2-\phi_2\pi_1)
\label{eq13-23}
\eea
where $\pi = \partial{\cal L}/\partial\dot\phi$ is the canonically conjugate
momentum.  Eq.\ (\ref{eq13-23}) is like the cross product of the vectors $(\phi_1,\phi_2)$ and
$(\pi_1,\pi_2)$.

Now we would like to deduce the potential between quarks from the operator 
$(\mathbb{D}\cdot\bbnabla)^{-1}\nabla^2(\mathbb{D}\cdot\bbnabla)^{-1}$ in
(\ref{eq13-22}).  Although $\mathbb{D}\cdot\bbnabla$ cannot be inverted in closed
form, it can be expanded in powers of the gluon field, which corresponds to weak
coupling.  We get
\bea
{1\over \mathbb{D}\cdot\bbnabla}\,\nabla^2{1\over \mathbb{D}\cdot\bbnabla}
&=& {1\over\nabla^2} + 2\,{1\over\nabla^2} \mathbb{A}^\times_{\ \!\bigcdot}\bbnabla
	{1\over\nabla^2}\nn\\
	&+& 3\,{1\over\nabla^2}\,\mathbb{A}^\times_{\ \!\bigcdot}\bbnabla\,
	{1\over\nabla^2}\,
\mathbb{A}^\times_{\ \!\bigcdot}\bbnabla\,{1\over\nabla^2}\nn\\
	&+& \dots
\label{eq13-24}	
\eea
The first term corresponds to the Coulomb potential, as in (\ref{eq13-13}).

\footnote{The remainder of this section was added to my submitted draft by 
RPF.}  If the following terms are evaluated to 2nd order we get
a strong attraction.  In $\beta_0 = 11-\sfrac23 n_f$\footnote{The beta function has
not yet been introduced; this will come in the next lecture.} it contributes
$+12$ units---most of the confining effect.  Thus the instantaneous Coulomb
interaction probably rises with distance.  The charge densities of transverse 
(``real'') gluons
and quarks are $(\rho + \mathbb{P}^\times\mathbb{A})$ and their interaction via
${\cal H}_I$, via $(\mathbb{D}\cdot\bbnabla)^{-1} \cong \nabla^{-2}$, 
to second order in ${\cal H}_I$, makes a vacuum polarization of the normal sign,
contributing $-1-\sfrac23 n_f$ to $\beta_0$.

This leads to an interesting model of a string connecting heavy quarks.  
Because of
the rapid rise (with $r^2$)\footnote{This seems to be a slip since the
equations imply a linear potential between charges, leading to a constant
force, as is usually understood.
It can be derived using dimensional analysis, by evaluating the Green's function $\nabla^{-2}$ at large
separations.}  of the force between charges, unbalanced color charges
$\rho + \mathbb{P}^\times\mathbb{A}$ at any distance are intolerable.  Suppose we
start with a red quark at $r=0$ and say anti-red far away to the right.  By
creating a dipole gluon $\bar R G$ within this range where opposite colors are
tolerable we cancel redness at larger transverse distances.  But then the $G$ end
of the gluon is unbalanced, so another gluon dipole $\bar G B$ forms (the energy
for which coming from the decrease in $G$ energy).  This continues until we meet
the final quark.  Thus we have a state of superpositions of color arrangements,
\be
\includegraphics[width=0.35\textwidth]{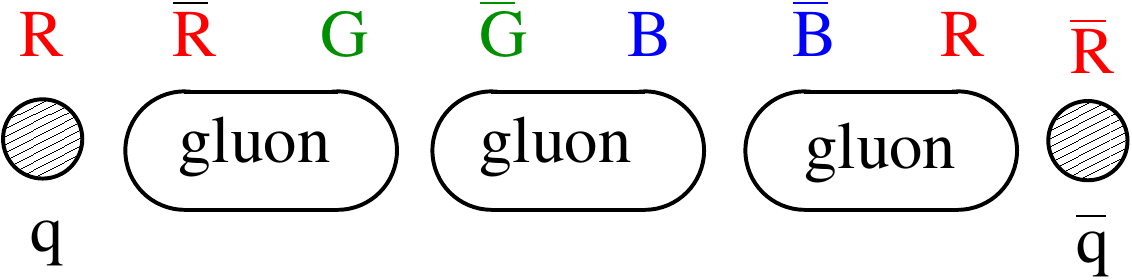}\nn	
\ee
In the transverse region away from the string, all the colors cancel and there
is not much energy in the Coulomb-like quantum chromostatic field.

We shall have a more general and precise discussion of these strings later on in
the course, and will see if this is a useful viewpoint of a string.
(Such pictures have also been discussed by Greenstreet.\footnote{RPF  means
Greenberg, who wrote several papers about quark confinement.})

\section{Interlude (1-5-$xy$)}
\label{interlude}

[From the audio tape.  RPF begins with some remarks about the research project.]
\it
First, last term we added an assignment to write something up; I've given some of
them back already and here are the rest.  I've written a lot on them but it doesn't
mean that I've corrected everything.  Mainly I'm trying to suggest some ways of
looking at things, which is not meant as a criticism necessarily; well it could be
a criticism, but very often a great deal of what I'm writing is not a criticism.
I think this is a very useful way of teaching, and I would like to do the same
thing this term, at the end of this term, having another paper to hand in on the
first day of exam week at the end of the term.  

And this paper this term will have
to be better than the paper of the last term.  You remember that I decided on that
rather late; I only gave you a few weeks.  So I said it doesn't have to be too
good.  So everybody passed. On some of the papers I was rather surprised and
disappointed and thought that you were unable or didn't have the time to do your
subject justice, and I made some remarks that you should talk to me about it or
something like that; that's only one or two so don't worry.  Sometimes a person
isn't prepared or doesn't have the background or the focus needed, and it's 
worth finding out earlier rather than later.  

The other possibility is that you've got some kind of a block somewhere, a
misunderstanding of what it's all about.  And it's very surprising \rm[noise]\it\
\dots it looks very simple to me, but it's only a block.  I know that because I've
had that experience once myself, and I understand it.  I also used to do tutoring,
and I've discovered many blocks like this \dots and straightened the person out
who had the wrong idea.  I'll give you an example.  My own block occurred when I
studied solid geometry in high schoool.  I was pretty good at math as you might
guess, and I thought I'd have a good time in solid geometry.  The class started and
I didn't understand anything.  The guy would ask questions, and I couldn't figure
out what the hell the answer was, and moreover something that I thought was usually
pretty dopey would come out as the right answer.  It would proceed like this and
 I was getting foggier and foggier for about a week.  All of a sudden, 
{\bf thank God}, it suddenly hit me.  I understood.  They were drawing things like
this, parallelograms, overlapping.  Then sometimes lines that would come out
\rm[connecting the parallelograms]\it; the question was were they perpendicular to
another line like that \dots \rm[laughter]\it.  I was looking at every one of the diagrams
as though it were a plane diagram, but it was a three-dimensional picture.
So when I finally figure it out the teacher told me ``that's why we call it solid
geometry, you idiot!'' \rm[laughter]\it.  I couldn't figure out the theorems and
equations and relations because I was looking at the diagrams flat.  That's a
block.

Another example was a student that a lady came to me and told me that her son was
very good in math but was needing some help in geometry, he was in high school.
So the first thing I did was to ask him some questions, like if this is a rectangle
and this is the diagonal, how far is it from here to here?  It's the same, he says;
and then a few more questions, showing that he had a first-rate intuition about
geometry; there was no problem about understanding what geometry was about. 
But to
make a long story short, he had a double block.  One: they had in his course in
high school that you write what they call proofs.  You write something and then you
write the reason.  Given: \dots and so on and so on. The way he thought was, and
he didn't understand, was how you knew what to write on each line.  He thought
there was some logical process to know what to write on each line.  And I had to
explain to him, no, what you did was you first had to figure out how to prove it.
And then after you figured out how to prove it you wrote the proof down.  Okay,
that was stupid, but that's the way they did it.  

Another block was, he didn't
realize the rule that the theorems that you were allowed to use in proving, the
things you could write on the right-hand side as reasons, always had to be
statements that were earlier than the thing you were trying to prove.  For example,
no, you can't use the Pythagorean theorem here because you haven't gotten there
yet.  Sounds dumb, but it's just artificial conventions of human beings 
\rm[long section where RPF is speaking away from the microphone]\it.

Same thing happens in algebra, where people have great difficulty
because it's not realized and it's not explained to them that $x$ is
used in two ways to represent a number; when you have a problem like
this \rm[writing on board]\it.  The problem is $x$ is some special number
and you have to find it, and when you have a statement like this \rm[more
writing]\it that's an entirely different use; there it means it's true no
matter what $x$ is.

So these blocks may be the cause of some \dots if you look at your papers and find 
some remark \dots let's talk
about it \dots a chance to maybe figure out if you have some 
difficulty. Alright?  Most of you have no such difficulty.

Is everything alright?  Do you want me to give more problems during the year,
or is it okay if we do these final exam papers, because they're very useful,
I think \dots  If you have some objections, come and tell me because I know 
it's been \dots

\rm
\section{Scale dependence (1-5-$xy$)}
\label{lec-1-5-$xy$}

So far in this course we have emphasized perturbation theory as the main tool
for making theoretical predictions, but it is important to keep in mind that
the path integral is not limited to this treatment.  For example, lattice gauge
theory, which we will discuss in some detail, is a way of computing observables
nonperturbatively.  And I will bring up some other possible ideas for going
beyond perturbation theory.  That will occupy us in the second part of this
term.  For the first part, we will continue to explore aspects of perturbation
theory.

I would like to spend some time in this lecture on the topic of running coupling
constants, which must be handled carefully in order to avoid confusion.
In fact, there is quite a great potential for confusion in this subject,
and an apparent complexity, because of the lack of agreement about the 
best conventions for carrying out the renormalization of the couplings,
and also some misguided suggestions that the running coupling should be 
defined in terms of some particular processes.

\subsection{Measuring couplings}

But before we discuss the running, we should try to understand what are
the most efficient ways of experimentally determining the values of
the couplings in the QCD Lagrangian,
\be
	 {1\over g_0^2}\tr FF + \sum_i\bar q_i(i\slashed{\partial}+m_0^{i})q_i 
\ee
where I have summed over the quark flavors, $i=u,d,s,c,b$ (and presumably $t$,
although it has not yet been observed).  The subscript $0$ means that these are
the bare couplings, that would coincide with the physical values if we were to
make predictions only at tree level, but which of course will differ once we
start to include loops.  So we have at least six Lagrangian parameters, 
\be
	g_0^2,\ m_0^{u},\ m_0^{d},\ m_0^{s},\ m_0^c,\ m_0^b 
\ee

In principle, we could determine all of these by measuring six independent
observables, since generically each one would constrain different combinations
of the parameters.  But in practice we usually focus on one thing at
a time, and try to choose an observable that is most sensitive to the quantity
of interest.  For example, to compute $m_0^b$, we could initially estimate it
as approximately half the mass of the $\Upsilon$ meson 
(bound state of $b\bar
b$), since we know that $\Upsilon$ gets most of its mass from the quarks and not
the gluons.  This would give $m_b^0\sim 5\,$GeV.   Similarly we can estimate 
$m_c^0$ as being about half of the mass of the $\psi$.  
 Of course the $\Upsilon$ and $\psi$ do
get some of their mass from the gluons, and so we could try to improve on these
estimates by doing a bound-state calculation to take that into
account---but that would depend on other parameters, namely $g_0^2$.

To determine $g_0$, we could try to compute the mass of a particle like the
proton, which is believed to get most of its mass from the gluons, since 
$m_0^u,\,m_0^d\ll m_p$.  However we then encounter the
problem that we don't know how to compute $m_p$; it is very far from being a
perturbative calculation, and we would have to rely on the lattice, which for
the present looks hopeless, although maybe someday it will be feasible.  

Instead, to measure $g_0$ it is more practical to look at a process involving
higher energies, much higher than the quark masses, so that the measurable
quantity depends very weakly on the $m_0^i$; in this way we can disentangle the
dependences and determine $g_0$ independently.  At arbitrarily high energy
$E\gg m_0^i$ this should become an increasingly good approximation since the
amplitudes will depend only on the ratio $m_0^i/E$.
 
Let me start with a process that at first looks like it will not help us, 
since it does not seem to depend on the QCD coupling $g_0$ at all:
electroproduction of quarks: 
\be
\includegraphics[width=0.1\textwidth]{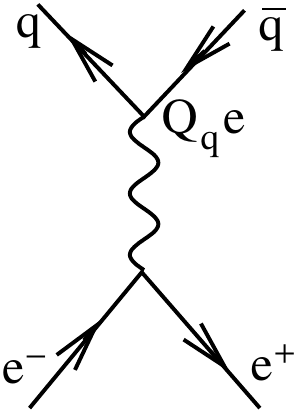}\label{15eeqq}	
\ee
Of course the real physical process does depend on $g_0$ because what we actually
 observe is not the
quarks, but rather $e^+ e^-\to$ hadrons, and the hadronization process
depends strongly on $g_0$.  But this is not the kind of $g_0$ dependence we are
interested in because (as I will argue later in the course) it is more
characteristic of the low-energy scales of the hadron masses and it is not a
perturbatively computable process.  Instead, we want to ignore the details of 
hadronization and pretend that the quarks are produced freely.  This is actually
a much better approximation than one might at first imagine, for the simple
reason that hadronization mainly occurs {\it after} the quarks have been produced.
Therefore it cannot affect the production cross section (at least at energies that
are not too close to resonances), which is the quantity
that we {\it can} compute perturbatively, and which we can measure more or less
cleanly despite the complications of hadronization.

Hence the electroproduction process does not depend on $g_0$ at all, at the
leading order in couplings; I am going to take advantage of a subleading effect
to suggest a way of measuring $g_0$.  But first let's look at the leading 
contribution, in the high-energy limit where we can ignore all the masses.  
We see that the calculation for producing quarks is hardly different from
that for producing $\mu^+\mu^-$.  The only differences are the charges of the
quarks, and the fact that quarks come in three colors.  Consider $e^+e^-\to
u\bar u$; the charge of $u$ is $+2/3$, so we can relate the quark and lepton
cross sections as 
\be
	\sigma_{u\bar u} = 3\, \left(2\over 3\right)^2\, \sigma_{\mu^+\mu^-}\nn
\label{eq15-2}
\ee
The factor of 3 is commonly understood as coming from the number of quark
colors: you can produce either a red quark, or a blue or a green.  This is
actually a cheat: in reality you can only produce a color-singlet combination
\be
	{1\over \sqrt{3}}\left(R\bar R + B\bar B + G\bar G\right)
	\equiv {3\,a\over\sqrt{3}} = \sqrt{3}\,a
\label{eq15-3}
\ee
In this formula, $a$ represents the amplitude for producing a quark of a
definite color, that we assumed could occur when we multiplied its
cross section by $3$.
We see that the amplitude of the color singlet state that is actually produced
is bigger by a factor of $\sqrt{3}$, and so this is the proper explanation of the 
factor of 3 in the cross section.  But either way of thinking about it gives
the right answer.

Similarly for producing $d\bar d$, we have
\be
	 \sigma_{d\bar d} = 3\, \left(-{1\over 3}\right)^2\, 
	\sigma_{\mu^+\mu^-}\nn,
\label{eq15-4}
\ee
and we can continue this to include all the higher mass quarks if
the center-of-mass energy is sufficient to produce them.

The interesting quantity to measure and compare to predictions is the
ratio
\be
	R = {\sigma_{\rm total}\over \sigma_{\mu^+\mu^-}}
\label{eq15-5}
\ee
whose contributions, at high enough $Q^2$, we can read off from 
equations like (\ref{eq15-3},\ref{eq15-4}).  The experimental data look
like this:
\be
\includegraphics[width=0.45\textwidth]{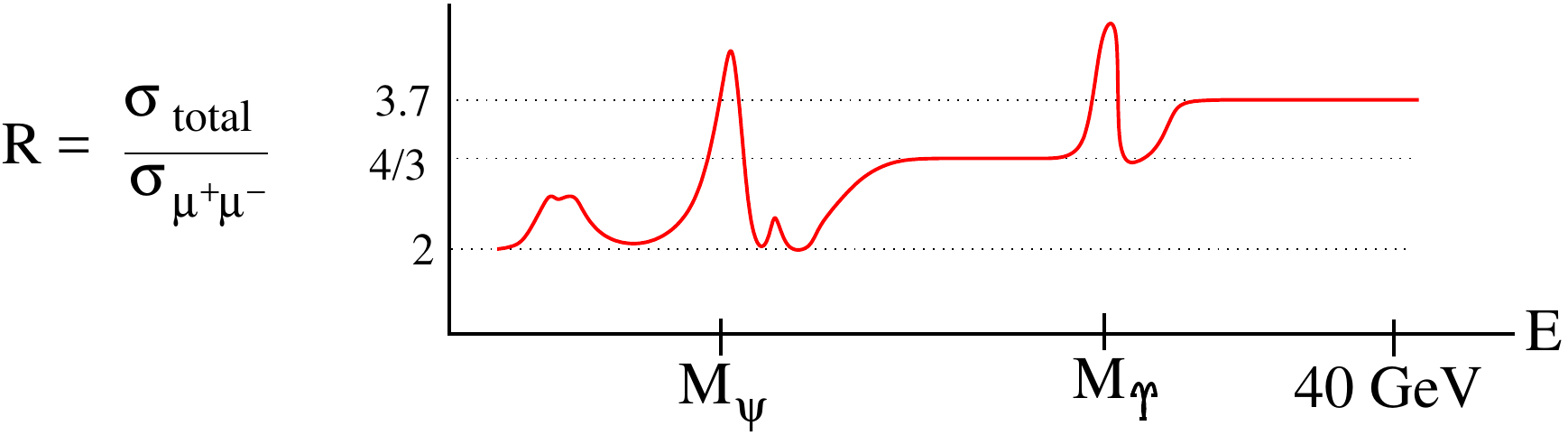}\nn	
\ee
At each quark threshold, a resonance for the corresponding bound state occurs, 
giving rise to the peaks and dips.  These are the hadronization complications
that we would like to avoid.  In between the thresholds, we observe the simple
behavior that we can understand within our approximations: $R$ is just flat, and
it is counting the number of quarks that can be produced, weighted by their
electric charges (squared).

Now this is all very beautiful, but it doesn't yet help us to determine the
QCD coupling.  For that, we should consider a higher-order process,
where a gluon gets radiated from one of the quarks, or exchanged:
 \be
\includegraphics[width=0.3\textwidth]{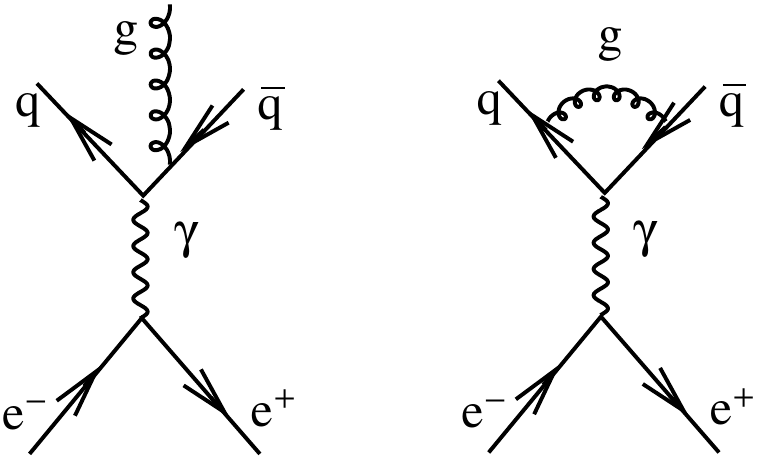}\label{eq15-6}	
\ee
One can show that at high energies, these processes are also independent
of the quark masses, and they lead to a multiplicative correction of the 
leading order prediction $R_0$, 
\be
R = R_0\left(1+ {\alpha_g\over\pi}+\dots\right)
\label{eq15-7}
\ee
where $\alpha_g = g^2/4\pi$.  Then if we can measure $R$ well enough, the
deviation between $R$ and the lowest order prediction $R_0$ will
determine $\alpha_g$.  So this is a possible way of measuring $g_0$.
It is not quite as direct as we would like, since it depends on a relatively
small correction, of order a few percent, to the basic quantity that is
insensitive to $g_0$.  Can we do better?

A more direct approach would be to observe the gluon that is emitted.  That of
couse is impossible, just like for the quarks, since they are all colored
objects.  Instead, we observe the jets of hadronized particles emerging from
those primary particles,  
\be
\includegraphics[width=0.4\textwidth]{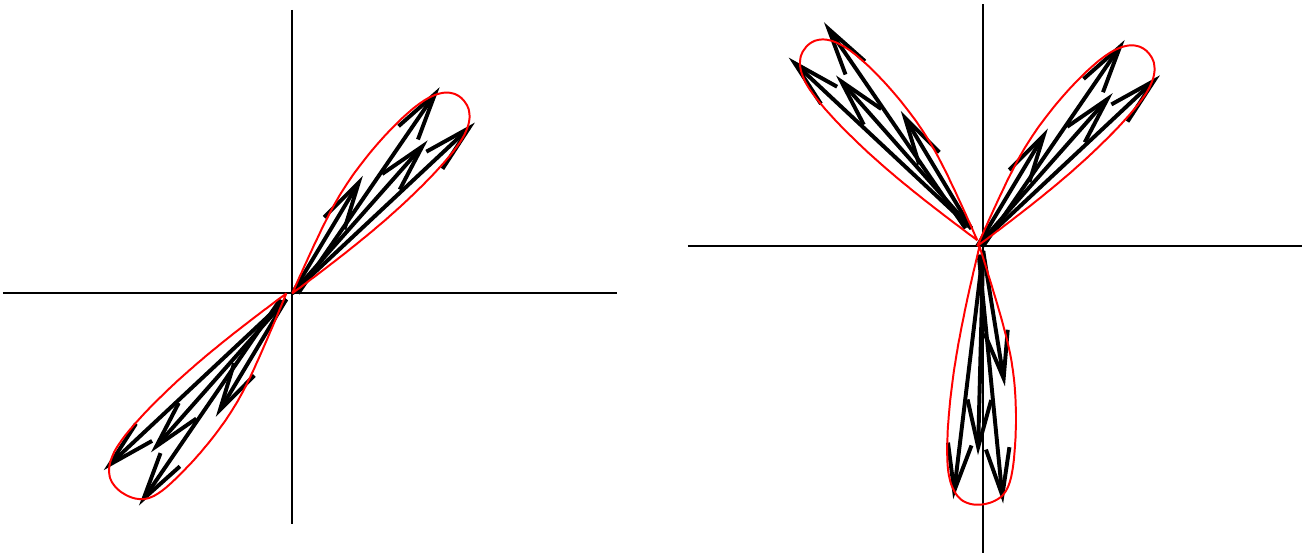}
\label{eq15-8}	
\ee
where I have drawn the momentum vectors of the individual hadrons and enclosed
them with an envelope to indicate the jet.  At high energies, the jets are
well-defined, with lengths (in momentum space) of order the center-of-mass
energy $E$, and widths of order the QCD scale $\sim 1\,$GeV.  The two-jet event
corresponds to the leading order process (\ref{15eeqq}), while the three-jet
event comes from the first diagram in (\ref{eq15-6}), assuming the gluon was
radiated at a large enough angle to produce a distinct jet. 

As we will discuss in a later lecture, it could happen that the gluon is too
soft, is radiated nearly in the same direction as the quark, and its
hadronization products get lumped in with the quark jet.  But at high enough
energies, there is a significant probability for the gluon jet to be
well separated from that of the quarks.  It will never be as probable as the
soft gluon emission, because the  extra quark propagator in the radiative
diagram would like to be on shell, which favors the soft emission, but at high
energies there is the competing effect from the available phase space to make
the hard gluon jet observable.  

Of course there is no definite dividing line
between hard and soft, so this is somewhat less clean in practice than the
idealized pictures shown in (\ref{eq15-8}); we might confuse some of the hadrons
that came from the gluon as being  associated with the quark jet, or vice versa,
leading to errors in the estimates of the total momenta of the respective jets. 
These errors become relatively smaller at high energies, where the soft jet contributions
become relatively less important.  The important point is, this is an observable that depends directly
on $\alpha_g$, so it is a more senistive determination than using the ratio $R$. 
In the following discussion, I will continue to focus on high-energy processes 
that allow us to neglect the quark masses and focus on the coupling $g_0$.

\subsection{Ultraviolet divergences}
You may have noticed something peculiar about the diagrams I drew in 
(\ref{eq15-6}): they mix up different orders of perturbation theory.
And of course there are other loop diagrams not shown, like the self-energy
correction to the quark from a gluon loop.  As you probably know from a
previous course on electrodynamics, the tree dia\-gram with the gluon emission
{\it cannot} properly be separated from this self-energy correction: the
infrared divergence of the soft gluon emission is canceled by a similar
divergence in the self-energy diagram, when it interferes with the tree diagram.
So that is one reason we cannot avoid considering the loops.  But a more
important one, for the present discussion, is that the loops will modify our
predictions at high energies, which I have stressed is the best regime for
comparing predictions to experiment.

And as you know, these loop contributions are problematic because they diverge
at high virtual momenta, so we need to introduce the procedure for cutting off
the divergences.  A typical kind of integral that we have to deal with
(continuing in our approximation of neglecting masses) is
\be
	\int {d^{\,4}k\over k^2 (p-k)^2}\,,
\label{eq15-9}
\ee
which by counting powers of momenta in the numerator and denominator
is logarithmically divergent.  A naive way of regularizing this, which 
later we will see does not quite work in the case of gauge theories, is to
modify the propagators by taking
\be
	{1\over p^2}\to \lim_{\Lambda\to\infty}
	{1\over p^2} - {1\over p^2-\Lambda^2}
\label{eq15-10}
\ee
This will render (\ref{eq15-9}) finite, in the intermediate step before taking
the limit $\Lambda\to\infty$, and yields a divergent term going as $\ln\Lambda$.
This divergence can be absorbed by redefining the coupling $g_0$, before taking
the limit.

Let's recall how this works in electrodynamics, 
for electron-electron scattering.  First consider the
low-energy, large distance limit, corresponding to scattering in the
Coulomb potential $e^2/r$.  The tree-level scattering amplitude is 
\be
\raisebox{-0.6cm}{\includegraphics[width=0.1\textwidth]{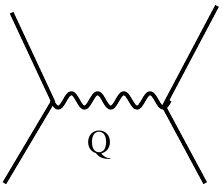}}
\sim {4\pi e^2\over Q^2}
\label{eq15-11}	
\ee
I want to consider the $Q\to 0$ limit first, since in this limit the
charge $e$ becomes the familiar constant value $e_{\rm exp}\cong \sqrt{1/137}
\cong 0.09$.\footnote{Notice RPF's unconventional normalization of the
coupling, that we have seen before and which he sometimes abandons in favor
of the usual one in later lectures.}
This is the physical value, which is not the same as the bare value $e_0$ that
is needed to cancel the divergences from loops, such as
\be
\includegraphics[width=0.15\textwidth]{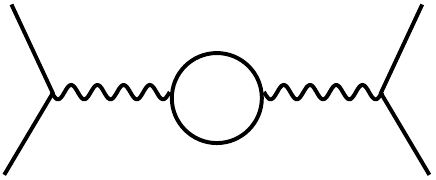}
\label{eq15-12}	
\ee
One can show that the divergences from this loop (plus the others not shown)
can be canceled by defining $e_0$ such that 
\be
	{1\over e^2_{\rm exp}} = {1\over e_0^2(\Lambda)} + {1\over
3\pi}\ln{\Lambda^2\over Q^2}
\label{eq15-13}
\ee
When we originally wrote the Lagrangian for QED, we thought that $e_0$ was
a parameter very close in value to the measured charge; now we see that it
is not a fixed parameter, but rather a function of the cutoff.  In the end nothing can
depend on $\Lambda$, since we are taking the limit $\Lambda\to\infty$.
What about the dependence on $Q^2$?  We thought that $e^2_{\rm exp}$ is supposed
to go to $0.09$ as $Q^2\to 0$.  In reality, we should have something like $Q^2+
m_e^2$ in the log; then this makes sense as $Q^2\to 0$.  The expression
(\ref{eq15-13}) is valid for $Q^2 \gg m_e^2$.

Of course there is one peculiarity in QED: we {\it can't} take the
$\Lambda\to\infty$ limit!  There is some scale $\Lambda \sim m_e\, e^{3\pi\times
137/2}$ at which $e_0$ diverges, and we cannot make sense out of the theory 
beyond this point since
$e_0^2$ becomes negative.  This is the Landau pole.  In practice, it is such a
high scale, far greater than the mass of the universe, that we don't care: the residual dependence of amplitudes on
$\Lambda$ is negligible.  It is more of an aesthetic shortcoming.  But a very
nice feature of QCD is that the analogous correction to $1/g_0^2$ has the
opposite sign, so there is never any Landau pole; instead $g_0$ goes to zero
as $\Lambda\to\infty$.  

In this course I will discuss three different methods of regularization,
which are summarized as follows:
\bea
1. && \hbox{ Pauli-Villars: }\qquad{1\over k^2}\to{1\over k^2} - {1\over
k^2-\Lambda^2}\nn
\eea
As I mentioned, this is not quite right as written; we need to be more careful
to avoid spoiling gauge invariance; we will come to that later.
\bea
2. && \hbox{Dimensional regularization:}\nn\\ 
&& \alpha_g \hbox{\ becomes dimensionful when\ } d\neq 4;\nn\\
 && \alpha_g\to \alpha_g\, \mu^{d-4}\nn
\eea
where we analytically continue amplitudes in the number $d$ of spacetime
dimensions, and 
\bea
3.\ &&\hbox{Lattice regularization:}\qquad\qquad\qquad\phantom{.} \nn\\
&& \hbox{has a dimensionful parameter: }\nn\\
 &&a \leftrightarrow 1/\Lambda \nn
\eea
where we have approximated spacetime as a discrete lattice, with sites
separated by the lattice spacing $a$, that plays the role of $1/\Lambda$
relative to the first method.  Any of these approaches leads to equivalent
results when we take the continuum limits, $\Lambda\to\infty$, $\epsilon\to 0$,
$a\to 0$.  Each one of them has its own different expression for how the
bare couplings depend on the cutoff, but the physical predictions that they make
to a given order in perturbation theory are the same.  To fully define the
theory we will also have to discuss gauge fixing.

Let me continue to work in the $\Lambda$ cutoff scheme for now; we will come
back to the other regulator methods in later lectures.  One can show that there
is a general form for the $\Lambda$ dependence of the bare coupling, which
we will derive later on,
\bea
	{1\over g_0^2(\Lambda)} &=& \beta_0\,\ln {\Lambda^2\over M^2}
	+ {\beta_1\over \beta_0}\ln\left(\ln {\Lambda^2\over M^2}\right)\nn\\ &
+& c_0 + \left(
{c_1\over\ln\Lambda^2/M^2}+\dots\right)\,.
\label{eq15-14}
\eea
Here $\cdots$ denote terms falling with $\Lambda$ even faster than
$1/\ln\Lambda^2$, that do not concern us in the large-$\Lambda$ limit.
The interesting observation is that the constants $\beta_0$ and $\beta_1$
are the same in different regularization methods, if we make the appropriate
identifications $\ln\Lambda^2\to 1/\epsilon$ or $\ln\Lambda^2 \to \ln(1/a^2)$
to translate between them.  There is an arbitrary scale $M$ appearing, known as
the renormalization scale.   Notice that any change $M^2\to e^{-\gamma}M^2$ 
can be absorbed into a change of the constants $c_0$ and $c_1$ (for $\Lambda\gg M$): 
\bea
	{1\over g_0^2} &\to & {1\over g_0^2} + \underbrace{\beta_0\gamma} +
\underbrace{\beta_1\,\gamma\over\beta_0}\,\ln{\Lambda^2\over M^2} + O\left(\ln^{-2}
{\Lambda^2\over M^2}\right)\nn\\
&& \qquad\ \ \delta c_0\quad\ \  \delta c_1
\label{eq15-15}
\eea
Thus the $c_i$ are not universal at all, and they also depend on the method of
regularization.  Their values get fixed, within the given method, by comparing
some prediction in which $g_0^2$ appears to its corresponding measured value.
The essential fact is that it should not matter which observable we choose;
any quantity that is sufficiently sensitive to $g_0$ will suffice.  And once that
is done, we can use the same formula to predict other observables, and use these
to test the theory.

You may wonder why I bother to write the $c_1$ term at all, since it is
irrelevant as we take the continuum limit.  There may be some practical situations
where we are not able to take this limit, notably the lattice, where it 
is computationally prohibitive to do so.  The best we can do there is to
take $a$ to some small value.  To get the best description of the data at two
different values of $a$, we would need to keep the $c_1$ term, so as to properly
compare predictions at different values of $a$, to see whether we think 
$a$ is small enough to trust our predictions.

It is worth remarking on the observation that for QCD, $g_0\to 0$ as we remove
the cutoff.  Does that mean that perturbation theory should always work
in QCD?  No, it depends on the scale of the physical problem.  This is because
the limit $\Lambda\to\infty$ is the same as the limit $M\to 0$ in eq.\ 
(\ref{eq15-14}).  When we compute an observable quantity at a physical
scale $Q^2$, the amplitude will have factors going as $\ln\Lambda^2/Q^2$.
When we combine these with the coupling $g_0^2$, the cutoff dependence
disappears, and the log becomes $\ln M^2/Q^2$, which diverges as $M\to 0$.
Thus we can't necessarily take the continuum limit while keeping perturbation theory under
control; it only works if $Q^2$ is larger than the characteristic scale of
QCD, $\lambda_{\rm Politzer}^2 \simeq (1\,{\rm GeV})^2$, where perturbation theory starts to break down.

Related to this, another possible source of confusion is the arbitrariness
of the choice of $M$ in (\ref{eq15-14}).  I showed that $M$ is perfectly
arbitrary since we can absorb any change of $M$ into a redefinition of
the $c_i$ constants.  That is true, but in practice, for a given application,
some choices of $M$ will be more advantageous than others.   Generally, if we
are going to use an observation at a particular scale $Q^2$ to fix the value
of $g_0^2$, then it makes sense to choose $M^2\sim Q^2$, to get rid of
large logarithms $\ln M^2/Q^2$, that would also appear in the higher-loop
diagrams that we are not including in our calculation.  This choice reduces
the error coming from these higher-order diagrams.  This argument also shows
why it would not make sense to try to take $M^2$ all the way to zero:
perturbation theory in QCD is breaking down at scales 
$Q^2\lesssim \lambda_P^2$.  So we should not take $M<\lambda_P$.  It is
sometimes convenient to {\it define} $M$ by imposing a choice like $c_0=0$
on the arbitrary constant, to make eq.\ (\ref{eq15-14}) look nicer.  Then
$M$ takes on a physical significance (within a particular regularization 
method), that can be called the scale where QCD is becoming nonperturbative.
In this convention, $M=\lambda_P$, which has been determined to be around 200\,MeV.

Because the coupling only runs logarithmically, it takes considerable experimental
effort to reach energies where convergence of the perturbation expansion
improves dramatically.  And on the lattice, it takes enormous numerical effort. 
Consider that on a 4D lattice, by cutting the lattice spacing in half, we
compound the computational problem by a factor of $2^4=16$, yet the coupling has
decreased by a factor of only $1/\ln 4 \cong 0.7$.  Thus it seems nearly
hopeless to achieve quantitative accuracy on the lattice, given the limitations
of computers, and it would be nice to come up with another nonperturbative
method that is not just brute force.

\section{The renormalization group (1-7-$xy$)}
\label{1-7-$xy$}
Last time we saw that the bare coupling could be expressed in a way
that I will rewrite in the form
\bea
	{1\over g_0^2(\Lambda)} &=& 
	\beta_0 \ln{\Lambda^2\over\lambda_P^2} 
	+ {\beta_1\over \beta_0}
     \ln\left({\ln\Lambda^2\over\lambda_P^2}\right) +{\rm const.}
\nn\\ &+&
{\cal O}\left({1\over\ln(\Lambda^2/\lambda_P^2)}\right)
\label{eq16-1}
\eea
where $\lambda_P$ is short for a scale I am calling 
``$\lambda_{\rm Politzer}$,'' and the last term, that vanishes as the
cutoff goes to infinity, is scheme-dependent.   If we choose $g_0$
 in this way, for a given method of regularization, physics is
unchanged from scheme to scheme and as the cutoff varies. If we rewrite the second term on the
right-hand side of (\ref{eq16-1}) as
$(\beta_1/\beta_0)\ln\left( 1/(\beta_0g_0^2)\right)$, the terms that vanish as 
$\Lambda\to\infty$ are cleaner and have no $\ln\ln$ coefficients in
the numerators of $1/\ln\Lambda^2$.\footnote{This statement is justified
in eq.\ (\ref{eq16-3}).} 

Now we can differentiate,
\be
	{d\left(1\over g_0^2(\Lambda)\right)\over
	d\left(\ln{\Lambda^2\over\lambda_P^2}\right) } = 
	\beta_0 + {\beta_1}\,g_0^2 + \beta_2\,g_0^4 +
\dots
\label{eq16-2}
\ee
This is closely related to the beta function, 
\be
\beta(g) = -{dg_0\over d\ln\Lambda} =  \beta_0\, g_0^3 + \beta_1\, g_0^5 + \dots
\ee
Here $\beta_0$ and $\beta_1$ are scheme-independent, but $\beta_2$,
$\beta_3, \dots$ are scheme-dependent and they affect only the 
${\cal O}(1/\ln\Lambda^2)$ terms.  As I explained in the last lecture,
we must choose the constant in (\ref{eq16-1}) in order to define 
$\lambda_P$; so we made a specific choice.  We define the constant to
be zero in the $\overline{\rm MS}$ scheme of dimensional regularization, and we can do
likewise in Pauli-Villars.  This defines $\lambda_P$.
It is an arbitrary choice which depends on the regularization method.
Making this choice, we can write an exact formula
\be
	{1\over g_0^2(\Lambda)} = \beta_0 \ln{\Lambda^2\over\lambda_P^2}
	+ {\beta_1\over \beta_0}\ln\left({1\over
\beta_0\,g_0^2}\right)
\label{eq16-3}
\ee
which is an implicit {\it definition} of $g_0^2(\Lambda)$.  It satisfies the
differential equation
\bea
	{d\left(1\over g_0^2(\Lambda)\right)\over
	d\left(\ln{\Lambda^2\over\lambda_P^2}\right) } &=& 
	\beta_0 + {\beta_1}\,g_0^2 + {\beta_1^2\over\beta_0}\,g_0^4 +
\label{eq16-4} \dots\\
	&=& {\beta_0\over 1 - {\beta_1\over\beta_0} g_0^2}\qquad
	\left({\beta_0 = 11-\sfrac23 n_f\ \ \ \ \!\!\atop \beta_1 = 102 -
\sfrac{38}{3} n_f}\right)\nn
\eea
where $n_f$ is the number of quark flavors.  Notice that $\beta_2$ and higher coefficients
are all determined by $\beta_0$ and $\beta_1$ in this definition of $g_0(\Lambda)$.
For large $\Lambda$, the
different choices one could make for these higher coefficients are not important.

As I mentioned in a previous lecture, in QED we have a problem when we
try to do the analogous thing.  Let $e^2_{\rm th}$ denote the
physical, observed value of the coupling.  Then
\be
{1\over e^2_{\rm th}} = {1\over e^2_{0}(\Lambda)} + {1\over 3\pi}
	\ln\left(\Lambda^2\over m^2\right)
\label{eq16-5}
\ee
where we have a plus sign instead of minus between the two terms.
Hence	
\be
{1\over e^2_{0}(\Lambda)} = {1\over e^2_{\rm th}} - {1\over 3\pi}
	\ln\left(\Lambda^2\over m^2\right)
\label{eq16-6}
\ee
As $\Lambda$ gets large, ${1/e^2_{0}(\Lambda)}$ becomes negative, and we
lose unitarity.  This means we can't push QED to arbitrarily high energies,
at least not greater than
\be
	{\Lambda^2\over m^2} \sim e^{137}\,.
\label{eq16-7}
\ee
Grand unification is one solution to this problem.\footnote{In my
notes I have parenthetically, ``(Even just electroweak theory seems to
fix it.)'' perhaps referring to the modification of the QED beta function
by the other standard model interactions.}

\subsection{Measuring $g^2$}

Recall that we could measure $g^2$ through the ratio
\be
	R = {\sigma_{e^+e^-\to {\rm had.}}\over \sigma_{e^+e^-\to
\mu^+\mu^-}};\qquad (R-1) = {g_0^2\over \pi} + a\, g_0^4 + \cdots
\label{eq16-8}
\ee
at high energy.  We wanted something that didn't depend on the mass of
the quarks, another example being $qq$ scattering at high energies.  In this way
we could concentrate on determining $g_0^2$.  We would also like our
observable to be dimensionless, such as $Q^2\sigma$
since the cross section $\sigma$ has dimensions of (length)$^{-2}$.
To avoid extraneous dependences, we can imagine keeping the 
scattering angles fixed as we increase the energy scale.

In general, our calculation of this quantity will depend on $g_0^2$, 
$\Lambda$ and $Q$; call it $D_{\rm theory}(g_0^2,\,\Lambda,\,Q)$.
At first we consider $g_0^2$ to be independent of $\Lambda$---it's
just a parameter.  Now we want the physically measured value of 
the observable
\be
	D_{\rm phys.} = D(g_0^2(\Lambda),\,\Lambda,\,Q)
\label{eq16-9} 
\ee
to be independent of $\Lambda$ as $\Lambda\to\infty$ (and of course it
must have a well-defined limiting value).  Recall that the coupling
is dimensionless, which you can see from the action,
\bea
	S &=& \int d^{\,4}x\,{1\over g_0^2} F_{\mu\nu}F_{\mu\nu}\\
\label{eq16-10} 
	&=& 
	\int d^{\,4}x\,{1\over g_0^2} \left(\partial_\mu A_\nu - \partial_\nu A_\mu
	- A_\mu^{\,\times} A_\nu\right)^2\nn
\eea
and remembering that $[A] = 1/L$ hence $[F] = 1/L^2$.  So $[F^2] = 
1/L^4$ and since the action is dimensionless, $[g_0^2] = 1$: the
coupling is dimensionless.  Therefore $D_{\rm theory}(g_{00}^2,\,\Lambda,\,Q)$
can only depend on the dimensionless ratio $\Lambda/Q$ if we keep
$g_{00}^2$ fixed.\footnote{The subscript on $g_{00}^2$ is to 
distinguish the coupling $g^2_{00}$ that is considered to be independent
of $\Lambda$ from $g_0^2(\Lambda)$.}\ \  
Hence we can write $D_{\rm theory}$ in the form
$D_{\rm theory}(g_{00}^2,\,\ln\Lambda^2/Q^2)$.  Similarly, our prediction
for the physically observed value must take the form
\be
	D_{\rm phys.} =
D\left(g_0^2(\Lambda),\,\ln\Lambda^2/Q^2\right)
\label{eq16-11}
\ee
and it {\it cannot depend on $\Lambda$}.  This tells us how it depends
on $Q$ for large $Q$.  

Suppose we ignore the $\beta_1$ term in $g_0^2(\Lambda)$, and the 
$g_0^4$ term in $(R-1)$.  Then
\bea
\label{eq16-12}
	D_{\rm theory}\left(g_{00}^2,\,\ln{\Lambda^2\over Q^2}\right) 
	&=&
	{g_{00}^2\over \pi}\\
	D_{\rm phys.}\left(g_0^2(\Lambda),\,\ln{\Lambda^2\over Q^2}\right) 
	&=& {g_0^2(\Lambda)\over\pi} = {1\over
\pi\beta_0\ln{\Lambda^2\over\lambda_P^2}}\nn
\eea
But this depends on $\Lambda$!  The problem is that we didn't go to
the next order:
\bea
	D_{\rm theory}\left(g_{00}^2,\,\ln{\Lambda^2\over Q^2}\right) 
	&=&
	{g_{00}^2\over \pi} + g_{00}^4\left(\beta_1\ln{\Lambda^2\over Q^2} +
a_1\right)\nn\\
\label{eq16-13}
	D_{\rm phys.}\left(g_0^2(\Lambda),\,\ln{\Lambda^2\over Q^2}\right) 
	&=& {g_0^2(\Lambda)\over\pi}\\
	 &=& {1\over
\pi\beta_0\left(\ln{\Lambda^2\over\lambda_P^2}-\ln{\Lambda^2\over
Q^2}\right)}\nn\\
&=& {1\over
\pi\beta_0\ln{\Lambda^2\over\lambda_P^2}} + {\ln\,\Lambda^2/Q^2\over
	\pi\beta_0 \,\ln^2{\Lambda^2/\lambda_p^2}}\nn\\
     &+&\cdots
\nn
\eea
where $\cdots$ represents terms of order $g_0^4$.  The next order
terms in $D_{\rm theory}$ must be
\be
{g_{00}^6\over\pi}\left(\beta_0^2\ln^2{\Lambda^2\over Q^2} 
+ 2 a_1\ln{\Lambda^2\over Q^2} +
a_2\right)\nn
\ee
Terms of the form $(\beta_0\ln\Lambda^2/Q^2)^n$ are the leading logs.
The fact that you can sum them is all due to demanding that $D_{\rm
phys.}$ does not depend on $\Lambda^2$.  If we do the sum then we get
\be
	{1\over \pi\beta_0\ln {Q^2\over\lambda_P^2}} + {a_1\over
	\pi \left(\beta_0\ln {Q^2\over\lambda_P^2}\right)^2} +\cdots
\label{eq16-14}
\ee
where the first term comes from the leading logs, the second from the
next-to-leading logs, and so on.  Now we could write
\be
	(R-1) = {\alpha(Q)\over\pi} + {a_1\over\pi}\,\alpha^2(Q)
+\cdots
\label{eq16-15}
\ee
where
\be
	\alpha(Q) = {1\over \beta_0\ln{Q^2\over\lambda_P^2}}
\label{eq16-16}
\ee
This shows that $\alpha(Q^2)$ operates like a coupling constant that
depends on energy.  We did this by neglecting $\beta_1$.  If we keep
$\beta_1$, we get
\be
	{1\over \alpha(Q^2)} = \beta_0\ln{Q^2\over\lambda_P^2} 
	+ {\beta_1\over\beta_0}\,\ln\left(1\over
\beta_0\,\alpha(Q^2)\right)
\label{eq16-17}
\ee
Evaluating it at different energies $Q$ in GeV, using $\lambda_P = 0.2\,$GeV, 
gives\footnote{This table, which RPF apparently computed himself, was not given
in the lecture, but it appears in his private notes for this lecture.}
\begin{center}
\begin{tabular}{|c|c|c|c|c|c|c|c|c|c|}\hline
$Q$      & 1 & 3 & 5 & 10 & 30 & 50 & 100 & 300 & 1000\\ 
\hline     
$n_f$    & 3 & 3 & 4 & 4 & 5 & 5 & 6\,? & 6\,? & 6\,???\\
\hline
$\alpha$ & $0.43$ &$0.26$ &$0.23$ & $0.19$ & $0.16$ & $0.15$ & $0.14$ & $0.12$ & $0.11\pm 0.01$\\
\hline
\end{tabular}
\setlength\tabcolsep{6pt} % default value: 6pt
\label{tab2}
\end{center}

\subsection{Renormalization group equations}

Now we are ready to derive the renormalization group equations, using
the fact that physical quantities cannot depend on $\Lambda$.  
$d D_{\rm phys.}/d\Lambda=0$ implies that 
\be
	\left.{\partial D\over\partial g_{00}^2} \right|_{g_{00}^2 =
g^2(\Lambda)} \times{d g^2(\Lambda)\over d\ln(\Lambda^2/\lambda_P^2)}
	-{\partial D\over\partial \ln Q^2} = 0
\label{eq16-19}
\ee
where
\bea
\label{eq16-20}
	{d g^2\over d\ln\Lambda^2} &=& - g\,\beta(g)\\
	&=& -\left[ \beta_0 g^4(\Lambda) + \beta_1 g^6(\Lambda)
	+ \beta_2 g^8(\Lambda) \cdots\right]\nn
\eea
Suppose we have worked out the theoretically predicted value
\be
	D_{\rm th}\left(g_{00}^2,\ln{\Lambda^2\over Q^2}\right)
	= b_0 + b_1 g_{00}^2 + b_2 g_{00}^4 + b_3 g_{00}^6 + \cdots
\label{eq16-21}
\ee	
where the $b_m$ may depend on $t\equiv \ln\Lambda^2/Q^2$.  Let 
$b'_m = db_m/dt$.  Then the RG equation says that
\bea
	b'_0 &+& b'_1\, g_{00}^2 + b_2'\, g_{00}^4 + b'_3\, g_{00}^6 \nn\\
	&=& \left[\beta_0\, g_0^4(\Lambda) + \beta_1\, g_0^6(\Lambda) 
	+ \beta_2\, g_0^8(\Lambda) \cdots\right]\nn\\
	&\times&\left[b_1 + 2 b_2 \,g_{00}^2 + 3 b_3\,
g_{00}^4\cdots\right]
\label{eq16-22}
\eea
This tells us that 
\be
	b_0' = b_1' = 0\,;
\label{eq16-23}
\ee
hence $b_0$ and $b_1$ are constants, that we can calculate
theoretically; call them $c_0$ and $c_1$.  Moreover 
\bea
	b_2' &=& c_1\,\beta_0\quad\implies\quad b_2 =
c_1\beta_0\ln{\Lambda^2\over Q^2} + c_2\nn\\
	b_3' &=& (b_1\beta_1 + 2b_2\beta_2) = 
	c_1\beta_1 + 2 c_1\beta_0 \ln{\Lambda^2\over Q^2} +
	2c_2\beta_0\nn\\
	&\implies& b_3 = c_1\beta_0\ln^2{\Lambda^2\over Q^2} +
	(2 c_2\beta_0 + c_1\beta_1)\ln {\Lambda^2\over Q^2} + c_3\nn\\
\label{eq16-24}
\eea
{\it etc.}\ \    Then
\bea
	&&\!\!\!\!\!\!\!\!\!\!\!\!D_{\rm phys.}\big|_{g_0^2(\Lambda)} = 
	c_0 + c_1 g_0^2(\Lambda) + 
	\left(c_1\beta_0\ln {\Lambda^2\over Q^2}
	+c_2\right) g_0^4(\Lambda)\nn\\
	&+& \left(c_1\beta_0 \ln^2{\Lambda^2\over Q^2}
	+(2c_2\beta_0 + c_1\beta_1)\ln{\Lambda^2\over Q^2}
	+ c_3\right) g_0^6(\Lambda)\nn\\
	&+&\cdots 
\label{eq16-25}
\eea
At each order $n$ in perturbation theory, a new constant term $c_n$ arises.
But this new constant is typically less important than the preceding terms
appearing in the coefficient of $g_0^{2n}$, that come with higher powers of logs.
In other words, $c_1$ determines all the leading logs, $c_1$ and $c_2$ all the
next-to-leading logs, {\it etc}.  
We can simplify (\ref{eq16-25})  because we know it doesn't depend on $\Lambda$.
Organizing it in terms of the $c_i$ coefficients we get
\be
	D_{\rm phys.} = c_0 + c_1\,\alpha(Q^2) + c_2\,\alpha^2(Q^2) + 
	c_3\,\alpha^3(Q^2) + \cdots
\ee
where $\alpha(Q^2)$ satisfies
\be
	{d\alpha(Q^2)\over d\ln Q^2} = \beta_0\,\alpha^2 +
\beta_1\,\alpha^3 + \beta_2 \alpha^4 + \cdots
\ee
If one could measure $D$ at such a high energy that $\alpha(Q^2)$ was
small and the series converged rapidly, it would provide way to 
determine $\lambda_P \sim 200\,$MeV.

\section{Renormalization: applications (1-12-$xy$)}
\label{lec17}
Renormalization is a confusing subject, and one factor contributing
to the confusion is the proliferation of different conventions.
I am guilty of this by my preferred normalization of the gauge
coupling, which is not the same as that of the rest of the world.
 For your convenience,
let me translate some previous results into the more conventional form,
where the coupling and its associated fine-structure constant are related as
\be
	\alpha = {g^2\over 4\pi}
\label{eq17-1}
\ee
Then the running of the renormalized coupling, (\ref{eq16-17}), takes the form
\be
	{4\pi\over\alpha(Q^2)} = \beta_0\,\ln{Q^2\over\lambda_P^2} + 
	{\beta_1\over\beta_0} \ln {4\pi\over \beta_0\,\alpha(Q^2)}
\label{eq17-2}
\ee

Now let's review what we learned in the previous lecture, concerning the 
utility of this expression, that sums up the leading logarithmic dependences in
the perturbation expansion.  Namely, we can take an amplitude computed at tree
level, and replace its $\alpha$ dependence by eq.\ (\ref{eq17-2}), to resum the
most important subclass of loop contributions to all orders, which improves the
tree-level prediction.  Furthermore, we can extend this to
subleading contributions.  Suppose we computed an amplitude at one-loop order
and found a result going as 
\be
	{\cal M} = \alpha_0 + \alpha_0^2\left(\ln{\Lambda^2\over Q^2} + c\right)
\label{eq17-3}
\ee
The term with $\ln Q^2$ is already contributing to the leading logs that
we obtain by replacing $\alpha_0\to\alpha(Q^2)$ in the lowest order
contribution.  Thus the correct way to incorporate the next-to-leading logs is
to replace  
\be
{\cal M} \to \alpha(Q^2) + c\,\alpha^2(Q^2)
\label{eq17-4}
\ee

I have avoided some of the complications of renormalization so far
by only discussing gauge invariant, physical quantities.  We could also
consider the renormalization of more general quantities like Green's
functions
\be
	\langle A(x_1)\cdots A(x_n)\rangle\, .
\label{eq17-5}
\ee
Then it is not sufficient to talk about only the renormalization of the
couplings, but also the wave function renormalization, that contributes to 
anomalous dimensions in the scaling of such a Green's function.  I am not
going to discuss these kinds of issues, but there are many references that
do so, for example \underline{Renormalization} by John Collins,
Cambridge University Press, 1984.  If you are only interested in physical,
measurable quantities, these complications can be avoided.

One further point pertaining to the previous lecture is about the dependence
of the beta function on the number of flavors.  We have noticed that one
of the nice features of QCD is its good behavior in the ultraviolet.  
This assumes there aren't too many flavors of quarks.  There are five or six
that we already know about, but there could be more---nobody knows why there
should only be three families.  But probably there are not 17 flavors of
quarks, which is the critical number that would ruin the good UV behavior of
QCD.

\subsection{Power counting of divergences}

Before embarking on explicit calculations of loop diagrams, it is 
enlightening to understand the general structure of divergences of
the theory, and you are probably already familiar with this, but
I would like to review it nevertheless.  Consider some rather complicated
diagram like
\be
\raisebox{-0.96cm}{\includegraphics[width=0.1\textwidth]{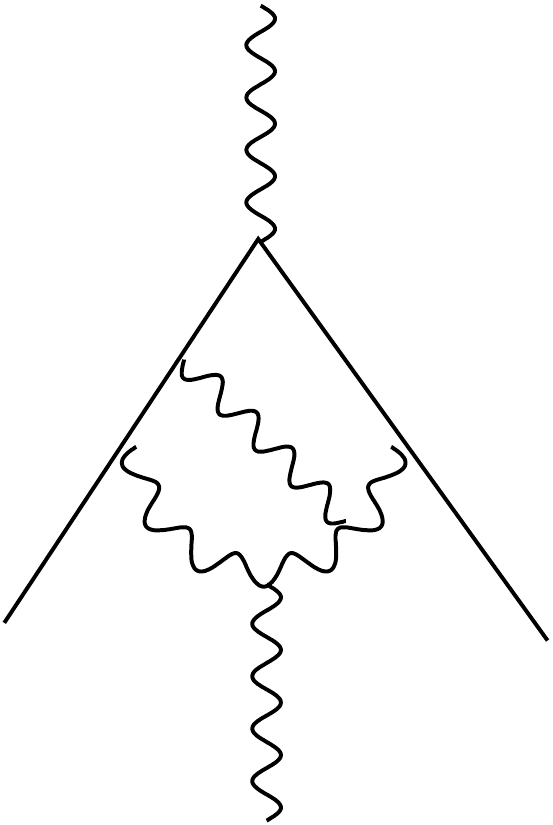}}
\ \sim \int {d^{\,4}k\,d^{\,4} p\over (p^2,\ k^2,\ p\cdot k\,  + \cdots)^7}
\times f(p,\ k, \cdots)
\label{eq17-6}
\ee
We count 7 propagators, schematically indicated by the denominator
of (\ref{eq17-5}), 8 powers of momentum from the integration measure,
and a numerator $f$ that goes like (momentum)$^5$ from rationalizing the
fermion propagators and counting the 3-gluon interactions.  So according to power counting, this goes as
(momentum)$^{-1}$ and is therefore superficially convergent.  We call this
exponent $N_d$, the superficial degree of divergence.  It tells us that a given
diagram generically behaves as
\be
\begin{array}{ll} N_d=0,& \hbox{log divergence}\\
	N_d = 2, & \hbox{quadratic divergence}\\
	N_d < 0, & \hbox{converges}\end{array}
\label{eq17-7}
\ee
We call it ``superficial'' because it is possible to construct exceptional
cases in which there is a divergence even if $N_d<0$.  This would be the case if
the integrals somehow factorized into a product of one that was highly
convergent times another that diverged.  More typically however, we will see
that this $N_d$ often {\it over}estimates the degree of divergence, as a 
consequence of gauge symmetry.

Now it would be
rather tedious to have to do this kind of counting for every possible diagram
that may arise, but fortunately we don't have to.  There is a beautiful
topological relation that does it for us, solely in terms of the numbers of
external lines of different kinds, independently of how complicated the diagram
is, such as the number of loops.  The relation is easiest to prove for a vacuum
diagram with no external legs.  It is a fact from topology that such a diagram
satisfies
\be
	\hbox{\# of Edges} + \hbox{\# of Vertices} = \hbox{\# of Faces} + 2
\label{eq17-8}
\ee 
This is a relationship between the number of loops, 
the number of vertices and the number of propagators.  

An easy way to arrive
at the result is to first notice that 
a vacuum diagram, by dimensional analysis, must
have $N_d=4$.  Imagine that we cut an internal quark line in a loop in such a
diagram, to add two external quarks.  We thereby remove one integration over loop
momenta and one fermion propagator, which reduces $N_d$ by 3.  On the other hand,
imagine adding an external gluon to some line on the diagram.  It creates an
extra fermion propagator, or else an extra gluon propagator with a coupling
proportional to momentum in the numerator.  Or we convert a triple-gluon
vertex with dimension 1 into a dimensionless 4-gluon vertex. In any case, 
we reduce $N_d$ by 1.  Therefore it must be that  
\be
	N_d = 4 - {\cal N}_g - \frac32{\cal N}_q,
\label{eq17-9}
\ee
where ${\cal N}_g$ (${\cal N}_q$) is the number of external gluon (quark) lines.

Another way of deriving (\ref{eq17-9}) is to use dimensional analysis for the
general case.  Let's illustrate this for some diagram with an arbitrary
number of external particles, representing a process in which particle 1
decays into $N-1$ final state particles:
\centerline{\!\!\!\!\!\!\!\!\!\!\!\!\!\!\!\!\!\!\!\!\!\!\!\!\!\!\!\!\!\!\raisebox{-0.96cm}{\includegraphics[width=0.1\textwidth]{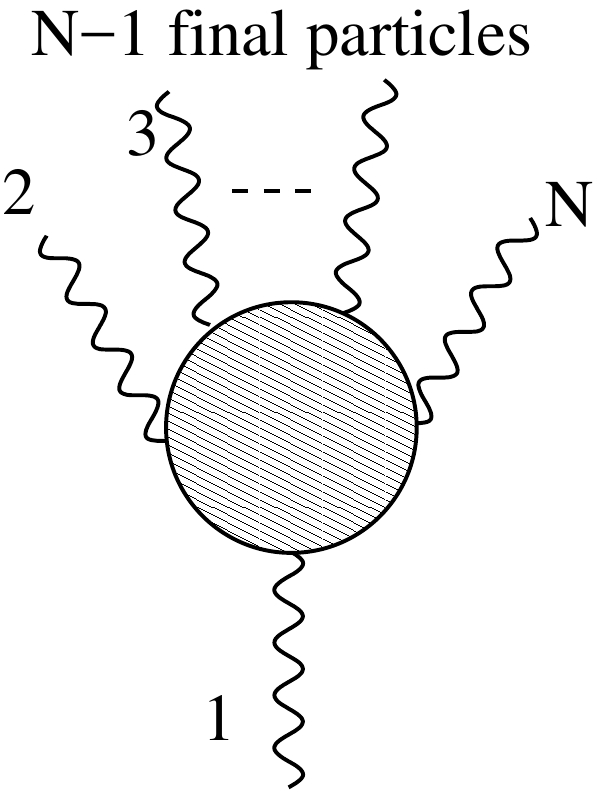}}
\parbox{0.3\textwidth}
{\bea &&\hbox{\!\!\!\!\!\!\!\!\!Rate of decays of 1 into $N-1$ particles:}\nn\\
d\Gamma &=& {1\over 2 E_1}|T|^2\prod_{i=2}^{N}
(2\pi)\delta(p_i^2-m_i^2)
{d^{\,4}p_i\over(2\pi)^4}\nn\\ &\times& (2\pi)^4\delta^{(4)}\left(p_1-\sum_{i=2}^N\,
p_i\right)
\label{eq17-10}
\eea}}
I have drawn them as though they are all external gauge bosons, but we will 
also discuss the case when some of them are quarks.  
 The shaded blob could contain any number of loops and internal
lines; it doesn't matter how complicated the diagram is.  Next to it I gave
the formula for the differential decay rate, that we know has dimensions of
mass or energy. Therefore we can determine the dimensions of the amplitude
$|T|$ that corresponds to the diagram.   By moving the factor $1/E_1$ to the
other side of the equation, we get 
\bea
\hbox{[energy$^2$]} &=& [|T|^2] \, [{\rm energy}]^{2(N-1)} \, 
[\hbox{energy}]^{-4}\nn\\
\hbox{hence\ \ \ \  [$|T|$]} &=& [E]^{4-N}
\label{eq17-11}
\eea
where $N$ is the total number of lines coming out, regardless of 
whether they are bosons or fermions.

If the amplitude $T$ had the same dimensionality as the amputated loop diagram, we would
conclude that $N_d = 4-N$, since the coupling $g$ is dimensionless, and so all dimensional
factors are associated with momenta.  In the case where all external particles were
bosons, this would be the right answer.  If some of them are fermions,
it is not right because the external fermions have spinors associated with them,
that have dimensions, and this makes the mass dimension of the diagram differ
from $N_d$.  But that is easy to correct for.  You recall that summing
the exterior product of two spinors over their polarizations gives the
projection operator $\sum u\bar u = \slashed{p}+m$; therefore each spinor
has dimension $1/2$.  This means we have to correct the previous result for
$N_d$ to read
\be
	N_d = 4 - {\cal N}_g - {\cal N}_q - \sfrac12 {\cal N}_q
\label{eq17-12}
\ee
in agreement with (\ref{eq17-9}).

I made a remark above, about gauge invariance, or possibly other symmetries,
 causing some diagrams to be more convergent than predicted by our formula for $N_d$.  The most
famous example is the case of ${\cal N}_g=2$, which are diagrams
contributing to the gluon vacuum polarization.  Gauge invariance tells
us that they should depend on the external gluon momentum $q$ as 
\be
	q^2\,\delta_{\mu\nu} - q_\mu q_\nu
\label{eq17-13}
\ee
This means that two factors of momentum that we counted toward the
degree of divergence are not loop momenta; instead they can be brought
outside of the integral, making it more convergent that we naively
estimated.  To remind ourselves of this possibility, we could add
an extra term $-{\cal P}$, so that 
\be
	N_d = 4 - {\cal N}_g - \sfrac32{\cal N}_q -{\cal P}
\label{eq17-14}
\ee
where ${\cal P}$ is the known power of the coefficient in front of the integral.

{\bf Exercise.}  Show that the power of couplings $g^P$ of an arbitrary 
diagram is given by 
\be
	P = 2\times (\hbox{\# of loops}) + {\cal N}_g + {\cal N}_q - 2
\ee
This is probably easiest to do in the usual normalization of
the fields, where 
the couplings appear in the conventional way, $g$ for the 3-particle
vertices and $g^2$ for the 4-gluon vertex. 

A fortunate consequence of the formula (\ref{eq17-14}) is that only a 
finite number of the different kinds of diagrams are divergent, when we 
classify them by their numbers of external lines.  Therefore we can make a
table to illustrate all the possibilities:
%\vskip-0.3cm
\begin{table}[h!]
\centering
\tabcolsep 2.5pt
\begin{tabular}{|c|c|c|c|c|c|c|}
\hline
${\cal N}_g$ & 1 & 2 & 3 & 4 & 0 & 1\\
\hline
${\cal N}_q$ & 0 & 0 & 0 & 0 & 2 & 2\\
\hline
$N_d$          & 3 & 2 & 1 & 0 & 1 & 0\\
\hline
${\cal P}$     &   & 2 & 1 & 0 & 1& 0 \\
\hline
\raisebox{0.cm}{${\hbox{typical}\atop\hbox{diagrams}}$} &
\includegraphics[width=0.03\textwidth]{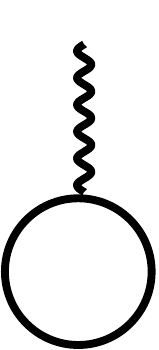}&
\includegraphics[width=0.03\textwidth]{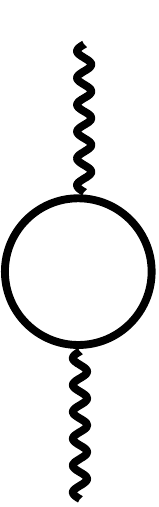}&
\includegraphics[width=0.033\textwidth]{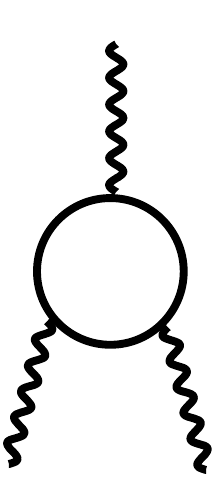} &
\includegraphics[width=0.033\textwidth]{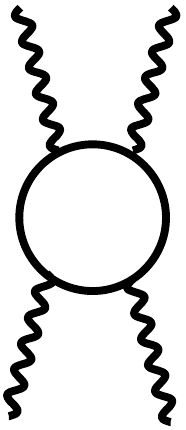} & 
\includegraphics[width=0.02\textwidth]{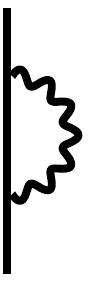}& 
\includegraphics[width=0.06\textwidth]{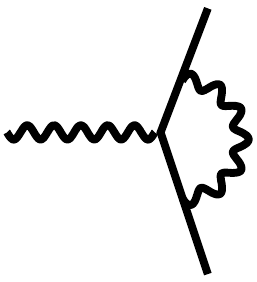}\\
 &
\includegraphics[width=0.03\textwidth]{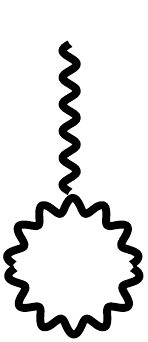} & 
\includegraphics[width=0.03\textwidth]{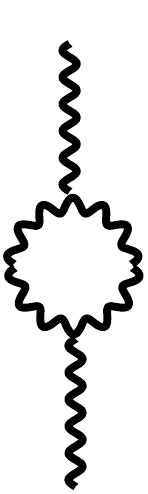} & 
\includegraphics[width=0.033\textwidth]{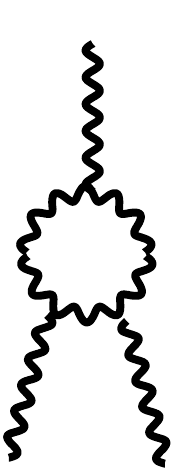} & 
\includegraphics[width=0.033\textwidth]{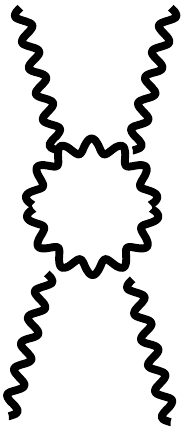} &
&
\includegraphics[width=0.06\textwidth]{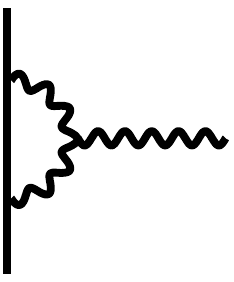}\\
\hline
\end{tabular}
%\caption{Superficial degrees of divergence, $N$, for graphs with
%${\cal N}_q$ external quarks and ${\cal N}_g$ external gluons.
%\label{tab1}}
\end{table}
\vskip-0.3cm
{\small Table I: Superficial degrees of divergence, $N_d$ (before accounting for ${\cal P}$), for graphs with
${\cal N}_q$ external quarks and ${\cal N}_g$ external gluons.  ${\cal P}$ is
the power of external momentum factors.}
\vskip0.3cm

The tadpole diagrams, with ${\cal N}_g=1$, provide another example of
our statement that symmetries can make a dia\-gram more convergent than
power counting would suggest,
\be
\raisebox{-0.25cm}
{\includegraphics[width=0.04\textwidth,angle=90]{tadpole}}\sim
\langle A_\mu\rangle\,,
\ee
This diagram is naively divergent with $N_d=1$, but in fact it vanishes.
One can think of it as an expectation value of the gluon field.  Such a thing,
if nonzero, would spoil Lorentz invariance, as well as gauge invariance.
And it would break discrete symmetries like C and P.

We already discussed the vacuum polarization diagram, the fact that
it is proportional to 
\be
\raisebox{-0.25cm}
{\includegraphics[width=0.04\textwidth,angle=90]{vacpole-g}}
\sim \delta_{\mu\nu}q^2 - q_\mu q_\nu 
\label{eq17-17}
\ee
We will derive this result later on.  It is an example of how
one must be careful about regulating the divergences from the
loops in a gauge invariant way.  
The behavior (\ref{eq17-17}) reduces its $N_d$ by 2, so that instead of being quadratically
divergent, it is only logarithmically divergent, but if the
regularization method failed to respect gauge invariance, it would be
afflicted with this more severe quadratic divergence. 

Moreover, symmetry prevents the diagram 
\be
\raisebox{-0.cm}
{\includegraphics[width=0.05\textwidth,angle=90]{3g-loop}}
\ee
from being linearly divergent as its $N_d=1$ would suggest.  The only 
kind of Lorentz-invariant loop integrand one could write, consistent with
this power counting, has the form $p\cdot X/(p^2)^2$, schematically,
where $p$ is the loop momentum and $X$ represents an external momentum
or polarization vector.
But $p_\mu/(p^2)^2$ is odd under $p_\mu\to -p_\mu$, so its integral
must vanish, as long as the regularization procedure does not introduce
any pathology that would spoil this reasonable expectation.  And we can 
also argue that it must vanish by Lorentz symmetry, since $\int d^{\,4}p\,
p_\mu/(p^2)^2$ would define some preferred direction in spacetime if it were
nonzero.  Therefore the 3-gluon amplitudes are also only logarithmically
divergent.  In fact gauge invariance provides yet another reason this must be
so: we know that the three-gluon interaction comes with a power of 
external momentum,
from its Finemensch rule, and this explains why the actual form of the integrand
must be $q_\mu/(p^2)^2$, giving $N_d=0$.

There is one possible caveat to the gauge invariance argument that should be
kept in mind however.  There is no guarantee that individual diagrams will be
gauge invariant; only the sum of all diagrams contributing to a given process at
a given order must necessarily be gauge invariant.

\subsection{Choice of gauge}

Before we embark on explicit calculations of loop dia\-grams,
I wanted to discuss the relative advantages of some choices of gauge relative to
others, when it comes to defining the gluon propagator.  In our earlier
discussion of gauge fixing, lecture \ref{sect12}, we discussed a particular
class of gauges $\eta\cdot A = 0$, involving an arbitrary four vector
$\eta_\mu$.  This is actually not the most convenient one for doing perturbative
calculations, even though it was conceptually appealing.  For one thing, 
it spoils Lorentz invariance temporarily, although these terms must cancel
out in the end.  

A simpler choice would be the Lorentz gauge 
$\partial_\mu A_\mu=0$.  Let us recall how 
the Faddeev-Popov procedure would work in this case.  The gauge-fixed path integral
takes the form
\be
	Z = \int e^{i\int \frac14(\partial_\mu A_\nu-\partial_\nu A_\mu -
A_\mu^{\,\times} A_\nu)^2} \,\Delta(A)\, 
	\delta[\partial_\mu A_\mu]\,{\cal D}A_\mu\,,
\ee
omitting for simplicity the quarks.  We rewrite the determinant $\Delta(A)$
as a path integral over ghost fields.  We take advantage of the fact that
this determinant does not change at all if we impose a slightly
different choice of gauge, $\partial_\mu A_\mu=f(x)$, with an arbitrary function
$f$.  Therefore we are free to do a weighted average over the path integral,
\be
	Z \to \int{\cal D}\!f\,e^{\frac{i}{2}\!\int\! f^2 d^{\,4}x}\, Z
\ee
Then the delta functional gets rid of $\int {\cal D}f$, and we are left with
a new term $\sfrac12(\partial_\mu A_\mu)^2$ in the Lagrangian, that allows the propagator
to be defined.  To see this, consider the modified equation of motion for
the gauge field, including a source term, 
\be
	\square A_\nu - \cancel{\partial_\nu \partial_\mu A_{\mu}} = S_\nu
\ee
and notice that the crossed-out term is removed by the new gauge-fixing term,
and therefore we may invert the $\square$ operator and solve for the
gauge field,
\be
	A_\nu = {1\over \square} S_\nu = {1\over k^2} S_\nu\,,
\ee
which shows that the propagator is simply $\delta_{\mu\nu}/k^2$ in this
gauge.  That obviously simplifies many perturbative computations,
compared to the axial gauge propagator.

Of course nothing obliges us to choose $e^{\frac{i}2\int f^2 d^{\,4}x}$ as
the weighting factor.  One can equally well take 
$e^{\xi\frac{i}2\int f^2 d^{\,4}x}$ with some arbitrary number $\xi\neq 0$.
This yields a more general class of covariant propagators of the 
form 
\be
	P_{\mu\nu} = {1\over k^2}\left(\delta_{\mu\nu} - \eta{k_\mu k_\nu\over
k^2}\right)
\ee
where $\eta$ is related in some simple way to $\xi$.\\
{\bf Exercise.}  Find the relation between $\eta$ and $\xi$.\\
The choice $\eta=0$ is known as Finemensch gauge.
Another very convenient choice is $\eta=1$, the Landau gauge.
It has the property of being transverse, $k_\mu P_{\mu\nu}(k) = 0$,
which leads to some simplifications in loop calculations.  For example
it greatly reduces the number of diagrams in the process we are going to 
consider next.\footnote{In the lecture RPF says it reduces the number from 17 to
4. Perhaps he had in mind that the number of terms in the 3-gluon vertex is
greatly reduced when taking only the transverse terms.  Also the ghosts decouple
in Landau gauge.}\ \ Nowadays you can find computer programs that will do the
symbolic algebra for you, for computing such diagrams.\footnote{Wolfram's SMP (Symbolic Manipulation
Program), the forerunner of
Mathematica, was in use at Pactech at this time.}   Nevertheless, it
is much easier to avoid mistakes if you can reduce the number of diagrams.

\subsection{Explicit loop calculations}

So far we have made numerous statements and expositions of a rather
general nature, without getting into the details of computing loop diagrams.
I would now like to go over some of those details, just to illustrate the
calculational techniques.  The example I will consider is the scattering of
two quarks.  To lowest order, as we know, it looks like
\be
\raisebox{-0.6cm}{\includegraphics[width=0.1\textwidth]{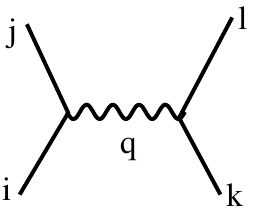}}
\sim {g_0^2\over q^2}\,{\lambda^A_{ji}\over 2}\,{\lambda^A_{lk}\over 2}
\ee
When we go to the next order, there are quite a few
 diagrams, including
\[
\includegraphics[width=0.45\textwidth]{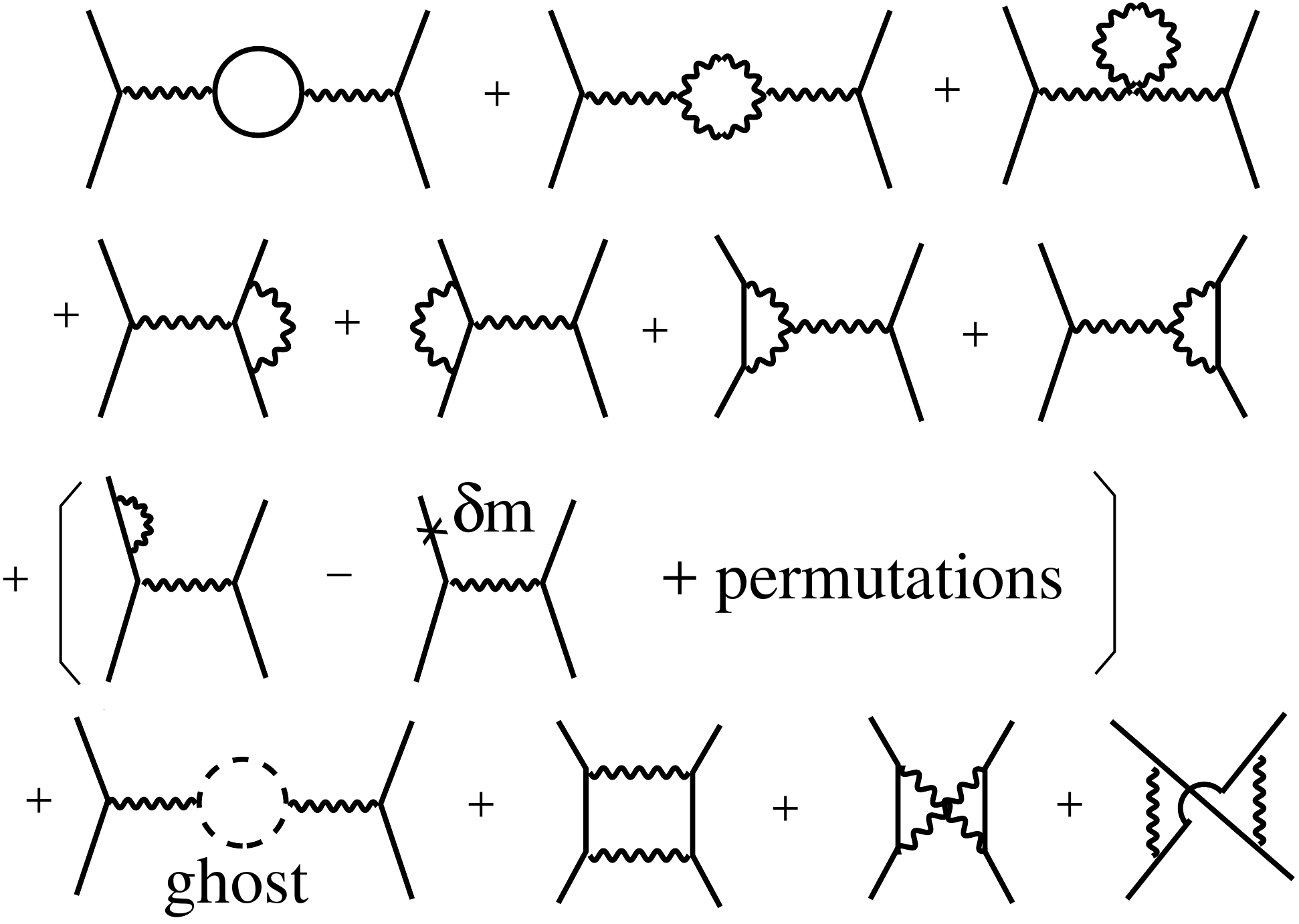}
\]
One simplification we can immediately make is to first isolate the primitive
divergences.  For example, consider the diagrams of the form
\[
\includegraphics[width=0.35\textwidth]{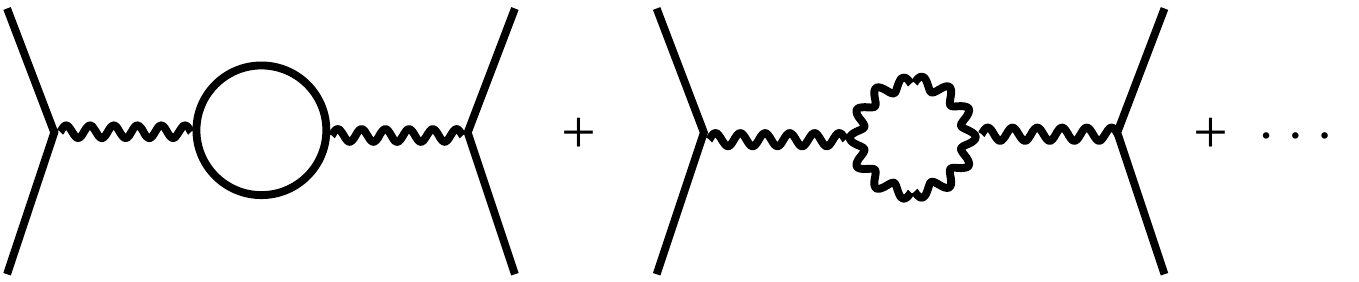}
\]
The interesting part of this calculation is the loop, not the external currents
nor the gluon propagators that connect them to the loop.  We might as well 
calculate the diagrams
\[
\includegraphics[width=0.4\textwidth]{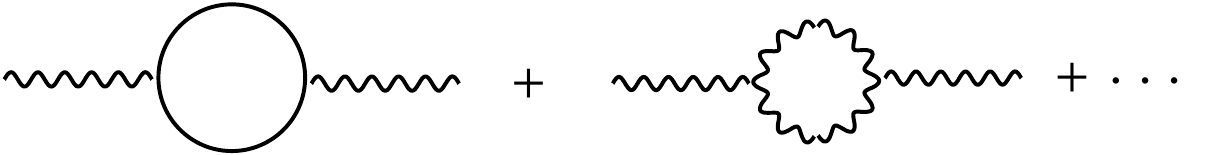}
\]
by themselves, since it is trivial to take that result and add to it all
the tree-level parts such as the gluon propagators and the spinors for the
external currents.

Now to illustrate the techniques, I am just going to compute the simplest 
and dullest of all of these, namely the quark loop contribution to the
vacuum polarization, 
\[
\includegraphics[width=0.2\textwidth]{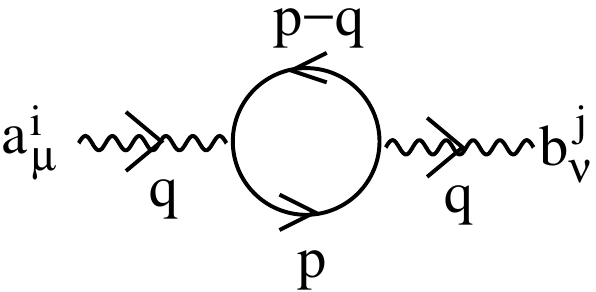}
\]
Once you understand the principles, it is just a matter of tedious effort
to compute the harder ones.\footnote{RPF mentions in the transcript that it is not
so straightforward to get a gauge invariant result for diagrams with gluon loops,
using Pauli-Villars regularization.} By applying the rules, we can write down the
expression for this diagram, 
\be
\label{eq17-25}
	g_{0}^2\left(-{\rm Tr}\int {d^{\,4}p\over (2\pi)^4}
	\,b^j_\nu {1\over \slashed{p} -\slashed{q}-m} {\lambda_j\over 2}
	\gamma_\nu {1\over \slashed{p} -m}{\lambda_i\over 2}
	\gamma_\mu a_\mu^i\right)
\ee
where I am taking $a^i_\mu$ and $b^j_\nu$ to be combined spin-color polarization
vectors for the external gluons, but if you prefer you could replace these
by spin polarization vectors $\epsilon_\mu$ and $\epsilon_\nu$, and consider
the gluon colors to be simply $i$ and $j$.  The trace here is a sum over all 
possible intermediate states, both spins and colors.  Therefore it is really
the product of 
two traces, one for the Dirac matrices and one for the color matrices.

Next, we should rationalize the quark propagators---multiply numerator and
denominator by $(\slashed{p}+m)$---and carry out the traces.
This looks like
\bea
\label{eq17-26}
&&\ \ 	\underbrace{\tr\left[{1\over \slashed{p}-\slashed{q}-m}\gamma_\nu
{1\over \slashed{p}-m}\gamma_\mu\right]}\,\ \ 
\underbrace{\tr\left(\lambda_i\lambda_j\over 4\right)}\\
&&
{{\rm tr}\left[(\slashed{p}-\slashed{q}+m)\gamma_\nu 
	(\slashed{p}+m)\gamma_\mu\over [(p-q)^2-m^2][p^2-m^2]\right]}
 \! 	\qquad{\delta_{ij}\over 2}\nn\\
&&= 2{\Big[(p\!-\!q)_\mu p_\nu + (p\!-\!q)_\nu p_\mu\! -\!\delta_{\mu\nu}
(p\!\cdot\!(p\!-\!q) - m^2)\Big]\over [(p-q)^2-m^2][p^2-m^2]}\delta_{ij}\nn
\eea
Notice that the gluon color is conserved by the loop.

Now we are left with the integral, that has exactly the same form as in 
QED.   There is a famous trick for combining the denominators, that 
I adapted from Schwinger by eliminating a step from his Gaussian integral
method,\footnote{In my notes at this point I have written ``stolen from 
Schwinger by eliminating a step from his Gaussian integral method.''}
\be
\label{eq17-27}
	\int_0^1 {dx\over [ax+b(1-x)]^2} = {1\over ab}
\ee
This allows us to combine the propagators into the form
\bea
	&&{1\over (p^2 - 2p\cdot q + q^2-m^2)(p^2-m^2)} = \qquad\qquad\quad\nn\\
	&&\qquad\int_0^1 dx\,{1\over[p^2 - 2 p\cdot qx + q^2 x - m^2]^2}
\label{eq17-28}
\eea 
so that the loop integral becomes
\bea
\label{eq17-29}
	\int_0^1 dx\!\!\!\!\!\! &&\int {d^{\,4}p\over (p^2 - 2 p\cdot qx + q^2 x -
m^2)^2} \\
	\times&&\left(2 p_\mu p_\nu - p_\mu q_\nu - p_\nu q_\mu -
	\delta_{\mu\nu}(p^2 - p\cdot q - m^2)\right)\nn
\eea
To further simplify it, we wish to complete the square in the denominator,
by shifting the integration variable by $p\to p+ qx$:
\bea
\label{eq17-30}
	\int_0^1 dx\!\!\!\!\!\! &&\int {d^{\,4}p\over (p^2 + q^2
x(1-x) - m^2)^2} \\
	\times&&\Big(2p_\mu p_\nu + (p\cdot q\hbox{\ terms})
	+2 q_\mu q_\nu\, x(1-x)\nn\\
	&-& \delta_{\mu\nu}(p^2 - q^2x(1-x) - m^2 + p\cdot q\hbox{\ terms\ }
	)\Big)\nn
\eea
Once it is in this form, it is not necessary to keep careful track of the
$p\cdot q$ terms in the numerator, since they are odd in $p$ and integrate to
zero.

\subsection{Regularization}

Everything so far seems perfectly innocuous and standard, but if we want
to be careful, you will notice that I have cheated.  The integral is divergent,
so how do we know that the step of shifting the integration variable
by $p\to p+ qx$ is legitimate?  It could conceivably change the result
in some unphysical way, unless we have carefully defined what we mean by
this integral.  To be rigorous, we must specify exactly how we are going
to regularize the integral, to cut off the ultraviolet divergence.  

The method I want to use in this lecture is the historical one, invented
by Pauli and Villars.  As I alluded earlier, it turns out to be too 
simplistic to change the propagator as in (\ref{eq15-10}).  This was attempted
by some of the early workers in the field, and it was found to spoil gauge
invariance.  Instead, one needs to apply this prescription to the {\it whole
amplitude}.  The consistent way is to replace the quark mass in the denominators
by $m^2\to m^2+ \Lambda^2$, and subtract the resulting expression from the
original amplitude.  The new integrand obtained in this way has good behavior
in the ultraviolet, and so the procedure of shifting the integration variable
is perfectly consistent, and moreover it preserves the gauge invariance, as we
will see.  And it is also consistent with our assumption that integrals like
\be
	\int d^{\,4} p\, {p_\mu\over p^2-m^2} = 0
\label{eq17-31}
\ee
should vanish, with the understanding that this is now a shorthand for the
fully regulated expression, where we have subtracted the corresponding term
with $m^2\to m^2+ \Lambda^2$.

The statement (\ref{eq17-31}) looks trivial, but we can use it to derive
a more interesting result, now that we are confident that shifts in the
integration variable are legitimate.  By shifting $p\to p-a$, we obtain
\bea
	\int d^{\,4} p\,
{p_\mu\over (p-a)^2-m^2}
	&=& a_\mu\int{ d^{\,4} p\,\over (p-a)^2-m^2}\nn
\eea
And then by differentiating with respect to $a_\nu$ and setting $a_\mu=0$,
we get the useful identity
\be
	\int d^{\,4} p\,
{\delta_{\mu\nu}(p^2-m^2)  - 2p_\mu p_\nu\over (p^2-m^2)^2} = 0
\label{eq17-32}
\ee
Incidentally, this could also be obtained more directly, using
\be
	\int d^{\,4} p\, {\partial\over\partial p_\nu}
	{p_\mu\over p^2-m^2} = 0\nn
\ee
which should be true for the integral of the derivative of anything.
We could be suspicious of such a statement in the unregulated theory, since
the surface term might fail to vanish, but it rigorously vanishes in the
regulated theory.

Let us now return to the calculation we started above, the computation of the vacuum 
polarization diagram.  We want to reorganize the numerator of (\ref{eq17-30})
so that it has one term in the same form as (\ref{eq17-32}):
\bea
 \int\!\! {d^{\,4}p\over (p^2\!\! +\!\! q^2
x(1-x)\!\! -\!\! m^2)^2}&&\!\!\!\!\!\!\! \Big[ 
	2p_\mu p_\nu \!\!-\!\!\delta_{\mu\nu}(p^2\!\! +\!\! q^2x(1-x)
	\!\!-m^2)\nn\\
	&& \!\!\!\!\!\!-2 q_\mu q_\nu x(1-x)\!\! +\!\! 2\delta_{\mu\nu} 
q^2x(1-x) 
\Big]\nn
\eea
This has the pleasing feature that the first term, which appears to be
quadratically divergent, actually vanishes.  We are left with the second
term, having the form (\ref{eq17-13}) that I said must arise as a
consequence of the gauge symmetry:
\be
	2(q_\mu q_\nu -\delta_{\mu\nu}q^2)\int_0^1 x(1-x)dx
\int {d^{\,4}p\over (p^2+q^2x(1-x)-m^2)^2}
\ee
This leaves the logarithmically divergent integral, whose evaluation I will
take up in the next lecture.

\section{Renormalization, continued (1-14-$xy$)}
\label{lec18}

To remind you, we were computing the one-loop corrections to quark-quark
scattering, and had noted that it is convenient to factorize the amplitude
with the gluon vacuum polarization correction 
in the form
\be
\raisebox{-0.4cm}{\includegraphics[width=0.1\textwidth]{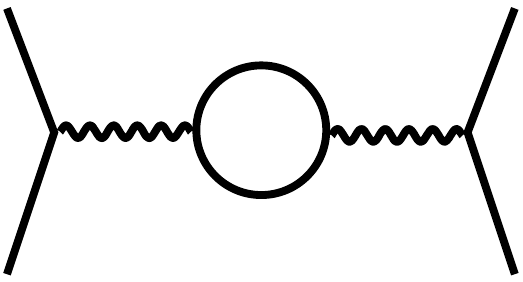}}
\  =\ 
J^i_\mu\, g_0^4 \, {1\over q^2}\,  B\,  {1\over q^2}\,  J^i_\mu
\label{eq18-1}
\ee
where $B$ represents the simpler diagram 
$\raisebox{-0.1cm}{\includegraphics[width=0.07\textwidth]{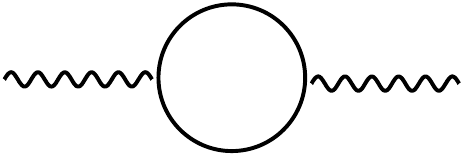}}$ with the
gluon lines amputated, and $J^i_\mu$ are the external quark currents, with
color indices $i$.  We found that $B$ takes the form
\bea
\label{eq18-2}
	B &=& 2(\delta_{\mu\nu}q^2 - q_\mu q_\nu) I,\\
	I &=& \int_0^1 dx\,x(1-x)\int {d^{\,4}p/(2\pi)^4
	\over [p^2 - m^2 + q^2 x(1-x)]^2}\nn
\eea

Because the currents are conserved, $q\cdot J=0$, we can simplify 
(\ref{eq18-1}) slightly,
\be
\raisebox{-0.4cm}{\includegraphics[width=0.1\textwidth]{qq-vacpol-q}}
\  =\ 
2 J_\mu^i\, g_0^4 \, {I\over q^2}\,  J^i_\mu
\label{eq18-3}
\ee
which has the same form as the tree-level contribution,
\be
\raisebox{-0.6cm}{\includegraphics[width=0.1\textwidth]{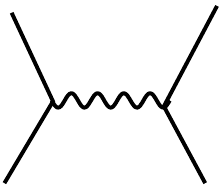}}
= J_\mu {g_0^2\over q^2}J_\mu
\label{eq18-4}
\ee
Written in this way, it is clear that the loop contribution can be expressed as
a change in the coupling constant, 
\bea
	g_0^2&\to& g_0^2 + 2 g_0^4 I\nn\\ &\to& {g_0^2\over 1 - 2 g_0^2 I}
\equiv g^2_{\rm eff}
\label{eq18-5}
\eea
where the first arrow indicates a result that is 
consistent to the order of perturbation theory at which we are working,
while the second one uses the hindsight of resumming the leading logs, 
that we discussed in the last lecture.

Now to do the logarithmically divergent integral $I$, I will continue to use
Pauli-Villars regularization, although later on we will introduce the
more elegant method of dimensional regularization.  Hence we must subtract
from $I$ the similar quantity with the modified propagator
\be
	{1\over (p^2 - (m^2+\Lambda^2) + q^2 x(1-x))^2}
\label{eq18-6}
\ee
Rather than directly subtracting, there is a nicer way to implement this,
by first  thinking of $I$ as a being a function of $m^2$, 
\be
	I(m^2) = \int_0^1 dx\,x(1-x)\int {d^{\,4}p/(2\pi)^4
	\over [p^2 - m^2 + q^2 x(1-x)]^2}
\label{eq18-7}
\ee
and then differentiating with respect to $m^2$.  Doing this makes the 
integral convergent, even without subtracting anything.  If we then integrate
$I'$ with respect to $m^2$, 
\be
	-\int_{m^2}^{m^2+\Lambda^2} I'(M^2)\,dM^2 = 
	I(m^2) - I(m^2+\Lambda^2)
\ee
the result is the subtraction we originally wanted to carry out.  The trick then
is to evaluate the convergent integral appearing in $I'(M^2)$, and postpone
doing the integral over $M^2$ until afterwards:
\bea
\int_{m^2}^{m^2+\Lambda^2}\!\!\!\!\!\!\!\!\!\!dM^2\int_0^1dx\,x(1-x)&&\!\!\!\!\!
\underbrace{\int{d^{\,4}p/(2\pi)^4\over [p^2+q^2x(1-x) -M^2]^3}}\nn\\
&&\!\!\!\!\!\!\!\!\!\!\!\!\!\!\!\equiv\int{d^{\,4}p\over (2\pi)^4} {1\over (p^2-L)^3}={1\over 32\pi^2 iL}\nn\\
\eea
In performing the integral, I have glossed over a few steps that I assume you are already
familiar with, from a previous course on quantum field theory, notably doing the
Wick rotation to avoid the poles from the $i\epsilon$ prescription, which
gives the factor of $1/i$.

Next we carry out the integral over $M^2$,
\bea
\int_{m^2}^{m^2+\Lambda^2}dM^2\!\!\!\!\!\!\!&&{1\over 16\pi^2 i}\,{1\over
M^2-q^2x(1-x)}\\
&=& {1\over 16\pi^2 i}\ln\left(m^2+\Lambda^2 - q^2x(1-x)\over
	m^2- q^2x(1-x)\right)\,.\nn
\eea
Since we are ultimately interested in the limit as $\Lambda\to\infty$, 
this can be simplified by ignoring the finite terms in the numerator
of the argument of the logarithm.  As we discussed before, the $\ln\Lambda$
divergence gets absorbed into the tree-level contribution
by redefining the bare coupling $g_0$.

To further simplify the discussion, I would like to consider momenta
such that $-q^2\gg m^2$, so that we can ignore the quark mass.  You might
be concerned that this could give rise to an infrared divergence from
the places where $x=0$ or $1$ when we perform the integral over $x$, but
because the integrand is a log, these singularities are integrable and lead to no
difficulty.  In this approximation, we have
\bea
 \int {dx\,x(1-x)\over 16\pi^2 i}\!\!\!\!\!\!\!\!\!\!\!&&\left[\ln\left(\Lambda^2\over
-q^2\right) -\ln x -\ln(1-x)\right]\nn\\
&=& \left(\frac16\ln\left(\Lambda^2\over
-q^2\right) - \frac{5}{18}\right)
\eea 
Remember that  $-q^2\equiv Q^2$ is positive, since $\vec q$ is the momentum transfer in 
the electron-electron scattering.  The integral of the logarithm can be
done using integration by parts.

Now we have evaluated the integral $I$, and we can put it back into the
expression (\ref{eq18-3}): 
\bea
\label{eq18-12}
\!\!\!\!\!\!\!\!\!\!\!\!J{1\over Q^2}J\left[\rule{0cm}{7mm} g_0^2 \right.&+& \left.{g_0^4\over 16\pi^2}
	\left(\Delta\beta_0\left(\ln{\Lambda^2\over Q^2}\right) + a\right)
\right]\\
\hbox{where\ } \Delta\beta_0 &=& -\frac23 n_f\hbox{\qquad and\ } a = \frac{10}9
\, n_f
\label{eq18-13}
\eea
where $n_f$ is the number of quark flavors having mass less than $Q^2$;
otherwise our approximation $Q^2\gg m^2$ is not valid.  Here $\Delta\beta_0$
is just the quark contribution to $\beta_0$; the full $\beta_0$ gets an 
additional contribution of $+11$ from the gluon loop, that we are not
calculating here.

We can now see more explicitly how the renormalization of the bare
coupling is derived, which absorbs the dependence on the cutoff arising
from the loop diagram.  Consider the effective coupling defined in
(\ref{eq18-5}), 
\bea
g_{\rm eff}^2 &=&	{g_0^2\over {1-{g_0^2\over 16\pi^2}\beta_0\ln{\Lambda\over Q^2}}}
\nn\\
&=& {16\pi^2\over \left(16\pi^2\over g_0^2\right) - \beta_0\ln{\Lambda^2\over
Q^2}}\,.
\label{eq18-14}
\eea
The second line makes it clear how $g_0^2$ must depend on $\Lambda$ in order
that $g_{\rm eff}$ be independent of $\Lambda$, 
\be
	{16\pi^2\over g_0^2(\Lambda)} \equiv {1\over \hat g_0^2(\Lambda)}
	= \beta_0\ln{\Lambda^2\over\lambda_P^2} + c_1
\label{eq18-15}
\ee
where I have inserted a renormalization scale $\lambda_P$, since 
$g_0$ is a Lagrangian parameter that cannot depend on the 
external momentum $Q$.  As I previously mentioned, we are free to
choose a convention for defining $\lambda_P$ such that the arbitrary
constant $c_1$ vanishes, if so desired.

\subsection{Effective Lagrangian perspective}

Although we imagined that the value
of $g_0^2$ has been fixed in this example by comparing to a particular
observable, the scattering cross section for two quarks, 
it is important to emphasize that  once
$g_0^2(\Lambda)$ has been determined, it is now valid for the study of
any process; we do not need to define a separate $g_0^2(\Lambda)$ for
every different observable.  One way to understand this is from the fact that
there is a finite number of primitively divergent diagrams in the theory,
repeated in this table,
\[
{\includegraphics[width=0.5\textwidth]{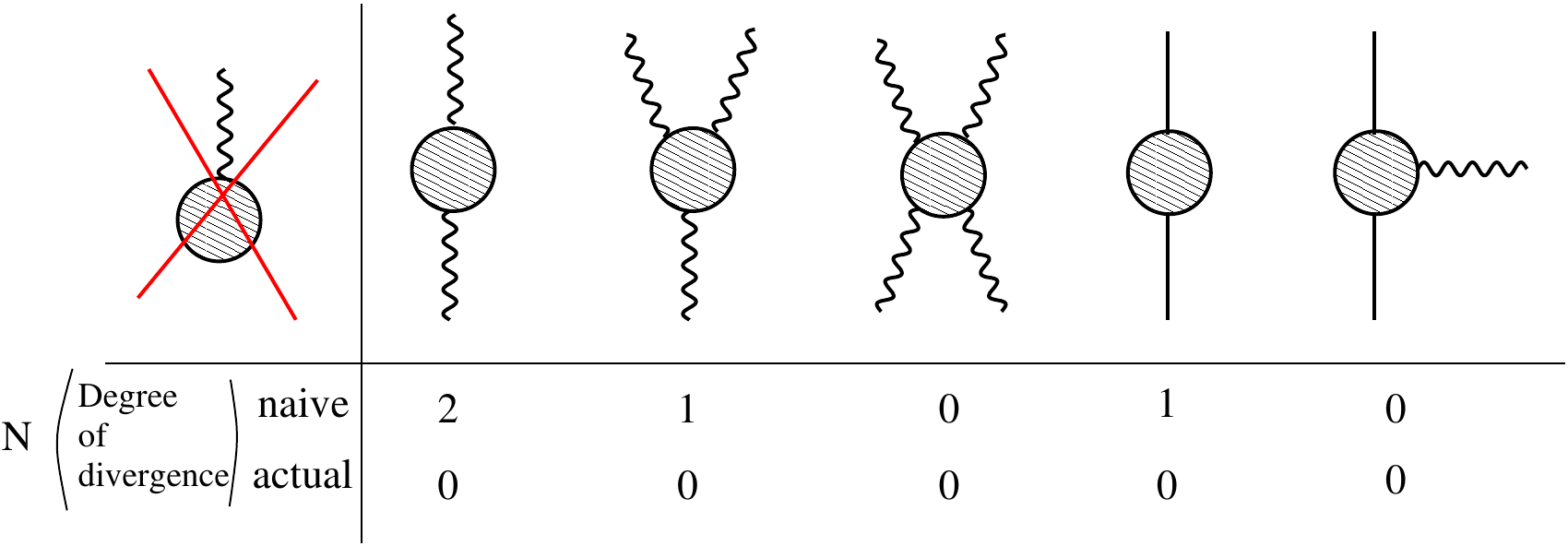}}
\]
You recall that the tadpole diagram vanishes.  The renormalizability of 
the theory implies that all the divergent diagrams can be related to
an effective Lagrangian contribution that has the same form as the 
bare Lagrangian.  So for example, all the divergent diagrams with only external 
gluons must correspond to terms in the Lagrangian in this manner:
\bea
\centerline{\includegraphics[width=0.35\textwidth]{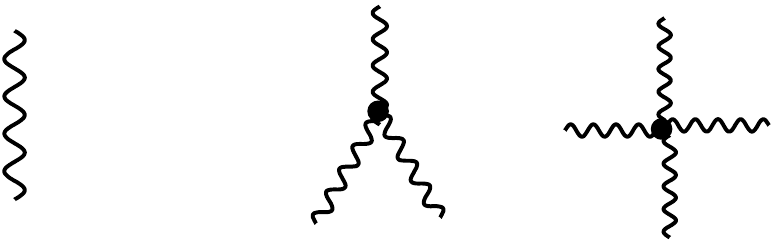}}\nn\\
(\partial_\mu A_\nu - \partial_\nu A_\mu)^2\qquad
A_\nu^{\,\times} A_\mu^{\,\bigcdot}\partial_\mu A_\nu\qquad\ \  
A_\nu^{\,\times} A_\nu^{\,\bigcdot} A_\mu^{\,\times} A_\nu\nn\qquad
\eea
{\it A priori,} we would have to renormalize three different
parameters to absorb the divergences.  But because of gauge invariance,
we know that they must organize themselves into the form 
$\tr\, F_{\mu\nu}F_{\mu\nu}$, that depends only on the single parameter
$g_0$.  Therefore these divergences are related to each other in 
such a way that  they can all be absorbed by the renormalization
of the single parameter $g_0$.  This is perhaps easiest to see in the
convention where we keep $g_0$ out of the field strength definition and
put it as a prefactor in the Lagrangian, 
$(1/g_0^2)\tr\, F_{\mu\nu}F_{\mu\nu}$.  Then the three divergences indicated
above would all contribute to the shift in the effective Lagrangian that goes as
\be
	\ln\Lambda^2 \, \tr\, F_{\mu\nu} F_{\mu\nu}
\ee
This of course assumes that the regularization method did not spoil gauge
invariance.  Otherwise we would have to fudge the results to make this
work out.

A similar argument applies to the terms with external quarks.  
The self-energy diagram requires us to renormalize an additional
quantity, the quark mass.  But the vertex correction does not need anything new;
gauge symmetry guarantees that the same renormalization of $g_0$ as needed
for the gluons also suffices for the coupling to quarks.

To make this clearer, let's rephrase these statements in the context
of the path integral.  We could imagine, if we were sufficiently adept,
being able to carry out the integral over quarks for a fixed gauge 
field background,
\bea
\int {\cal D} A_\mu\, e^{i{1\over g_0^2}\int F_{\mu\nu} F_{\mu\nu}}&&
\!\!\!\!\!\!\int {\cal D}\psi {\cal D}\bar\psi\, e^{i\int\bar\psi
(i\slashed{D}-m)\psi}\nn\\
	=\int {\cal D} A_\mu\, e^{i{1\over g_0^2}\int F_{\mu\nu} F_{\mu\nu}}
	&&\!\!\!\!\!\!\underbrace{\det\left(i\slashed{D}-m\right)}\\
&& G(A) \sim e^{i\ln\Lambda^2 \int F_{\mu\nu} F_{\mu\nu}}\nn
\eea
giving a functional determinant of the covariant Dirac operator, that I am
calling $G(A)$.  Even though we can't calculate $G(A)$ exactly, from
perturbation theory we know that it has a divergent contribution like I have
indicated, which has the same form as the tree-level gluon action.  We can
then rewrite the bare coupling in terms of a renormalized coupling, which
is finite as $\Lambda\to\infty$, plus a correction designed to cancel the
$\ln\Lambda$ divergences.  In the convention I have chosen above, it would
be easier to think of it as a correction to $1/g_0^2$:
\be
 {1\over g_0^2} ={1\over g_{\rm ren}^2} + \delta\left(1\over g_0^2\right) 
\ee
These extra terms labeled  as $\delta(1/g_0^2)$ are known as the
{\it counterterms}.

There is an interesting consequence of the running for the convergence
properties of the loops when we start to think
about the higher-order contributions, such as
\[
{\includegraphics[width=0.35\textwidth]{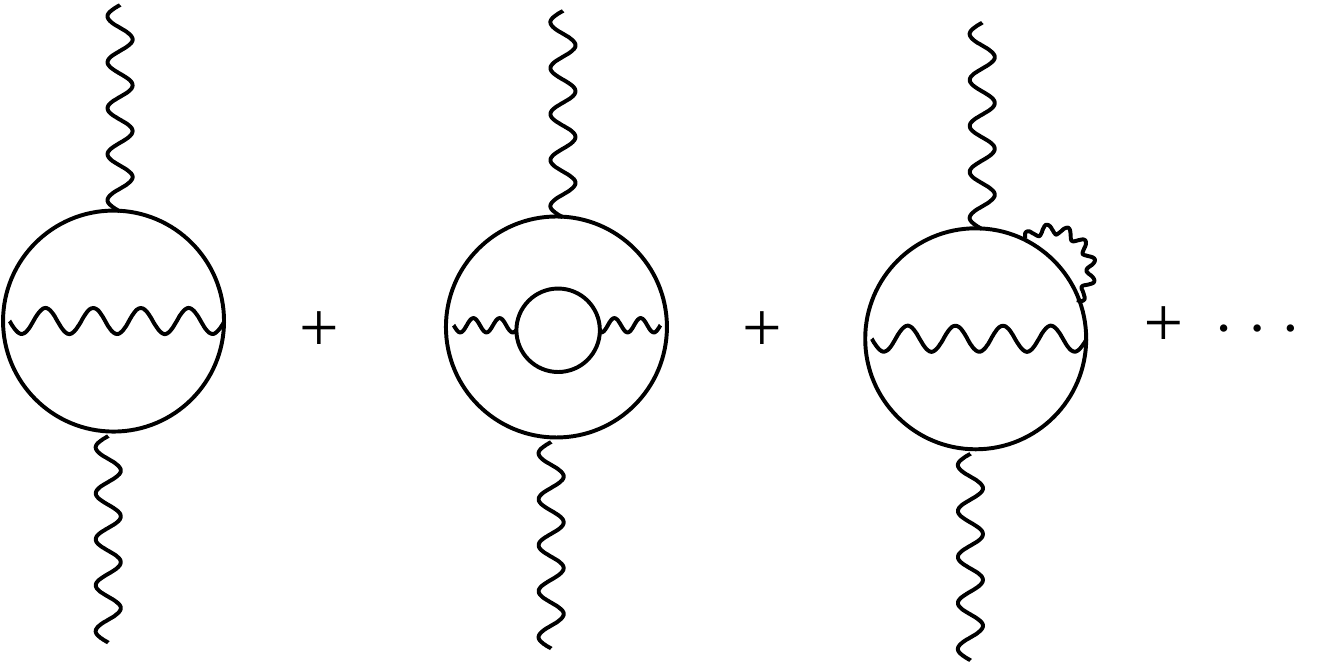}}
\]
One can think of such diagrams as though they were at one order
lower in perturbation theory, but 
constructed from propagators that have already been dressed at one 
loop,
\be
\raisebox{-0.25cm}{\includegraphics[width=0.25\textwidth]{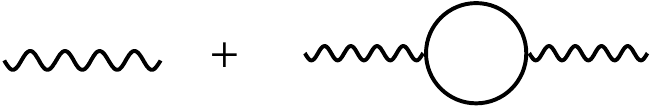}}
\ +\ \dots\ \sim\  {1\over Q^2\ln
Q^2}
\ee
An integral that is normally considered to be logarithmically divergent would
instead behave like  
\be
\int^\Lambda {dp\, p\over p^2 \ln p^2} \sim \ln(\ln\Lambda^2)
\label{eq18-21}
\ee
This is still divergent, but more mildly so.  And going to higher order,
one could get higher powers of logs in the denominator, which would make
the integral converent,
\be
\int_c^\Lambda {dp\, p\over p^2 \ln^2 p^2} \sim 
\int_c^\Lambda {d\ln p^2 \over \ln^2 p^2}  \sim {1\over\ln c} - {1\over
\ln\Lambda}
\label{eq18-22}
\ee
This could be understood as a consequence of replacing the vertices at the ends of the
dressed propagator by the dressed vertices, that  behave as $\alpha(Q)\sim 1/\ln(Q^2)$ to
give this more convergent behavior.

\subsection{Misconceptions}

I would like to discuss a different viewpoint of the running coupling, that
you may encounter in the literature, and that I consider to be misguided.
The idea is to choose some physical amplitude---suppose for simplicity that
at tree level it is linear
in $\alpha$---and to consider it as a function of $Q$.  One could then 
{\it define} a running coupling $\ul{\alpha}(Q^2)$ to be exactly determined by
this physical observable.  An example would be the amplitude for the scattering
of two quarks, to all orders in perturbation theory,
\be
\raisebox{-0.8cm}{\includegraphics[width=0.05\textwidth]{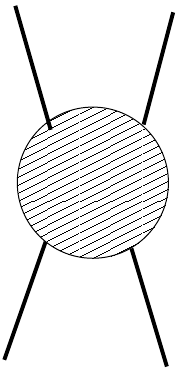}}
\  \equiv \ {\ul{\alpha}(Q^2)\over Q^2}
\label{eq18-24}
\ee
However there would be no such thing as perturbation theory in regard
to this particular process, since there is nothing to expand in:
$\ul{\alpha}(Q^2)$ is the exact result.  Another example is the correction $R-1$ for the process $e^+e^-\to$ hadrons that we discussed
previously.  It gives rise to a different definition of the coupling,
call it 
\be
	\ul{\ul{\alpha}}(Q^2) = \pi(R-1)
\ee
If we were to compare these two definitions, we would find out that they
approximately agree, and a useful way to compare them would be by
differentiating and trying to reconstruct the beta function.   We would
find that both definitions satisfy equations of the form
\be
	{d\ul{\alpha}\over d\ln Q^2} = \beta_0\ul{\alpha}^2 + 
\beta_1\ul{\alpha}^4 +
\beta_2\ul{\alpha}^6 + \dots
\ee
and that the first two coefficients $\beta_0$, $\beta_1$ agree for both
definitions.  But beyond that, the remaining coefficients are in general
different between the two definitions, and unrelated to the values of $\beta_0$, 
$\beta_1$.
Contrast that to the definition I made,
\be
	{d\alpha\over d\ln Q^2} = {\beta_0\alpha^2\over
1-{\beta_1\over\beta_0}\alpha^2} = \beta_0\alpha^2 + \beta_1\alpha^4 +
{\beta_1\over\beta_0}\alpha^6 + \dots
\ee
where the higher coefficients are all determined.  In our procedure, the
amplitudes for the two processes have to both be calculated perturbatively
in $\alpha$.   This is how physics should work: we have a definite theory that
is independent of the process and we predict the observable from it.  The theory
should not be predicated on one particular process or another.

Adding to the confusion caused by such proposals is the misconception that 
there is a momentum-dependent coupling constant in the Lagrangian.  
As I explained before, the $Q$ dependence in $\alpha(Q)$ is just a shorthand
to remind us how the loop-corrected amplitude depends on $Q$, that we can
deduce by replacing $\lambda_P\to Q$ in the definition of $g_0(\Lambda)$.
But the actual coupling that goes into the Lagrangian is $g_0(\Lambda)$, which
does not depend on $Q$.

\subsection{Dimensional regularization}

Let me finish by giving a preview of how the logarithms arise in dimensional
regularization, that we will discuss in more detail in the next lecture.
Because the action must be a dimensionless quantity, when we continue the
dimension of spacetime to some value $D = 4-\epsilon$, the coupling that was
dimensionless in $D=4$ is no longer so. For this discussion 
 I will adopt the field normalization where $F\sim \partial A + [A,A]$ and the
coupling constant appears in front of $F^2$.  Because of this form of $F$, there is
no choice but to say that the dimensions of $A$ are $[A] = 1/[L] = [M]$, and therefore
the action looks like
\bea
	{1\over g^2}\!\!\!\!\!\!\!&&\underbrace{\int F_{\mu\nu}^2 \,d^{\,D}x}\nn\\
	&& \hbox{dimensions of M$^{4-D}$ = $[M^\epsilon]$}
\eea
Therefore to make the action dimensionless, we must have 
\be
	[g^2] = [M^\epsilon]
\ee
To make this explicit, it is convenient to relate $g$ to a dimensionless coupling
$g_0$ and a mass scale that I suggestively call $\lambda_P$,
\be
	g^2 = g^2_0 \lambda_P^\epsilon
\ee

Now imagine redoing the calculation that led to eq.\ (\ref{eq18-5}) using
dimensional regularization.  The result takes the form
\bea
\label{eq18-31}
  g^2 + g^4 Q^{-\epsilon}\left (2\beta_0\over\epsilon\right)
&\to& {1\over {1\over g^2} - {2\beta_0 Q^{-\epsilon}\over \epsilon}}\\
&\approx& {1\over {2\beta_0\over\epsilon}\lambda_P^{-\epsilon} \nn
	- {2\beta_0\over\epsilon}Q^{-\epsilon}}
\eea
in which the $\ln\Lambda^2$ divergences of Pauli-Villars get replaced
by $1/\epsilon$ poles.  Here  I have resummed the leading logs and
observed that necessarily $g_0^{-2} = 2\beta_0/\epsilon + {\rm const.}$, in order to 
cancel the divergence.  Then we notice that 
\be
  {2\over\epsilon}(\lambda_P^{-\epsilon} - Q^{-\epsilon}) = \ln(Q^2/\lambda_P^2) +
O(\epsilon)
\ee
just like the outcome of the cutoff method of regularization.

\section{Renormalization (conclusion); Lattice QCD (1-19-$xy$)}

I would like to add something to our previous discussion concerning the
renormalization group equation.  Recall that the definition
\be
\label{eq19-1}
{16\pi^2\over \beta_0 g^2(\Lambda^2)} - {\beta_1\over\beta_0^2}\ln {16\pi^2\over
\beta_0 g^2(\Lambda^2)} = \ln{\Lambda^2\over\lambda_P^2}
\ee
tells us (in an implicit way) how the coupling $g$ must depend on the cutoff
$\Lambda$.  Now imagine some physical process, whose amplitude ${\cal M}$---for definiteness
I will pick an
example where it is dimensionless---we compute
from the theory with the cutoff, using the  coupling $g_0$.  ${\cal M}$ is generically
a function of $g_0$, some momenta which for simplicity I will 
represent by a single scale $Q$, and the cutoff, of the form
\be
\label{eq19-2}
	{\cal M}_{\rm theory}\left(g_0^2,\ln{\Lambda^2\over Q^2}\right)
\ee
The explicit $\Lambda$ dependence goes away, if we replace $g_0\to g(\Lambda)$ 
using the coupling defined by eq.\ (\ref{eq19-1}).  In perturbation theory, we would find that
\bea
	{\cal M}_{\rm theory} &=& g_0^2 + g_0^4\left(\beta_0\ln{\Lambda\over Q} +  
	c_1\right) +\dots\nn\\
	&=&  \alpha(Q^2) + c_2\alpha^2(Q^2)
\label{eq19-3}
\eea
Now since ${\cal M}_{\rm theory}\left(g^2(\Lambda),\ln{\Lambda^2/ Q^2}\right)$
is independent of $\Lambda$, we are free to set $\Lambda^2 = Q^2$.  Then
the physical amplitude is
\be
	{\cal M}_{\rm phys} = {\cal M}_{\rm theory}(g^2(Q^2),0)
\label{eq19-4}
\ee
where $g^2(Q^2)$ is {\it defined} by eq.\ (\ref{eq19-1}), or explicitly
\be
\label{eq19-5}
{16\pi^2\over \beta_0 g^2(Q^2)} - {\beta_1\over\beta_0^2}\ln {16\pi^2\over
\beta_0 g^2(Q^2)} = \ln{Q^2\over\lambda_P^2}
\ee
and now the physical amplitude is
\bea
\label{eq19-6}
	{\cal M}_{\rm phys} &=&  g_0^2(Q^2) + c_1 g_0^4(Q^2)\left(
0 +  
	c_2\right) +\dots\nn\\
\eea
Comparing to (\ref{eq19-3}), we can see that 
\be
\label{eq19-7}
	{g_0^2(Q^2)\over 16\pi^2} = \alpha(Q^2)
\ee
since
\bea
\label{eq19-8}
	{\cal M}_{\rm phys} &=&  \alpha(Q^2) + c_2 \alpha^2(Q^2) +\dots
\eea
This makes clear what is the correct interpretation of the running coupling constant that
I was criticizing in the previous lecture.  The $Q^2$ dependence is not present
in any fundamental coupling in the Lagrangian, but rather it arises from taking
advantage of the $\Lambda$-independence of the physical amplitude and using our
freedom to set $\Lambda$ equal to the relevant scale of the process, to get rid
of the log.\footnote{The preceding sentences are not in my notes, but seem to be
the logical connection to the previous lecture.}\ \     Of course, this is a consequence of solving
the renormalization group equations, but this point of view seems to me simpler
and more intuitive than the RG equations.

\subsection{Lattice QCD}

Now we will move on to the main subject of this lecture, which is a comparison
of different kinds of cutoff schemes, including the lattice and dimensional
regularization.\footnote{RPF called it ``dimensional
renormalization'' in his lectures, but would have adopted the more common
terminology for the version to be published.}\ \   As we have mentioned before,
the path integral
\be
\label{eq19-9}
	\int e^{iS}{\cal D}A =  \int e^{i{1\over 4 g^2}\int 
	F_{\mu\nu} F_{\mu\nu} d^{\,4}x + \dots}\, {\cal D} A
\ee
is meaningless until a UV cutoff is introduced.  I find that the lattice is 
the most physically satisfying way of accomplishing this.  We approximate
spacetime as a lattice, and discretize all the field variables and notions
of differentiation; for example 
\be
\label{eq19-10}
	(\partial\phi)^2 \to
{1\over\epsilon^2}\big(\phi(x)-\phi(x+\epsilon)\big)^2
\ee
K.\ Wilson invented this technique especially for solving QCD on the computer.
In this framework, the gauge field $A_\mu$ is a connection that relates the
relative orientations of the color frames at neighboring points on the lattice; these
can be different from each other by an arbitrary SU(3) rotation.  Hence it is
natural to regard $A_\mu$ as living on the links connecting neighboring lattice
points, rather than sitting on the lattice points themselves.  Consider two
neighboring points, labeled by 1 and 2, and define the link variable
\be
\label{eq19-11}
	U_{12} = e^{i\int_1^2 A^i_\mu {\lambda^i\over 2} dx_\mu}
\ee
where the path is a straight line connecting lattice points 1 and 2.  The
quarks, on the other hand, live on the sites.  In this formulation, the $U_{ij}$
become the dynamical variables rather than $A^i$.

Now we need to formulate the action in terms of the link variables.  Of course
it has to be gauge invariant.  An invariant quantity must involve a
closed path in configuration space, 
\bea
\label{eq19-12}
\raisebox{-2cm}{\includegraphics[width=0.2\textwidth]{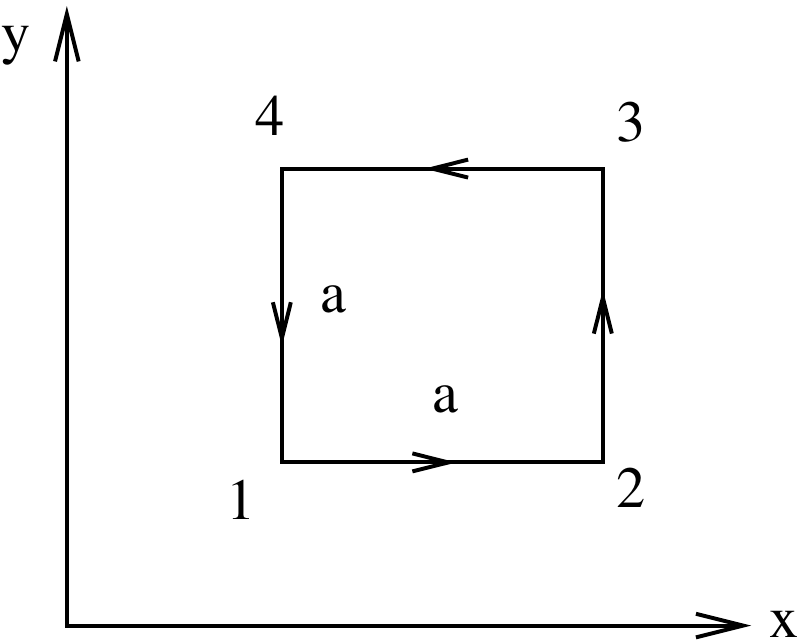}} &:&
U_{14} U_{43} U_{32} U_{21}\\[-1.5cm]
&=& 1 + i F_{xy}^i {\lambda^i\over 2} a^2 + O(A^2)\nn
\eea

\bigskip
\medskip
The small $a\times a$ planar region together with this product is known as a 
plaquette.  $U_{14} U_{43} U_{32} U_{21}$ is
invariant under gauge transformations at any of the interior points $4,3,2$,
but to make it also invariant at site 1, we must take the trace:
\be
\label{eq19-13}
	\tr\left[ U_{14} U_{43} U_{32} U_{21}\right] = \tr\, 1 + 0 + O(A^2)
\ee
The $O(A^2)$ term is the interesting part.  Notice that
\bea
\label{eq19-14}
	U &=& 1 + i\sum b_i {\lambda^i\over 2} + \dots \nn\\
	\bar U &=& 1 - i\sum b_i {\lambda^i\over 2} + \dots\nn\\
\tr\,\bar U U &=& \tr\, 1 + \sum b_i b_i \sfrac12 + \dots 
\eea
But since $U$ is unitary, it must be that all the terms of $O(b^2)$ cancel
out, implying that we need to also keep track of such terms in the individual
$U$ matrices.   It turns out that the relevant ones are
\bea
\label{eq19-15}
	U &=& 1 + i\sum b_i {\lambda^i\over 2} -\sfrac14\sum b_i b_i + \dots \nn\\
	\bar U &=& 1 - i\sum b_i {\lambda^i\over 2}  -\sfrac14\sum b_i b_i+ 
\dots
\eea
When we keep these second-order terms in the expansion of the $U_{ij}$ variables in
the plaquette, and expand the result to $O(A^2)$, the resulting expression is
the finite-difference version of the free part of the gauge kinetic term,
\be
\label{eq19-16}
	F_{\mu\nu}^a F_{\mu\nu}^a\, a^4
\ee
where $\mu\nu = xy$ in the example shown in (\ref{eq19-12}).  Summing on all
plaquettes gives a sum on $\mu,\nu$ as well as the sum over all locations,
resulting in the action
\be
	S = {2\over g^2}\sum_{\rm plaquettes}\big(\tr[UUUU] -\tr\,1\big)\,,
\label{eq19-17}
\ee
where the additive constant is unimportant.  We must be careful about the
relative orientations of the $U$'s on the links, as indicated by the arrows in
(\ref{eq19-12}), to respect gauge invariance.    This is not indicated
explicitly in (\ref{eq19-17}) but it should be kept in mind.  The factor of
$a^4$ in (\ref{eq19-16}) represents the integration measure, since
\be
\label{eq19-18}
	\sum F^2\,a^4 \to \int d^{\,4}x\,F^2
\ee
in the continuum limit.\footnote{I have in parentheses the question ``take real
part of $\tr[UUUU]$?'' in my notes. The answer is yes; for a nice review from this
era, see J.\ Kogut, 10.1103/RevModPhys.51.659.}

Next we turn to the interaction of gluons with quarks.  How do we represent
$\bar\psi \slashed{D}\psi$?  Consider 
\be
\label{eq19-19}
	\sum_{\rm links} \bar q(x) \gamma_\mu\left[q(x+\mu) - q(x)\right]
\ee
using the shorthand that $x+\mu = x + a \vec e_\mu$, in terms of a lattice
unit vector $\vec e_\mu$ that points in the $\mu$ direction.  But this is
not yet gauge covariant, and it is missing the integration measure factor.
Instead, take
\be
\label{eq19-20}
	i a^3\sum_{\rm links} \bar q(x) \gamma_\mu\left[U_{x,x+\mu}q(x+\mu) - q(x)\right]
\ee
In fact, the second term $\bar q(x)\gamma_\mu q(x)$ can be dropped, since it
has no effect on the dynamics, but just contributes an overall phase to the path
integral.\footnote{I have elaborated here on what is written in my notes:
``because it is just a number (?)''}

Let's look more closely at the gauge invariance of the gluon kinetic term.  One can transform
the color axes independently at each lattice site.  Suppose we transform site
1 by the SU(3) matrix $\Lambda(1)$, and similarly at site 2 by $\Lambda(2)$.
One finds that $U_{21}$ changes by 
\be
\label{eq19-21}
\raisebox{-0.3cm}{\includegraphics[width=0.1\textwidth]{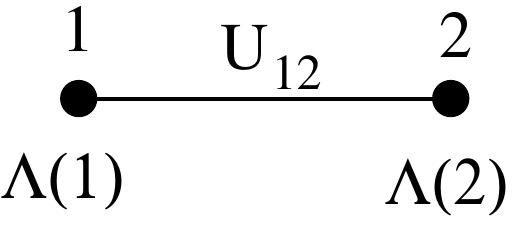}}\qquad 
U_{21} \to \Lambda^\dagger(2)\,
U_{21}\, \Lambda(1)
\ee
Therefore the action is invariant because
\bea
\label{eq19-22}
\tr\left[ \dots U_{32}\underbrace{\Lambda(2)\bar\Lambda(2)} U_{21}\dots \right]
 &=& \tr\left[\dots U_{32}U_{21}\dots\right]\nn\\
 1\qquad\qquad\quad\  && 
\eea
This reasoning also shows how the quark kinetic term is invariant.

However we must still specify the form of the path integral measure for the 
link variables.  Since $U$ is an SU(3) transformation, a gauge invariant measure
is required.  This is known to mathematicians as the Haar measure.
Since SU(3) is a compact group, $\int dU$ is finite at a given site, unlike the
usual measure $\int dA$.  We will not go into the mathematical details of the
Haar measure here, but it is intuitively similar to the more familiar
integration measure for rotations ($\sin\beta\, d\alpha\, d\beta\,
d\gamma$ for SO(3) in terms of Euler angles). 

Just like for a Pauli-Villars cutoff $\Lambda$, we need to find out how
$g$ must vary with the lattice spacing $a$ in order for the theory to give
results that are independent of $a$ in the continuum limit, $a\to 0$. 
Technically, this is not so easy to do as for the $\Lambda$ cutoff, but 
conceptually, it can be carried out in the same way.  We would have to work out
perturbation theory on the lattice to do it properly.  Less rigorously, one
can expect that $g(a)$ has a similar structure to our previous expression for 
$g(\Lambda)$ if we identify $\Lambda = c/a$ for some constant.  Then the task 
becomes determining the correct value of $c$.  Another way of thinking about it
is in terms of the renormalization scale we called $\lambda_P$ in eq.\
(\ref{eq16-1}).  We could define $\Lambda = 1/a$, but we would find that some
other value of $\lambda_P$ is needed to describe the same physics on the lattice
relative to the $\Lambda$ cutoff.  

To my mind, the lattice is the most concrete and least mysterious of all
regulators, and it is the most physically sensible.  But---and now I'm going to
speculate---it seems like we are missing something by having to rely on these
rather ad hoc schemes for defining our theories.  It's comparable to Leibnitz
and Newton inventing the integral calculus, which also looks like taking the
continuum limit of a lattice,
\be
\label{eq19-23}
	\lim_{h\to 0} \sum_m f(x + mh)\, h = \int f(x)\,dx\, .
\ee
The value of the integral does not depend on the machinery of cutting off
the small scales and taking the limit, so one need not be preoccupied with 
the details of exactly how to discretize, like we are doing with
all of our different regulator schemes for the path integral.  We know how to
do the ordinary integrals directly.  Similarly we believe that the path integral
has some kind of intrinsic meaning that does not depend on the cutoff scheme, but
the difference is that we can't avoid that whole discussion, and the dependence
on details of whether we use this kind of cutoff or that kind of cutoff.  This
makes me dream, or speculate, that maybe there is some way, and we are just
missing it, of evaluating the path integral directly, without having to make
this detour into the machinery of renormalization, since we know that the
physics has to be independent of it in the end.

\subsection{Dimensional regularization}

Previously we already discussed dimensional regularization in a preliminary way.
Here I would like to do it in somewhat more detail.  The basic observation is
that an integral like 
\be
\label{eq19-24}
\int {d^{D}\!p\over (p^2-m^2)\left((p-q)^2 - m^2)\right)}\hbox{\ converges for 
	$D < 4$}\,.
\ee
Hence we define $D = 4-\epsilon$, and take the limit $\epsilon\to 0$ in the end.
An integral like (\ref{eq19-24}) gives a pole $1/\epsilon$.  This idea was
used to great advantage by K.\ Wilson to understand phase transitions in
statistical mechanics, and it was applied to gauge theories by `t Hooft and 
Veltman:
\begin{trivlist}
\item
K.\ Wilson, Phys.\ Rev.\ D7, 2911 (1973)
\item
G.\ `t Hooft and M.J.G.\ Veltman, Nucl.\ Phys. B44, 189 (1972)
\end{trivlist}

{\bf Exercise.}  Prove the following statements in $D$ dimensions:\\
1. $g^2$ has dimensions of $E^{4-D}$.\\
2. Defining $\eta_g$ and $\eta_q$ as the number of external gluon or quark
lines, and $L$ as the number of loops, the power of $g$ for any diagram is
$2(L-1) + \eta_g + \eta_q$.\\
3. The dimension, in powers of mass, of the matrix element (amplitude) $|T|$  is
$d= D+N - DN/2$ where $N = \eta_g + \eta_q$.\\
4.  The dimension of the integral for a diagram is $d = 4-\eta_g - \sfrac32\eta_q
-L(4-D)$.

Now I would like to consider the defining properties of the
momentum space integrals in $D$ dimensions, to show that they can be evaluated
rigorously and with no ambiguity.  There are four basic properties.

\begin{trivlist}
\item 1. $\int f(p)\, d^{D}\!p$ is \underline{linear}:
\be
\label{eq19-25}
\int[\alpha f + \beta g]\, d^{D}\!p = \left(\alpha\int f + \beta\int g\right)\, d^{D}\!p
\ee
Among other things, this means that we can use Fourier and Laplace transforms 
in $D$ dimensions.

\item 2. \underline{Shifts of the integration variable} are allowed,
\be
\label{eq19-26}
\int f(p+a)\,d^{D}\!p = \int f(p)\, d^{D}\!p
\ee
where $a$ is a constant vector.

\item 3. \underline{Scaling}:
\be
\label{eq19-27}
\int f(\alpha p)\, d^{D}\!p = {1\over \alpha^D} \int f(p)\, d^{D}\!p
\ee

\item 4. \underline{Normalization}:
\be
\label{eq19-28}
	\int e^{-p^2/2}\, d^{D}\!p = (2\pi)^{D/2}
\ee
since $\int du\, e^{-u^2/2} = \sqrt{2\pi}$.
\end{trivlist}

We can define Fourier transforms in the usual way,
\bea
\label{eq19-29}
	\int d^{D}\!p\, f(p)\, e^{2\pi i p\cdot x} &=& \phi(x)\nn\\
	\int d^{D}\!p\, \phi(x)\, e^{-2\pi i p\cdot x} &=& f(p)
\eea	

Next let us evaluate some integrals.  Notice that
\be
\label{eq19-30}
	\int e^{-{\alpha\over 2} p^2}d^{D}\!p =
\left(\sqrt{2\pi\over\alpha}\right)^D
\ee
Then we can use the shift property to find that
\be
\label{eq19-31}
	\int e^{-{\alpha\over 2} p^2} e^{ip\cdot x} d^{D}\!p =
\sqrt{2\pi\over\alpha}^D\, e^{-x^2/2\alpha}
\ee
Other integrals can be generated from this one by differentiating with
respect to $\alpha$.

{\bf Exercise.}  Prove that 
\be
	\int e^{-{\alpha\over 2} p^2} (a\cdot p)(b\cdot p)\, d^{D}\!p  = 
	X (a\cdot p)
\ee
and find $X$.  Further prove that
\be
	\int e^{-{\alpha\over 2} p^2} p^2\, d^{D}\!p  = 
	D X
\ee
Another useful integral is
\bea
 	\int {d^{D}\!p \over (2\pi)^D} {(p^2)^R\over [p^2-C]^M} &=& 
{i(-1)^{R+M}\over (4\pi)^{D/2} C^{R-M+D/2}}\\
&\times& {\Gamma(R+D/2)\,\Gamma(M-R -D/2)\over \Gamma(D/2)\,\Gamma(M)}\nn
\eea
The reader is invited to derive this one as well.

\section{Dimensional regularization, continued (1-21-$xy$)}

Another very useful class of integrals is that where the integrand 
 $f$ depends only on the magnitude of $p$.  (Imagine that we have already
Wick-rotated to Euclidean space.) Then
\be
\label{eq20-1}
	\int f(p) \, d^{D}\!p  = C_D \int f(\rho) \rho^{D-1}d\rho
\ee
To determine $C_D$, we can consider the case where $f = e^{-\rho^2\alpha/2}$,
since we already know the value of this integral from (\ref{eq19-30}). 
Comparing with (\ref{eq20-1}), 
\bea
	\sqrt{2\pi\over\alpha}^D &=& C_D \int_0^\infty e^{-\rho^2\alpha/2}\,
\rho^{D-1}\,d\rho\\
	&=& C_D \int_0^\infty e^{-u\alpha/2}\, u^{{D-1\over 2}} {du\over
2\sqrt{u}}\nn\\
	&=& C_D \left({2\over\alpha}\right)^{D\over 2}\sfrac12 
	\underbrace{\int_0^\infty e^{-u\alpha/2}\, 
u^{{D\over 2}-1}\, du}\nn\\
&& \qquad\qquad\qquad\qquad\Gamma(D/2)\nn
\eea
hence
\be
	C_D = {2\pi^{D/2}\over \Gamma(D/2)}
\ee
{\bf Exercise.}  Find a general formula for $\int d^{D}\!p\, f(p^2,a\cdot p)$.
Hint: relate it to $\int e^{-\alpha p^2/2 + i p\cdot x}\, d^{D}\!p$.

To combine denominators, we can use formulas like\footnote{In my notes there
is a question mark over the $=$ sign and a parenthetical note to check the
formula, probably a caution from RPF.  It is correct.}
\be
	\int_0^1 dx\, {x^{m-1} (1-x)^{n-1}\over [ax + b(1-x)]^{m+n}} = 
	{\Gamma(m)\Gamma(n)\over \Gamma(m+n)}\, {1\over a^m b^n}
\ee
Then 
\bea
&&\int {d^{D}\!p \over \left((p-q)^2 - m^2\right)(p^2-m^2)} \\
&&\qquad = \int_0^1 dx \int d^{D}\!p {1\over \left(p^2 - 2p\cdot q x + q^2 x -
m^2\right)^2}\nn\\ 
&&\qquad	= \int_0^1 dx \int d^{D}\!p' {1\over \left(p'^2  + q^2 x (1-x) -
m^2\right)^2}\nn\\ 
&&\qquad	= \int_0^1 dx\, C_D \int_0^\infty d\rho\,\rho^{D-1}\,
{1\over \left(\rho^2 + q^2x(1-x) - m^2\right)^2}\nn
\eea

\underline{Claim}: 
\be
	\int d^{D}\!p \, f(p^2)(a\cdot p)(b\cdot p) = 
	{a\cdot b\over D}\int d^{D}\!p\, f(p^2)\,p^2
\ee
Similarly,
\bea
&&	\int f(p^2)(a_1\cdot p)(a_2\cdot p)\cdots (a_n\cdot p)\, d^{D}\!p\\
&& \qquad = \#\big[(a_1\cdot a_2)(a_3\cdot a_4) + \dots\big]
	\int d^{D}\!p\, f(p^2) p^n\nn
\eea
where $n$ is assumed to be even.

As we showed before, dimensional regularization reproduces the logarithms
that we get from Pauli-Villars regularization, through the combination of the
$1/\epsilon$ poles with terms like $(p^2/\mu^2)^\epsilon$.  One shortcoming
however is in the definition of chiral theories, since it is not clear how to
define $\gamma_5$ in $D$ dimensions, nor correspondingly the totally antisymmetric
tensor $\epsilon_{\alpha\beta\gamma\delta}$.  This of course is not a problem
for QCD where parity is conserved.

A novel potential use of dimensional
regularization, which is not usually considered, is that it could provide more than
just a method
for evaluating divergent loop integrals: it is also possible to use it to define
a quantum field theory nonperturbatively in $4-\epsilon$ dimensions. I spent
some time thinking about this, but was not able to get anything interesting out
of it.\footnote{In the next lecture RPF expands on this,
presenting a way to formulate quantum field
without recourse to the path integral or canonical quantization.}

\subsection{Physics in $D$ dimensions}

Inspired by dimensional regularization, it is interesting to try to
formulate a more complete picture of what physics would look like in an
arbitrary number of dimensions.  We start by imagining a linear vector space
in $D$ dimensions, with vectors $x$, $y$, {\it etc.}\ \ Given any two such
vectors, linearity implies that
\be
	\alpha x + \beta y \hbox{\ is also a vector}
\ee
We need a scalar product, a number associated with every pair of vectors,
\be
	x\cdot y \hbox{\ maps (vectors)}^2 \to \mathbb{R}
\ee
It must be linear,
\be
	(\alpha x + \beta x')\cdot y = \alpha\, x\cdot y + \beta\, x'\cdot y
\ee 
and associative,
\be
	x\cdot y = y\cdot x
\ee

Now suppose we had four such vectors.  We could construct the object
\be
	(a\cdot b)(c\cdot d) = a_\mu c_\nu b_\mu d_\nu
\ee
where the indices are no longer numbers taking on discrete values, but rather
markers telling us which vectors should be dotted with each other.  It is
just an alternative notation.  In this case, what would it mean to write an 
expression like
\be
 	n_\mu = a_\mu c_\nu d_\nu 
\ee
in which we have an index that is not contracted?  Such an equation makes
sense if we interpret it to mean that 
\be
	x_\mu n_\mu = x_\mu a_\mu\, c_\nu d_\nu\hbox{\ for all\ }x\,.
\ee
We can also think of $n_\mu$ as being a linear map from vectors into
$\mathbb{R}$.  And we can generalize this to several uncontracted indices,
like
\be
	m_{\mu\nu} = a_\mu b_\nu + c_\nu d_\mu
\ee 
which means that $m_{\mu\nu}$ maps pairs of vectors into $\mathbb{R}$, {\it
etc}.  Hence there is no need for the indices to take on discrete values 
as they would in an integer number of dimensions.

\underline{Contraction.}  In addition to operating on two vectors, $m_{\mu\nu}$
can be contracted on its own indices, $m_{\mu\mu}$, which is just a number 
in $\mathbb{R}$.  For the above example, it is obviously $a\cdot b + c\cdot d$.
However there is a special bilinear mapping,
\be
	\delta:\ (x,y)\to x\cdot y
\ee
that we call $\delta_{\mu\nu}$ in discrete dimensions.  Then
\be
	\delta_{\mu\nu} y_\nu = y_\mu
\ee
is a consistent definition of $\delta_{\mu\nu}$.  Now
$\delta_{\mu\mu}$ is a pure number, that we are free to choose.  Let us make 
the {\it definition}
\be
	\delta_{\mu\mu} \equiv D
\ee  
Obviously, there is no restriction that $D$ should be an integer.

Next let's consider how calculus should work.  We can consider a nonlinear
mapping, $F(x):$ vectors $\to\mathbb{R}$, such as 
\be
\label{eq20-19}
	F(x) = {x\cdot x\over (1+(a\cdot x)^2)^2}
\ee
What is the derivative of $F$?  We define
\be
	D_c F \equiv \lim_{\epsilon\to 0} {F(x + \epsilon c) - F(x)\over
\epsilon}
\ee
where $c$ is some vector.  This is the directional derivative, along the
$c$ direction.  It is  a linear function of $c$, so it must be of the form
$c\cdot$(something).  Therefore we define
\be
\label{eq20-21}
	D_c F \equiv c\cdot \nabla F = c_\mu \nabla_\mu F
\ee
This specifies the $\nabla_\mu$ operator, independently of $c$, since 
(\ref{eq20-21}) must be true for any $c$.  For the example (\ref{eq20-19}),
\be
c\cdot \nabla F = {c\cdot\nabla\, x\cdot x\over \left[1+(a\cdot x)^2\right]^2}
- {2 x\cdot x\over \left[1+(a\cdot x)^2\right]^3}\,2 a\cdot x (c\cdot\nabla)
(a\cdot x)
\ee
as usual.  Now
\be
	(c\cdot\nabla)(x\cdot x) = \lim_{\epsilon\to 0} {(x+\epsilon c)\cdot
	(x+\epsilon c) - x\cdot x\over\epsilon} = 2 c\cdot x,
\ee
so $\nabla_\mu(x\cdot x) = 2x_\mu$, as expected, and $(c\cdot\nabla)(a\cdot x)
= a\cdot c$, giving $\nabla_\mu (a\cdot x) = a_\mu$ and $\nabla_\mu x_\nu = 
\delta_{\mu\nu}$.

Next we construct the Laplacian, by considering successive derivatives:
\be
	a_\mu b_\nu \nabla_\mu \nabla_\nu F = (a\cdot\nabla)(b\cdot\nabla)F
\ee
This allows us to isolate $\nabla_\mu\nabla_\nu F$, which is a tensor
$T_{\mu\nu}$, whose contraction gives the Laplacian.  For example,
\bea
	\nabla_\mu\nabla_\mu\, x\cdot x &=& 2\delta_{\mu\nu}\nn\\
	\nabla^2 (x\cdot x) &=& 2\,\delta_{\mu\mu} = 2D
\eea

So far, all of our results look completely reminiscent of their counterparts in
integer dimensions.  But the concept of orthogonal subspaces leads to a
novelty.  Suppose we have a vector $a$ such that $a\cdot a\neq 0$.  For any 
vector $x$ we can define the part that is orthogonal to $a$ as
\be
	x' = x - {a\cdot x\over a\cdot a}\, a
\ee
so that $a\cdot x' = 0$.  Then any vector can be written as a piece proportional
to $a$ plus a piece in the orthogonal direction.  Furthermore the dot product of
two vectors splits into
\be
	x\cdot y = x'\cdot y' + {a\cdot y\, a\cdot x\over a\cdot a}
\ee
If we take away the external vectors, this gives the definition of the Kronecker
delta of lower dimensionality, living in the subspace orthogonal to $a$,
\be
	\delta_{\mu\nu} = \delta_{\mu'\nu'} + {a_\mu a_\nu\over a\cdot a}
\ee
with $\delta_{\mu'\nu'}$ defined in the orthogonal subspace.  Its trace is
\be
	\delta_{\mu'\mu'} = D-1
\ee
This procedure can be repeated to get subspaces of successively lower
dimensionality, which of course terminates if $D$ is an integer, but does not if
$D$ is noninteger.  It means that we can construct infinitely many directions 
that are all mutually orthogonal in the noninteger case.

{\bf Exercise.}  Suppose that we have two vector spaces of dimension $D_1$ and
$D_2$, and vectors $x_1$, $y_1$, $x_2$, $y_2$ respectively defined in the two
spaces.  There is a rule for combining them to make a vector of dimension $D_3$,
such that
\bea
	x_3 &=& x_1 \oplus x_2,\quad y_3 = y_1 \oplus y_2 \implies\nn\\
	x_3 + y_3 &=& (x_1+y_1)\oplus(x_2 + y_2)
\eea
Prove that $D_3 = D_1 + D_2$, and 
\be
	\int F(p_3)\, d^{D_3}p_3 = \int F(p_1,p_2)\,  d^{D_1}p_1\,  d^{D_2}p_2
\ee

\section{Physics in $D$ dimensions, conclusion (1-26-$xy$)}
We have not yet discussed how to extend the Dirac algebra to 
arbitrary dimensions.  Let's consider the gamma matrices.  For any vector
$a_\mu$, we can associate a quantity
\be
	\slashed{a} = a_\mu\gamma_\mu
\ee
that has the property
\be
	\{\slashed{a},\slashed{b}\} = 2\, a\cdot b
\ee
Therefore $\{\gamma_\mu,\gamma_\nu\} = 2\,\delta_{\mu\nu}$ and 
$\slashed{a}^2 = a\cdot a$, where $\delta_{\mu\mu} = D$ as usual.

For the trace properties, I like to define two traces, that are normalized
differently from each other.  When acting on the unit matrix in the Dirac space
they give
\be
	{\rm tr}[1] = 1;\qquad {\rm Tr}[1] = D\,,
\ee
so ${\rm tr}[X] = (1/D) {\rm Tr}[X]$.  They satisfy the usual property
\be
	{\rm tr}[\slashed{A}\slashed{B}] = {\rm tr}[\slashed{B}\slashed{A}]
\ee
from which you can derive that ${\rm tr}[\slashed{a}\slashed{b}] = a\cdot b$
and ${\rm tr}[\slashed{a}]=0$.  In fact, by using the property
\be
	{\rm tr}[\slashed{a}\slashed{b}\slashed{c}\slashed{d}]
= {1\over 2D}\left({\rm Tr}[\slashed{a}\slashed{b}\slashed{c}\slashed{d}]
+ {\rm Tr}[\slashed{d}\slashed{c}\slashed{b}\slashed{a}]\right)
\ee
and its generalization to an arbitrary number of gamma matrices in the products, one can
demonstrate that the trace of any odd number of gamma matrices vanishes.

The Dirac equation in $D$ dimensions can be written as
\be
	i\slashed{\nabla}\psi - V(x\cdot x)\gamma_0\psi = E\psi
\label{eq21-6}
\ee
for a spherically symmetric potential.  It can be solved exactly; I have carried this out.\\
{\bf Exercise.} Reduce the problem (\ref{eq21-6}) to a conventional ordinary differential
equation.

It turns out that the sum of two spaces of dimension $1/2$ is not an ordinary
1-dimensional space.  We would expect that 
\be
	(x\cdot y)^2 = (x\cdot x) (y\cdot y) \hbox{\ in 1 dimension}\,.
\ee
But this property does not hold when you construct it from two half-dimensional
spaces, along the lines of the exercise at the end of the previous lecture.\\
{\bf Exercise.}  Show that a $D=0$ space constructed from two spaces of equal and opposite
dimension has nontrivial properties.

Newtonian mechanics is quite straightforward in $D$ dimensions. We define a 
time-dependent vector $x(t)$ and impose the principal of least action, with
the action
\be
	S = \int\left({m\over 2}\left({dx\over dt}\cdot {dx\over dt}\right) -
	V(x)\right)\, dt
\ee
Here I do not insist on a central potential; for instance $V(x)$ could have the
form
\be
	V(x) = f(x\!\cdot\! x) + \sum a_i\!\cdot\! x \, g(x\!\cdot\! c) + \dots\,,
\ee
for example $V(x) = (a\cdot x + x\cdot x)/(1 + (x\cdot x)^2)$.  You can prove
that orbits stay in the same plane---this is just conservation of angular
momentum---for a spherically symmetric potential; it's very dull.

One can solve the wave equation in noninteger dimensions,
\be	
	\nabla^2\phi = -\kappa^2\phi;\quad \phi = \phi(x)
\ee
where $\nabla^2 = \bbnabla\cdot\bbnabla$, as we discussed in the last lecture.
It can be done using the Fourier transform, which works in $D$ dimensions, with $\phi = e^{ik\cdot x}$
as usual.  If there is a source $s$, so that
\be	
	\nabla^2\phi = -\kappa^2\phi + s
\ee
then by Fourier transforming one can solve
\be
	\tilde\phi(p) = {\tilde S(p)\over p^2-k^2}
\ee
and also do the inverse Fourier transform to obtain $\phi(x)$ in position 
space.

In quantum chromodynamics we would like to be able to generalize the path integral to $D$
dimensions,
\be
	Z = \int e^{iS}\prod {\cal D}A(x),\quad S = {1\over 2 g^2}\int
E_{\mu\nu}\cdot E_{\mu\nu}
\ee
with $E_{\mu\nu} = \nabla_\mu A_\nu \dots$.  Here $A_\nu(x)$ is just a vector
field defined on the space of vectors, both in $D$ dimensions.  Therefore there
is no difficulty in defining the action.  However the measure is problematic
 because the spacetime is too abstract in $D$ dimensions.  In particular,
we don't know how to construct a lattice if $D$ is not an integer.  

This motivates us to consider a different formulation of quantum field
theory.  Imagine some super\-functional that I will denote by
$\{F[A]\}$, that acts on functionals $F[A]$ of the field in the same way as
the usual normalized path integral, for the cases of integer dimensions where we know
how to define it:
\be
	{\int e^{iS} F[A(x)] \prod{\cal D}A\over \int e^{iS} \prod{\cal D}A}
\equiv \{F[A]\}
\ee
Even though the definition of the left-hand side is not obvious for noninteger
dimensions, we can generalize its known properties in integer dimensions to
obtain a functional differential equation,  that defines our mysterious
super\-functional.  Namely, we know that the path integral is invariant
under a change of variables $A_\mu \to A_\mu +\epsilon \alpha_\mu(x)$, that
I take to be infinitesimal.  Moreover, the measure by itself is invariant under
this trivial shift.  
Recall that the functional derivative 
${\delta F[A_\mu(x)]/\delta A_\mu(x)}$
is defined by
\be
	\delta F = \int \epsilon \alpha_\mu(x)
{\delta F[A_\mu]\over \delta A_\mu(x)}\, d^D\!x
\ee
It follows that 
\be
	\left\{{\delta F[A_\mu]\over\delta A_\mu(x)}\right\} + 
	i\left\{ F\, {\delta S\over\delta A_\mu(x)}\right\} = 0
\label{eq21-16}
\ee
This equation is equivalent to the path integral, but more fundamental
since it extends to the case of noninteger dimensions, and it can be used
as our starting point.  It is a statement of Schwinger's action principle,
from which one can derive the usual perturbation expansion, but it can also
serve as a nonperturbative definition of the theory.

One issue we have glossed over is the signature of the spacetime metric. This is
a discrete choice in integer dimensions, and it is not perfectly clear how to
deal with it in arbitrary dimensions.  Should the fractional difference in the
dimension be spacelike or timelike?  In a more extreme case, what would the world look like if the
metric of spacetime was
\be
	\delta_{\mu\nu} = \left(\begin{array}{cccc}-1 & & & \\
                                                    &-1& & \\
                                                    &  & 1 & \\
                                                    &  & & 1 \end{array}
	\right)\ ?
\ee
It might be advantageous to find some kind of geometrical description of events
to answer this.

Although it is a somewhat different issue than noninteger dimensions, one could
also think about derivatives of fractional order.  We know that conventional
derivatives have the form
\bea
	{df\over dx} &=& \lim_{\epsilon\to 0} {f(x) - f(x-\epsilon)\over \epsilon}
	\\
	{d^2\!f\over dx^2} &=& \lim_{\epsilon\to 0} {f(x) - 2f(x-\epsilon) +
f(x-2\epsilon)\over \epsilon^2}\nn
\eea
What about a derivative of order $1.5$ or $1/2$?   I will illustrate the 
correct generalization for the $1/2$ order:
\be
	{d^{1/2}\!f\over dx^{1/2}} = \lim_{\epsilon\to 0} 
{f(x) - \sfrac12f(x-\epsilon) +
\sfrac18 f(x-2\epsilon) - \dots\over \epsilon^{1/2}}	
\ee
where $-1/2$, $1/8$ \dots are the binomial coefficients from expanding
$(1+x)^{1/2}$.  This turns out to be a valid procedure.  And it has an inverse:
you can find half-order integrals as well, in an analogous way.

Returning to dimensional regularization, I wanted to show in slightly 
more detail how the
$Q^2$
dependence of the coupling comes out from the perturbation series in that 
method.   Here I will set $D=4+\epsilon$ and take $\epsilon\to 0$ at the end.
Recall that in our $\Lambda$ cutoff scheme, the form of an amplitude would look
something like
\bea
g_\Lambda^2 &+& g_\Lambda^4\left(\beta_0\ln{\Lambda^2\over
Q^2}+c\right)+\dots\nn\\
	&=& \alpha(Q^2) + \alpha^2(Q^2)c + \dots
\eea
In dimensional regularization, we must introduce an arbitrary mass $m_0$ to make
the coupling dimensionless:
\be
	g^2 = g_0^2\, m_0^{-\epsilon} \equiv \alpha_0\, m_0^{-\epsilon}
\ee
The renormalization of the coupling constant now becomes
\bea
	{4\pi\over \beta_0\alpha(Q^2)} &-& {\beta_1\over \beta_0}\ln \left(
{4\pi\over \beta_0\alpha(Q^2)}\right)\\ &=& {4\pi\over \beta_0\alpha_0} + 
	{2\over\epsilon} + \ln{Q^2\over m_0^2} + \gamma_E - \ln 4\pi\nn
\eea
where $\gamma_E = 0.5772\dots$ is Euler's constant.  Nothing physical depends
on $m_0^2$, nor on $\epsilon$.  By choosing the arbitrary normalization of $m_0$
appropriately, we can cancel the $2/\epsilon$ pole (the minimal subtraction MS scheme),
or the pole along with the $\gamma_E - \ln 4\pi$ terms (modified minimal subtraction, 
$\overline{\rm MS}$ scheme).

\subsection{Scattering at high $Q^2$}

We have seen that at high $Q^2$, the effective coupling of QCD is supposed to 
become small.  But this by itself does not guarantee perturbation theory is necessarily
very good, since the numerical coefficients of the expansion might turn out to
be large.  In particular, there is always a low-energy effect mixed in with any
high-energy process because of hadronization of the final state particles.  This
part of the process is taking place at scales where the coupling is definitely not small and
the calculation is not perturbative.  It might seem like after all our wonderful
efforts of using renormalization to improve the perturbative predictions at high
energy, we could get foiled by these low-energy effects.  However all is not
lost, because we can separate these two phenomena from each other in a more or
less clean manner.  If we can argue that the details of hadronization are
independent of the high energy scale $Q^2$, then this separation can be done
quantitatively.

Recall that in $e^+e^-$ collisions that produce $q\bar q$, we will see 
two hadronic jets,
\[
\includegraphics[width=0.3\textwidth]{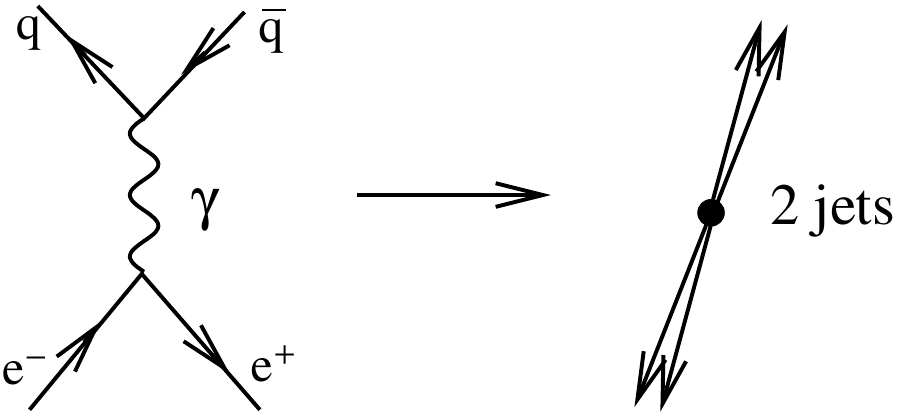}
\]
or if a gluon is radiated at a large enough angle, we will see three jets,
\[
\includegraphics[width=0.3\textwidth]{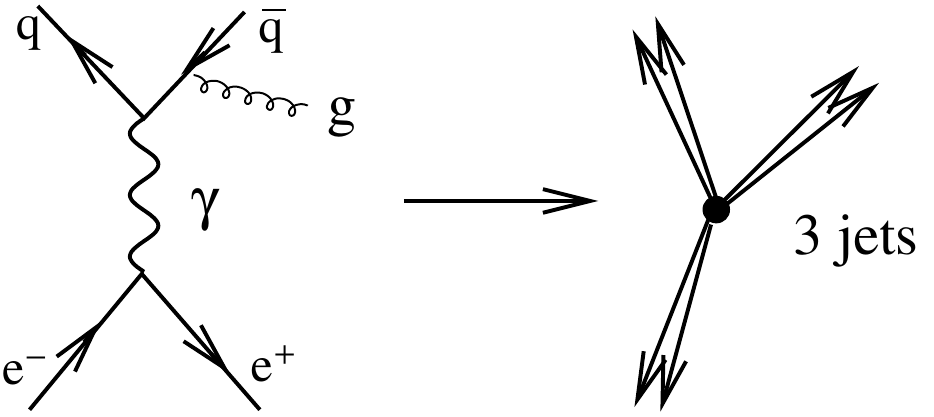}
\]
At low $Q^2$, the third one is likely to be too soft to appear as a distinct 
jet.  In this case the envelope of momentum vectors would appear to be a
cylinder with a small bump for the third soft jet
\[
\includegraphics[width=0.2\textwidth]{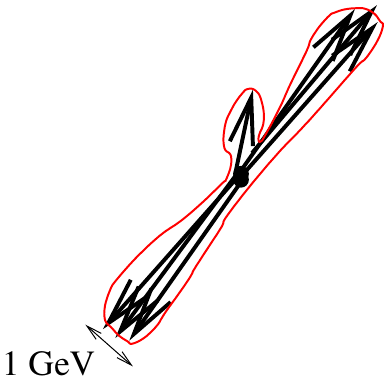}
\]
At high $Q^2$, the cigar gets relatively narrower, and the jets become better
defined.  The gluon jet starts to become more distinct as $Q^2$ increases,
like bringing a picture into focus.  And as $Q^2$ continues to increase, greater
numbers of jets start to appear.

We can measure the distributions of hadrons in jets at low $Q^2$, described by
what are known as fragmentation functions.  Although we don't know the
fragmentation function for gluons, we can make some educated guess.  The
important point, related to my claim above, is that these functions do not
change appreciably with $Q^2$, which leads to the factorization phenomenon that
I described.  These distributions take the form
\be
	c\, {dp_z\over E}\, f(p_\perp)\, d^{\,2}p_\perp
\ee
where $p_z$ is the momentum along the jet axis, and $p_\perp$ is the transverse
momentum (shown as having a spread of $\sim 1\,$GeV in the picture above).
To the extent that $f(p_\perp)$ is flat, within the jet, you will notice that
this distribution is Lorentz invariant.

A useful quantity for characterizing particles in the jets is rapidity.  Consider the
quantity $E-p_z$, where $E = \sqrt{p_z^2 + p_\perp^2 + m^2}$ is the energy of 
a particle in the jet.  Under a boost in the
$z$ direction, this becomes $\gamma(E + v p_z) - \gamma(p_z + v E) = 
\gamma(1-v)(E-p_z)$: it changes multiplicatively.  Similarly $E + p_z \to
\gamma(1+v)(E+p_z)$.   Therefore the rapidity
\be
	w = \ln{E+p_z\over E-p_z}
\ee
changes by
\be
	w\to w + \ln{1+v\over 1-v}
\ee
where $v$ is the boost parameter.  It is related to $w$ by
\be
	v = \tanh w
\ee
and in terms of $w$ a boost takes the form
\bea
	E' &=& E\cosh w = {E\over \sqrt{1-v^2}},\nn\\
	p' &=& p\sinh w\,.
\eea

It turns out that the distribution $f(p_\perp)$ is not quite flat over its
region of support, but instead goes as
\bea
	f(p_\perp) &\sim& \alpha(p_\perp^2)\, {d^{\,2}p_\perp\over p_\perp^2}
	\nn\\
	&\sim& d\ln p_\perp^2\over \ln p_\perp^2\nn\\
	&\sim& d\ln\ln p_\perp^2\,.
\eea
Hence it is {\it nearly} constant, the log of a log.   This small deviation from
flatness has been observed and provides a confirmation of QCD.

\subsection{Sphinxes}
[The following figures appear without explanation in my notes.]
\[
\includegraphics[width=0.475\textwidth]{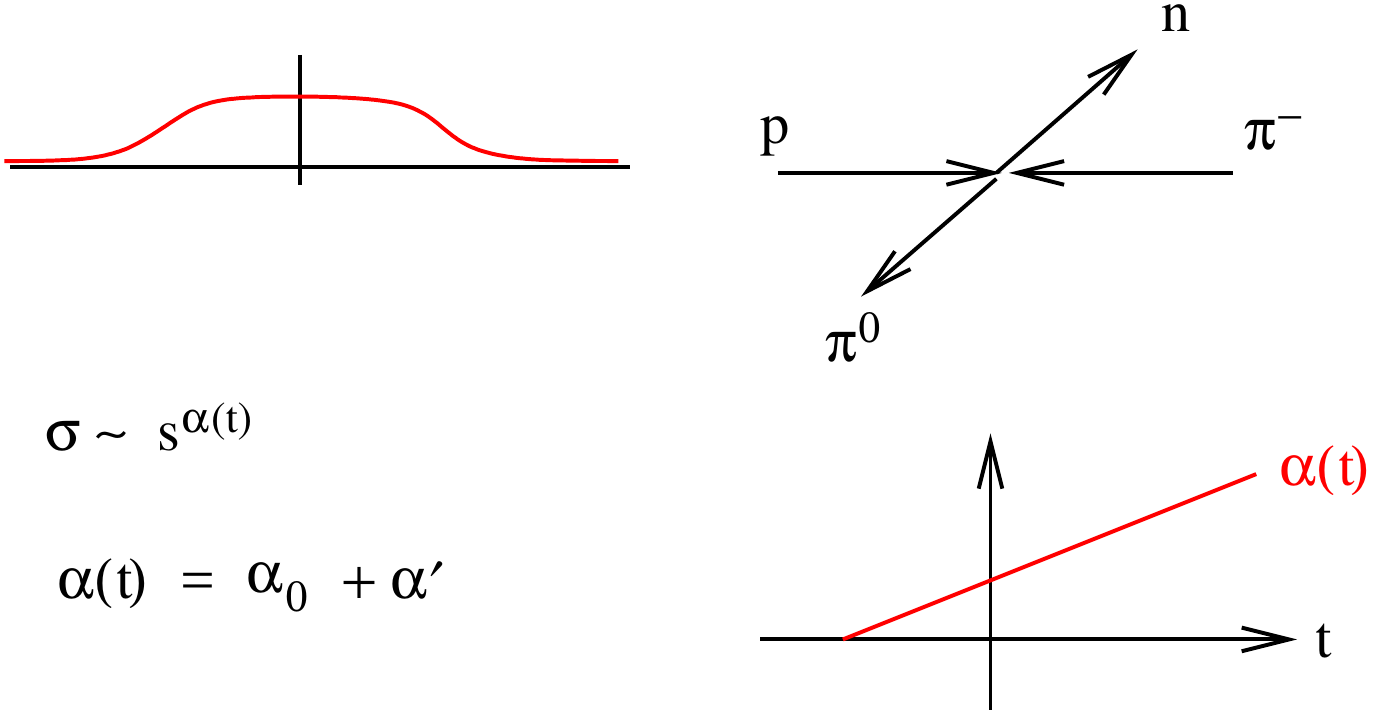}
\]

\section{Final lecture (1-28-$xy$)}

\subsection{Schwinger's formulation of QFT, continued}

I would like to come back to the alternative formulation of quantum field
theory that I started to discuss last time, eq.\ (\ref{eq21-16}).  To understand
in more detail how to use it, let's consider the simpler example of a scalar
field theory, where it takes the form
\be
\left\{{\partial F\over\partial\phi}\right\} + i\left\{F\, {\partial
S\over\partial\phi}\right\} = 0,
\ee
and now
\be
	S = \sfrac12\int\left((\nabla\phi)^2 -\mu^2\phi^2 -\lambda\phi^4\right)
	\,d^{\,4}x\,.
\ee
Remember that we are free to choose any functional $F$; it is instructive to
take
\be
	F[\phi] = e^{i\int \sigma(x)\,\phi(x)\, d^{\,4}x}\,.
\ee
From it, we can generate Green's functions by taking functional derivatives
$\delta^n/\delta\sigma^n$.  Then, with $n=1$,
\be
	i\Sigma(x)\equiv \left\{ i\sigma(x)\,F\right\} = -i\left\{F\,
(-\nabla^2-\mu^2)\phi\right\}
\ee
to zeroth order in $\lambda$.   If we define $f(x) = \{F\,\phi(x)\}$, we
see that
\be
	i(\nabla^2 +\mu^2)f = i\Sigma(x),
\ee
which can be solved to get
\be
	f = \left(\nabla^2 + \mu^2\right)^{-1}\Sigma(x)
\ee
This of course is just the propagator acting on $\Sigma$.  Fourier transforming
to momentum space, it reads
\be
	f(p) = -{1\over p^2-\mu^2}\, \Sigma(p)
\ee
To define what happens at the pole, we need to make an $i\epsilon$ prescription,
as usual.

Now we can write
\be
	\left\{\phi(x)\,F\right\} = \int dy\left\{ I(x-y)\,\sigma(y)\,
	e^{i\int \sigma\,\phi\,d^{\,4}x}\right\}
\ee
where $I(x-y)$ is the propagator in position space.  Hence
\be
	\left\{\phi(x)\,e^{i\int \sigma\,\phi\,d^{\,4}x}\right\} = 
	\int dy\,I(x-y)\,\sigma(y)\,
	\left\{ e^{i\int \sigma\,\phi\,d^{\,4}x}\right\}
\ee
or
\be 
	{\delta\{F\}\over \delta\sigma(x)} = \int I(x-y)\,\sigma(y) dy\,\{F\}\,.
\label{eq22-10}
\ee
We can integrate this to get
\be
	\{F\} = e^{\frac12 \int\sigma(x)I(x-y)\sigma(y)\,dx\,dy}\,.
\ee
This is all at the level of free field theory so far.  I leave it as an exercise
for you to show that eq.\ (\ref{eq22-10}) generalizes to
\be
	{\delta\{F\}\over \delta\sigma(x)} = \int I(x-y)\,\sigma(y) dy\,
\left[1-2\lambda\left(-i{\delta\ \ \over\delta\sigma(y)}\right)^3\right]\{F\}
\ee 
in the presence of the interaction.  This can be solved perturbatively, or
perhaps if you are clever enough, in some nonperturbative fashion.  Obviously,
there is no very sensitive dependence on the number of spacetime dimensions in
this formulation, so it could serve as a nonperturbative definition of the
theory in $4-\epsilon$ dimensions.

\subsection{Parton model; hadronization}
Now I would like to come back to some things that we started to discuss earlier
in the course, during the first few lectures.  Remember the parton picture of
the proton,
\[
\centerline{
\includegraphics[width=0.15\textwidth]{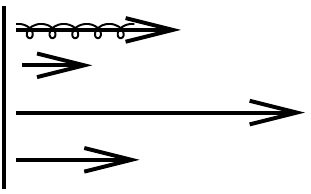}}
\]
where the quarks and gluons have various momentum probability distributions
inside the proton, that we denoted by $u(x)$, $d(x)$, {\it etc}.  I had mentioned a conceptual problem, the fact that even
if we knew the wavefunction of all the constituents for a proton at rest,
this is not sufficient for determining $u(x)$, $d(x)$, \dots.  The problem has
to do with how the wave function transforms under a boost, 
\[
\centerline{
\includegraphics[width=0.35\textwidth]{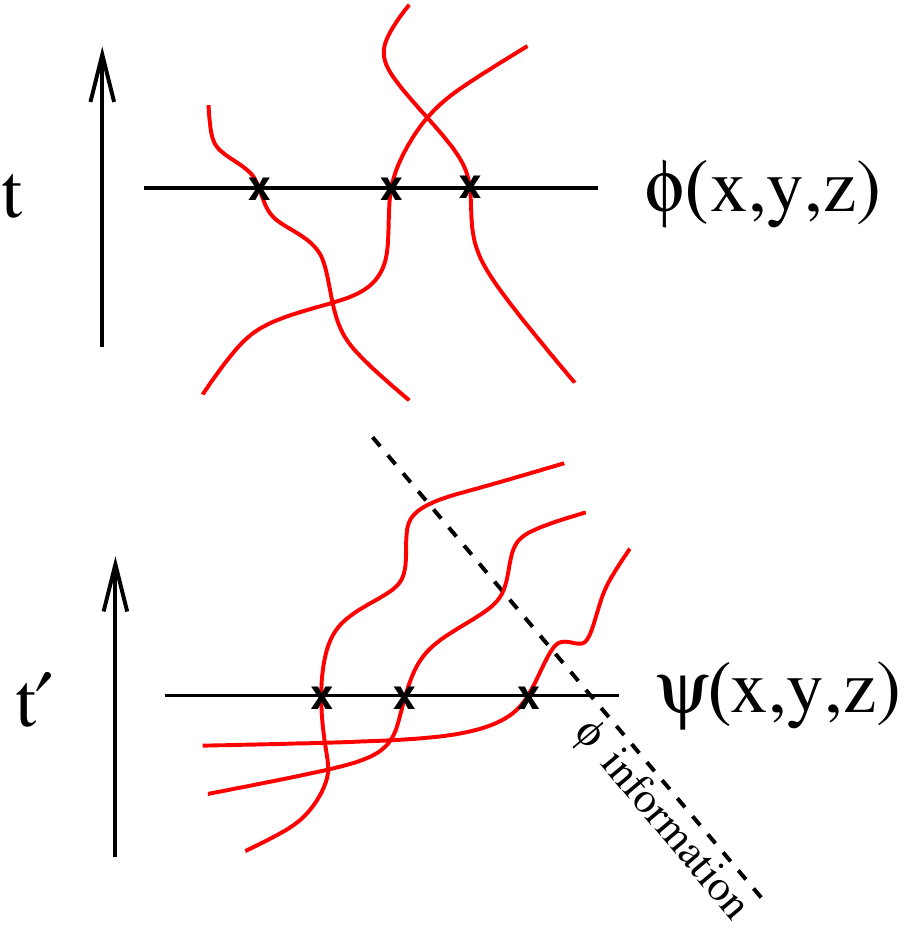}}
\]
In one reference frame, at $t=0$, the wave function for three quarks at
respective positions $x,y,z$ is $\phi(x,y,z)$, while in some boosted frame, at
$t'=0$, it is $\psi(x,y,z)$.  It is a nontrivial task to get $\psi$ from $\phi$,
since we need to solve the Schr\"odinger equation to propagate the quarks
forward in time.  As we saw before, this is complicated by the fact that the
concept of the wavefunction is not relativistic.  

But once we know the
distribution functions at high energies, it turns out that they don't change 
very much as you go to even higher energies.  We can determine these functions
by doing proton-electron scattering,
\[
\centerline{
\includegraphics[width=0.25\textwidth]{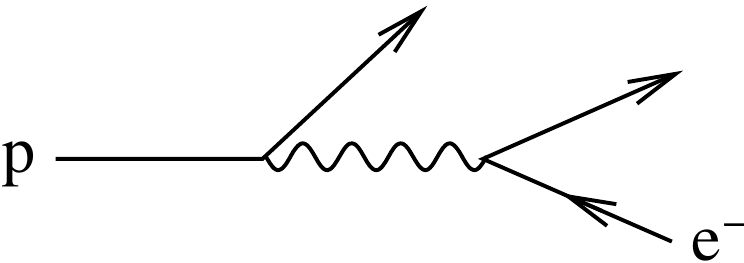}}
\]
The reason that the distributions continue to change at higher energies is that
the kinematics are not so simple as in this diagram:  in reality, gluons are
radiated, in particular in the forward direction, where we don't see them as 
distinct jets.  In fact as we increase $Q^2$, gluons are more likely to be
emitted, which leads to $u(x)$ depending on $Q^2$ and not just the momentum
fraction $x$.  This gives a correction to the naive parton picture, that
neglected such effects.  Qualitatively, the correction looks like
\[
\centerline{
\includegraphics[width=0.45\textwidth]{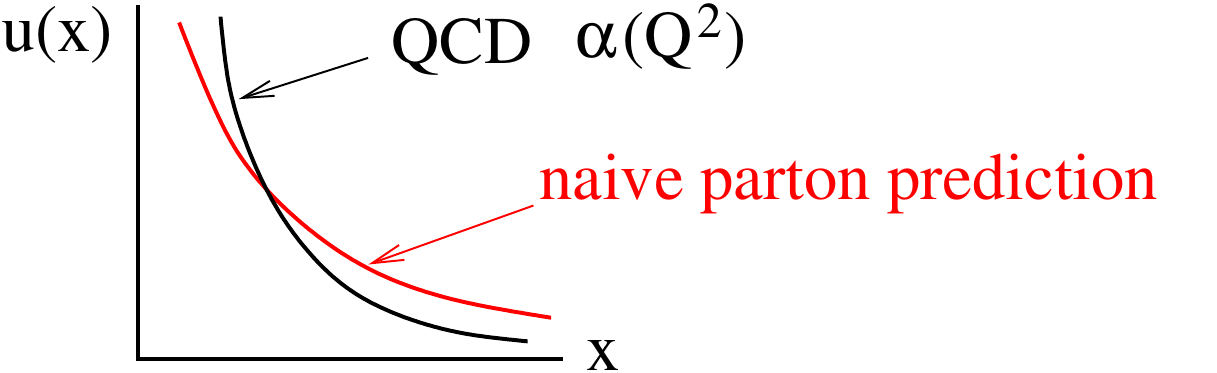}}
\]
You might think that gluon emission should decrease at high $Q^2$ because of the
running of $\alpha(Q^2)$, but it turns out that the increase in phase space
outweighs this effect.  It is similar to the infrared/brehmsstrahlung problem in
QED.

Consider electroproduction of quarks, $e^+ e^-\to q\bar q$.  Many
soft gluons will be emitted, but they all get lumped in with the quark jets, and are
of little consequence {\it a priori}.  Our concern is how this description may
change as a function of $Q^2$. 
\[
\centerline{
\includegraphics[width=0.25\textwidth]{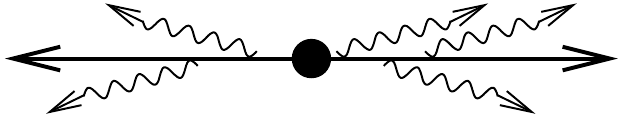}}
\]
Recall the fragmentation functions, $D^h(z,Q^2)$, that tell us the distribution
of the momentum fraction $z$ carried by hadron $h$ in the jet.  We would like to
know how it depends on $Q^2$.  Since the main dependence on $Q^2$ comes from the
coupling, and this log dependence is also tracked by the cutoff, we can infer
that
\bea
	{dD\over d\ln Q^2} &=& {\delta D\over \delta g^2_\Lambda}
	\, {\partial g^2_\Lambda\over \partial\ln\Lambda^2}\nn\\
	&=& {\delta D\over \delta\alpha(Q^2)}\,{\partial\alpha(Q^2)\over
 	\partial\ln Q^2}\,,
\eea
or, defining $\tau = \ln Q^2$, 
\be
{dD\over d\tau} = {\delta D\over\delta\alpha}\, \beta_0\alpha^2\,.
\ee

So we need $\delta D^h/\delta\alpha$, which arises at first order in
perturbation theory, and is related to the probability for emitting a gluon
(or possibly a $q$-$\bar q$ pair) 
that carries away some fraction $(1-y)$ of the momentum:  
\[
\centerline{
\includegraphics[width=0.35\textwidth]{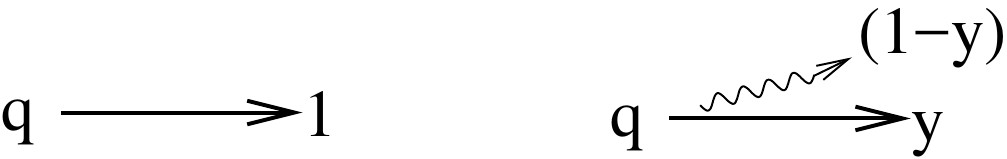}}
\] 
This leads to the evolution equation
\be
{\delta D\over \delta \alpha} = \int {\cal P}(y)\, D^h(z/y,Q^2)\,{dy\over y}
\ee
where ${\cal P}(y)$ is the probability of emitting a gluon that leaves the
quark with momentum fraction $y$, if it was normalized to be 1 initially.
One can show that
\be
	{\cal P}(y) = {2\over 3\pi}\,\ln{Q^2\over m^2}\,\left(1+y\over
1-y\right) + \dots
\ee
where the $\dots$ represent terms with weaker dependence on $Q^2$.

{\bf Exercise.} 
Consider $e^+e^-\to q\bar q$ with the kinematics indicated below:
\[
\centerline{
\includegraphics[width=0.2\textwidth]{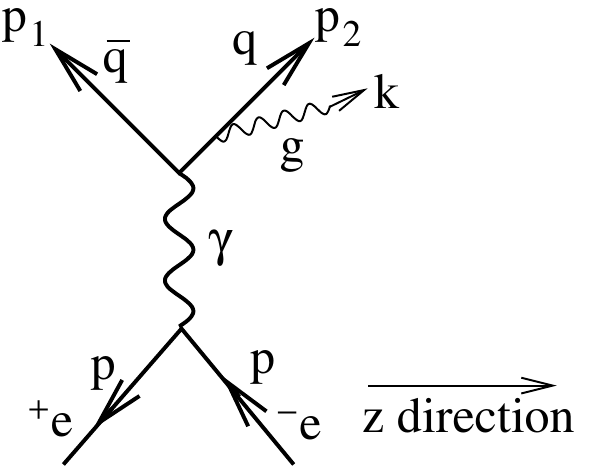}}
\] 
Show that the
probability to emit the gluon is proportional to 
\bea
P &\sim& \alpha\, {\epsilon_1^2 + \epsilon_2^2 + p_{1z}^2 + p_{2z}^2\over
\epsilon_1\epsilon_2}\nn\\
&\sim& \sfrac43\alpha (\epsilon_1^2+\epsilon_2^2)
\eea
where $\epsilon_i$ is the energy of quark $i$, and the second expression is the
result from integrating over the quark directions.

We can write
\be
	{d D\over d \alpha} = \tau {\cal P}\cdot D + \dots
\ee
where ${\cal P}\cdot D$ denotes the convolution $\int {\cal P}D \,dy/y$, or defining
$\kappa = \beta_0\ln\tau$,
\be
	{dD\over d\kappa} = {\cal P}\cdot D\,.
\ee
Hence if we evolve from $Q_0^2$ to $Q_1^2$, the change in $\kappa$ is
\bea
	\Delta\kappa &=& \beta_0\ln{\tau_0\over\tau_1} \nn\\
	 &=& \beta_0\,\ln\left(\ln Q_0^2/\lambda^2\over \ln Q_1^2/\lambda^2
	\right)
\eea
This is a double logarithm, so the change is typically quite small.  Take for
example
\bea
	Q_0 &=& 6\,{\rm GeV}, \quad Q_0^2\cong 40\,{\rm GeV}^2,\nn\\
        Q_1 &=& 4000\,{\rm GeV}, \quad Q_1^2 \cong 2\times 10^7\,{\rm GeV}^2
\eea
This gives $\Delta\kappa\cong \sfrac12$.  To get a change as large as 
$\Delta\kappa = 1$, we would need to go to $Q_1 = 400,000\,$GeV!

So it is necessary to vary the energy quite dramatically to see any appreciable
change in the fragmentation function $D^h$.  And only by starting from rather
low energies will we observe much variation at all.  Notice that at such low
energies as 6\,GeV, many of our approximations that were appropriate for
high $Q^2$ are not very good.  The important point is that the details of hadronization
are indeed insensitive to $Q^2$ as long as $Q$ is well above the QCD scale.

\bigskip
\bigskip
{\bf Acknowledgment.}  GGT thanks Jezebel Moss for encouragement to complete this
project, Professor P.\ Schickele for 
useful comments on the manuscript, and Midge Finemensch for permission to 
make these notes public, except for the censored pages.

\bigskip
\appendix
Appendices \ref{A1}-\ref{A3} are the verbatim transcriptions from the audio
tapes of lectures 15, 17 and 18.  Appendix \ref{AppD} contains scans of RPFs
hand-written addenda and corrections to revised drafts of two lectures.  The remaining appendices
are material that RPF handed out to the class participants, some of them written
in his own hand.
\let\theequation\oldtheequation
\section{Transcription: Scale dependence (1-5-$xy$)}
\label{A1}
\it

Okay.  Now that we're starting the second term, we've formulated
several times, in many different ways, in different kinds of gauges,
the rules for perturbation theory; also the formulas in terms of path
integrals.  One of the purposes of path integrals is a statement of
the equations which is not strictly speaking necessarily simply perturbation
theory.  If there was some way to compute the path integrals, and
there is for instance numerically, there would be a scheme for making
calculations which would not rely simply upon the need for
perturbations.  Because of our limited ability in doing path integrals
until the present time, we're only pretty good in perturbation
theory.  We got so good at it from working with quantum
electrodynamics where the coupling constant is very small, and therefore we've
had lots of practice.  But please don't think that we have to do
everything by perturbation theory.

During this term we are going to talk in the first half of it about
perturbation theory and what we can learn of quantum chromodynamics\,\footnote{RPF says electrodynamics but it is clear he meant to say
QCD.} from perturbation theory. The second half of the term will be an
attempt to understand the behavior of this theory, the fact that it
confines quarks and so on, in some way by looking at the path
integrals, without actually expanding them in perturbation theory.  It
will not be mathematically accurate; it will be qualitative.  Say this
will get big, this will get small, I think this will be bigger than
that, and therefore this will happen.  You will be very dissatisfied
if you want precision. One of our problems as we're discovering,
right, is doing these things with precision.  The state we are in now
is one where we will have to discuss it in a qualitative way. So
that's all I can do, but we'll do that in the second part of the term.
In the first part of the term, we will find out how much we can do by
perturbation theory to test the theory.

So far we made only passing reference to the running of the coupling
constant. This subject requires some care to avoid confusion.  But I
must say that it would not be at all difficult, there would not be any
particular problem, and it is a very simple matter;  the confusion
comes because we can't calculate anything, and so we try to say as
much as we can without  calculating \dots  It's something like the
subject of thermodynamics, which appears to be quite complicated, but
if you use always the same variables, such as temperature and volume,
to represent the system, it's much simpler than if you suddenly say
now wait, let's suppose I want to plot this on entropy and pressure; it's
the  perpetual change of variables from one to the other that makes
the subject so complicated.

So there is a certain apparent complexity here, which is due to our inability, or 
indefiniteness, in choosing a method of calculation, or indefiniteness in deciding
what process to calculate.  
That makes it look a little complicated.  Let me make
believe, at first,  that we could calculate whatever we want.  Then the problem would be the
following.  It would be straightforward.  We would start with our theory, with
Lagrangian 
\be
	 {1\over g_0^2}FF + \bar\psi(i\slashed{\partial}+m_0)\psi \dots\nn
\ee
Let me put the constant and call it $g_0$.  And then in the part that has to do
with the quarks, with different flavors, there would be masses for the different
flavors.  So there would be a number of constants which are in the theory,
\be
	g_0^2,\ m_0^{u},\ m_0^{d},\ m_0^{s},\ m_0^c,\ m_0^b \nn
\ee
I'm putting a subscript 0 on them, which means those are the values
that we put into the equations; in case there's any question of what I
mean, that's what I mean: the numbers that we put into the original
Lagrangian to make the calculation. So this is perfectly definite. 
There's the $c$ quark; we may discover one day that there are others
\rm[quark flavors]\it\ so we might need a few other parameters, but for the
future.\footnote{The top quark had not yet been discovered
at this time, though its existence was not doubted.}\ \   So at the present time we've got these six numbers, the
parameters that we can put into the theory.  

Now suppose we start out
with this theory, we put some parameters in, and we compute something:
the mass of the proton, the mass of the pion, and so on.  If we
computed six quantities, and we could compute perfectly and the theory
were right and experiments were available for all those quantities,
then we could determine these parameters.  Then if I  computed a
seventh quantity, that would be a completely predicted quantity and we
begin to test the theory.  That's simple and straightforward and
that's all there is to it, {\bf except} \dots First of all, we can't
compute the mass of the proton, so we can't determine these
constants.  In view of that you also would know that it would be very
practical to compute some quantities rather than others.  

For instance, if you wanted to compute the mass of the $b$ quark, you
could probably do a pretty good job by trying to compute the mass of
the Upsilon, which would be pretty close to  the mass of two $b$'s, $2
m_0^b$.  We've got a very crude beginning for  $m_0^b$ without being
able to calculate.  We would get $m_0^b$ most likely  from masses of
the Upsilon and its excited states, and we learn about the interaction
strength somewhere, we correct for the excitation states
 that would make a very accurate value of $2 m_0^b$, which would
presumably be very close to one half of the mass of the
Upsilon---reasonably close to a half the mass of the Upsilon,
in other words to, uh, 5, 10 GeV \rm[speaking to himself]\it\ the mass
of the Upsilon is \dots  so it has to be 5 \rm[for $m_0^b$]\it.  Yes that's
right.  In the same way the mass of the $c$ could be determined as
being about 1.8.  Now, if we did the calculation more elaborately,
you see we would be picking out the light quantities; instead of having the
mass of the proton and the mass of the pion looking for some tiny
deviation which was due to the $c$ quarks, which are hardly affecting
either one of them, neither one of them, then to get the \rm[$c$ quark]\it\
mass, that's not the way to do it.

  So it's sensible to try to pick out physical
quantities that are more sensitive to particular parameters 
than to others.  Of course 
in principle you could
compute any old six quantities with infinite accuracy
and deduce all the parameters.    But it would be more practical to
choose six quantities that are more directly sensitive to the
masses.  I've already got rid of $m_0^b$ \dots.

Now we want to try to concentrate on things which depend---because
there's a great deal of interest in that quantity---physical
quantities that depend upon $g_0^2$, and are not very sensitive to the
others.  I believe that one is the mass of the proton, because the
mass of the $u$ and the $d$ \dots we have good evidence are very
small, and that the mass of the proton is not due to the mass of the
quarks inside, I mean at least not directly \dots it has to do with
the value of $g_0$.  It's hardly sensitive to $m_0^s$, but we could 
imagine that someday we could correct for that \dots
However at the present time it is hopeless to
compute the mass of the proton from these constants theoretically,
and therefore we can't determine $g_0$ from the mass of the proton,
even though someday we could.

Another kind of effects where I look for $g_0$ are those phenomena
that use high energy and which---so high that the  masses of these
things \rm[quarks]\it\ don't make any difference.  And insofar as these \rm[quark
masses]\it\ are involved, we can presumably compute their effect.
In other words we look for
processes in which we expect there would be a limit, that limit would
still exist for this process if the masses of these things went to
zero.

I'll give you some examples.  This is only to suggest things to look
at that we can calculate 
 that will help us to isolate parameters, in particular $g_0$.  One
interesting experiment is $e^+ e^-\to$ hadrons.  
\be
\includegraphics[width=0.15\textwidth]{eeqq}\label{eeqq}	
\ee
And the idea of that is if you do that at high energies,
the electron and positron annihilate and produce a photon, which you
can understand.  And then the photon produces
a pair of quarks, as a sort of initial disturbance.  We have an operator
$Q_f\bar\psi_f\gamma_\mu\psi_f$ for each quark flavor, and this operator starts by
generating a pair of quarks.  Now what happens after that is of course that this
\rm[quark]\it\ maybe radiates a gluon, the gluon splits into quarks, and they combine
together and they make $\pi$'s and they make $K$'s and you get a big splash of 
junk.  The total cross section for doing this is the chance that we got started,
so to speak.  You can calculate the probability that we got these things started
by just figuring that they're free, because the energy is so high.  And then after
a while they scratch their heads and say, ``hey, I'm not supposed to be able to 
come out, I've gotta do something else,'' but they're already there.  I don't know
if you feel this intuition very well, I get it myself but I don't know how to 
express it; that at high enough energies, when we start this process and then after that
\dots 

Let's put it this way: suppose you did this, and you thought one day
that you made a $\rho$.  The next time you realize that you're not
going to see the $\rho$  but the $\rho$ actually disintegrates into a
pair of pions.  Well the fact is, by the time you got to the $\rho$,
and then it went into $\pi$'s, it isn't going to change the total
cross section; whether the $\rho$ does or doesn't disintegrate 
doesn't make any difference to the total rate.  The ultimate things
that happen to these objects are not much affected by things that
happen late, and therefore at low energy, and therefore involving the
masses of the quarks and so  on.  So therefore the total cross section
shouldn't involve the masses of the quarks. It shouldn't involve
anything in QCD, there's no coupling constant at all \rm[the QCD
coupling]\it\ because this rate to produce this pair of quarks that act
like free particles, we calculate it directly.

In the same way we could compute the rate to produce a pair of $\mu$'s
and calculate that pure electrodynamic thing and call it
$\sigma_{\mu\mu}$.  Then I could calculate the probability of
producing hadrons here \rm[in the diagram (\ref{eeqq}) with hadronization
of the quarks]\it; the cross section
would be the cross section for producing $\mu$'s---which also doesn't
depend much on the mass of $\mu$---we take a very high $q^2$ so the
masses don't make a difference; then we would have the cross section
for making, let's say, $u$ quarks.  Then the charge is $2/3$ for a $u$
quark and the cross section goes as 
\be
	\sigma_{u\bar u} = 3\, \left(2\over 3\right)^2\, \sigma_{\mu^+\mu^-}\nn
\ee
to produce $u$ quarks; so this would be the probability of producing
$u$ quarks.  We can produce the $u$ quarks in three colors: red, green
and blue, and therefore the thing is multiplied by 3.  Is there any question about
that?  Yeah? \rm[a question is asked, inaudible]\it\  No sir, because this is
an electromagnetic phenomenon I'm talking about, this is a \rm[photon]\it\
not a gluon, so this coupling is not QCD.  Any other questions?
You had me for a minute \rm[laughter]\it.

This {\bf three}, I say you can produce any color but that's 
a lot of nonsense,
you can't produce any color, because you have to go into a singlet
state.  So you're going to go into the state
\be
	{1\over \sqrt{3}}\left(R\bar R + B\bar B + G\bar G\right)
	= {3\,a\over\sqrt{3}} = \sqrt{3}\,a\\
\ee
Now let's say the amplitude to go into the red-anti-red state, what
I first calculated over there, let's call that amplitude $a$.  Then 
the amplitude to go into blue-blue would also be $a$, and the
amplitude to go into green-green would also be $a$, so the amplitude
to go into this state would be $3a$ times $1/\sqrt{3}$---this is a
normalized state---which is $\sqrt{3}a$.  And the rate is $3\,a^2$.
In other words, three times the rate of making $R\bar R$.  You can
fake it if you like, {\it i.e.} sloppily, that there's the same chance
for red quarks or green quarks or blue quarks, therefore I add
them---multiply by three.  Or realize that you don't produce that
state at all but you produce a superposition.  But you come out with
the same answer.

So that's for $u$ quarks.  But then we might produce $d$ quarks also.
And by the same method of thinking,
\be
	 \sigma_{d\bar d} = 3\, \left(-{1\over 3}\right)^2\, 
	\sigma_{\mu^+\mu^-}\nn,\hbox{\ \it etc.}\nn
\ee
And now, we might produce $s$ quarks.  Now $s$ quarks are not very
heavy, so if an $e^+$-$e^-$ experiment is done at several GeV, or
10 GeV or something like that, then that's another one third \rm[writes
on board]\it\ \dots
And then, if we have enough energy to get above the $c$, we produce
$c$ quarks \rm[writes on board]\it\ and then maybe the $b$; depends, but if
we have a total energy between say 9 and 10, probably 9.65 \dots
Then there's another factor yet; for the ratio of the total cross
section to the cross section of the $\mu$, we get a curve 
\be
\includegraphics[width=0.45\textwidth]{Ree}\nn	
\ee
which is an interesting thing, it bobbles around, there are bumps and
things for the $\rho$ meson \dots  And as you get to high energies,
here's 3.6, it comes here and there's a wonderful resonance \dots
and makes the $\psi$ and then some particles, and then the background
comes up here, and over here it starts to make the $\Upsilon$ \dots
until we have 30 or 40 GeV, and we don't see the resonance we
expected for the $t$ quarks.  But it does show more or less constant
as long as you're in a region where you're not at the same order as a
mass of a new kind of quark that you can make.  And it rises from
these various plateaus with these numbers \rm[gotten by adding the
squared charges from the previous calculations]\it\ \dots

At any rate, that's {\bf marvelous}, but that doesn't determine {\bf
any} of those constants---too bad.  However, maybe the idea if we do
this at high energy, and worry ourselves about the $b$
quarks and the $c$ quarks and the interactions, we might be able to
get a little accuracy this way.  There is, of course, an interaction;
the trick is \dots that it doesn't involve the masses.
 \be
\includegraphics[width=0.3\textwidth]{eeqq-rad}\label{eeqq-rad}	
\ee
\dots and therefore that should be a good approximation.  

On the other
hand there is the possibility that we can calculate what happens
with the possible emission of a gluon \dots It could be that emitted a
gluon.  Or it could be that there were interaction forces between
these \rm[quarks]\it\ by the exchange of a gluon.  Now it is not quite as
obvious that the effects of these things will not depend on the
masses of the particles; but calculations by putting masses in show
that it really doesn't.  And that there is a correction that now
involves, as you would like, the coupling constant $g$ which we
discussed.  And what happens is, that we get that the same theoretical
ratio \rm[as we discussed before]\it\ is multiplied by a correction,
\be
R = R_{\rm pure\ free\atop particle}\left(1+ {\alpha_g\over\pi}+\dots\right)
\ee
which is proportional to the coupling constant.  I'm going to use
$\alpha_g$\footnote{Noise interferes, but I believe he says that
$\alpha_g = g^2$, with the usual factor of $4\pi$ absorbed into his
unconventional definition of $g$.} \dots strong
interactions, put the $g$ just to remind you for gluons \dots
plus higher terms.  This, then, would be a way, if we could measure
accurately enough, to determine the $\alpha_g$, and therefore $g_0$.
It would be most sensitive to $g_0$.  \rm[Question from me: what was that
subscript you put on $R$, those words?]\it\  The words say, ``pure free
particle'' theory.  This is the real ratio, corrected to the first
degree for quantum chromodynamics.  You can make a power series
expansion in the coupling constant, the first term of which is
$\alpha_g/\pi$.  And this then is a way to determine a quantity which
is particularly sensitive to $g_0^2$ and which is presumably not
sensitive to the other masses, although we do have to do a little 
work
to get rid of these things, we make corrections for these things,
depending on what region of the graph you want, we correct for that
mass \dots
 and we can do a fairly good job of
correcting for the masses \dots

So that's one way of getting a quantity which depends on $g$ and it
would be a possible thing.  The trouble with it is it's a correction
to an experiment which gives 99---\rm[pauses to think and correct
himself, in undertone]\it\ no this is I think this is 5\% of the
total---95\% of the answer, that doesn't depend on quantum
electrodynamics \rm[QCD]\it\ at all, and you've only got a 5\% correction,
it's not very easy, you can't do it very well.  So it would have been
a nice thing, and it would have been a nice experiment, so if we're
talking {\bf ideally} that would have been a place to look for
something to calculate and to measure to get something that's
practically dependent on $g_0$.

Alright, now there's another thing that's observed.  According to
this model, here, when we knock these two quarks out, and they're
going very fast, and then they just tear out and I don't know what,
radiate gluons, and do all kinds of things, they fall apart and make
whatever strings there are \dots  and what happens is that if we look at the 
{\bf momenta}--- 
\be
\includegraphics[width=0.4\textwidth]{jets}\nn	
\ee
---I draw it in a plane because it's three-dimensional, the momenta,
what we get is {\bf thousands} of hadrons, lots and lots, most of them
pions.  Okay.  And if we plotted the momenta, and I'm only going to
plot it in two dimensions instead of three, we find that they're all distributed 
in sort of a---at least if $q^2$ is very large---in a kind of
a long ellipse, which is much longer than it is wide; this is 
\dots
the order of, well if you added all the momenta \dots conservation of
energy, it's a rather big number \dots but they have a certain width, the width is of order of
a half a GeV \dots  
Now, suppose you try to calculate the chance that this happens, and
that this thing goes off \dots you remember what I do, I have to
calculate this correction to this diagram \rm[RPF is apparently
explaining how the three-jet configuration on the right arises from
radiation of a gluon from one of the quarks as in the previous diagram
(\ref{eeqq-rad})]\it\  and I add that to the rate of this \rm[the 2 jet]\it.

Now in this case I could look at this one \rm[the 2-jet]\it\ and one would
notice that there's a small chance that what happens, looking down on
the plane, is that geometrically there's a momentum like that; you've
got one quark coming out over here in this direction, and this quark
starts out this way, if you want, if I employ the virtual diagram.
What you see, though, is a quark coming this way, and a gluon going
that way, the total momentum of which balances this \rm[the other jet]\it.
So if we didn't see this region here, it would seem \dots

That can happen, at wide angle, you can ask for wide-angle gluons
coming out.  What we see experimentally is that from time to time, 
it doesn't look like this \rm[presumably the 2-jet diagram]\it, it looks
more like this \rm[the 3-jet diagram]\it.  And if this is sufficiently \dots
\rm[probably referring to the hardness of the gluon jet]\it\  and stuck out
\dots then we can interpret it as being this \rm[the 3-jet picture]\it.  
Why I have to say that is of course, if it's not sufficiently obvious,
it might be that \rm[the 2-jet picture]\it\ with a fluctuation.  I mean if
these two are close enough together, how can you tell the difference?
You can't.  So you'd have to take the case where there's a pretty good
angle, which turns out to be a low chance.  It's low because, you see,
when these two open out \dots \rm[RPF explains on the board that in this case
where the gluon is hard, the extra intermediate quark propagator is
carrying large momentum, which suppresses the amplitude]\it.  So this
happens rarely, but we can see it.  And although it does, from the
phenomenological point of view does involve soft masses and so on in
determining whether we get $\pi$'s or $K$'s and how many, we can at
least count how many jets we get, and estimate these momenta fairly
well.  

The only uncertainty is whether we should include a particular
particle here; is that part of this jet or part of that jet?  So there
is some sloppiness in it.  The sloppiness will become less as the
energy of the experiment is increased.  There's some sloppiness, but
we can do a pretty good job of guessing that these things come through
this \rm[2-jets]\it, and by measuring their rate, we get a pretty good idea of 
{this} \rm[3-jet]\it\ rate, that has an $\alpha_g$ directly, because the
amplitude for this process has a coupling constant in it, and this
rate has a coupling constant squared, so this is a direct measurement
of $\alpha_g$ \dots

Later on in the course therefore we're going to calculate that \dots
 and compare it to experiment to try to determine the coupling
constant.  Okay?  I've therefore illustrated two examples, and there
are others, \dots that seem to isolate experimental data and seem to
be able to measure, that we can roughly calculate.  I say ``seem to be
able to measure''  because we have all these little uncertainties, so
it's nothing exact, but a pretty good measurement and a pretty good
calculation by which we can determine $\alpha_g$ today and prove our
claims \dots  So that is a perturbative effect of something that
depends mostly on $\alpha_g$.

But what I wanted to explain mainly, the main thing I wanted to
explain, is that there {\bf are} physical processes, for which we can
say that we don't need to know the masses of the other particles. 
That the process has the same limiting value, the same probability,
whether these are all zero or not.  It's not so easy to correct for
them; it depends on the energy.  Let us, since we've demonstrated
more or less that there are such processes, let us assume that there
are some physical data, which involve just that \rm[$g_0^2$]\it\ and not
these \rm[the quark masses]\it, because we make corrections for this, and
there are physical phenomena for which we believe, theoretically, the
phenomena will still exist as the masses went to zero, of the quarks, and it doesn't
involve any other length \rm[scale]\it.

So we could imagine those things, those are special kinds of data that don't
involve the masses of the quarks, and without saying so, from now on
I'm talking about that kind of data.  When I talk about a physical
quantity, I'm going to suppose it's that kind of a quantity, okay?
Not something like the mass of the $\pi$, which depends on the mass of
the $u$, or the mass of the $K$, or the difference between the
$\Lambda$ and the proton, which is certainly dependent on the mass of
the $s$.  Those are not the kind of thing that I want to talk about
here.  Alright?  \dots This will focus our attention.

Now we're ready to go, huh?  No! Another complication sets in to 
\dots  The theory doesn't make sense.  The perturbation
theory gives infinities \dots the theory diverges.  You all know that.
I don't have to prove it to you; we'll discuss it all later in detail.
We're going to go back over all this and do it.  I'm describing where
we're going to go and what we're going to see.  So the theory has
infinities.  In the case of electrodynamics, which you studied, 
and as you know what this does in that case is that
there is some kind scheme for cutting off
all the integrals that are divergent, in other words 
\be
	\int {d^{\,4}k\over k^2 (p-k)^2}\,. \nn
\ee
Then this is divergent logarithmically because you've got four $k$'s
down here and four $k$'s up there.  And what people do is they say
that the propagator for instance for a photon is replaced by
subtracting from it what you would get if the propagator had a mass
and then taking the limit as the mass goes to infinity,
\be
	{1\over p^2}\to \lim_{\Lambda\to\infty}
	{1\over p^2} - {1\over p^2-\Lambda^2}
\label{pv}
\ee
and obtain results which have logarithms in them; you have cutoffs,
actually, and what do you do with these infinities?  What we do with
these infinities is the following.  If we discuss for example the
scattering of two electrons by a photon at large distances, where the 
potential is $e^2/r$, or the scattering amplitude is 
\be
\raisebox{-0.6cm}{\includegraphics[width=0.1\textwidth]{g-exchange}}
\sim {4\pi e^2\over Q^2}
\label{eetree}	
\ee
If we were to compute this, at very low $Q$, and long distances,
then this \rm[the coupling
$e$]\it\ is an experimental
number.  And now we discuss that this experimental number is not the
same as the number $e_0$ that I would put into the theory, right away,
because there are virtual diagrams 
\be
\includegraphics[width=0.15\textwidth]{loop-corr}\nn	
\ee
where this may make a pair and the pair may annihilate, and like
that, and this has got that kind of a divergence in it, and corrects
this \rm[(\ref{eetree})]\it.   What we've discovered is that \rm[taking his time to
write it from memory]\it
\be
	{1\over e^2_{\rm exp}} = {1\over e_0^2} + {1\over
3\pi}\ln{\Lambda^2\over Q^2} \nn
\ee
Yes.  Now what we're supposed to do is to make $\Lambda$ go to
infinity and this gets some kind of nonnegative result.  $Q$ is 
the momentum transfer, which is supposed to be small.  But what we say
is, when we do the theory with a cutoff, we change the theory, because
the theory by itself doesn't mean anything.  So we use a particular
cutoff, and we take an $e_0$ such that the physics is independent---is
correct, agrees with experiment.  We choose a cutoff $\Lambda$ and an $e_0$ so
that it agrees with experiment.  And now if we change the cutoff, we 
change the $e_0$ so that it continues to agree with experiment.  
In this case, we could change the $e_0$---if we change the $\Lambda$
we could change the $e_0$ so that this \dots 

It's
very important, the right sign, and I get the right sign by  
\rm[long pause while RPF checks that the
sign of the running is correct]\it. That's 137 \rm[$1/e_{\rm exp}^2$]\it.\footnote{Recall that RPF 
prefers to normalize the gauge couplings as $\alpha = 1/e^2$.}\ \ 
 As $\Lambda$ gets very big, this
\rm[the log]\it\ can get bigger than 137, so this \rm[$1/e_0^2$]\it\ would have to
go negative which makes no sense.  This theory really doesn't work, 
because it means this would have to go negative; I must be missing a
sign\dots In any case, to encounter this problem you need
to put in such a large cutoff that the logarithm is $3\pi\times 137$
which is more than 1000.  Then we would need $\Lambda \sim e^{500}\,Q$,
which is a mass greater than the mass of the universe.  So there is
no practical problem.  So that's why we repeat all these calculations,  
without ever getting into any trouble, in practice.  In practice,
we never have to take the $\Lambda$ so large to get a good accuracy
\dots
From a theoretical standpoint that seems satisfactory.

Alright.  So what the trick is, and the point is, is that when we're
putting our cutoff down, we're changing the theory.  And when we use
different kinds of cutoffs, we're using different theories. However,
as it turns out that by putting different $g_0$'s in, we  can get the
same physics from the different theories.  For each theory you have to
have its $g_0$, for each $\Lambda$ there has to be an  $e_0$ \dots

But what happens in electrodynamics \rm[QCD]\it\ is better, because
the sign is the other way around, and this is what happens.  We could
ask, for any experimental data \dots we could ask the following thing. 
\dots 
So we first have to modify the theory, to make it work.   Really, we
have to define a process of calculation.  There are several methods.
One is the method I mentioned before \rm[(\ref{pv})]\it,\\
\smallskip

$\begin{aligned}
\ \ 1.\ \qquad\qquad{1\over k^2}\to{1\over k^2} - {1\over
k^2-\Lambda^2}\nn
\end{aligned}$\\
\smallskip

\noindent plus some tricks to keep gauge invariance.  It can be done, it's not
very good about that without those tricks for gauge invariance.
\dots 
\rm[Tape was changed here.  RPF is describing dimensional regularization.]\it
\bea
2. && \hbox{ dimensional regularization. }\nn\\ 
&\alpha& \hbox{\ becomes dimensionful when\ } N\neq 4;\nn\\
 &\alpha& = \alpha_s (\Lambda_D)^{4-N}\nn
\eea
\dots as I will show you, physics with a fractional number of
dimensions.  So we calculate with dimension $N$.  And then we write
$N\neq 4$.  As it turns out, when the number of dimensions is $N$, 
the coupling constant, which I'm going to write as $\alpha$ instead of
$g^2$, the coupling constant $\alpha$ has a dependence on a scale,
an energy scale; that is, it's not a \rm[dimensionsless]\it\ number, it's not
a pure number as it is in four dimensions; it has dimensions.  So if I
put a dimension, say $\Lambda_D$ or something which is some kind of a
length, to a certain power, namely $N-4$, then this thing in front
\rm[$\alpha_s$]\it\ will be a constant, as we will vary $\Lambda$ \dots

Third method: we replace space and time---spacetime---by a lattice of
points in spacetime.  And to define what to do on that lattice, which
is analogous to the Lagrangian here, and I will discuss that, that's
called the lattice model.
\bea
3.\ &&\hbox{Lattice model. } \nn\\
&& \hbox{Has a dimensionful parameter: }\nn\\
 &&a = {1\over \Lambda_a}\nn
\eea
The lattice model has a dimension to the lattice, how small it is.
The dimension $a$ corresponds to an energy---I'm going to talk about
energies---$1/\Lambda_a$; this is the
spacing of the lattice, the lattice spacing.  Later on in the course,
I will discuss both of these methods, not so much this one, but I'll
talk about this one too.  

Oh by the way!  This method is not defined
yet, we also have to say in what gauge we do it in.  There's an axial
gauge, there's a $\partial_\mu A_\mu = 0$ gauge, which propagator,
whatnot.  They're all
variations on a theme.  All I'm trying to say is, all these things
are mutilations of our beautiful scheme.  Mutilations which we have to
make, because otherwise it's all meaningless.  However, we expect the
following.  Let's take the lattice model; it's the easiest to
understand.  Surely, if the number of points taken is sufficiently
fine, we'll get a damn good representation \dots \rm[he compares to
numerical algorithms for approximating differential equations as
finite-difference equations]\it\ if we don't get enough accuracy we make
the lattice smaller.  For any particular size of lattice, there are
artifacts---errors---if you want to talk about something this big and
the lattice is that big \dots  The size of the physical phenomenon
here would be something involving the reciprocal of the momentum
cubed \dots  Then the $q$ \rm[momentum]\it\ is the reciprocal of the wave
number, and we would like to get our constant \rm[$a$]\it\ smaller than the
wave number, or we'll never be able to represent that \dots
But in order to get good accuracy, if we want to make it really good,
we should make it still smaller, because as we all know, the more fine
we have the lattice, the better the representation.  This is some 
sort of limit, that if we took the lattice fine enough, we should get
more and more perfect agreement with everything.

However \dots so we would like these $\Lambda$'s---or in this case
it's a question of how fast you come to the limit $N=4$, and in this
case it's a question, surely putting something like that \rm[the
wrong-sign propagator in Pauli-Villars regularization]\it\ is going to change the physics if $\Lambda^2$
is not enormous compare to $Q^2$.  Right?  The propagator is
different---you changed it.  But you don't think you changed it much.
So, we have the idea, and it turns out to be right, that if we have a
phenomenon at a certain scale $Q$, and if we take $\Lambda$'s
much bigger than $Q$, and we consider different $\Lambda$'s, we can
always find a constant \rm[bare coupling]\it\ to put in that gives the same physics.
At first sight you might not have realized that you'll have to change
the $g_0$.  You would have thought that's going to be fixed, but the
theory with its divergences shows that in the same way as in electricity
there are going to be logs \dots

And so what happens is, when we ask ourselves, how to choose---let's
pick a datum to analyze, any one of them \dots  We imagine we could
calculate it.  Then we ask how to choose the $g_0$---the other
constants are involved, but we suppose that we select the datum as 
something sensitive to $g_0$ \dots \rm[RPF is illustrating this on the
board, but I do not have it copied in my notes]\it\ \dots so the physics
is always the same.  Or the physical datum comes out always the same.
The datum agrees with the theory \dots.  With one datum we can always
make it agree with the theory by adjusting for a given $\Lambda^2$ by
picking out the $g_0^2$.  For another datum, we might not get perfect
agreement, if the $\Lambda^2$ wasn't large enough, right?  Because the
theory is a little dopey at low $\Lambda^2$.  If it was a lattice,
the lattice scale is too big.  So we really want this in the limit of
very, very large $\Lambda$.  We aren't interested in the formula for
a $\Lambda$ of the same order as $Q$ or $6Q$ or something like that.
That's not the problem, because things  will work for one datum but
the other one won't work, because the theory isn't right on the scale
of \dots  Is there any question about this idea?  I've tried to
explain the idea, and I hope that if you don't catch on tell me what's
bothering you, and we'll straighten it out now. 

\rm[Question from a
 student, inaudible]\it\  That's right, we're assuming that this
thing is going to work.  Yes, this business of choosing a lattice is
analogous to the usual one, choosing a lattice for, say, doing the
diffusion equation.  And you expect it to become more and more
accurate as you make the lattice finer.  The only complication is that
we have to keep changing the $g_0$ as we do so, in order to keep the
physics the same.  But the presumption is that we {\bf will} be able
to do that, that there will be a definite limit.  That's an assumption
which is, well I don't know whether people can claim they have proved
it, but it seems to be true.  Okay?

Anyway we want the physics to be the same when we change the
$\Lambda^2$, but also, by the way, when we change the method \rm[of
regularization]\it.  Like I said \rm[referring to a previous illustration on
board]\it, various $\Lambda^2$'s and various methods; I've labeled the
$\Lambda$'s differently for different methods, but it's the same idea.
And I'm now going to tell you the answer \dots and later on I'll prove
it \dots Is there another question?  Okay, the answer is that the
value of $g_0^2$ that you have to choose has the following expansion
in $\ln\Lambda$:
\bea
	{1\over g_0^2} &=& \beta_0\,\ln {\Lambda^2}
	+ {\beta_1\over \beta_0}\ln\left(\ln {\Lambda^2}\right)\nn\\ &+& c + \left(
{a\over\ln\Lambda^2}+\dots\right)
\label{betafn}
\eea
where the further terms that are small \rm[falling with $\ln\Lambda$]\it;
these don't interest us.  We're supposing that $\Lambda$ is big enough
\dots  $\beta_0$ and $\beta_1$ are computable and known.  $c$ is
arbitrary; it's where we have the room to choose, to make the thing 
fit the data.  That's the constant we choose to make it fit the data,
so $c$ is arbitrary, you can have any constant there, and it's chosen
to fit the data.  But the formula, and what you have to choose,
depends on the method---by the method I mean whether you use the
lattice method or dimensional renormalization or what method of 
ultraviolet \dots you use.  $a$ \rm[in (\ref{betafn})]\it\ also depends on 
the method; $\beta_1$ and $\beta_0$ do not depend on the cutoff
method.  

\rm[Question about the arguments of the logs not being
dimensionless.]\it\  Yes, now that's
ridiculous, isn't it?  Very good, $\Lambda$'s are energies.  And I've written
logs that aren't any---ha ha ha \dots  Well, the old professor can fix
that, we'll divide by $\rm{(1\,GeV)}^2$:
\bea
	{1\over g_0^2} &=& \beta_0\,\ln {\Lambda^2\over {\rm GeV}^2}
	+ {\beta_1\over \beta_0}\ln\left(\ln {\Lambda^2\over {\rm
GeV}^2}\right)\nn\\ &+& c + \left(
{a\over\ln\Lambda^2}+\dots\right)
\label{betafn2}
\eea
But why ${\rm1\,GeV}^2$?  We'd better contemplate that some other joe 
will come along and
use ${\rm 10\,GeV}^2$ in there.  So we'd better contemplate what happens if
instead of writing with ${\rm 1\,GeV}^2$ underneath with $M^2$ underneath,
before somebody else comes along, he likes to write his thing
different.  He would write this.  And I would like to explain to you
why it doesn't make any difference.  Because that makes it look that
it is still more arbitrary, that we have another thing \dots we don't.
And I'll show you why.  We might have a different constant \rm[$c'$]\it, that's the
clue:
 \bea
	{1\over g_0^2} &=& \beta_0\,\ln {\Lambda^2\over M^2}
	+ {\beta_1\over \beta_0}
  \ln\left(\ln {\Lambda^2\over M^2}\right) + c' + {a'\over
\ln{\Lambda^2\over M^2}}+\dots\nn\\ 
&=& \beta_0\ln\Lambda^2 \!+ {\beta_1\over \beta_0}
\ln\underbrace{\left(\ln\Lambda^2\!\!-\!\ln M^2\right)}\!+ (c'\!\!-\!\beta_0\ln
M^2)\dots\nn\\
&&\qquad\qquad\qquad
\ln\Lambda^2\left(1-{\ln M^2\over\ln\Lambda^2}\right)\nn\\
&=& \beta_0\,\ln\Lambda^2 + {\beta_1\over \beta_0}\ln\ln\Lambda^2
+\left(c'-\beta_0\ln M^2\right)\nn\\
&+& \left(a'-{\beta_1\over\beta_0}\right){1\over\ln\Lambda^2}+\dots
\label{betafn3}
\eea
You see that all I did was change the constant.  If I had used ``1''
in here \rm[for $M$]\it, I've got a certain constant, when I made them fit
the data.  If I had used 10 or $M$, I'd get a different constant,
that's all.  So I'm still okay, right?  The $M$ makes it look as if
there is another parameter, which adds to the confusion of this damn
thing.  Because when you hear about choosing $M$ and choosing
$\Lambda$ and choosing $g_0$, there's no choosing $M$, really: it
doesn't make any difference.  It's just a question of the definition
of the constant when you go to fit the data.

Now the part that I hadn't finished is here; you notice that as
$\Lambda^2$ goes to infinity \dots \rm[RPF explains that the 
regularization-dependent term $a'$ becomes negligible as the cutoff is
removed.]\it\  So I must take my $\Lambda^2$ big enough that this term
doesn't amount to anything.  So everything's okay, and that answers
your question about the units.  Is that alright?

Sir? \rm[Question from me: Is this equation just the two-loop
approximation?]\it\ Yes---no, it's exact!  \rm[me again: there aren't more
logs of logs?]\it\  Oh yeah, maybe down here, there's log log, this times
log log, stuff like that.  But always smaller, okay?  No, there's no log log
log, no.  There's no log log log.  I'll explain to you why.  It may
be wrong in that there may be a term---I'm not sure, okay, like
$\ln\Lambda$ times $\ln\ln\Lambda$, or something like that, which is
still smaller than this one, but not much.  And things like that, but
these are all dropping out as $\Lambda$ goes to infinity.  So, I
should say, terms of this order or smaller are going to drop out,
that's what this curly line means.  Okay?  All we have to do is to
take the $\Lambda^2$ very large and then we can do that.

Alright.  And that's the formula how then we believe, that if we do
that, choose the $g_0^2$ so it's equal to this and adjust the constant
$c$, we can fit the first physical datum.   And then a second physical
datum \dots what with the same thing, with the same constant, you
should get a fit to experiment, provided that we've chosen the
$\Lambda^2$ large enough that this \rm[the $a$ term]\it\ is small enough that
everything's okay.   Alright.  Are there any questions?  That's all
there is to it.  That's all there is to what we call the way the
theory's supposed to work. 

Review.  The theory diverges.  In electrodynamics that's really
serious \rm[the Landau pole problem]\it\ but we don't pay any attention to it
for practical reasons, $e^2$ is sufficiently small \dots  In quantum
chromodynamics because of the opposite sign,  there isn't any real
difficulty \rm[because of asymptotic freedom]\it.  We {\bf can} choose the
$g$ in terms of the method that we use to make the cutoff in such a
way that we would expect that the physical data agree with
experiment.  That is what the theory is.  The theory is, strictly
speaking, not the Lagrangian which we wrote down or the path integral
we wrote down, but the path integral plus all this crap about how to
make a cutoff, plus this baloney about how we have to choose $g$.
And then we should take the limit as $\Lambda$ goes to infinity to get
the most accurate result.   {\bf That's} the theory.  The theory of
quantum chromodynamics is {\bf not} defined by the Lagrangian alone.
To put it another way, you cannot say to a mathematician ``hey, here's
my Lagrangian \dots, figure out the consequences,'' because you
haven't told him the full physics of what you intend to do, which
is---because if you give him that he'll find out that the answer is
infinity---it doesn't make any sense.  The true theory is the
Lagrangian plus a cutoff scheme, plus a proposition as to how the
$g$'s go, so that the results will be independent of $\Lambda$, the
cutoff scale, as the cutoff scale gets sufficiently fine.  And this
has to be the way to do it.

Now you can find this out by perturbation theory, of course.  You'll
notice that as $\Lambda$ gets very large, the $g_0^2$'s are very
small.  When the $g$'s are small we can compute everything by
perturbation theory, and that's the way you computed the $\beta_0$ and
the $\beta_1$.  Well why didn't we compute $c$?  Because you can't;
it's an arbitrary constant.  And why didn't we compute $a$?  Because
that depends on the method.  We can compute it for each method, but it
doesn't do us any good.  I'll tell you why.   This is a signal that
there will be errors of $1/\Lambda^2$ \rm[$1/\ln\Lambda^2$]\it\ in the end,
because we haven't taken a fine enough lattice spacing, using the
example of the lattice.  Because the lattice doesn't really represent
the continuum.  And this is a kind of measurement error.  So it's no
use to compute this ``$a$\!'' accurately for a given scheme \dots

Yes? \rm[Question: $\Lambda$ is something you choose arbitrarily?]\it\  No,
we try to make it as large as possible. \rm[student: We try to make it as large as
possible, but we can {\bf choose} it to be as large as possible.]\it\  Yes,
that's true. \rm[student:  Okay, it seems like we can always choose
$\Lambda$ large enough to make $g$ small enough that we can use
perturbation theory]\it\ That's correct \rm[student: and then we can use perturbation
theory on any problem.]\it\  That's correct.  \rm[student, not satisfied:
I have been told that \dots ]\it\ But the series diverges \dots for
processes with small $Q$, small momentum transfer, the corrections
to the propagator get bigger and bigger \rm[student: even
when $g$ is small?]\it\  What happens is, when you take this exchange
between two quarks \dots you get corrections \dots say a loop of
gluons or something, which modify the propagator between two quarks.
Now this thing, when you calculate it, involves something like
$\ln \Lambda^2/Q^2$, when you calculate it.  This interaction will
now will have a term like this with an extra $g_0^2$.  Although the
$g_0^2$ is small, the $\ln\Lambda^2$ is undoing it, and you get a
finite \dots which isn't small.  It gets to be small in its effect 
if $Q^2$ were big
enough, as we will learn next time, but if you ask the question at
low $Q^2$, it just doesn't work.  The divergences of the perturbation
theory undo the smallness of the \dots \rm[bare coupling in the UV]\it. Okay?

Alright.  Well the difficulty with the perturbation theory is not that
it doesn't exist; it's that you can't sum it.  We don't know how.
Sometimes we can sum some terms, but we can't do a very good job,
we have to think about it, rather than calculate it, even though with
sufficiently large  $Q^2$, the effect of $g$ is small, for smaller
$Q^2$ the effects are bigger \dots

\rm[The following apparently refers to eq.\ (\ref{betafn2}) or
(\ref{betafn3}).]\it\
I'll be putting a mass squared here from time to time, maybe \dots
you'll appreciate that it doesn't represent an independent choice of
\dots It does---there is a way of making it look like an independent
choice.  Obviously, there is \dots suppose that we finally fix the
data and worked it out and determined $c$.  And somebody could find
an $M$ so that this canceled out.  And then he could say that this
formula for $1/g_0^2$ is exactly this.  And there's no constant and
the other constant is $M$.  And there's all these different ways of
representing the same thing, which causes a tremendous amount of
confusion to a lot of people, and I'm sorry for that, because we
really \dots calculate it, so we don't know the $M$; some people come
out with $0.2\,{\rm MeV}$ or something like that\footnote{RPF
meant to say $0.2$\,GeV, referring to the scale $\Lambda_{QCD}$ that
he usually denotes as $\lambda_P$.}
and say that's what it is, $0.5\,{\rm MeV}$, whatever.  So they don't know
it well enough, so we don't have the numbers accurate enough that we
can do one or another of these things once and for all and
be done with it, so we have to kind of leave all these balls in the
air, as to which way you would prefer to write it---whether to
choose an $M$ and say that it's the constant \rm[$c$]\it\ 
I want to determine, or to say the constant I'm going to choose is
zero and it's the $M$ I want to determine.  So you'll hear different 
people saying different things, but you have to understand that they're all
equivalent.  It always makes it a little easy to do it on the blackboard
because I've prepared the lecture, but then you have stop and think
are they really equivalent \dots or you'll forget how I did that.

Are there any other questions about the idea?  As you can probably
see, because of the logarithms, you might expect $1/g_0^2$ is not
really converging, and if we are working at $100\,{\rm GeV}$ or something like
that, and you wanted to change $\Lambda$ to change the logarithm in
order to get something that \dots it's damned hard!  \dots
And therefore something like the lattice \dots They have a technical
thing; they can do some calculations on lattices \dots The lattices
really aren't small enough to get a good answer  \dots  They're as
small as they can make them and still do the calculation, because of
the number of \dots that are available \dots But when they try to 
make the $\Lambda$\footnote{RPF means the lattice spacing} smaller to
be more accurate, they need a lot more computer time, because in four
dimensions if you decrease the lattice by one half, you have sixteen
times as many points to compute.  Sixteen times as much work.   But
even changing the lattice by $1/2$ is only changing $\Lambda$ by 
a factor of two, and $\ln\Lambda$ doesn't do much.  And so it is very
difficult, in fact I would say virtually impossible, to make the
numerical calculation practical \dots limitations of computers.
To make the numerical calculations with greater accuracy \dots
So I think we have to study this theory, not only to figure out
an analytic way \dots to understand well enough what happens at short
distances \dots so that we have a better way of computing that is less
sensitive to this brute force scheme that they're now using.
During this course I will discuss all the numerical calculations,
and more of these methods, and everything else; this is just an introductory lecture
to explain where we have to go.

\section{Transcription: Renormalization: applications (1-12-$xy$)}
\label{A2}
\rm[A student is asking whether RPF is going to explain the
correspondence between dimensional regularization and the cutoff to
which he has been referring so far.]\it\  Yes, I am.  I worked it out the
other day and it's very simple; I understood it \rm[RPF says something to
the effect that he might get some details wrong here since he is going
by memory]\it\  The point is that in dimensional analysis \rm[regularization]\it\
the coupling constant has a dimension,  so you represent the physical
coupling constant as $g$ times some dimension---some energy, which
corresponds to our $\Lambda$---to a power of 4 minus the number of
dimensions.  Now when you do an integral over correction terms in
perturbation theory, all of those integrals converge if the number of
dimensions is less than 4, and so the corrections to the coupling
constant---that's what I'm trying to get straight \dots  Now as the
$d$ approaches 4 it turns out---let's say we write
$d=4-\epsilon$---then there's an $\epsilon$ down here as you approach 
\dots  It ends up that you're trying to work out something like 
$(1/\epsilon) \times \Lambda^{-\epsilon}$ which gives you $1/\epsilon$
times $1 - \epsilon\ln\Lambda$ \dots I was trying to get that straight
just before I came \dots I couldn't figure it fast enough \dots
but there's a direct correspondence \dots

We have been talking about the renormalization of the coupling
constant in a kind of abstract way, and as usual at the beginning of
each lecture  I have to fix up some minor things \dots in order to
make everything I said consonant with the outside world, all the
equations in which I wrote $g^2$, in all those the $g^2$ should be 
replaced by $g^2/16\pi^2$; then
\be
	\alpha = {g^2\over 4\pi}
\label{eqA2-1}
\ee
This is the real world.  This is me.\footnote{Apparently RPF made a
side-by-side comparison of the two notations, but I only copied the
``real world'' version in my notes.}\ \   My $\alpha$ was $\pi g^2$
times $4\pi$, which is not good to do.  This is the right thing;
first you substitute this, then you substitute that \dots  All the
equations are changed with the appropriate positions of the $4\pi$'s.
There was one equation that we chose to define $\alpha(Q^2)$, and that
was $1/\alpha(Q^2)$, which now becomes 
\be
	{4\pi\over\alpha(Q^2)} = \beta_0\,\ln{Q^2\over\lambda_P^2} + 
	{\beta_1\over\beta_0} \ln {4\pi\over \beta_0\,\alpha(Q^2)}
\label{eqA2-2}
\ee
The equations I'd written before didn't have the $4\pi$'s.  Alright?

Now I will just remind you of what we discovered, that when we did
perturbation theory to  order $g_0^2$ to some process, that's all the
order we worked out.  And this is replaced ultimately by
$\alpha(Q^2)$ \dots that's in perturbation theory; it becomes this
if you would sum the leading logs, because you always know they come
in common.  It can't be $g$, because that's $\ln\ln\Lambda$ \dots
nothing depends on the cutoff \dots  In other words, when you do first
order perturbation theory, you simply replace the $g_0$'s by
$\alpha(Q)$, and you get a much more accurate result.  You've
included all the higher order leading logs.  If you try to do it to
next order in the coupling constant, it gets a little more
complicated.  You get
\be
	g_0^2 + g_0^4\left(\ln{\Lambda^2\over Q^2} + a\right)
\label{eqA2-3}
\ee
---I'm not going to try to get my $4\pi$'s right---and we know already
there was one of these \rm[in the argument of the $\ln$]\it\ in there, that has to be there, and that's just
to get the right coefficient.  And then there will be some constant
\rm[$a$]\it\ that has to be worked out when you do the second order
perturbation.  If that's the case, then this turns into 
\be
	\alpha(Q) + a\,\alpha^2(Q)
\label{eqA2-4}
\ee
So the way to do second order perturbation theory, is after you do it,
take away the logarithmic term and just look at the constant term, and
the constant term is the coefficient of the second order term in
$\alpha$.  And it will be a little more complicated with the 3rd order
term.  But we can work it all out.  And it tells you, in other words,
from the perturbation expansion \dots to write it not in terms of 
$g_0^2$ but in terms of $\alpha(Q)$.  It's the effect of summing the
leading logs, that are evidently going to come in, although we haven't
worked it out for the higher terms.   This is a much harder \dots

So in fact therefore people say that the coupling constant is
dependent on $Q^2$ because of the running---that's what running means
\dots I just want to say it again, it looks complicated but it's
relatively simple; at first order you replace $g$ \dots at second
order \dots.

There's many papers and places where you can read about this process
of renormalization; I mentioned a \rm[version of ?]\it\ the renormalization
group equations which will look much simpler than any ones you'll see
anywhere, unless of course  I have it physically right and have done
it nicely. The problem is making sure that the quantity that we're
dealing with depends \dots And that the quantity we're dealing with 
is physical and doesn't involve something like just a 
Green's function, an expectation of $\phi$ at one point and $\phi$
at another point.  Because the $\phi$'s--wave functions, or field
operators---also shift their coefficients with various $c$'s and so
on.  If you deal with a physical quantity that you measure, you don't
have any of that stuff.  For example, if we have an expectation of 
vector potentials
\be
	\langle A(x_1)\cdots A(x_n)\rangle
\nn
\ee
at one point times another point or something like that, then you have
to watch out that these vector potentials are also changing their
definition as we change the coupling constant.\footnote{In my notes I
have written that RPF is talking about anomalous dimensions here.}\ \   
So in the
renormalization group equations for Green's functions \dots
well it looks much more complicated, but it really isn't that much more
complicated, it's just the physical ideas were adequately 
described \dots it's always a good idea to stick to physical 
questions \dots So a lot of \dots book or any paper on
renormalization, you find it enormously more complicated than anything
you've seen; they've added all kinds of extra stuff.  It also has a
lot about the history where people tried this and that and did this
and did that and proved it this way and proved it that way
\dots  The subject looks worse than it is. Okay. So try to look it up
and  \dots  that I cheated you somehow in describing it \dots  I did 
find a nice book called \underline{Renormalization} by John Collins,
Cambridge University Press, 1984.  
I can't read it all, it's too complicated for me.  Now there's 
{\bf thousands} of references on renormalization \dots

Now that's the end of last time---I'm always fixing up the lecture
before.  Oh, there's one more \rm[thing to fix up]\it; I've forgot the
$\beta_1/\beta_0$ here \rm[eq.\ (\ref{eqA2-2})]\it, and I'd better tell 
you where the $\pi$'s are \dots The fact that this is positive, 
if it was only about electrodynamics and didn't have the solution to
write the conclusion to be negative; it also suggests, the fact that 
it's positive means that everything will work \dots wonderful theory
\dots \rm[RPF refers to the Landau pole problem of QED]\it\  However if the
number of flavors is more than something like 17, then we're in
trouble.  So most people believe there is not 17 flavors of quarks.
We only know of five so far.  It might be six, since people like to
fill out the symmetry with $t$ quarks.  But there is no theoretical
reason to say that there isn't another group of three, like $u$, $d$,
$s$---($u$,$d$), ($s$,$c$)--strange and charm, and you have beauty and
truth or something \dots \rm[inaudible, some other word starting with
``t,'' drawing laughter from the class]\it.  And maybe there's $x$ and
$y$.  And maybe there's $w$ and $v$---we don't know; there's no
understanding as to why there's more than one family, or why it stops at
three families \dots

Okay, well, \dots the problem is how do we actually calculate
something \dots and perhaps this whole problem \dots and you'll see it
all coming out \dots So that's what I'm about to do.  But even there,
before I do a particular problem, I want to do something about
guessing where the divergences are going to come.  We all know from
doing quantum electrodynamics and other field \dots I'm assuming you've
taken a course in field theory \dots  comes out to be divergent
because of \dots  And so let's try to find out when we're going to get
\dots  So if you were to take a very complicated diagram to calculate,
some crazy thing \dots there's a quark, quark, gluon, gluon \dots
it doesn't make any difference \dots something like that \dots
\be
\raisebox{-0.96cm}{\includegraphics[width=0.1\textwidth]{2loop}}
\qquad \int {d^{\,4}k\,d^{\,4} p\over (p^2,\ k^2 + \cdots)^6}
\times f(p,\ k, \cdots)
\label{eqA2-5}
\ee
And then such an object will end up with two integrals over momenta,
one for this loop and one from this loop, four-dimensional integrals
\dots Therefore we need to integrate over eight variables, and in the end
the question is does it diverge?  The real question is, will there
be---what kind of formula are we going to integrate?  There will be 
various $k^2$'s, minus this and that, propagators, maybe there'll be 
six propagators.  I don't mean there's a sixth power of the
propagator, I mean there are six of these kind of things in a row.
Perhaps \dots  So I'm not going to worry about whether it's $p$ or
$k$.  But there will be from the gradients in the couplings, up here
\rm[the numerator $f$]\it\ there will be some $k$'s and $p$'s.  And then the
question is when we go to do these integrals, we will get a
divergence, a logarithmic divergence or \dots depending on 
how many powers are down here and how many powers are up here.  If
there's more powers downstairs than there are upstairs, then it will
be a convergent integral.

So what we have to do is count how many powers there are upstairs and
downstairs.  And that means looking at all these couplings and seeing
if there are gradients in them, taking two powers for every propagator
of gluons, one power for every propagator of quarks \dots Now I think
there are no gradients in the coupling of a gluon to a quark and so
on.  And we get all \dots and have a big counting job.  And it will
depend on the structure of  the diagram.  And now, for a miracle. 
There are very many relationships between the way diagrams are
constructed, and what kinds of topology you can take.  Of course one
of the typical theorems of topology is that \dots I don't know why
that should be relevant for this because it doesn't have to be a
planar diagram, that \footnote{In my notes I have written that this
assumes a vacuum diagram, no external legs.} 
\be
	\hbox{Edges} + \hbox{Vertices} = \hbox{Faces} + 2
\label{eqA2-6}
\ee 
In other words, there's a relationship between the number of loops, 
the number of vertices and the number of propagators \dots
But what I'm trying to say is that the number of loops, the number of
junctions, the number of couplings, and all this stuff are not
completely independent of each other, but they're related to each
other.  And those relationships turn out to mean that I can make this
count and tell you the answer in a very nice way, it's very simple.
The net power of any integral, the number of numerator over the
denominator---in this case for example, one, two, four and four is
eight \dots there are twelve down here, let's say there's one more
here, that would be fourteen, so that's a net of minus two, so that's
convergent.\footnote{The answer for the diagram in
(\ref{eqA2-4}) should be $-1$}\ \  The net power of all the momenta,
which is $N$, has this property:
\be
\begin{array}{ll} N=0,& \hbox{log divergence}\\
	N = 2, & \hbox{quadratic divergence}\\
	N < 0, & \hbox{converges}\end{array}
\label{eqA2-7}
\ee
and the wonderful thing is, no matter how complicated the diagram,
the formula is that 
\be
	N = 4 - {\cal N}_g - \frac32{\cal N}_q,
\label{eqA2-8}
\ee
four minus the number of gluon lines coming in from the outside,
minus $3/2$ the number of quark lines coming from the outside, period!
It doesn't make any difference how it's all structured in there.
That's an entertaining thing; you can play around and try to prove it
to yourself \dots by actually counting things up and showing various relations
of the number of intersections and junctions and three-point
couplings.  See, for example, this relation between the number of 
junctions and the number of lines, because each line has two ends,
so you know, take the number of lines divided by two, it's going to
tell you how many junctions there are.  At any rate,  this ends
up as being true, which is most remarkable.  Well now I'm going to
prove \dots

\rm[Question from student: \dots superficial degree of divergence?]\it\ 
Yes, yes, yes, superficial degree of divergence.  It is often called
a naive counting divergence because, what could happen, is that this
whole problem, turns out there are momenta up here, but they're not
the momenta of the integrand that you have to integrate over, but they
might be the momenta of the outside lines---let's call $q$ the typical
momenta coming in, that we're not integrating over.  Then
dimensionally, from the point of view of the number of energy terms, it's the
same dimension, but the integration is more convergent.  \rm[Student:
Or what could happen is that we have two integrals, over $k$ and $q$;
the $q$ integral is very convergent, and the $k$ integral is
divergent.]\it\  That could happen, but it usually doesn't.  \rm[Another
student: but anyway this is a worst case, this counting?]\it\ Yes.

Let me explain how I did this.  A way of looking at it, one way of
a direct count, the most obvious way \dots  Here's another way.
This, we could say, is part, it's a diagram for some process.  It's
a piece of a lot of terms that are going to be added together to
produce a matrix element $T$ for a process.  Now since they're going
to be added together, they all have the same dimension, so the
dimension of this is the same as the dimension of $T$.  But in the
case of $T$, we have various rate formulas.  Let's take an example.
We have a single particle going into \dots I don't care how, and 
disintegrating into $N-1$ particles, one coming in and $N-1$ coming
out.  Then we say that the rate at which this happens goes like this:
\centerline{\!\!\!\!\!\!\!\!\!\!\!\!\!\!\!\!\!\!\!\!\!\!\!\!\!\!\!\!\!\!\raisebox{-0.96cm}{\includegraphics[width=0.1\textwidth]{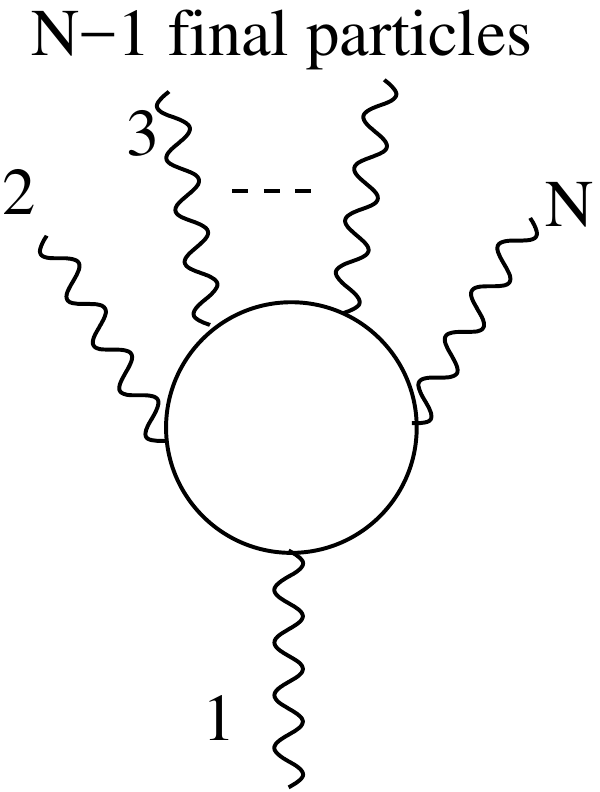}}
\parbox{0.3\textwidth}
{\bea &&\hbox{\!\!\!\!\!\!\!\!\!Rate of decays of 1 into $N-1$ particles:}\nn\\
d\Gamma &=& {1\over 2 E_1}|T|^2\prod_{i=2}^{N}
(2\pi)\delta(p_i^2-m_i^2)
{d^{\,4}p_i}\nn\\ &\times& (2\pi)^4\delta^{(4)}\left(p_1-\sum_{i=2}^N\,
p_i\right)
\label{eqA2-9}
\eea}}
including the $(2\pi)$'s---which have nothing to do with dimensions,
but I'm being accurate for a change---that's very rare.\footnote{The
factors of $(2\pi)^{-4}$ for each $d^{\,4}p_i$ are missing}\ \   That's
the exact formula for the rate.  There's a similar formula for the
cross section, but that's with two particles coming in; let's just
take the case of one.

What I'm going to use it for is to determine the dimensions of $T$.
By the way, in this case it's very important that the coupling
constant has no dimensions.  We're going to have coupling constants,
$g$, $g$, $g$, $g$ all over this thing, and those $g$, $g$, $g$, $g$'s
is not going to make any difference to the dimension.  So all I have to
do is find the energy dimension of this thing, and thereby obtain the
maximum degree to which it could possibly diverge.  Okay, now the
rate is one over the lifetime of that object, and which is therefore
an energy,
\be
[\hbox{Rate}] = \left[{1\over\tau}\right] = [\hbox{energy}]
\ee
and multiplied by this energy \rm[moving $E_1$ to the left-hand side]\it,
we get 
\be
\hbox{[energy$^2$]} = [|T|^2] \, [{\rm energy}]^{2(N-1)} \, 
[\hbox{energy}]^{-4}\nn
\ee
Okay?  And so we find out that
\bea
	[|T|^2] &=& [E]^{8-2N}\hbox{ or}\nn\\
	\hbox{\rm[$|T|$]} &=& [E]^{4-N}
\label{eqA2-11}
\eea
where $N$ is the total number of lines coming out.  And that's what
this formula \rm[(\ref{eqA2-8})]\it\ was supposed to be, only it's slightly
\dots Because we then can prove---you have to watch out, we talked
about the dimension of the integral that we're going to get when we 
do this.  Well that's not quite the same, because whenever a quark
comes in, there's a spinor for that quark, and that spinor has a
dimension.  So the integral \rm[the loops inside the diagram, with
external wave functions removed]\it\ is not the same as the dimension of 
$T$.  But you have \rm[the dimension of]\it\ the integral is four minus
the number of gluons minus the number of quarks, which is what I have
there \rm[(\ref{eqA2-11}]\it, that's $T$, but for each of the quarks there 
was a spinor, which has a dimension of $1/2$, and so the integral
doesn't quite have the same dimension as the $T$, 
\be
	I = 4 - {\cal N}_g - {\cal N}_q - \sfrac12 {\cal N}_q
\label{eqA2-12}
\ee

To remind you, that a quark or spinor has a dimension of $1/2$,
you remember that if you're going to sum this over spins, when you
sum this over spins and there are quarks in it, what you do is you
say, oh I know, I'll get some kind of a matrix element, and then I put
a projector $\sum u\bar u = \slashed{p}+m$ in it when I sum over
spins.  That means when you sum over spins, you put an extra energy
in.  So the dimensions of the $T$ when summed over all the spins 
has these factors, one of these for every quark.  That was $|T|^2$,
therefore half an energy for each quark, so that's where this \rm[last
term of (\ref{eqA2-12})]\it\ comes from.

And now finally if you have some good reason to know that the final
answer \dots for some reason you know, gauge invariance might be such
a reason, that you know the final answer must have zero divergence
\rm[in terms of contracting external momenta with the amplitude]\it, for 
a gluon with momentum $q$ that it has to come in this way, 
\be
	q^2\,\delta_{\mu\nu} - q_\mu q_\nu
\ee
so that it will automatically give that $q^\mu$ on that is zero.
So let's say we know there must be a factor like this in front.
Then of course we know the dimension of the integral is that much
smaller.  So the thing to do to remind yourself of that 
is to take $-p$,  
\be
	N = 4 - {\cal N}_g - \sfrac32{\cal N}_q -p
\label{eqA2-14}
\ee
where $p$ is the known power of the coefficient in front.  What I mean
by that is the power of external momenta.  If you say well I'm going
to take the worst case, then $p$ is zero.  We'll consider $p$ zero;
I'm going to talk about the cases where it isn't \dots  Alright?

Now let's---oh, I had noticed something, that I copied \dots scrap of
paper; I'm not going to remember all of it, therefore I can't
guarantee it, so you might like to try to prove it.  I also got interested in what the
order in $g$ is for a given diagram.  So I did all my algebra \dots
to find the order. 
And the particular way that I worked out the order was the 
conventional way, in which the fields that I usually use are replaced
by $g A$ so that the action looks like $(\partial_\mu A_\nu - 
\partial_\nu A_\mu - g[A_\mu,A_\nu])^2$---that's not my conventional
way.  That's the way I want to calculate the order \dots
When I define it that way, and I find the order of $g$, I find the
following rule, 
\be
	2\times (\hbox{loops}) + {\cal N}_g + {\cal N}_q - 2
\ee
\dots to give you an exercise \dots I don't guarantee it, because I
found it on an envelope without checking it.  In case you find that
useful, maybe you could disprove it or prove it; it would be
interesting to try.  

Let's find out what kind of diagrams diverge.  And to be specific in 
drawing the diagrams, I'm going to draw the lowest order in each kind.
The lowest order in $g$, not the lowest divergence \dots  Then we can 
represent our diagram by telling how many gluons there are, and how
many quarks there are.  Here's a little table, and for each case I'm
going to draw a diagram to illustrate it.  Now if there are no lines
coming in, so the diagram has no external lines, well we never have to
calculate it \dots So we start with one gluon.  And that's a thing
that looks like this: the gluon's coming along \dots you can have a
loop of gluons, or you can have a loop of quarks; that's a typical
diagram.  I just draw one typical diagram of this kind.\footnote{In my
notes I had added the extra examples.}\ \  There would be no electron
lines coming in from the outside, and this number \rm[$N$]\it\ would
be three \dots
\begin{table}[h]
\centering
\tabcolsep 2.5pt
\begin{tabular}{|c|c|c|c|c|c|c|}
\hline
${\cal N}_g$ & 1 & 2 & 3 & 4 & 0 & 1\\
\hline
${\cal N}_q$ & 0 & 0 & 0 & 0 & 2 & 2\\
\hline
$N$          & 3 & 2 & 1 & 0 & 1 & 0\\
\hline
\raisebox{0.cm}{${\hbox{typical}\atop\hbox{diagrams}}$} &
\includegraphics[width=0.03\textwidth]{tadpole}&
\includegraphics[width=0.03\textwidth]{vacpole-q}&
\includegraphics[width=0.033\textwidth]{3g-loop} &
\includegraphics[width=0.033\textwidth]{4g-loop} & 
\includegraphics[width=0.02\textwidth]{self-en}& 
\includegraphics[width=0.06\textwidth]{vertex-corr}\\
 &
\includegraphics[width=0.03\textwidth]{tadpole-g} & 
\includegraphics[width=0.03\textwidth]{vacpole-g} & 
\includegraphics[width=0.033\textwidth]{3gg-loop} & 
\includegraphics[width=0.033\textwidth]{4gg-loop} &
&
\includegraphics[width=0.06\textwidth]{vertex-corr-g}\\
\hline
\end{tabular}
\caption{Superficial degrees of divergence, $N$, for graphs with
${\cal N}_q$ external quarks and ${\cal N}_g$ external gluons.
\label{tab0}}
\end{table}
\rm[Question from student: you mean quarks?]\it\  Yes, always I mean quarks,
not electrons.  I say ``electrons'' and I say ``photons,'' but I mean
gluons when I say ``photons,'' and I mean quarks when I say
``electrons.'' \ \ 

Now the next case would be that there were two gluons, and there was
something going around.  It could be a quark.  Or if you prefer, to
make it more interesting, make it a gluon, I don't care.  Because
these are typical, I'm only illustrating.  If there are two gluons
coming in, as you know, now the divergence is 2.  Well you can keep
this going.  Now we've got three gluons, with something going around
here, these are just to illustrate the idea, and there is no quarks
coming, the divergence is 1.  Or there could be four, \rm[RPF makes sound
effects as he draws the legs]\it\ beep, beep, beep, beep, \rm[and the loop]\it\
loo-loo-loo-loo-loo; now we have four
of these, and the divergence is 0.  And now if I put five, then it
gets convergent.   And I stop now; I'm only interested in the
divergences. 
 
So I start now, over again, this time putting in some quark lines.
Now you can't just have one quark line because of the conservation
of quarks, so the first case you would get would be something like
that, it would be no gluons coming in, and two quark lines coming in,
and that's \rm[$N=$]\it\ 1.  And then you could have two quark lines coming
in and one gluon line, and you get 0.  Now the next thing would be
with four quarks, but that's already convergent.  And that's the end.
Those are the only diagrams---sorry the only types of
diagrams---that will bring divergences.  

Mind you, this diagram here,
this is lower order, it's plus any internal complications, don't
forget.  That is, the same divergence will appear to occur if I draw
a diagram like this.\footnote{I did not copy the diagram, but the
reader can imagine adding loops to the lowest order diagrams.}\ \ 
It's the same divergence.  And that, due to that theorem \rm[eq.\
(\ref{eqA2-12})]\it\ \dots

Now we discuss \rm[in more detail]\it\ the individual terms \dots
Well if you have a gluon coming in \dots the vacuum here \dots
\be
\raisebox{-0.25cm}
{\includegraphics[width=0.04\textwidth,angle=90]{tadpole}}\sim
\langle A_\mu\rangle\,,
\ee
a kind of expectation of the vector potential.  The vector potential
could be in any direction in the vacuum, and it averages to zero.
So this is physically zero, and we never calculate it, because of 
symmetry.  The symmetry is this: we can make $A\to -A$, the theory is
unchanged;  not true, not true \dots a little more subtle way; there
is a symmetry in here.  You change the sign of \dots something like
that; anyway there's no direction you can \dots the gluon, so there's
no expectation for the mean gluon field.  So there's no term, you don't
have to worry about it now.  In fact we never calculate it  \dots
there's never any \dots

Now ordinarily we would expect this
\be
\raisebox{-0.25cm}
{\includegraphics[width=0.04\textwidth,angle=90]{vacpole-g}}
\sim \delta_{\mu\nu}q^2 - q_\mu q_\nu 
\ee
to be a quadratic divergence,  from what it says here \rm[in the
table]\it.  It turns out that the conditions of gauge invariance in the
case of quantum electrodynamics for instance, and also in quantum
chromodynamics, means that there has to be such a factor 
\rm[$q^2 - q_\mu q_\nu$]\it, and I'm not
going to prove it now, I'm just telling you about this, that gauge
invariance makes that $p=2$ \rm[in eq.\ (\ref{eqA2-14})]\it.  And this
turns this to a log divergence.  In other words we overcount \dots
it's not as bad as we think.   However if we go to calculate it, if
we're not very careful with the cutoff, suppose that the method of
cutting off doesn't guarantee gauge invariance, then we can easily get
a quadratic divergence \dots we screwed up, okay?   But if you do it
right, so that you don't lose the gauge invariance in the cutoff
process,  then you can show that this will only be log.

Now the divergence of the first power, 
\be
\raisebox{-0.cm}
{\includegraphics[width=0.05\textwidth,angle=90]{3g-loop}}
\ee
really never occurs.  Because if we would have--how could we have it?
We could have $p$ dot something, like $q$ or something, and then you
would have $p\cdot q/(p^2)^2$ or something.  This would be \rm[in the
denominator]\it\ spherically symmetrical, and this would be \rm[in the
numerator]\it\ lopsided.  If I change $p\to-p$ in the integral I get the
same thing with the minus sign.  The mean value of $p_\mu$, a single $p$,
integrated over all directions, is zero.  So for that reason, this
``1,'' here and here both \rm[in the 3rd column of the table]\it\ the 1
is really equivalent in the end to 0; it's only log divergent.
Although it looks like it's linearly divergent, the linear pieces
average out, provided that your cutoff isn't lopsided, alright?
\dots If you have a reasonable cutoff that's symmetrical, you only get
log divergences.  And these \rm[the terms with superficial degree of
divergence 0]\it\ are log divergent.  And so it turns out that this whole
mess, in practice, this whole thing, are all log divergent, 
if the cutoff has any degree \dots 

\rm[Question from student: does gauge invariance reduce the degree of
divergence of some of the diagrams]\it\ There's this \rm[the vacuum
polarization] [student: but not the 3-gluon diagram?]\it\  Yes actually,
this cuts this down \rm[by]\it\ 1 \dots The ``1'' has to be an external
momentum.  This will produce an effect that is proportional to the
original coupling, which is like this.  That comes from an $A A$
cross a curl $A$.\footnote{RPF means
$\bbnabla_\times\vec{\mathbb{A}}\cdot
\vec{\mathbb{A}}_\times^\times\vec{\mathbb{A}}$; see eq.\
(\ref{eq8-18}).}\ \   So there's one gradient that comes on the external 
line.  So it always turns out that this ``1,'' in order to keep the
dimensions right, ends up as some external line \rm[momentum]\it.  So if 
we want the momentum of the incoming particle to be in front \dots

Yes.  \rm[Another student: So you do get gauge invariant couplings.]\it\  
Something tells us that there will be an extra gradient in front. But
it's also true that the mirror symmetry of the \dots the same result. 
Alright?  Another question?  \rm[Shouldn't gauge invariance  also reduce
the logarithmic divergence in the \dots]\it\  Yes and no.  That's \dots
more difficult because there's no procedure.  This is  gauge
invariant, but it's coupled directly with $A$ and we don't see any
extra gradients.  So there's no way to decrease the apparent power. 
The only way to make it smaller is by having momentum come in front. 
When you have two quarks---talking about the last diagram---two quarks
and a gluon, you're going to imitate a term with a quark, and a quark
or an antiquark, and a gluon, and there's no gradient, so it comes out
that it's logarithmic.  

Now it is possible  to choose a gauge, by the
right choice of gauge, you can make any of these damn integrals zero
\rm[in terms of the degree of divergence]\it, at the expense that the others
change \rm[?]\it\ \dots 
\footnote{I don't understand the claim, since the propagator
(\ref{eq12-8}) (with $\mu^2\to -k^2$) has the same power-counting properties as usual.}
You
have to be careful to compute a complete physical process always.
You want to make absolutely sure to compute something, the total
answer of which is gauge invariant.  Then you can't really say---it is
possible---you remember all the different propagators we had for the
gluon?  Well it depends on what propagator you use, whether you use a
propagator with $k\cdot\eta/k^2$ or whether you use the propagator
with $\delta_{\mu\nu}$, or still another propagator which is
interesting.  I wanted to mention this before \dots  When we were
doing the \dots business, with this gauge for instance; suppose we
started with this gauge $\partial_\mu A_\mu=0$, and we come out and we
have to do
\be
	\int e^{i\int \frac14(\partial_\mu A_\nu-\partial_\nu A_\mu -
A_\mu^{\,\times} A_\nu)^2} \,\Delta(A)\, 
	\delta[\partial_\mu A_\mu]\,{\cal D}A_\mu
\ee
and I'm not going to write the quark business in.  
Then we had in addition the statement that there was a determinant 
$\Delta(A)$, and this produced ghosts, it was represented by ghosts,
and on top of that was a delta function of $\partial_\mu A_\mu$.
And then we integrate over all $A$.  But I suggested that we get 
exactly the same $\Delta(A)$ here if you try to make this
$\delta[\partial_\mu A_\mu-f]$, and that the answer was independent of that
\rm[$f$]\it.  And then suggested further that you multiply by $e^{-f^2/2}$
or something \rm[times]\it\ ${\cal D}f$.  And the result of that was to bring
up a $e^{-(\partial_\mu A_\mu)^2/2}$---I'm just outlining what I did.
In order to eat, in the square of this \rm[$(\partial_\mu A_\nu-\partial_\nu A_\mu)^2$]\it\ 
the divergence pieces, and then I could
show you that the equation of motion which was
\be
	\square A_\nu - \cancel{\partial_\nu \partial_\mu A_{\mu}} = S_\nu
\ee
At any rate this term \rm[slashed out]\it\ didn't appear at all, and
therefore
\be
	A_\nu = {1\over \square} S_\nu = {1\over k^2} S_\nu
\ee
So we got the propagator $\delta_{\mu\nu}/k^2$.  Now the interesting
thing is what happens if you put a different number here \rm[in the gauge
fixing Lagrangian]\it, $a(\partial\cdot A)^2/2$.  And I'll just leave it as
an exercise, because if you put a different number in there, you get a
propagator of this form,
\be
	{1\over k^2}\left(\delta_{\mu\nu} - \eta{k_\mu k_\nu\over
k^2}\right)
\ee
where $\eta$ is not the same as $a$, I can't remember exactly;
$\eta$ is something like $a/(1\pm a)$.  Ah, it should be when $a=1$
this disappears; when $a$ goes to infinity, this should go to 1.
Because $a=\infty$ brings us all the way back to here; this is
Gaussian, such a tightly Gaussian, 
it's equivalent to a delta function.  And it says you calculate
everything exactly when $\partial_\mu A_\mu = 0$.  And that's a
propagator like that \rm[$\delta_{\mu\nu}-k_\mu k_\nu/k^2$]\it.  
Now you see that if you take a $k_\mu$ of that, you get zero
automatically, because the divergence of $A_\mu$ is always exactly 
zero.  Well, this general propagator when $\eta=1$ is called Landau's
propagator; when $\eta=0$ it's called the Finemensch propagator. And
other $\eta$'s are possible too; I call it to your attention because
it's interesting \dots the effects of this going to zero \dots
you've got to be careful what propagator you use \dots what sizes you
get for the different \dots It's only when you have a gauge-invariant
quantity that you get an answer that does not depend on the \dots
propagator you use.

\rm[Some of the lecture was lost during the change of tapes.]\it\ 
\dots Nowadays it's possible to do all these diagrams and all these
calculations on machines, programs for algebra \dots programmed
specifically for working on these dia\-grams and integrals involving
quark data, 
and therefore it gets to be no big deal.  You choose a propagator
\rm[gauge]\it, you turn on the switch, and it does all the 17 diagrams.
Whereas by hand, you are happy to discover that by using Landau's
gauge, you only have four diagrams; remember I had 17.  This would 
be useful \dots without \dots machines.  The reason that four diagrams
is better than 17 is mainly, it's impossible to do anything without
making mistakes, when you have too many pieces \dots Alright?

I am now going to calculate, at last; let us talk about the
scattering of two quarks.  To lowest order we already know 
that it looks like this
\be
\raisebox{-0.6cm}{\includegraphics[width=0.1\textwidth]{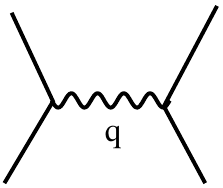}}
\sim {g_0^2\over q^2}
\ee
and we computed it.  I don't remember what we got for the color \dots
but it involves a $g^2$ and a one over $q^2$ \dots  That's the lowest
order.  Obviously this is $g_{00}$ \dots now it's going to be $\alpha$
when we're done \dots

Okay, now we get the next order, the $g^4$.  
We're trying to get the
next order in $g$ \dots So what we do---this is
the lowest order---next order, we have a lot of possibilities, 
\includegraphics[width=0.45\textwidth]{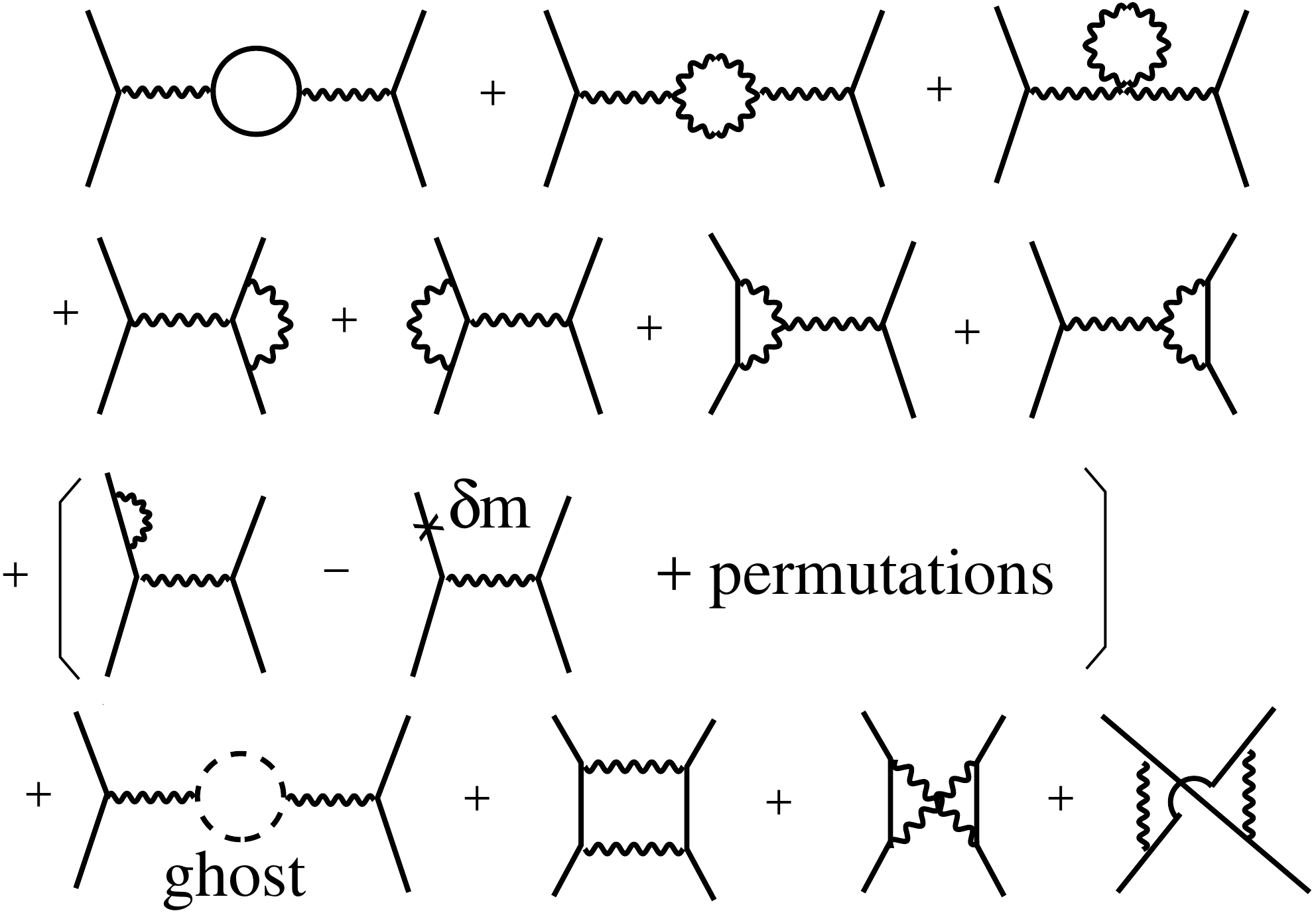}
Could be that you had a quark loop here, coming out this way.  Could
be you had a gluon loop coming out here.  Could be that you had a
gluon loop like that---what? Yes, with the four gluons that couple
there; I believe that's zero when you work it out, but I'm not \dots
It's obviously simple.  Then there's a diagram that looks like 
this \rm[vertex correction]\it\ plus one on the other side.  Then there's a diagram that looks
like this \rm[3-gluon vertex correction]\it---I wish the blackboard went a 
little bit further, I'll draw it up here, a diagram that looks like
this \rm[quark self-energy correction and mass counterterm]\it\ \dots
subtract the effect of this, this thing is divergent, the correction
to the quark mass, you just have to subtract it.  Alright now I've
drawn all the diagrams except the mirror image of this one, the mirror
image of this one, this \rm[self-energy]\it\ could be here or here or here
\dots  Alright?  Now any other diagrams you can think of?
\rm[Student mentions the ghost.]\it\  Ah yes, I'm sorry, the ghost. 
Important, important, the ghost, the ghost. Very vital.  Ghost, ghost,
ghost, ghost. Thank you, yes.  Anyone else think of some
oversight?  \dots\footnote{I realize that the box diagram is missing 
a few minutes later in the lecture and interrupt RPF.}

Alright.  Now the situation is, a lot of people will like to compute
this 
\[
\includegraphics[width=0.35\textwidth]{qq-vacpol}
\]
and say, well this is just a gluon going along, so don't do all
this, just do this,
\[
\includegraphics[width=0.4\textwidth]{vacpole-q2}
\]
on the gluon \dots \rm[Student (me): there's a box diagram with \dots]\it\ Just a
moment.  Just let me finish this part. Just do these \rm[now including the
gluon loop contribution to the vacuum polarization]\it.  You can't,
because this depends on what kind of gauge propagators you use, it's
not a gauge-invariant process \dots  You have to finish it by having
quarks make the gluons \dots that's why I made it more complicated.
The answer is that depending on what kind of propagator you use, you
get different answers for that.  It's only when you put the whole mess
together that you get an answer that's independent of what propagator
you use.  Somebody was going to say something.  \rm[Student: there's a
box diagram too, where you've got two gluons connecting the quark
propagators.]\it\   Of course, sure there is.  Yes.  And crossed, right?\,\footnote{There are many other crossed diagrams missing.}
I thought we'd find something \dots 

What we can do is do one as simple as possible, and then say the rest of
them, you know the rules for putting this in and the rules for putting that
in, they follow the same kind of \dots the labor is enormous \dots how to sum
over the colors, various things \dots If you had to do it, you would do it.
Or find a book that will do it, okay?  But in order to understand  the nature
of the results that we're going to get, I'm going to take only one case.  I'm
going to take the dullest and the simplest case, this one,
\[
\includegraphics[width=0.2\textwidth]{vacpol-q3}
\]
\dots The others teach you a little bit, but \dots you've probably
done this \dots  So here we go.  Alright? \dots I'm going to have
something going around here; let's say that the momentum coming 
in here is $q$, and this has a momentum $p$ let's say, and then since
the momentum coming in is the same as the momentum going out \dots
If you don't know what I'm doing, then it's because you didn't take a
course that's supposed to be a prerequisite for this, having to do
with perturbation field theory, and you're going to have to learn it.
Alright?  If you have taken such a course, this will be very boring,
and I'm sorry, but I'll go as fast as I can, and hope that you \dots
will stop me and ask some questions \dots  Alright, now here we would
have a quark going around, and we would have a coupling in here, which
is a $\gamma_\mu$ in the direction of the polarization of this, so
let's say that $a$ is a vector polarization, and let's say $a^i$ is
the color of the gluon.  So this would be $\gamma_\mu$ and it would be
multiplied by $a_\mu^i$.  But there would also be a $\lambda_i$ matrix
for the color matrix that this couples with,
\be
	g_{00}^2\left(-{\rm Tr}\int {d^{\,4}p\over (2\pi)^4}
	\,b^j_\nu {1\over \slashed{p} -\slashed{q}-m} {\lambda_j\over 2}
	\gamma_\nu {1\over \slashed{p} -m}{\lambda_i\over 2}
	\gamma_\mu a_\mu^i\right)
\ee
Now the next thing happens is that this damn thing propagates, 
the quark propagates around to here, and the fact that they're
propagating, one over $\slashed{p}$ minus the mass of this quark.
Sir: \rm[student: Do you mean to have $\lambda_i/2$?]\it\  Yes sir, I do, I
do.  Yes, thank you.  Then comes this baby which I'll call $b$, and we
can say the polarization is $\gamma_\nu$, we're going to be
multiplying by $b_\nu$ and by $j$.  And then we'll also have a
$\lambda_j$, for the color business there, and then we'll have---I
didn't leave enough room did I?---well maybe I just about did,
I have the propagator $p$ dagger minus $q$ minus $m$---oh, I forgot the
2 again, I shouldn't do that.  And I have two $g$'s for the couplings
at either end.  And those are the kind of $g$'s that I was tagging
$g_{00}$ \dots

Of course, that looks like this is second order, but I didn't put the
rest of these two lines on \rm[the external quarks]\it, and when I did that
there would be two more $g$'s, and this is important.  The first
diagram which got two $g$'s and the second diagram has four $g$'s
\dots   Now, I have to sum over everything.  I have to sum over all
the possibilities for the momentum of the quark loop, and I have to
sum over the colors, and the spinors.   Well, first the spin.  When
you come around and come back, you've got all these matrices, and it
becomes a trace.  So really it should have been a matrix trace, a
trace for the gamma matrices,
\bea
\label{eqA2-25}
	\underbrace{\tr\left[{1\over \slashed{p}-\slashed{q}-m}\gamma_\nu
{1\over \slashed{p}-m}\gamma_\mu\right]}&\,&
\underbrace{\tr\left(\lambda_i\lambda_j\over 4\right)}\\
\underbrace{{\rm tr}\left[(\slashed{p}-\slashed{q}+m)\gamma_\nu 
	(\slashed{p}+m)\gamma_\mu\over [(p-q)^2-m^2][p^2-m^2]\right]}
 \!\!\!\!\!&& 	\qquad{\delta_{ij}\over 2}\nn\\
= \Big[(p-q)_\mu p_\nu + (p-q)_\nu p_\mu &-&\delta_{\mu\nu}
(p\cdot(p-q) - m^2)\Big]/\dots
\nn
 \eea
\dots In addition, we'll have a matrix operator for the color, and
then it carries all the way around here and goes to this color, and 
then it
carries all the way around so we get another kind of a trace, this 
time on the colors.  So this is a trace on the gamma matrices, you
might say, and then there's another \dots trace on the color indices,
of $\lambda_j \lambda_i$ over 4.  And this one is easy, we know the
color of this---without the $1/4$---is $2\delta_{ij}$, so that's
$\sfrac12\delta_{ij}$.  So the first thing it tells us of course is
that the color of the quark that comes out must be the same as the
one that went in.  That's the conservation of color; if it's a
red-antiblue gluon here, it will be a red-antiblue gluon there \dots

This particular trace \rm[the Dirac trace]\it, the famous way of handling
that, is to multiply numerator and denominator by
$\slashed{p}-\slashed{q} + m$ and by $p$ dagger plus $m$,
 and then in the denominator, you'll have the rationalized thing
there, $(p-q)^2-m^2$ and $p^2-m^2$, and then there will be this factor
we had before, $\delta_{ij}$ over 2.  Alright? \dots Any question up
to here?  In fact it's the same as the corresponding correction in 
electrodynamics, except there's some slightly different number \dots
from the colors.  This trace here that I've written can be directly
simplified; this trace is exactly the same as this dot this plus the
other way around \dots minus \dots it's all memory \dots 

So the net result is that effectively I have to do this integral, an
integral that you must have learned about when you were doing
electricity, QED, so I turn over here to do the integral.  But before
I do the integral, I look at it, and notice how it diverges.  We can
see by counting there's 2 $p$'s in the numerator, there's 4 $p$'s in
the denominator, there's 4 $p$'s in $d^{\,4}p$.  That cancels but
you've still got 2 so it's quadratically divergent.  In exact
agreement with the prediction.  But, now the trickery, the method for
doing this like this, is to have a whole list of integrals \dots
a list like this---I'll do the integrals in a minute---but things like
this,
\be
	\int {d^{\,4} p\over (2\pi)^4}{1\over (p^2 - m^2)^3} =
	{1\over 16\pi^2 i m^2}
\ee
or something; I'm not chasing that, alright?  This permits you to do
anything that has powers in the denominator, by integrating over \dots
But this \rm[(\ref{eqA2-25})]\it\ has two different kinds of powers in the
denominator, so there is an invention for putting that together,
which runs like this: 
\be
	\int_0^1 {dx\over [ax+b(1-x)]^2} = {1\over ab}
\ee
The integral from 0 to 1 $dx$ \dots and this you can verify directly.
And therefore if you take the product of two pieces, you can write it as
one denominator.\footnote{Here RPF makes a joke that elicits laughter,
something like ``Classic, right?''}\ \   I copied it from Schwinger,
actually, I cheated. He had another way of doing it which was
extremely clever. Involving a Gaussian integral.  I noticed that I
could eliminate one step \dots and make it look \dots \rm[laughter]\it.\footnote{Perhap RPF is saying something like ``and make it look like
my own idea.''}

So therefore \dots
\bea
	&&{1\over (p^2 - 2p\cdot q + q^2-m^2)(p^2-m^2)} = \nn\\
	&&\qquad\int_0^1 dx\,{1\over[p^2 - 2 p\cdot qx + q^2 x - m^2]^2}
\eea 
Alright?  Alright, \dots what we're going to replace that thing by,
and so we would have this integral from 0 to 1 $dx$, to be done,
later, and then we'll have the integral $d^{\,4}p$, and then we have
something like \dots 
\bea
	\int_0^1 dx\!\!\!\!\!\! &&\int {d^{\,4}p\over (p^2 - 2 p\cdot qx + q^2 x -
m^2)^2} \\
	\times&&\left(2 p_\mu p_\nu - p_\mu q_\nu - p_\nu q_\mu -
	\delta_{\mu\nu}(p^2 - p\cdot q - m^2)\right)\nn
\eea
Alright.  Now we have to do this \rm[$p$]\it\ integral.  But we can't, because
it's divergent.  The first thing that's a good idea to do, always, is
to shift the $p$, let me shift $p$, let $p\to p+ qx$.  Then we have
\bea
	\int_0^1 dx\!\!\!\!\!\! &&\int {d^{\,4}p\over (p^2 + q^2
x(1-x) - m^2)^2} \\
	\times&&\Big(2p_\mu p_\nu + (p\cdot q\hbox{\ terms})
	+2 q_\mu q_\nu\, x(1-x)\nn\\
	&-& \delta_{\mu\nu}(p^2 - q^2x(1-x) - m^2 + p\cdot q\hbox{\ terms\ }
	)\Big)\nn
\eea
I just completed the square here.  And then there are some terms with
$p\cdot q$, I don't know what they are \dots they are linear in $p$,
you'll see why I don't care about them in a minute \dots
Anything linear in $p$ that I didn't bother to write out, the reason
is that when I integrate over all directions of $p$, the plus and
minuses are going to go out, and since I'm going to get nothing from
them, I didn't bother to be careful.  Alright?

Now I have cheated.  Why?  We made a mistake.  Because, the integral
is divergent; I didn't tell you how to make it convergent.  Therefore
this business of shifting $p$ by $qx$---how do I know that the
integral---that you can shift $p$ by $qx$ and get the same result?
I must insure that my method of renormalization, whatever the hell I'm
going to use, has that property that if you shifted the $p$, it would
be alright, okay?  So that's one thing.  The second is, what about the
integral over $p$ at zero?  That's right, if the method of
renormalization is round, okay, but round against what?  Against the
shifted $p$?  If you say it's round against this $p$, it's not round
against the shifted $p$.  You know what I mean, symmetric in both
directions.  It's cheating.  So only if I made a method of
renormalization, I mean a method of cutting off the integral,
specific, and carefully, can I really do these things which I've been
doing.  

Now it happens to turn out that one method that's been
invented for this kind of loop of quarks or electrons is this: you 
subtract the same expression with a larger value for the mass of
the electron.  In other words, the method of renormalization, I
should have at least specified before I made those steps.  And the 
method that I want to use is going to be this one.  You take the value
and then take with $m^2$ replaced by a bigger $m^2$, which I like to
write as $m^2\to \Lambda^2 + m^2$, and subtract.  That's the scheme.
That scheme, it so happens, will permit the steps which I did of
shifting.  It was all right.  But it's very easy to slide off the
wagon and make operations which are not quite right until you specify
the right way \dots You see, this method of subtraction maintains the
gauge invariance for this diagram.   Because if this were electricity,
everything is exactly the same except for a number, and if this is
electricity, if this is any mass whatever, this is a gauge invariant
integral, and the gauge invariance is maintained by subtracting the
same expression.  To show you that it isn't necessarily obvious what
to do,  an early worker in the field first proposed or tried to
subtract from this propagator  the propagator with a different mass.
In other words, to use for the propagator 
\be
{1\over p^2 - m^2} \to {1\over p^2 - m^2}-{1\over p^2 - (\Lambda^2 +m^2)}
\ee
That doesn't work.  That doesn't keep the gauge invariance.  Pauli and
Villars pointed out that if you subtracted the whole thing, the whole 
closed \dots \rm[amplitude]\it\ then you maintain the gauge invariance.  So
you see it's easy to \dots if you don't maintain the gauge invariance 
\dots  electrodynamics \dots Pauli-Villars \dots  and that's the way
if you do this then all these things are legitimate.  Okay?

Now I'll show you a very interesting---is that okay, you had a
question?  It's nerve-wracking, alright.  But it's been straightened
out, in the case of an electron. I want to point out, that when we 
come to the second diagram here, with the gluon, it's more subtle and
more complicated.  And the exact way to do it is very hard, and it
took a lot of finagling around to get it right, when I first tried it.
But I was able to guess and push and hammer.  In the meantime, another
method of cutoff was invented, which is called the dimensional
renormalization, invented by Wilson and 't Hooft \dots for which 
gauge invariance and covariance in space in four dimensions
are automatically maintained in \rm[$d$ dimensions]\it, 
not a chance of losing it, so that it's a good scheme.  We don't have
to have the old-fashioned hammer tricks.  Here we know a good trick
that will work, and I wanted just to point out \dots  Now I would like
to point out---I think we could almost see---Now I want to show you
something.  Suppose we had a method of renormalization, that we knew
was going to be symmetric, and everything is going to be alright.
Then, I claim this integral, 
\be
	\int d^{\,4} p\, {p_\mu\over p^2-m^2} = 0
\label{eqA2-32}
\ee
if we could renormalize, to make it finite anyway, would have to be
zero because of the asymmetry.  Provided we had a good method to 
protect the asymmetry, and subtracting this thing with a different
mass obviously does that.  That's it, no problem.  Now differentiate
both sides of this \dots  Now what I want to prove, let's see \dots
If you differentiate inside the integral with respect to \rm[$p$ sub]\it
$\nu$, you'll get 
\be
	\int d^{\,4} p\, {\partial\over\partial p_\nu}
	{p_\mu\over p^2-m^2} = \int d^{\,4} p\,
{\delta_{\mu\nu}(p^2-m^2)  - 2p_\mu p_\nu\over (p^2-m^2)^2} = 0
\label{eqA2-33}
\ee
It's not hard to prove, by the same kind of symmetry, one way or
another, you might not like the way I did it, that this is also zero.

\rm[Student: what did you differentiate?\footnote{I reconstructed the
previous equation from the tape rather than my notes; apparently it
was less explicit, leading to this question.}]\it\  I differentiated the integrand
with respect to $p_\nu$.  Isn't it legitimate to differentiate the
integrand if you're going to integrate it back anyway?  Well, if you don't
like it this way, then another way to do it is to shift $p$, and
differentiate with respect to the shift in $p$.  You start out with 
some kind of  thing like
\bea
	\int d^{\,4} p\,
{p_\mu\over (p-a)^2-m^2}
	&=& a_\mu\int{ d^{\,4} p\,\over (p-a)^2-m^2}\nn
\eea
which is just this thing \rm[(\ref{eqA2-32})]\it\ with $p$ substituted with
$p-a$.
And then differentiate both sides of this expression with respect to
$a_\nu$; then put $a_\mu=0$.  Alright?

I write this particular thing by putting it all under the same 
denominator, and putting this here \rm[arranging the numerator so that
the first term is like in (\ref{eqA2-33})]\it:
\bea
 \int\!\! {d^{\,4}p\over (p^2\!\! +\!\! q^2
x(1-x)\!\! -\!\! m^2)^2}&&\!\!\!\!\!\!\! \Big[ 
	2p_\mu p_\nu \!\!-\!\!\delta_{\mu\nu}(p^2\!\! +\!\! q^2x(1-x)
	\!\!-m^2)\nn\\
	&& \!\!\!\!\!\!-2 q_\mu q_\nu x(1-x)\!\! +\!\! 2\delta_{\mu\nu} 
q^2x(1-x) 
\Big]\nn
\eea
So we have this sort of general statement that it is a displaceable
method of doing things.  You can think of it as a shift of the 
origin; the method of renormalizing doesn't have anything to do with
it.  The question is, if this were automatically zero.  And then if we
look at this we expect that \dots we have just the right combination:
we have $2 p_\mu p_\nu$ and $\delta_{\mu\nu}$ times the
denominator---watch out on that sign! \rm[The $q^2$ term in the numerator
has the opposite sign to that in the denominator.]\it\  This part's okay
but this is wrong.  Well I'll fix it; I'll put plus and I'll make it
minus 2 \rm[in the second line]\it.  Alright, that's not wrong.  Now this
thing, times the delta, plus this one, go to zero \dots by the
argument about the way to do it \dots in the integral \dots

The result of that is that all of this can be \dots the net result of 
the whole thing, altogether, is 
\be
	2(q_\mu q_\nu -\delta_{\mu\nu}q^2)\int_0^1 x(1-x)dx
\int {d^{\,4}p\over (p^2+q^2x(1-x)-m^2)^2}
\ee
and I'm close to the end of the hour \dots I wanted to subtract---this
is only a logarithmically divergent thing---and I subtract the upper
limit and so on. It's going to introduce something that goes like the logarithm of 
this cutoff divided by some pole mass which is practically the $q^2$
of the quark.  It's a little more complicated than that and I'll
finish it next time.  What I'm trying to say is that we can get to
this thing in front, and that is necessary for gauge invariance, 
because $q$ dot \rm[the prefactor]\it\ is zero \dots  The current is
conserved.

Now let me, since it's just a few more minutes to ten, just to
remind you \dots I'll explain that this means that the vector
potential that is coming in and out \dots finish the job next time
\dots and also next time I'll start to explain \dots dimensional
renormalization, which is so handy, because the old-fashioned way you
had to do a lot of thinking, trickery, to make sure you didn't screw
up \dots invariance.  

\section{Transcription: Renormalization, continued (1-14-$xy$)}
\label{A3}

We were doing one of a number of diagrams that have to do with the
correction to scattering of two quarks.  The scattering of these two
quarks behaves as---$g$ is the coupling constant---something like
$g^2$ over $q^2$ in the first order.   Or that's sometimes called
second order because it's in $g^2$ \dots  First order in $g^2$.
That's straightforward.  We want to get corrections for it.  And the 
corrections appeared to be a large number of diagrams which I wrote
\dots and you'll see that there's a $g$ here and here and here and
here,
\be
\raisebox{-0.4cm}{\includegraphics[width=0.1\textwidth]{qq-vacpol-q}}
\  =\ 
J_\mu\, g_0^4 \, {1\over q^2}\,  B\,  {1\over q^2}\,  J_\mu
\ee
where $B = \raisebox{-0.1cm}
{\includegraphics[width=0.07\textwidth]{vacpol-q4}}$.
So this is going to be a correction which will be of the order 
$g_0^4$.  There will be two propagators $1/q^2$ here and $1/q^2$
there, and so on.  So it will be $1/q^4$ multiplied by an integral,
\bea
	B &=& (\delta_{\mu\nu}q^2 - q_\mu q_\nu) I,\\
	I &=& \int_0^1 dx\,x(1-x)\int {d^{\,4}p/(2\pi)^4
	\over [p^2 - m^2 + q^2 x(1-x)]^2}\nn
\eea
What we're going to discover is that this is---sorry, not by the 
integral, but by $B$, the bubble \dots I left out all the indices,
the colors and all that.  You can go away and discover that the colors
are all \dots multiplied by a certain integral \dots in this integral
there is a number of constants which I inadvertently, carelessly
dropped---twos and $\pi$'s and things which presumably you can
calculate \dots\footnote{I had added the $1/(2\pi)^4$ factor as an
afterthought in my notes.} 

Now, the currents \rm[$J_\mu$]\it, which I didn't write here, which are
operating here and here \dots this bubble, when acting on the current,
the $q$---the current is conserved---and the $q$ acting on the current
is zero, so that $q\cdot J=0$.  The bubble gives you $\delta_{\mu\nu}$,
which means that these two currents are in the same direction, and
this $q^2$ in the bubble eats one of these $q^2$'s.  So this thing
turns into 
\be
\raisebox{-0.4cm}{\includegraphics[width=0.1\textwidth]{qq-vacpol-q}}
\  =\ 
J_\mu\, g_0^4 \, {1\over q^2}\, I\,  J_\mu
\label{eqA3-3}
\ee
So therefore, it's the same form as this \rm[the tree level contribution]\it
\be
\raisebox{-0.6cm}{\includegraphics[width=0.1\textwidth]{g-exchange3}}
= J_\mu {g_0^2\over q^2}J_\mu
\ee
except we have $g_0^4$ times the integral up here instead of $g_0^2$
in the correction.  So the easy way to think about it is it's a
correction to this coefficient at the top,
\be
	g_0^2\to g_0^2 + g_0^4 I
\ee
Alright?  Are there any questions about that?

In getting to this form we did a little  hocus-pocus about correcting,
shifting origins and this and that, and talking about tricks to get 
rid of the quadratic divergence which originally arose. However we
still have a divergence and we have to talk about how to handle it.
And the particular rule that works for log divergences in 
quark loops is to subtract---the rule I'm going to use here, now---
is the method of subtracting the same
result for different masses.  But I will describe, perhaps today, but
later, the method called dimensional renormalization, which is to
change this $d^{\,4}p$ to   $d^{\,3.9}p$.  This is a function of only
$p^2$, so $d^{\,4}p$ is something like $p^3\,dp$, with a coefficient
that is $\pi^2$ or something, that depends on the number of 
dimensions; in three dimensions it's $p^2\,dp$ times some other number,
$4\pi$.  And in $d$ dimensions $d^{\,d}p$  is $p^{d-1}\,dp$ times
some coefficient involving Gamma functions of $d$ and so on
\dots  That's all there is to it---mainly that's all there is to
dimensional renormalization; to use $d=3.9$, and then $3.99$,
go to the limit.  Alright?  That's what it's about.  However it's very
pretty and I must have spent a lot of time because I enjoy it \dots

Anyhow, the way of renormalizing is to subtract---this is what we're
going to do now
\be
	{1\over (p^2 - (m^2+\Lambda^2) + q^2 x(1-x))^2}
\ee
\rm[from the integrand of $I$]\it.  That will make the results convergent
\dots  It would be---another way to make this subtraction is to 
consider this integral as a function of $m^2$,
\be
	I(m^2) = \int_0^1 dx\,x(1-x)\int {d^{\,4}p/(2\pi)^4
	\over [p^2 - m^2 + q^2 x(1-x)]^2}
\label{eqA3-7}
\ee
and then consider taking $I'(m^2)$, the derivative with respect to
$m^2$, and integrating that 
\be
	-\int_{m^2}^{m^2+\Lambda^2} I'(M^2)dM^2 = 
	I(m^2) - I(m^2+\Lambda^2)
\ee
Let's take the derivative with respect to $m^2$, but we're going to
have to have a variable for it, so let's call it $M$; integrate that
with respect to this $M^2$ \dots you can certainly do this \dots
This is a trick that I wanted to use
\dots converges \dots Well of course I need to
differentiate and then put it under the integral sign, calculate the
integral and then \dots 

\dots This particular method would produce this final result without
all those tricks about shifting and so on;  all those could be 
perfectly done by \dots and not notice all that stuff \dots
all that stuff that I did for quadratic divergences, it's also
taken care of \dots  I did that to show you how in cases where
there's some confusion, it is always possible to get an answer.
People were very clever to squeeze answers out of these \dots

The logarithmic divergences are always much easier to handle, much
less uncertain than quadratic and  higher divergences.  And when we
were doing this stuff for example we found that one of these things
would produce a logarithmic divergence directly.  So another thing to
do is to compute a process that we know produces a logarithmic
divergence, and then have no more trouble, and use gauge invariance
to get the terms in front.  But with dimensional renormalization, 
you don't need any guessing \dots

Okay, we're going to do it this way \rm[by Pauli-Villars subtraction]\it\ here.  So first
I'll differentiate that \rm[(\ref{eqA3-7})]\it\ and I will find myself needing to do the
integral
\bea
\int_{m^2}^{m^2+\Lambda^2}\!\!\!\!\!\!\!\!\!\!dM^2\int_0^1dx\,x(1-x)&2&
\underbrace{\int{d^{\,4}p/(2\pi)^4\over [p^2+q^2x(1-x) -M^2]^3}}\nn\\
&&\!\!\!\!\!\int{d^{\,4}p\over (2\pi)^4} {1\over (p^2-L)^3}={1\over 32\pi^2 iL}\nn\\
\eea
And now at the end when I'm all done I have to integrate with respect to
$M^2$ \dots Alright, so that's where I'm at, you see where I got that:
differentiate that with respect to $M^2$, the second power \dots now it's
got the third power.  Now of course this integral will present
no divergence because there are six $p$'s in the denominator and four in
the numerator; then this integral can be done.  We can do it in lots of
ways, and I'm not going to bother \dots Anybody who's ever done anything
in perturbation theory is always going to put this very same integral,
sooner or later.  It's the integral of $d^{\,4}p/(2\pi)^4$ divided by $p^2$
minus something, cubed, and it's equal to \rm[one over]\it\ $32\pi^2 i L$ or
something like that.  Factors of 2 or so I'm not going to \dots I can't
remember and I didn't bother to look it up  \dots

This then, this integral here then---I'm keeping these lines this way so
that you don't have to write that over and over and over---this piece
becomes
\bea
\int_{m^2}^{m^2+\Lambda^2}dM^2\!\!\!\!\!\!\!&&{1\over 16\pi^2 i}\,{1\over
M^2-q^2x(1-x)}\\
&=& {1\over 16\pi^2 i}\ln\left(m^2+\Lambda^2 - q^2x(1-x)\over
	m^2- q^2x(1-x)\right)\nn
\eea
Okay, now the idea is that $\Lambda$ is the cutoff, it's supposed to be
higher than any of this part with the mass, even $q^2$.  We only want this
theory as $\Lambda\to\infty$.  As  $\Lambda\to\infty$ then, relatively
speaking we can drop this \rm[$m^2- q^2x(1-x)$]\it.  Of course we get an infinite
answer because we have a divergence and that's where all the trouble began.
What happens to that infinity?  It's fixed by changing the coupling
constant $g_0$ with $\Lambda$, by making $g_0$ a function of $\Lambda$,
so that the variation of the first order takes away the \dots if you use
different $\Lambda$'s you use different $g_0$'s at the end.

I'm going to continue this calculation disregarding the masses.  Suppose
you have a large momentum transfer, and disregard the mass.  The purpose of
this is only to do the arithmetic; if you want to you can always do it 
with \rm[nonvanishing mass]\it\dots I'm going to disregard this just so to take
a simple example where $q^2$ is much larger than the $m^2$ \dots 
It looks like it's dangerous because when
$x$ is small, even if $q^2$ is large, maybe that \rm[the mass]\it\ is important,
but it turns out in the log it don't make any difference, but anyway you
do it with $m^2$, I 
don't want to do it with $m^2$.  I'm only illustrating, explaining what
comes up \dots  So I'm going to write this as
\bea
 \int {dx\,x(1-x)\over 16\pi^2 i}\!\!\!\!\!\!\!\!\!\!\!&&\left[\ln\left(\Lambda^2\over
-q^2\right) -\ln x -\ln(1-x)\right]\nn\\
&=& \left(\frac16\ln\left(\Lambda^2\over
-q^2\right) - \frac{5}{18}\right)
\eea 
We suppose that $-q^2$ is positive, actually.  It's going to be a momentum
transfer for the scattering \dots 

Now I have to integrate this and I'm almost finished, see?  So I have to
integrate $x(1-x)$.  That's a constant \rm[meaning that
$\ln(\Lambda^2/q^2)$ does not depend on $x$]\it\ \dots 
that's well within my power.  $x-x^2$ \rm[integrates to]\it\ $x^2/2 - x^3/3$
that would be the integrand, and if I put $x$ from 0 to 1---no the
integrand is the differential of that---I get $1/2$ minus $1/3$ is $1/6$.
It's going to come out to $1/6$.  Silly now, 
because I've lost the constant \dots so I have the logarithm, I'm just trying
to keep things that are relevant \dots  Now I have to do the log of $x$
times this, which I'll do by parts.  And I get $1/4$ minus $1/9$ actually,
when I do it by parts I get some kind of number here, something like
$5/18$, okay?  Coming from integrating those logs, which are easy to do,
and I'm sure you'll \dots  Yeah, $5/18.$  Anyway you get some constant.
And that's it.  

And that's the integral ``$I$,\!'' and it will have the $\Lambda$
there.  How are we going to look at the physics?  Assuming the physics
is right, and we're going to get an answer, now we put that back: this ``$I$\!''
goes back in here \rm[eq.\ (\ref{eqA3-3})]\it, right?  So, I'm just going to write
the coefficient of the term \dots
\bea
\label{eqA3-12}
\!\!\!\!\!\!\!\!\!\!\!\!J{1\over Q^2}J\left[\rule{0cm}{7mm} g_0^2 \right.&+& \left.{g_0^4\over 16\pi^2}
	\left(\beta_0\left(\ln{\Lambda^2\over \underbrace{-q^2}}\right) + a\right)
\right]\\
&& \qquad\qquad\qquad\quad Q^2\nn\\
\hbox{where\ } \beta_0 &=& -\frac23 n_f\hbox{\qquad and\ } a = \frac{10}9
\, n_f
\label{eqA3-13}
\eea
$g_0^2$ plus $g_0^4$ times the various numbers of $\pi$'s, which I have
recovered by looking at the answer in the book, alright?  Times a certain
constant \rm[$\beta_0$]\it\ times the logarithm of $\Lambda^2$ over $-q^2$
plus another constant, where for us, $\beta_0$ it turns out is $-2/3$
when we put it in this form.  Alright?
And the ``$a$\!'' for us is equal to---this is $5/3$ of that, $5/3$ minus $2/3$,
probably something like plus $10/9$---highly questionable \dots

Now first of all, to remind us all that $-q^2$ is positive, let's call it
$Q^2$.  So I'm going to make it like that.   This \rm[eq.\ (\ref{eqA3-12})]\it\ is all multiplied by
$1/Q^2$, and by the currents and so forth in the final interaction.  That's
what comes out, alright? \dots The reason I wrote it this way is that
there's going to be more contributions that come from the other diagrams
that we haven't worked out.  And I'm going to have to add those in
when I discuss \dots  

Oh---\rm[the factor of $n_f$ was not written in 
eq.\ (\ref{eqA3-13}) at first]\it\
there's more than one flavor of quark, and 
each flavor of quark makes a loop, and each one those is the same as this
one.  And insofar as $Q^2$ is large enough to neglect the mass of the
quark, insofar and therefore for the first few, certainly for the $u$ and
the $d$ and very likely for the $s$, maybe for the $c$ quark, there will be
a certain number of flavors that we would use, that contribute to this
formula; it would be the number of flavors whose mass is less than $Q^2$.
The flavors with masses higher than $Q^2$ are not much contribution.  The
flavors that are in between, then you just have to do the integral better
and so on.  So put the number of flavors \rm[in (\ref{eqA3-13})]\it.  Alright?

It looks like this answer means that the probability of scattering will 
depend upon the cutoff $\Lambda$---there it is, explicitly there.  And
therefore our original program, which was that we were going to calculate
somehow in quantum chromodynamics and make predictions, but we found our
theory was divergent, and what are we going to do?   Well we have a
 \dots But the trick is to arrange that it doesn't depend on the cutoff,
by supposing that $g_0^2$ is chosen, for each $\Lambda$ that you choose,
you must take a different $g_0$.  You must choose a $g_0$ which is a
function of $\Lambda$, chosen so that the answer to a physical question
does not depend on $\Lambda$.  We need a formula for that to do that.
Well this is very hard to figure out here by looking at this thing, what
kind of a shenanigans, how you're going to vary $g_0$ to get rid of this
$\ln$.  Of course this could also be written as $g_0^2$, and to the same
order as I have it here, it's convenient to write it this way,
\be
	{g_0^2\over {1-{g_0^2\over 16\pi^2}\beta_0\ln{\Lambda\over Q^2}}}
= {16\pi^2\over \left(16\pi^2\over g_0^2\right) - \beta_0\ln{\Lambda^2\over
Q^2}}
\label{eqA3-14}
\ee
There's one other term I forgot about, this $(16\pi^2)$, this is very much
like \dots That's equivalent to this approximation \rm[Taylor expanding to
first nontrivial order]\it; we're only worried about this log part now; the
$``a\!$'' is something else we have to work out a little more accurately. 
To this order, as far as the $\ln\Lambda^2$, it's like this \dots
which could also be written $1/g_0^2$, or I would prefer to put my 
$16\pi^2$ here, so perhaps there's a $16\pi^2$ here, minus
$\beta_0\ln\Lambda^2/Q^2$.  Now this \rm[$g_0^2$]\it, we can suppose depends 
upon $\Lambda$, and we have to make this \rm[the right-hand side of (\ref{eqA3-14})]\it\ not depend on 
$\Lambda$.  If we make $16 g_0^2$, if we arrange all the time that $16\pi^2$
over $g_0^2$, which is by the way $g_0^2/16\pi^2$ \dots old notation where
I had forgotten the $4\pi$ squared, so you can remember this in connection
with the previous lecture, without seeing that lousy $4\pi$ squared all the
time.  This, if we suppose this, which is defined as this, 
\be
	{16\pi^2\over g_0^2(\Lambda)} \equiv {1\over \hat g_0^2(\Lambda)}
	= \beta_0\ln{\Lambda^2\over\lambda_P^2} + {\rm const.}
\label{eqA3-15}
\ee
is equal to,
when we vary the $\Lambda$, we make sure that this is some kind of 
constant, well we should make this is $\ln\Lambda^2$ plus a constant.
And since it's a constant, we can put anything down here, that we want
\dots  This will cancel the $\ln\Lambda^2$, yes?  No, you've got to have a 
$\beta_0$ in here \rm[initially forgotten in the above formula]\it.  

So if we suppose our $g$'s are chosen like this, then 
we'll get an answer that is independent of $\ln\Lambda^2$; that's the
trick.  When we do different degrees of convergence with the cutoff,
and we change the cutoff, and we change $g^2$ that we use appropriately for
that cutoff, then we can arrange the whole thing so that it doesn't make
any difference where that cutoff is, that's the miracle of this theory,
and that's why we have a theory.  Because otherwise you would have
predictions that would depend on still another parameter which is the
cutoff,  where we have to write our theory with explicit formulas for
the cutoff \dots This way, we don't; we just have to say that it's going to
happen, and hope that it does, it's been proved that it does \dots Alright?

Well this is explaining the machinery that we're going to \dots
The thing is we haven't gone to the next order, $g_0^6$, and discovered
that this isn't quite enough; that there has to be a log-log term there.
I'll explain why there's a log-log term \dots

Yes? \rm[Question about the constant in conjunction with $\lambda_P$.]\it\  I don't know that
constant.  It's arbitrary, you choose anything you want.  Later on
we chose, specifically, to make Politzer's $\lambda$, to make that 
constant serve for a special method of cutting off dimensional
renormalization, which is not what I'm going to use, so the $\pi$, the $10/9$
and all that stuff is changed, because of the kind of handling of the 
integral.  And because it's a different---yes, if you had decided that the
method of cutting off was going to be the $\Lambda$ method, then in that
$\Lambda$, the way I did it, changing the masses of the quarks, that would
be enough \dots then a more convenient $\lambda$ Politzer might be to make
that zero \rm[referring to the constant in (\ref{eqA3-15})]\it.
\dots Anything would be alright.   Somebody has to make a choice somewhere.
\dots There's a choice to make that constant zero for dimensional
renormalization \dots  Let's take the case of the lattice.  You say that
the $\Lambda$ corresponds to the wavelength of the spacing.  So is $\Lambda$
\rm[equal to]\it\ 
$1/a$ or $2\pi/a$?  One guy does one way, another one another way; all he
does is change the scale of $\Lambda$.  So you need to put that as the 
log of $4\pi$ squared over here, or you can change the definition of this
\dots That's why I like to put the constant here at the end \dots
Any other questions?

\rm[Question: why does this procedure work for all processes, using the
same $g_0^2(\Lambda)$ for all of them?]\it\   That has been proved.  That we
had to assume, that the theory was \dots It's not obvious at all.  It turns
out that \dots it's not true only of the calculation at second order,
but at the next order also it's independent of the process.  Only beyond
that does it become dependent on the process.  And that's connected to the
discovery that the rate of change of the $1/g_0^2$ with respect to
$\Lambda$ computed as a series \dots Oh I shouldn't say it that way, I
should have used a real process.  The first two terms \dots how we know 
is something I didn't prove.  \rm[Student interjects, I think it is me:
Isn't the answer to that question the fact that you have a finite number of
\dots with primitive divergences?]\it\  Yes, we'll discuss that.  Thank you.
Yes, thank you.  That's good, let's discuss it.  
He's got the right answer \dots  We worked out the various divergences.
\[
{\includegraphics[width=0.5\textwidth]{prim}}
\]
This one \rm[the tadpole]\it\ vanishes.  Then there was divergences of this kind
\dots two things coming out, never mind what's in here; there would be
three gluons, remember this little table I made?   With the degree of 
divergence \dots  There was a table that said how many glues there were and
how many quarks there were on the outside lines, from which we calculated 
as 4 minus this \rm[number of gluon lines]\it\ times \rm[minus]\it\ $3/2$ that 
\rm[the number of 
quark lines]\it. And the result for $N$ here was 2, 1, 0, 1, 0, and everything
else convergent.  Now using gauge invariance you can show always that in
this case \rm[$N=2$]\it\ there have to be two $q$'s outside, in front \dots
and therefore this is really \dots 0, logarithmic divergence.  And this 
one \rm[the linear divergence]\it\ by symmetry again, a single power of
 momentum \dots this is also a logarithmic divergence, so is this.  So
altogether they're all logarithmic divergences, it's no big deal.  So 
these are the kinds of things that are divergent.  This \rm[the vacuum
polarization]\it\ will always have $q_\mu q_\nu - q^2 \delta_{\mu\nu}$, 
something like that.  Because of dimensions however, you see there must be,
the whole thing has dimension one so there has to be at least one $q$
sticking out in front \dots these things \rm[the two-point function]\it, insofar as they diverge, 
at high momentum, they must be numbers times these two $A$'s with $q$'s
in it.  And if you call the vector potential of this $A$, 
and \rm[of that]\it\ $A$, and you make the Fourier transform back again, in other
words, if we hadn't had those cutoffs and I've just got quarks in here---er
gluons in here---then this bubble \dots for having a $q^2$ in front, which
I illustrated here \dots double gradient on the $A$.  And because of 
gauge invariance, the result must be 
\bea
\centerline{\includegraphics[width=0.35\textwidth]{gauge-inv}}\nn\\
(\partial_\mu A_\nu - \partial_\nu A_\mu)^2\qquad
A_\nu^{\,\times} A_\mu^{\,\bigcdot}\partial_\mu A_\nu\qquad\ \  
A_\nu^{\,\times} A_\nu^{\,\bigcdot} A_\mu^{\,\times} A_\nu\nn\qquad
\eea
Likewise this term \rm[the three-point function]\it, 
involving three potentials, is equivalent to the effect of some direct
contact---the divergent piece---is equivalent to some kind of contact
which involves three $A$'s and one gradient.  But because of gauge
invariance, the only thing you can write that has that property is
this kind of thing, to go along with this one \rm[the kinetic term]\it.  And
furthermore, these four, has four $A$'s, and they will turn out to be
of this form.  Not only that, but the \rm[coefficients]\it\ of every one of these
things will be adjusted just right so that the combination of these things
with their coefficients, all the coefficients will be right, so that this
is equal to 
\be
	\ln\Lambda^2 \, F_{\mu\nu} F_{\mu\nu}
\ee
times a number, which involves this divergent log \dots  The log divergent
part looks like this.  So if I had computed this one or this one, I would
have gotten the same result.  The reason it has to have this form is gauge
invariance.  If I did not destroy the gauge invariance by the cutoff
method.  Now the particular cutoff method I used was forced to not spoil
the gauge invariance \dots

Likewise this thing \rm[the quark self-energy]\it\ is going to involve two 
$\psi$'s.  And this one \rm[the vertex correction]\it\ is going to involve two
$\psi$'s and an $A$.  This one \rm[the vertex correction]\it\  corresponds to
changing the coupling constant here, and this one \rm[the self-energy]\it\
corresponds to this---there  is a term $\bar\psi m \psi$ \dots  corresponds
to the idea of changing the mass of the quark.  Well I have just erased
something here that I think I need:  when I change $\Lambda$, I'll change $g$
because I'm going to suck that number into the original zero order
$(1/g_0^2)F_{\mu\nu}F_{\mu\nu}$ which is what I started with, and this
gives corrections---are going to produce corrections---to this thing times
$F_{\mu\nu}F_{\mu\nu}$, and I'm going to say ``Oh.  I could have started
with a $g$, I'll make this $g$ change to eat that number.''  
In other words $1/g^2$ will vary in such a way to eat that number.
And that's what those formulas \rm[(\ref{eqA3-15})]\it\ are for $1/g^2$.
They're just designed to eat these logs.  The gauge invariance enables
you to know that all of these are going to all go together \dots you just 
have to look to
higher order divergences, and find out this never stops \dots
but you have to show that you don't keep getting more and more in trouble.
Which I'll show you why \dots in a minute.

To make this even clearer, if I have to; to look at it another way.
We originally have to do an integral that looks like this.
Then there's another term, which I'll write as a factor $\bar\psi$
gradient dagger\,\footnote{RPF habitually says ``dagger'' to mean
what we call ``slash''} minus $A$, which I will write like this \dots
You integrate this over $\psi$ and also over $A$,
\bea
\int {\cal D} A_\mu\, e^{i{1\over g_0^2}\int F_{\mu\nu} F_{\mu\nu}}&&
\!\!\!\!\!\!\int {\cal D}\psi {\cal D}\bar\psi\, e^{i\int\bar\psi
(i\slashed{D}-m)\psi}\nn\\
	=\int {\cal D} A_\mu\, e^{i{1\over g_0^2}\int F_{\mu\nu} F_{\mu\nu}}
	&&\!\!\!\!\!\!\underbrace{\det\left(i\slashed{D}-m\right)}\nn\\
&& g(A) \sim e^{i\ln\Lambda^2 f(A)}
\eea
That's the kind of thing we're trying to do.  Now we can imagine first that
I had just done this integral \rm[over $\psi$]\it\ completely, it would be nice if
we could do it for arbitrary $A$; this is some terrifying functional 
of $A$.  \rm[question from student\footnote{Evidently RPF wrote $g(A)$ first
without identifying it as the functional determinant.}]\it\ Yes, yes, yes, exactly.  This is one
over---no, the determinant of the Dirac operator $i$ gradient minus $m$
minus $A$, with some color terms; yes, that's just what it is; it's the
same thing, we can't work this out either.  Now we can expand this by
perturbation theory, and try to make a calculation, and we discover that
this is \dots at very short distances, high momentum, at very short
distances there's some trouble.  The trouble comes from too many gradients
on top of each other, 
the propagators from this thing have delta functions in the origin, delta
functions on a line, 
two of them on top of each other, they strongly diverge.  So for very 
high frequency $A$, this function has to be discussed,  this is a little bit
wrong.  So we had to fix it a little bit.  It's still a determinant, we
just
fixed it by a cutoff.  When we fix it by the cutoff, we discover that this
thing is $e$ to the $i$ times $\ln\Lambda^2$ multiplied by another function
of $A$, of course you can always write it that way.  But insofar that this
only involves high frequencies, short distances, $A$ is at two points very
close together.  

And, if everything has been done right, since this
expression here \rm[the determinant]\it\ is gauge invariant, with respect to $A$,
because if I make a gauge transformation of $A$, and then fix up the $\psi$
so as to get the same answer, this has to be a gauge invariant expression 
\dots we discovered over there that it involved the gradient of $A$
squared.   But we know that it's gauge invariant, and therefore if I did 
it completely, I would get the whole string \rm[the three terms in $F^2$]\it, 
I could only get this; this is the only gauge invariant expression which
starts like that.  And that it starts like that is a statement of the forms
that we got---by the way, although we didn't notice it, but this $A$,
this kind of propagator, is just---see, if I put an $A$ on each side of
this, this becomes $A_\nu$ squared and two gradients. Well that's
$\partial_\mu$ squared.  Let's figure out which way \rm[to contract the
Lorentz indices]\it.  The other one is $q_\mu A_\nu q_\nu A_\mu$ \dots
Then assume an $A$ which is a plane wave and substitute it in this 
expression \rm[the gluon kinetic term]\it; you would get this kind of thing back
\rm[$\delta_{\mu\nu}q^2 - q_\mu q_\nu$]\it, so this is in fact the operation 
producing this kind of combination.  So we're getting the first term right.
Because we only looked at the two-gluon.  If we looked at the three-gluon,
we would be {\bf surprised} to discover that it produces 
$A$ dot $A$ to the cube 
\rm[$A_\nu^{\,\times} A_\mu^{\,\bigcdot}\partial_\mu A_\nu$]\it, which is just 
this combination.  And even {\bf more} surprised to discover that the
coefficient is the same $g_0$ exactly.  The surprise would disappear when
we realized that it has to be with a cutoff scheme that \rm[preserved gauge
invariance]\it\ \dots

And in the same way, when we go to integrate over $A$, we know that there's
a problem with the meaning of this thing \rm[the functional measure]\it; let's 
forget about it.  But we could imagine some kind of rule: stop the
integrations above a certain high frequency---unfortunately, that's not
gauge invariant---do a lattice.  Well let's say stop high frequencies,
forget about \dots one of these days---dimensional renormalization.  So you cut this off 
at high frequency; you say wait a minute, what if I cut it off at a
different high frequency?  Then I could say that the intermediate between
the medium high and the very high frequencies is what I'm integrating over
to see what happens if I cut it off at the medium frequency.  That will
produce a number of terms that will involve the logarithm of the very high
and the medium high frequencies.  And the coefficient of that log \dots
will again have to be gauge invariant and have the same kind of form
\dots \rm[change of tape]\it

\dots it's just pretty; the only place where there appear to be divergences
are just the places you need to make the simple form to be the same 
shape as the original one.  Sometimes the good way to look at it is that this
thing can be compensated by putting a term like this with a $g^2$.
The Lagrangian \rm[gets a]\it\ correction term, this can be put in by putting a
term like that times some $q$ times some number.  So the Lagrangian has a
correction term, those are called counterterms; in other words, if we
started with a Lagrangian, instead of saying $1/g_0^2$ exactly, times
$F_{\mu\nu}F_{\mu\nu}$, we say we're going to with a Lagrangian which has
already in it counterterms, this thing minus those numbers, which you're
gonna find out what they're gonna be, times $F_{\mu\nu}F_{\mu\nu}$, which 
are counterterms.  That's the Lagrangian that I started with, and then
we're going to have a cutoff at $\Lambda$.  And the cutoff at $\Lambda$ is
going to be equivalent to making corrections---divergences---well, they're
not divergent because we're cutting them off, which undo these, to get
something which is independent of $\Lambda$.  So if I write the real $g$
\rm[the renormalized value]\it
\be
 {1\over g^2} ={1\over g_0^2} + \delta\left(1\over g^2\right) 
\ee
this is equivalent to starting with some constant, call that $g_0$, the 
constant we started with, plus some counterterm which will depend on
$\Lambda$, and they're built in such a way as to compensate the divergences
that we get here.  The net result when we're all finished is just that
constant times $F_{\mu\nu}F_{\mu\nu}$, independent of the cutoff. And
that's where all those formulas come from, that talk about the $1/g^2$
being corrected by things that depend on $\Lambda$.  Alright?  Any other
questions?  It's just another way of describing the same thing.  But the
beauty of it is \dots the divergent terms are exactly right to reproduce 
the form of the Lagrangian, and therefore by changing the coupling constant
we can undo the $\Lambda$ dependence.  Alright?

Now I do have to complete the discussion, to discuss higher order
calculations \dots I only did this one \rm[the gluon vacuum polarization]\it\,
to lowest order---I didn't do everything, I didn't do these loops, but
let's suppose I had; well let's say this one, I don't care, look,
it doesn't make any difference, quarks or gluons \rm[in the loop]\it.  I'm not
going to worry about \dots  However, we just noticed when we were counting
divergences that now we're in trouble, because now after we did all that,
then we look and we find something like this,
and that table of calculations says that this will have the same divergence
as this and this.

By the way, there was a step in here that was very 
clever: taking this and putting it \rm[in the denominator of (\ref{eqA3-14})]\it\,
implies something about the higher orders---the leading logs, what that's
all about, is that I have not only taken this \rm[one loop]\it\ diagram,
\[
{\includegraphics[width=0.35\textwidth]{nested-loops}}
\]
but I've added this diagram, and this diagram \rm[two loops]\it\ and so on,
to get the sum of one plus $x$ plus $x^2$ plus $x^3$ plus $x^4$ \dots
is equivalent to $1/(1-x)$, and when I did this \rm[(\ref{eqA3-14})]\it, I was 
already predicting the higher terms, but I know where they're coming from,
obviously \dots So I've done all the single loops.

Now, so this looks as if it's going to produce another contribution to the
log, and so on, and the millions of diagrams---but it's not true.  Because,
I look at it this way, every time from now on that I see a gluon propagator,
I really should correct it, or could---can correct it, by putting a loop in
there like so, or two loops or three loops, and so on,
\be
\raisebox{-0.25cm}{\includegraphics[width=0.25\textwidth]{gluon-prop3}}
\ +\ \dots\ \sim\  {1\over Q^2\ln
Q^2}
\ee
I want to include all these loops. So what this ought to be, this line
now really means the propagation compensated, or corrected by these loops.
Now we found out that---I guess I should have emphasized, that when I make
this choice, substitute that back in here, I find this thing, $\beta$
times the log of $\Lambda^2$, minus this logarithm; this is the logarithm
of $Q^2$ \dots And therefore the effective propagation is not really
$1/Q^2$, but is really $1/(Q^2\ln Q^2)$ if the loops are included.  That's
slightly more convergent.  That goes down a little bit faster than $Q^2$.
So that these propagators are not strictly speaking $1/Q^2$, they're 
$1/(Q^2\ln Q^2)$.  By the way, we sometimes write 
that as $\alpha(Q^2)/Q^2$; we talked about that.  Anyway it's one over log,
so these divergences are no longer computed right by just saying things 
like $\int d^{\,4}p/p^4$ with zero powers left over, which is equivalent of
course to $\int d(p^2)/p^2$.  But the propagators are not $1/p^2$, they
have logs in them, and I have to tell you how many there are; the lowest
possibility is that it begins with $\ln p^2$,
\be
\int^\Lambda {dp\, p\over p^2 \ln p^2} \sim \ln(\ln\Lambda^2)
\label{eqA3-21}
\ee
Indeed, there are terms, you might have higher powers of logs \dots
but not worse, you have at least the log.  But you see what this looks
like, this differential log over log?  This is log log, less divergent.
So if I integrate this to some high frequency, this is still divergent,
but it's $\ln(\ln\Lambda^2)$.  And when I go to the next order, I'm going
to get log log.  Of course there can be terms with log squared here,
 but those will be convergent:
\be
\int^\Lambda {dp\, p\over p^2 \ln^2 p^2} \sim 
\int^\Lambda {d\ln p^2 \over \ln^2 p^2}  \sim {1\over
\ln\Lambda^2}
\label{eqA3-22}
\ee
In other words this integral \rm[(\ref{eqA3-21})]\it, this doesn't quite converge, 
but if I had more logs in there it would converge.  So that there are
little log-log divergences, no worse.  

So the next order terms, when you include the corrections in them from the 
lower order, does not produce the same divergence \dots You might say,
well how do I know that there are \dots maybe they're only log squared
or log cubed \dots $\beta_1$ terms \dots
\rm[Question from student: why did you have just two powers of the momenta
in the numerator of that integral?]\it\  Well it's a logarithmic divergence.
I don't care about \dots it's just to understand the log \dots
The differential log over log is log log, that's it.

Now, you say well now I'm going to go on to the third.  But it's no longer
true that it goes on to the next one, because the higher ones give more
powers of logs down here.  And more powers of logs down here, it converges,
it stops.  Thank God.   Now you say, well, it wasn't really log $p^2$,
it was that plus $\ln\ln p^2$, and I'll let you make the argument
that this doesn't change the divergence; it doesn't make it converge any
better \dots \rm[Question from student about absence of triple logs]\it\  w
Because the next order produces an integral like this \rm[(\ref{eqA3-22})]\it
\dots There's no way to isolate log-log.  This correction is this 
propagator, which is corrected by one over log.  If it's got a log-log
in it, it's additional.  In other words the corrections of $\beta_0$ log
plus $\beta_1$ log log.  So the correct thing to put in there, if it's
anything, is $\ln p^2$ plus $\ln\ln p^2$ at that order.  But that argument,
that doesn't make any difference compared to the $\ln p^2$ when you
calculated the divergence.   What happens in the higher orders, you get
more logs down here, but you get this to a higher power \dots there's no way to 
isolate this \dots

\rm[Question from student]\it\ This has to do with the behavior in terms of $Q^2$.
The propagator is $\ln Q^2$ \dots it's true that the correct formula 
has $\ln Q^2$ plus $\ln\ln$ terms; that doesn't \dots you just get rid of 
this log and subtract it from the \dots and try to get the log log
isolated; there's nothing that comes out 
\dots as you would like it, as you might imagine it.  \dots $Q^2 (\ln Q^2
+\ln\ln Q^2)$.  Of course there's no propagator that goes inversely to log
log.  That's not the way it goes.  It goes inversely to the sum.  And that
would be log log log.  You don't get this kind of a form, because you can't
isolate that piece \dots My argument here is very heuristic, but it does
work and it gives you an idea of why \dots this is the way the thing works
out \dots more or less why it works out \dots  any other question?

In order to add to your confusion \dots different conventions \dots
different ways of looking at it; some are better than others.  If you want
to read the literature, you have to read everybody's ideas.  Some are
better \dots  because they get rid of some confusion, so they straighten
something out.  Now if you learned only the way which  is all straightened
out \dots then you have some trouble reading literature in which something
is a little older or something which the guy is using some old-fashioned
idea; well I wouldn't say old-fashioned but less \dots then you couldn't
understand the subject completely.  What I described, I had tried to prove
differently than is general in any textbook---if it's in some textbook,
you don't think it's original, not at all \dots I did it myself---but I
have to tell you about what I consider a kind of mistake, okay?  Which is
very prevalent and it's all over the place.  It's {\bf not} the way I'm trying
to explain it.  Now the way I'm trying to explain it is that the coupling
constant---when you make a cutoff, you change the theory.  And the answers
in general appear to depend on the cutoff.  But it turns out they depend on
two things: the way you cut it off, $\Lambda$, and also the coupling
constant you put in.  But by the very wonderful situation we have
of renormalization, that when you adjust the coupling constant correctly,
when you change the cutoff, you change the coupling constant, you'll get
the same predictions in the long-range wavelength physics \dots And that's
what I wrote in the beginning, was a formula for how you have to change
\dots in order to make the  results independent of the cutoff. You change
the cutoff, you'd better change the \rm[bare coupling]\it\ \dots

Next.  It turned out that in many circumstances where you could expect the
mass of the quarks to be unimportant \dots that for such processes, the
behavior of the process could be worked out as a perturbation theory in 
$g$,
\bea
	&&c_1 \pi g_0^2 + g_0^4\left( b_1\ln{\Lambda^2\over Q^2} + a_1\right) +
\dots\nn\\ &=& c_1\alpha(Q^2) + a_1\alpha^2(Q^2) +\dots
\label{eqA3-23}
\eea
These coefficients depend on the logarithm of the cutoff and the momentum
of the operation, the process, some momentum associated with the process,
some definition of the momentum.  And this sum, these logs, could be summed
so this was written in the form of $\alpha(Q^2)$ plus $a_1 \alpha^2(Q^2)$
and so on, and this is for some physical process.  In other words, for 
some physical process, the calculations go like this.  For a different
process, all the coefficients would be different \dots there might be a 
$\pi$, so let's put one in here\,\footnote{I have the $\pi$ in
(\ref{eqA3-23}) crossed out in
my notes} \dots anyhow, it would go like this and it would be---so, 
therefore when you start to work perturbation theory there's a rule 
\dots work at second order you get the log, but you've already eaten that
when you made the substitution.  So this is a way I managed to write it, 
and it's perfectly okay, and the formula for how loops \dots describe \dots
the type of thing that's connected \dots which we wrote down \dots I'm just
repeating \dots thing I found out by \dots or rather not the best way to do
it.

Now let me tell you the wrong way, what I consider not as good a way. It's 
a way that you could have done it, but it's got annoyances in it.  It works
like this.  You start out to define an $\ul{\alpha}$.  Now in order to make sure
that this $\ul{\alpha}$ is not exactly the same as that $\alpha$, 
I'm going to put an underline on it, so you'll always know which one I'm
talking about.  Now define an $\ul{\alpha}(Q^2)$ {\bf by} a physical
process.  I'm going to give you examples \dots  For instance, we recalculate
the scattering of two quarks, to {\bf all} orders, exactly.  All orders
exactly, it's going to be written as this super-duper $\ul{\alpha}$
over $Q^2$,
\be
\raisebox{-0.8cm}{\includegraphics[width=0.05\textwidth]{blob}}
\  \equiv \ {\ul{\alpha}(Q^2)\over Q^2}
\label{eqA3-24}
\ee
and that's going to define $\ul{\alpha}$.  There would be no such thing as
finding a power series expansion of the coefficient, which is what I would
do there \rm[eq.\ (\ref{eqA3-23})]\it; \dots that's alpha, by definition, to all
orders.  Another thing would be, another way for example, when we talk about 
$e^+e^-\to$ hadrons, the ratio \rm[$R$]\it, and remember that the formula for that
ratio in the first order perturbation theory is $1+\alpha/\pi$, 
remember that?  Or rather $1+g^2/\pi$?  I'm not worried about the $4\pi$'s
in the definition of $\alpha(g^2)$,
that's a pain in the ass that I can't remember.  Let me {\bf define}
$\ul{\ul{\alpha}}(Q^2)$; this is another definition, let's put two lines,
it's another definition, it's identically equal,
\be
	\ul{\ul{\alpha}}(Q^2) = \pi(R-1)
\ee
But anyway, we could do this to all orders, this is defined to all orders.
So there's no such thing as a perturbation expansion for this to first
order and next order and next order \dots it's just a definition.  That
would be a possibility.  As it turns out, that up to the first two orders,
the formula for this $\ul{\alpha}$ and the formula for that
 $\ul{\ul{\alpha}}$ and my $\alpha$ all agree, in terms of the first log
and the log-log.  All of those formulas satisfy
\be
	{d\alpha\over d\ln Q^2} = {\beta_0\alpha^2\over
1-{\beta_1\over\beta_0}\alpha^2} = \beta_0\alpha^2 + \beta_1\alpha^4 +
{\beta_1\over\beta_0}\alpha^6 + \dots
\ee
And for every one of these definitions, for any one of these alpha bar
things, the $\ul{\alpha}$ satisfies exactly the same equation,
up to the fourth order, 
\be
	{d\ul{\alpha}\over d\ln Q^2} = \beta_0\ul{\alpha}^2 + 
\beta_1\ul{\alpha}^4 +
\beta_2\ul{\alpha}^6 + \dots
\ee
but the next order depends on the process.  These \rm[the coefficients of
the first two terms]\it\ don't.  In other words it depends on the process
needed to define \dots the $\ul{\alpha}$.  So for practical purposes
up to second order it doesn't make any difference, but if you want to make
things definite so that one guy can compare his results to the other, 
in higher order, they're all mixed up, because one guy is using one way,
another is using another way, because of the difference in processes.
You say what's any better, why not use a definite process?   Because
\dots Instead of using a definite process, I used a definite theory
\dots You'll notice this \rm[$\beta_2 = \beta_1^2/\beta_0$]\it\ is a special
choice, but it's definite, and this involves \dots independent of 
process.  And an advantage is, you don't have to compute this special
process \dots you've got it done \dots If you want to know some physics
then you have to compute.  And that's saying that you should really
calculate the power series for this process, in terms of $\alpha$, 
my $\alpha$ \dots  We should naturally expect to do each process 
separately as a perturbation expansion, instead of arbitrarily choosing
one \dots one is no better than the other \dots it all adds confusion to \dots

They then said that the physical coupling constant depends on $Q^2$,
but there's no definition, it depends on how you define it.  You could say
that the coupling constant depends on momentum squared, but \dots 
So, people talk about this as if it's a running coupling constant, 
but you can't put that into the Lagrangian, as a running coupling constant.
The only thing you can put into the Lagrangian is something that depends on
$\Lambda$, not on $Q$, so I was rather confused \dots  
You see how much confusion \dots in the
definitions; Politzer's $\lambda$ \dots we saw the equations depend on
which method of cutoff you use, and how you define the constant, is it zero or is
it Euler's constant times the log of $4\pi$ \dots
And on top of that, on top, I wanted to add the ambiguities that slipped in 
\dots to define alpha and it's not \dots 

Now in the electrodynamic world, there was a wonderful special process to find
the electric charge, which was unique, which is, let me evaluate the interaction
of the particles when they're very far apart---the very long wavelength coupling 
of photons to electrons.  In quantum chromodynamics you can't find any \dots
like that \dots For very long wavelengths \dots so we have no simple phenomenon
which \dots Any other questions?  Alright then.

Someone asked me last time how dimensional renormalization produces the same
results.  The answer is more or less the following.  You would make a process
in the scattering---first of all we have less than four dimensions.  We have
something like
\bea
	{1\over g^2}\!\!\!\!\!\!\!&&\underbrace{\int F_{\mu\nu}^2 d^{\,D}x}\nn\\
	&& \hbox{\rm dimensions of Energy$^{4-D}$}
\eea
and then  integrate with respect to $D$ dimensions of spacetime, $D$ is not 
four.  Now that means---and what about the dimensions?  In  $F_{\mu\nu}$ as you 
all know and must have written \dots there's the combination of $\partial A$
and $A$ $A$, and that means that $A$ is an inverse length or an energy.  
That's independent of dimension.  And that 
$F_{\mu\nu}$ is an energy squared.  And $F_{\mu\nu}$ squared is an energy to the
fourth.  A length is an inverse energy.  So this quantity would have dimensions
of energy to the $4-D$.  Therefore $g^2$ is not dimensionless.  $g^2$ has 
dimensions of energy to the $4-D$,
\be
	[g^2] = \hbox{\rm Energy$^{4-D}$} =  E^\epsilon
\ee
So here, one way is to just say, alright, I know that.
Another way is to write $g$ as some other constant times some particular 
length to the $4-D$, I'll call that epsilon:
\be
	g^2 = c_0 \lambda_P^\epsilon
\ee
\dots Now what happens is, if you do perturbation theory, $g^2$ plus  $g^4$
times an integral, same way as we did before.  Except those integrals because
they don't have $d^{\,4}p$ any more, they only have a $D$ integral, are less
divergent, in fact they converge.  So you can actually do the integral, and
there's no problem, and you find that the integral varies as $Q^{-\epsilon}$
\bea
\label{eqA3-31}
  g^2 + g^4 Q^{-\epsilon}\left (2\beta_0\over\epsilon\right)
&=& {1\over {1\over g^2} - {2\beta_0 Q^{-\epsilon}\over \epsilon}}\\
&\approx& {1\over {2\beta_0\over\epsilon}\lambda_P^{-\epsilon} \nn
	- {2\beta_0\over\epsilon}Q^{-\epsilon}}
\eea
minus $\epsilon$ because it's $D-4$ \dots  so you get this kind of a term and
\dots it's no problem, everything converges and it's fine.  But as we vary
$\epsilon$, we discover that the coefficient that we actually get, the
coefficient here, is a certain constant \rm[$2\beta_0$]\it\ over $\epsilon$.  That is,
if I did the calculation with different $\epsilon$, I'd get something that varies
with $\epsilon$ this way: the coefficient diverges as $\epsilon \to 0$.

I can go through all this usual stuff of rearranging the sum and \dots
when $Q$ is very large this is small; but the point is that aside from constants
which will be taken out \rm[RPF writes out the right-hand side of eq.\ 
(\ref{eqA3-31}) at this point]\it\ \dots and as $\epsilon$ approaches zero,
\be
  2(\lambda_P^{-\epsilon} - Q^{-\epsilon}) = -\epsilon \ln(\lambda_P^2/Q^2)
\ee
The point is, the theory is convergent.  It depends on $\epsilon$, and has a
very high coefficient as $\epsilon$ goes to zero, and in the limit produces 
logarithms, just like the logarithms you see, and we have to adjust the coupling
constant.  So it has the right behavior with $\epsilon$ \dots the same problem,
how does this $g$ depend on $\epsilon$?  \rm[Student: in dimensional
regularization, don't you want to take $\lambda$ to be close to the energy
scale of the \dots ]\it\ $\lambda_P$ \dots In the end yes, it's better, yes I think
that's the right thing to do \dots  So what we're saying is that the coupling
constant has the dimensions of the physical energies we're interested in
\dots but it turns out the strength varies inversely as $\epsilon$ \dots
\footnote{In my notes I have written ``So we have traded large dimensionful
$\Lambda$ for small dimensionless $\epsilon$.''}
\newpage

\begin{widetext}
\rm
\section{Revision examples}
\label{AppD}
Additions written by RPF to the revision of lecture \ref{sect12}.
\[
\centerline{
\includegraphics[width=\textwidth]{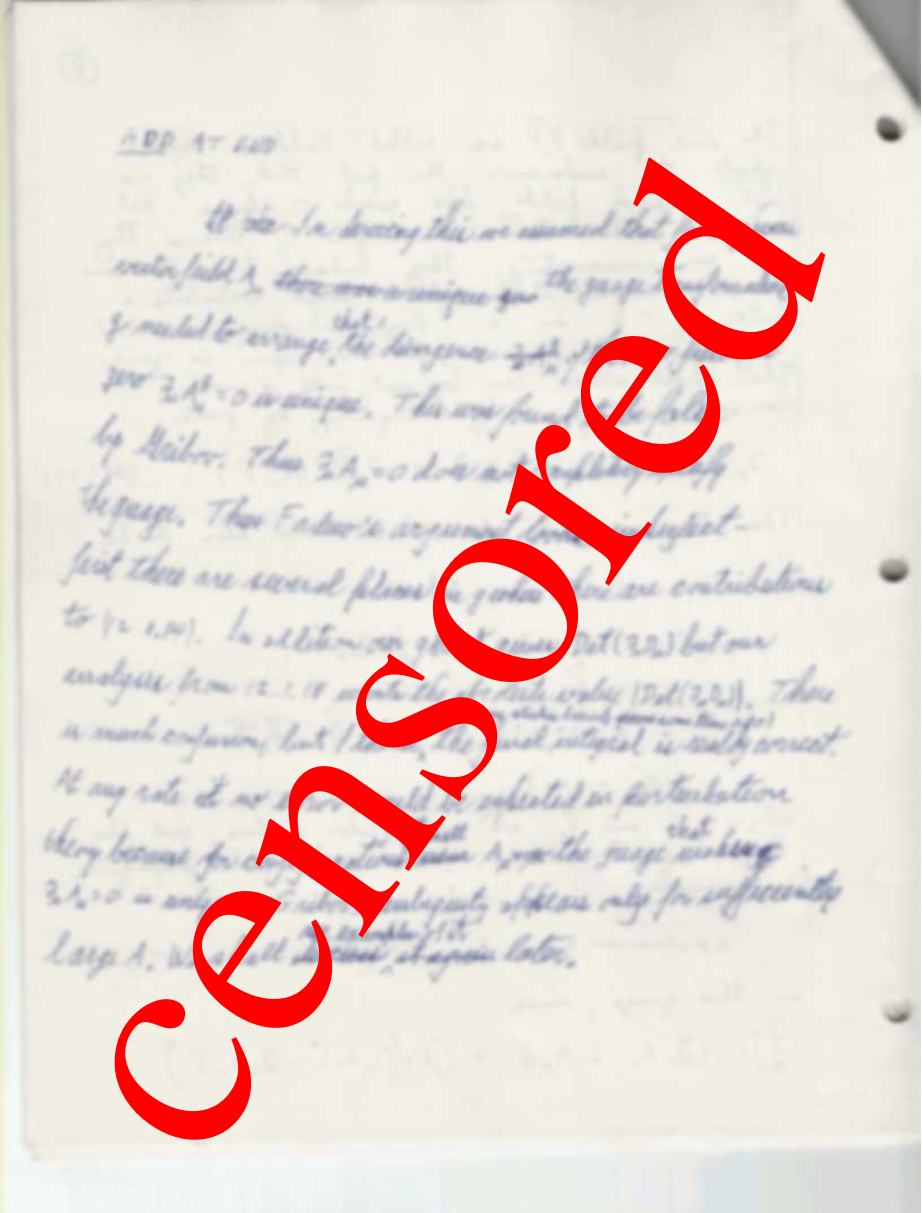}}
\] 
\newpage
Additions written by RPF to the revision of lecture \ref{sect13}.

\[
\centerline{
\includegraphics[width=0.95\textwidth]{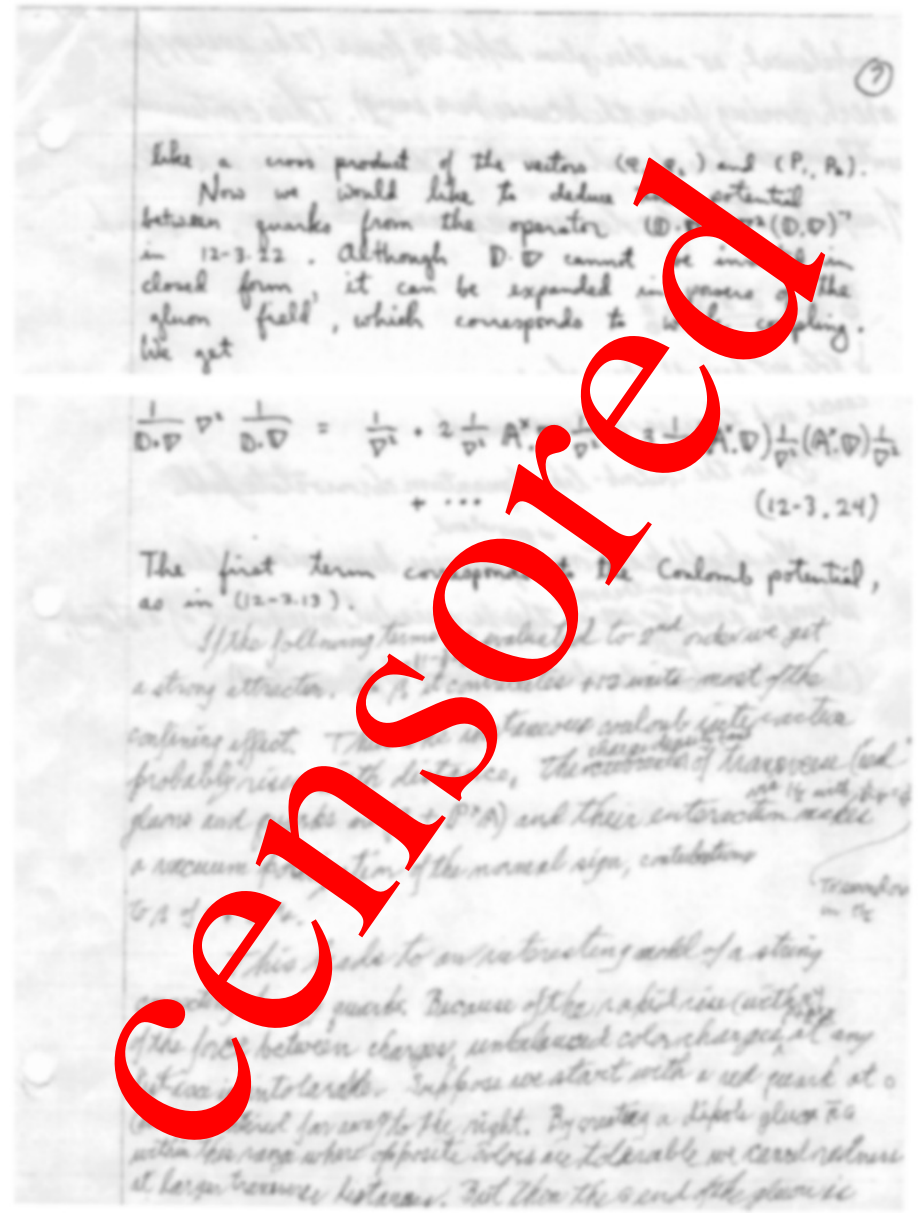}}
\] 
\[
\centerline{
\includegraphics[width=0.95\textwidth]{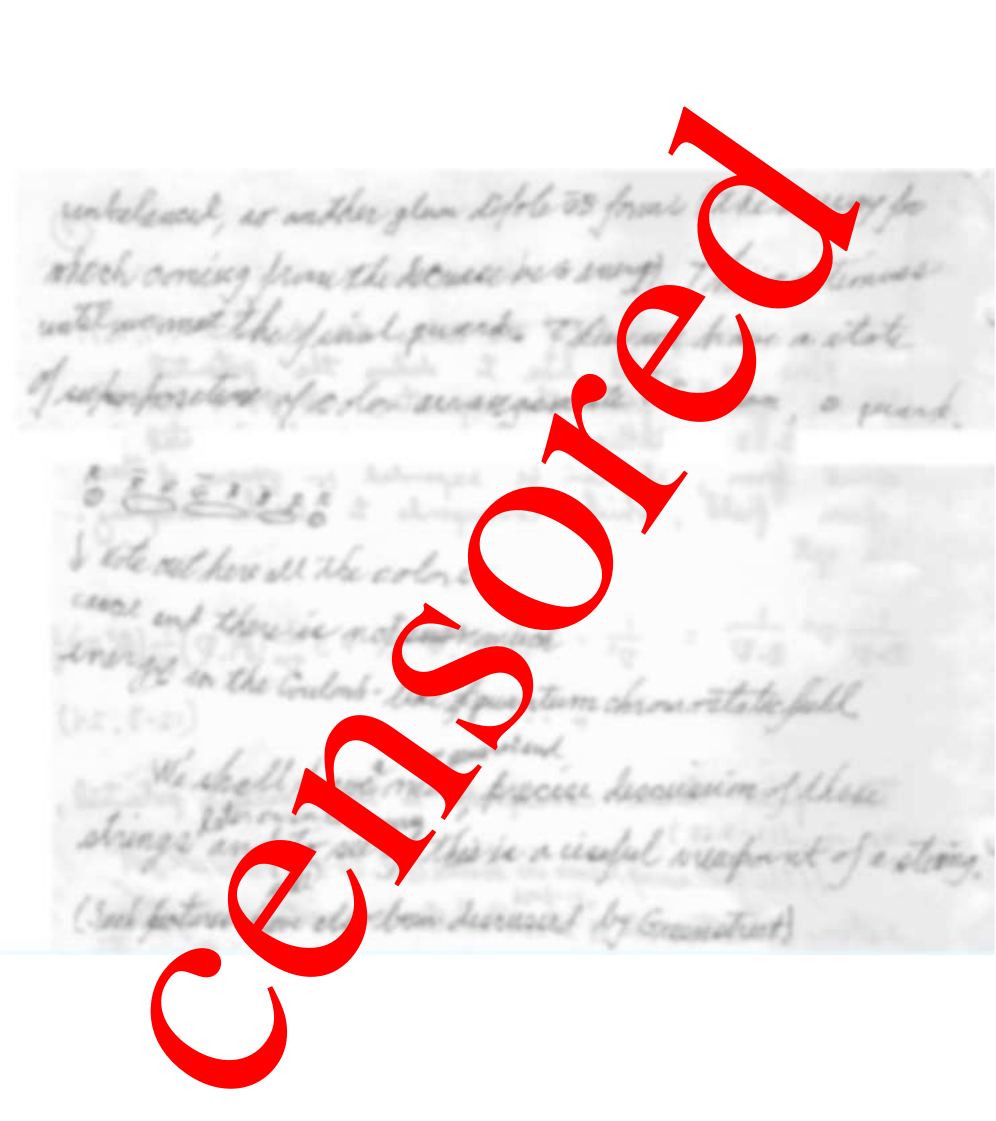}}
\] 
\newpage
There were sometimes also subtractions:
edits by RPF to the revision of lecture \ref{sect8}.
\[
\centerline{
\includegraphics[width=0.9\textwidth]{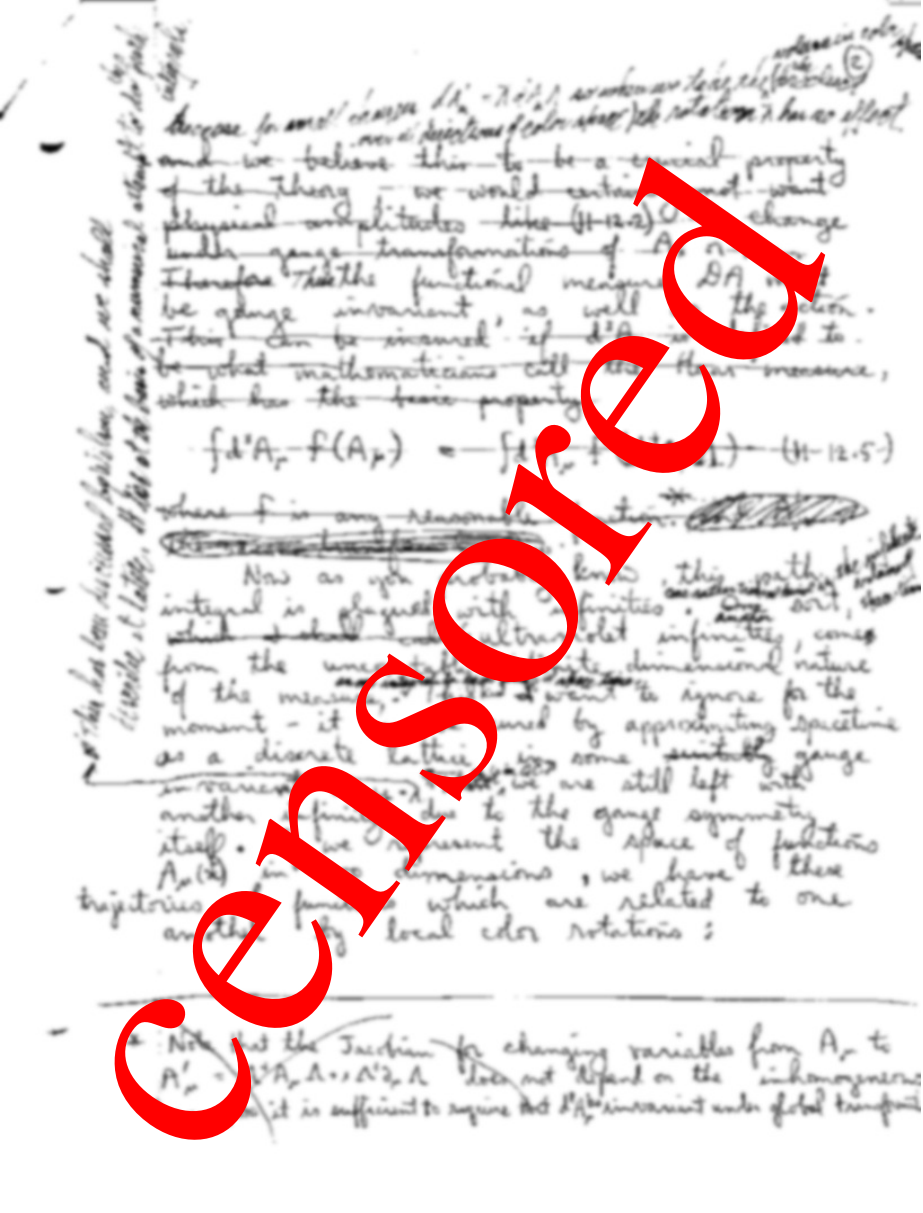}}
\] 

\section{Hadron masses and quark wave functions}
\label{AppE}
The following three pages were copied out of an unidentified textbook and 
handed out at the beginning of the course.  Handwritten corrections of
quark wave functions were added by me.
\[
\centerline{
\includegraphics[width=0.7\textwidth]{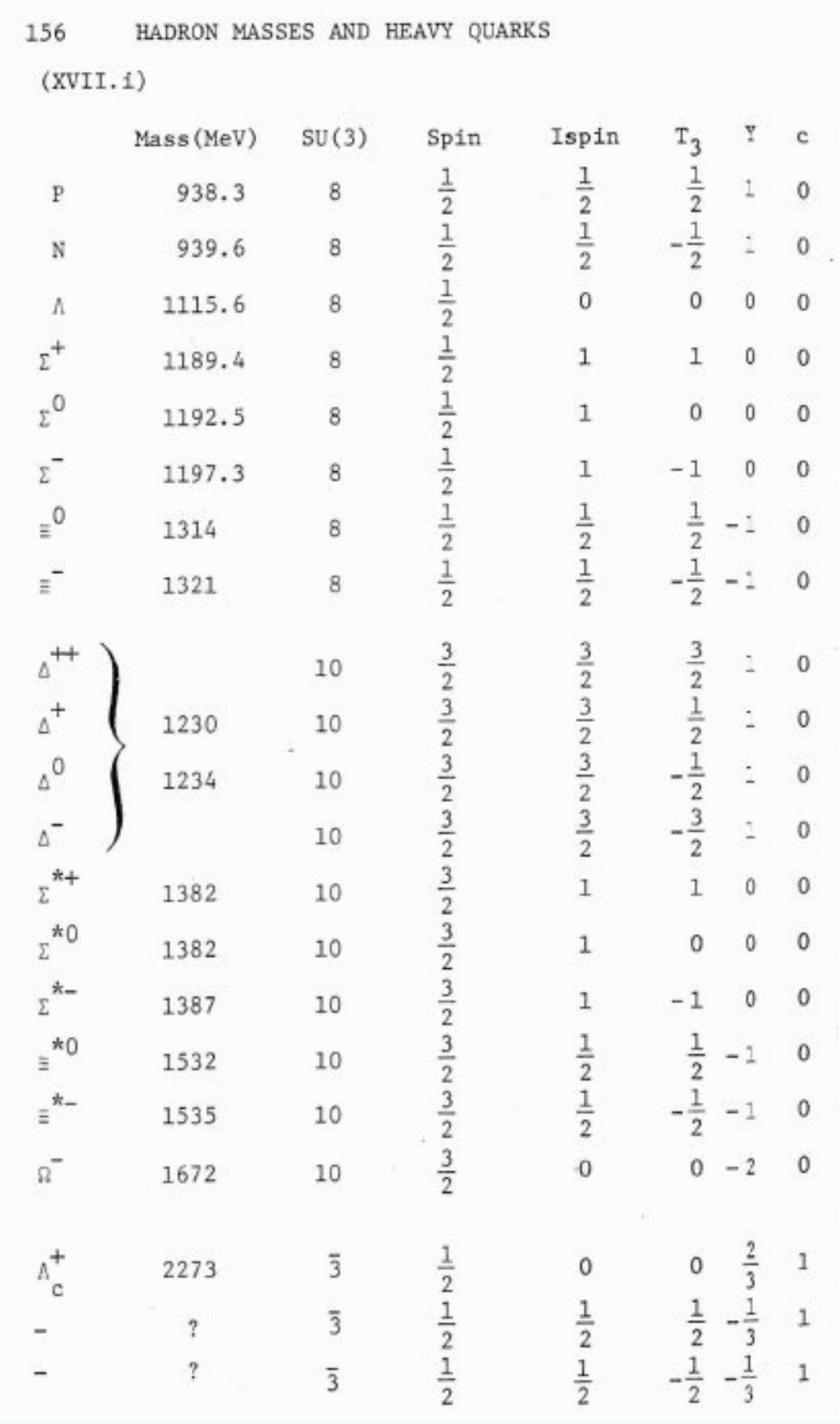}}
\] 
\[
\centerline{
\includegraphics[width=0.8\textwidth]{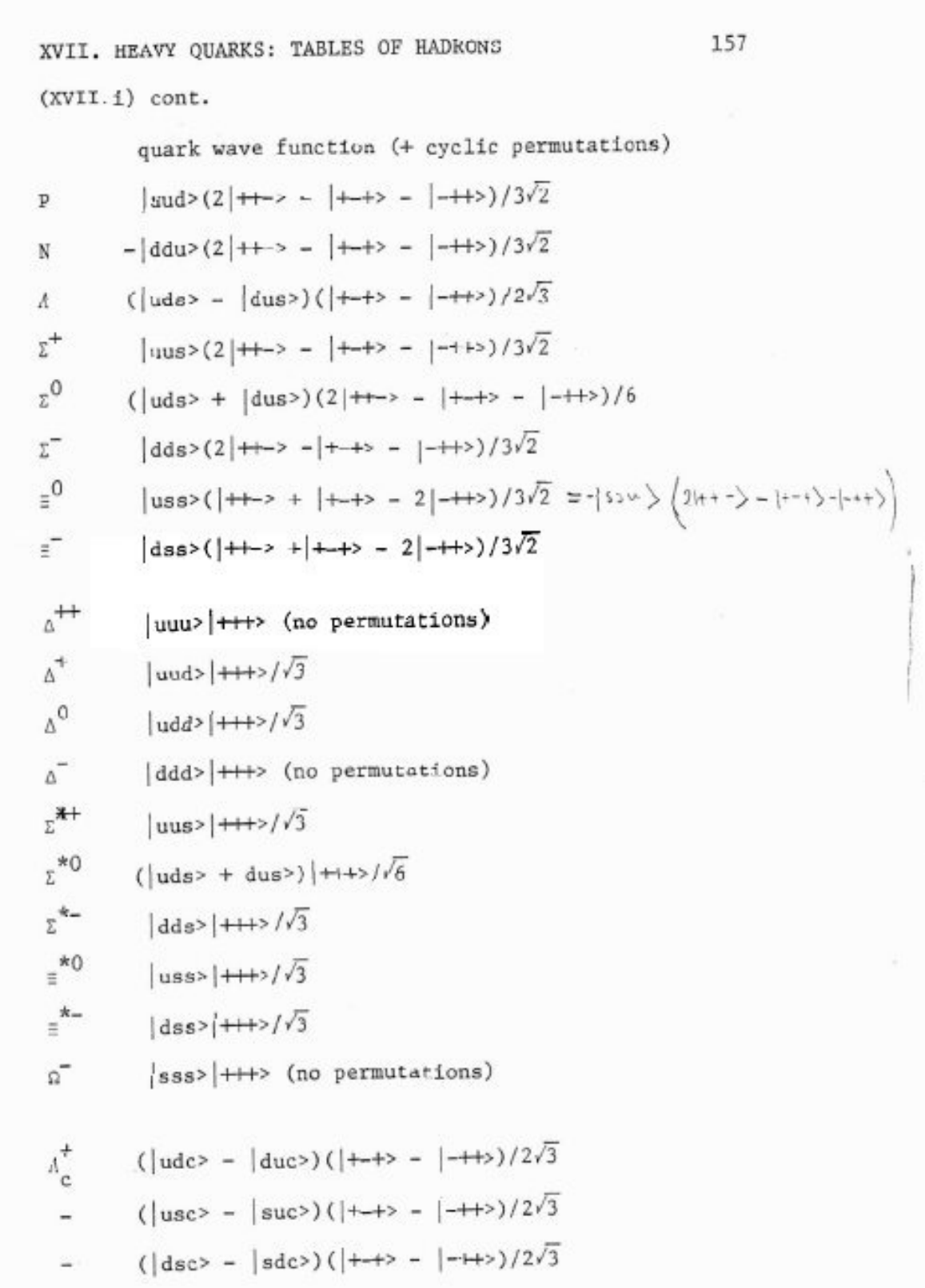}}
\] 
\[
\centerline{
\includegraphics[width=0.7\textwidth]{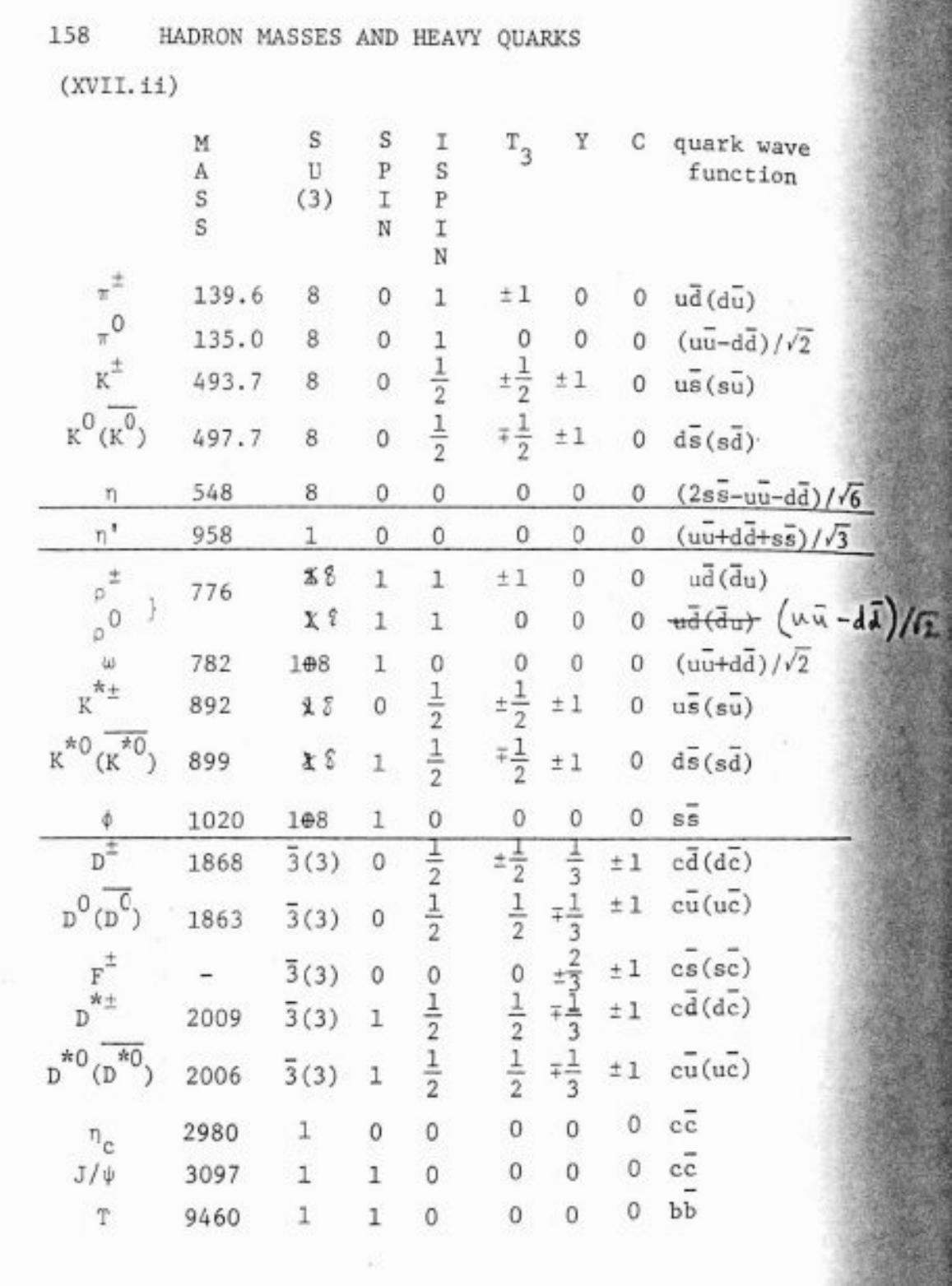}}
\] 

\section{Tables of hadrons}
\label{AppF}
These tables of meson and baryons were written by RPF.
\[
\centerline{
\includegraphics[width=0.9\textwidth]{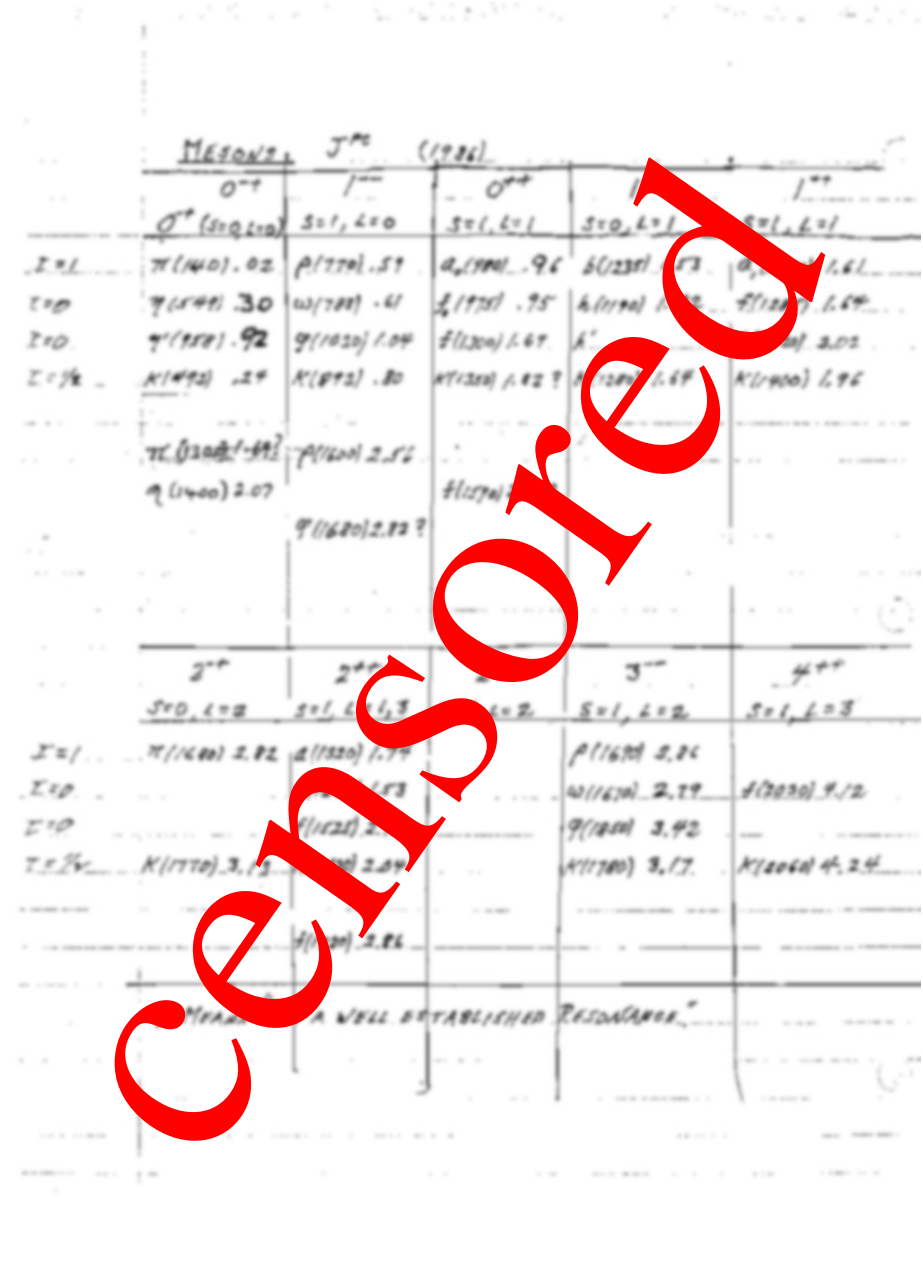}}
\] 
\[
\centerline{
\includegraphics[width=0.9\textwidth]{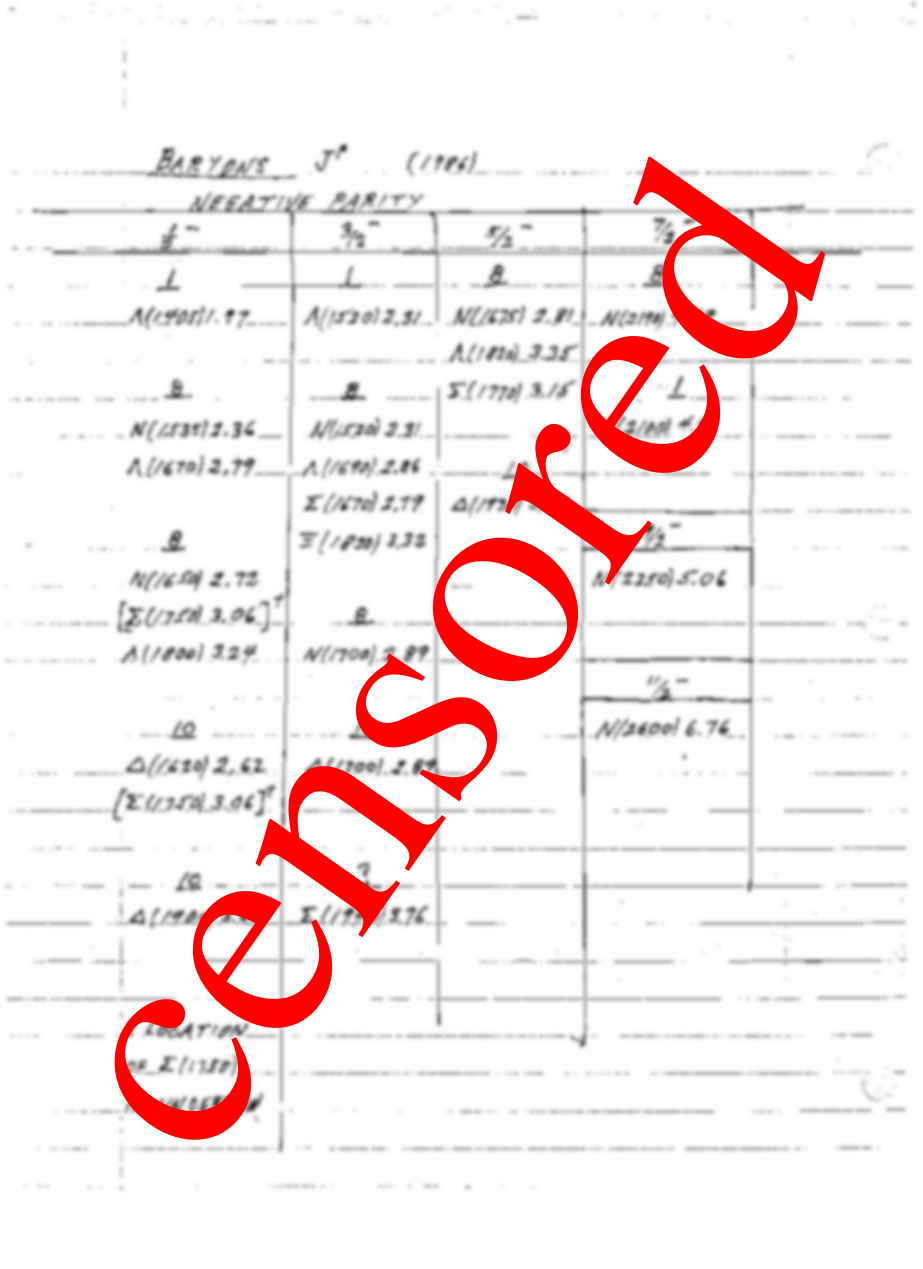}}
\] 
\[
\centerline{
\includegraphics[width=0.9\textwidth]{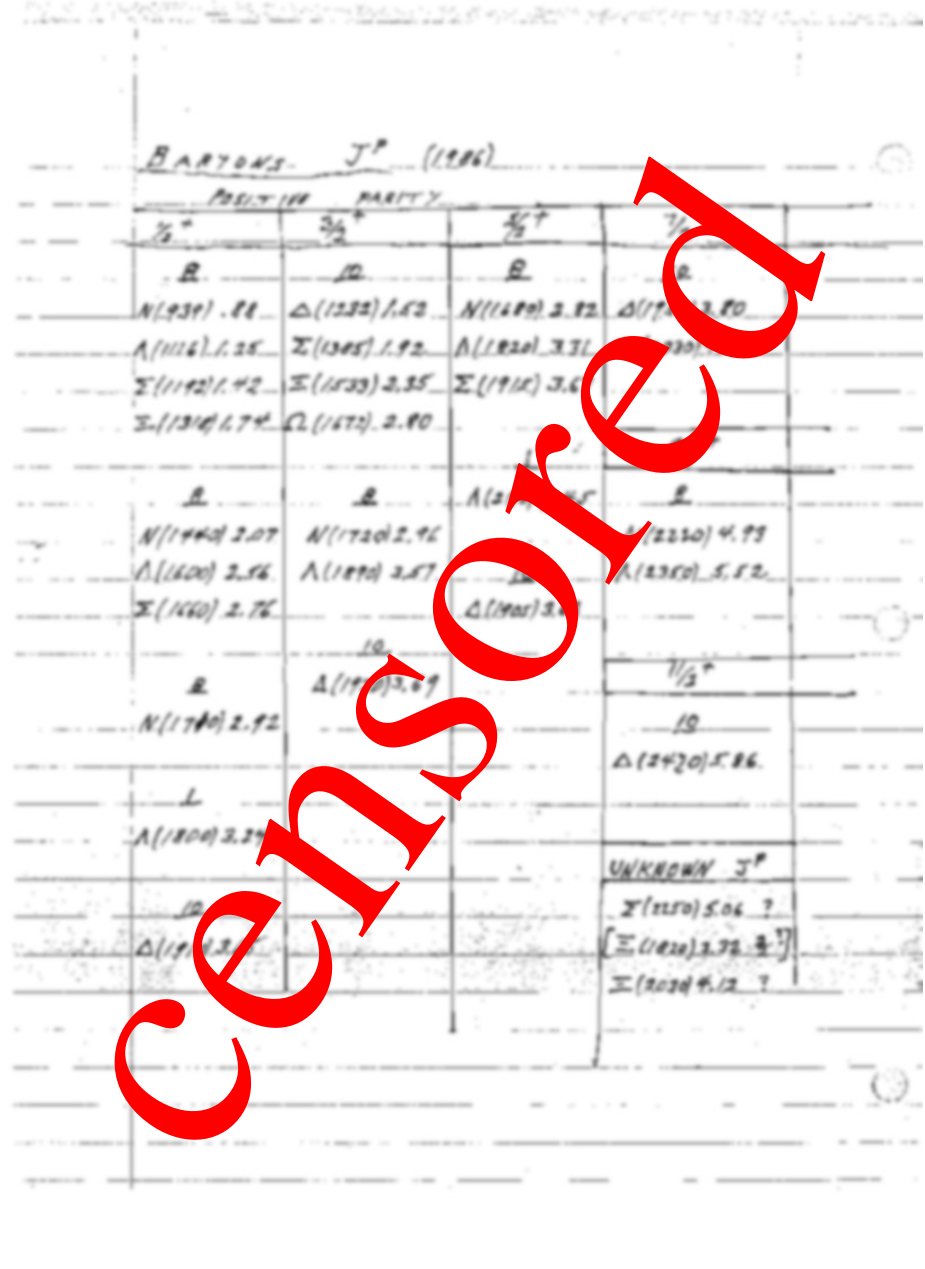}}
\] 

\section{Rules for amplitudes and observables}
\label{AppG}
These were also  hand-written by RPF.  Annotations in blue were made by me
at the time the course was given.
\[
\centerline{
\includegraphics[width=0.9\textwidth]{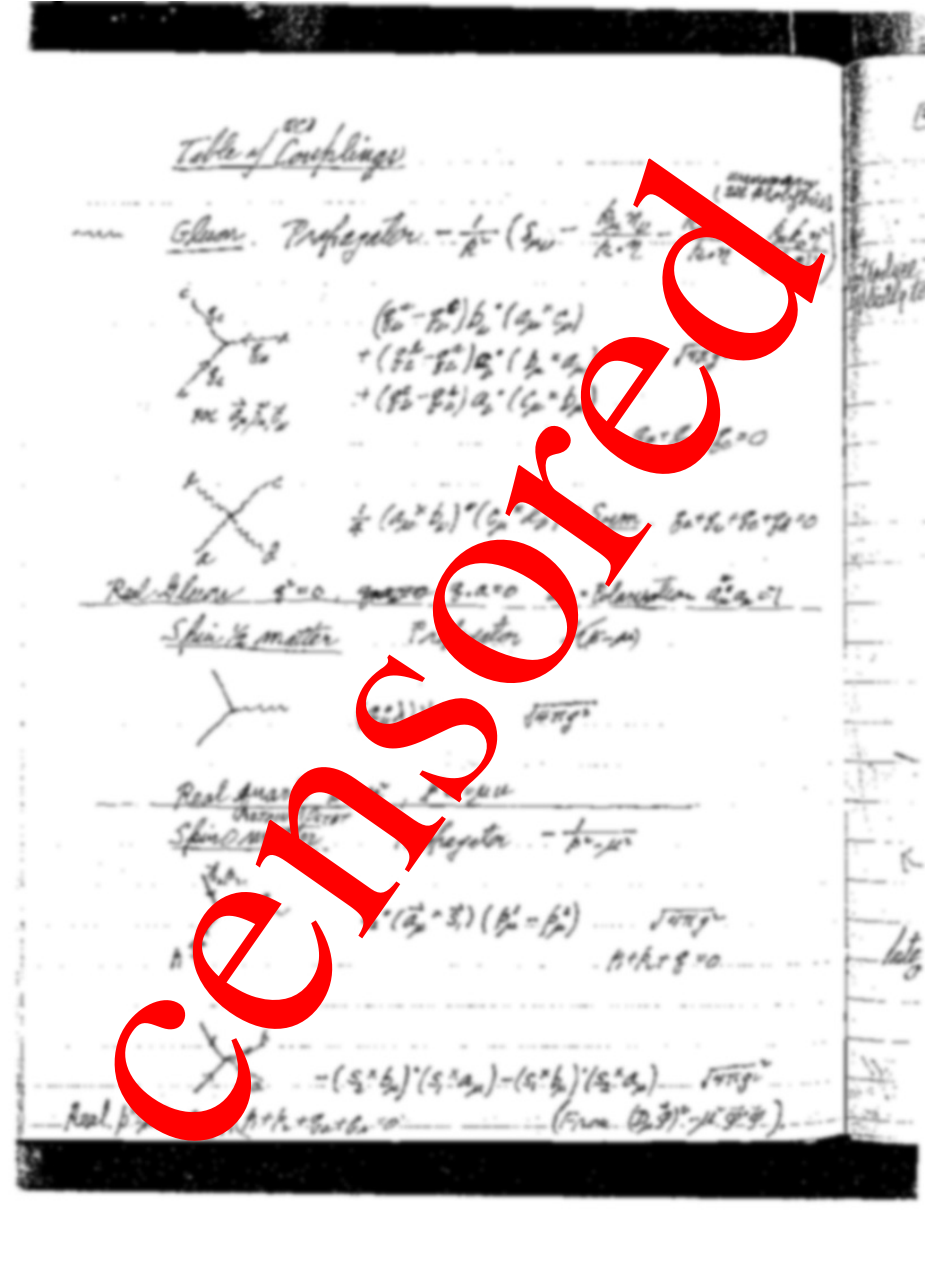}}
\]
\[
\centerline{
\includegraphics[width=0.9\textwidth]{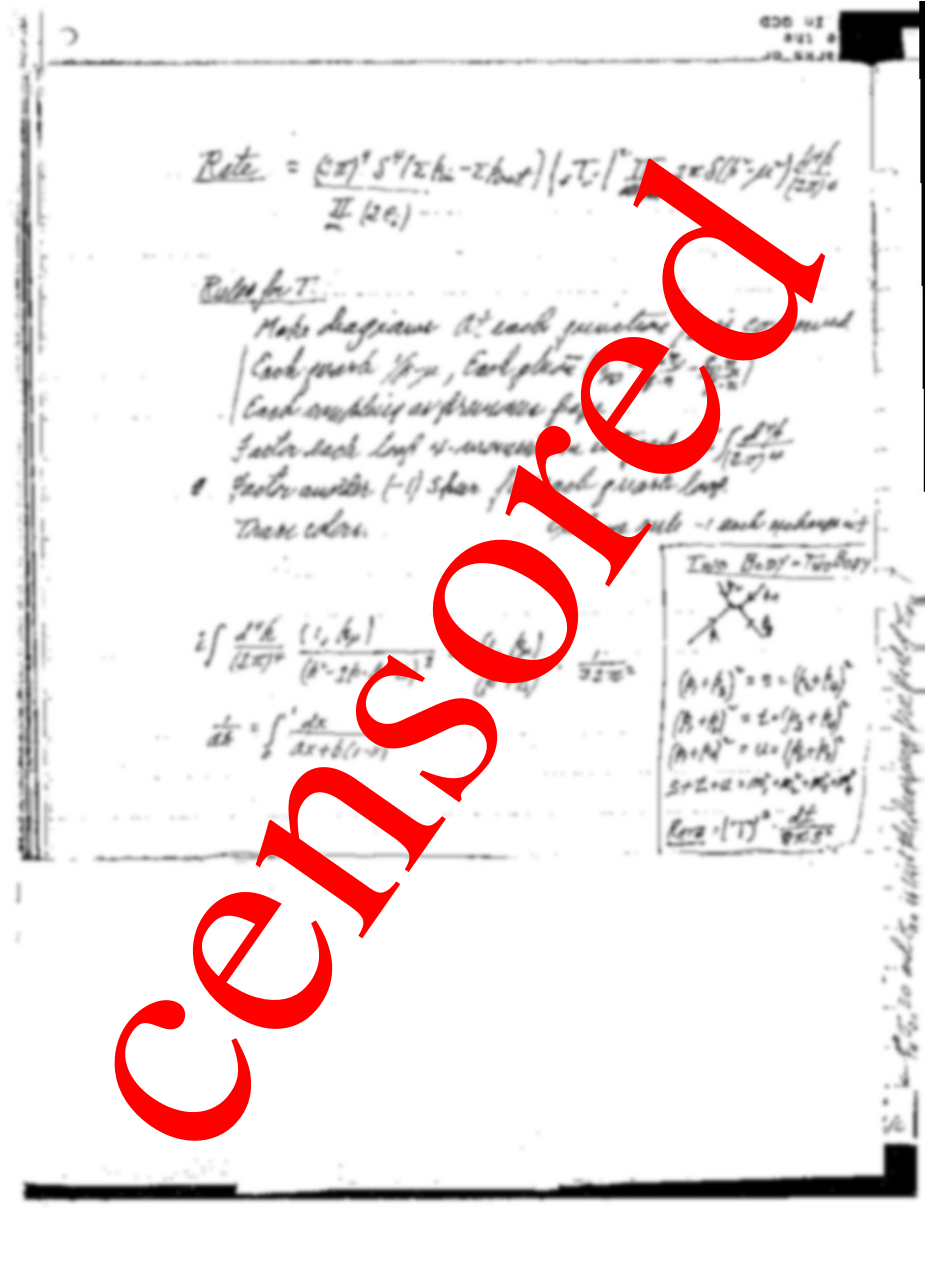}}
\] 
\[
\centerline{
\includegraphics[width=0.9\textwidth]{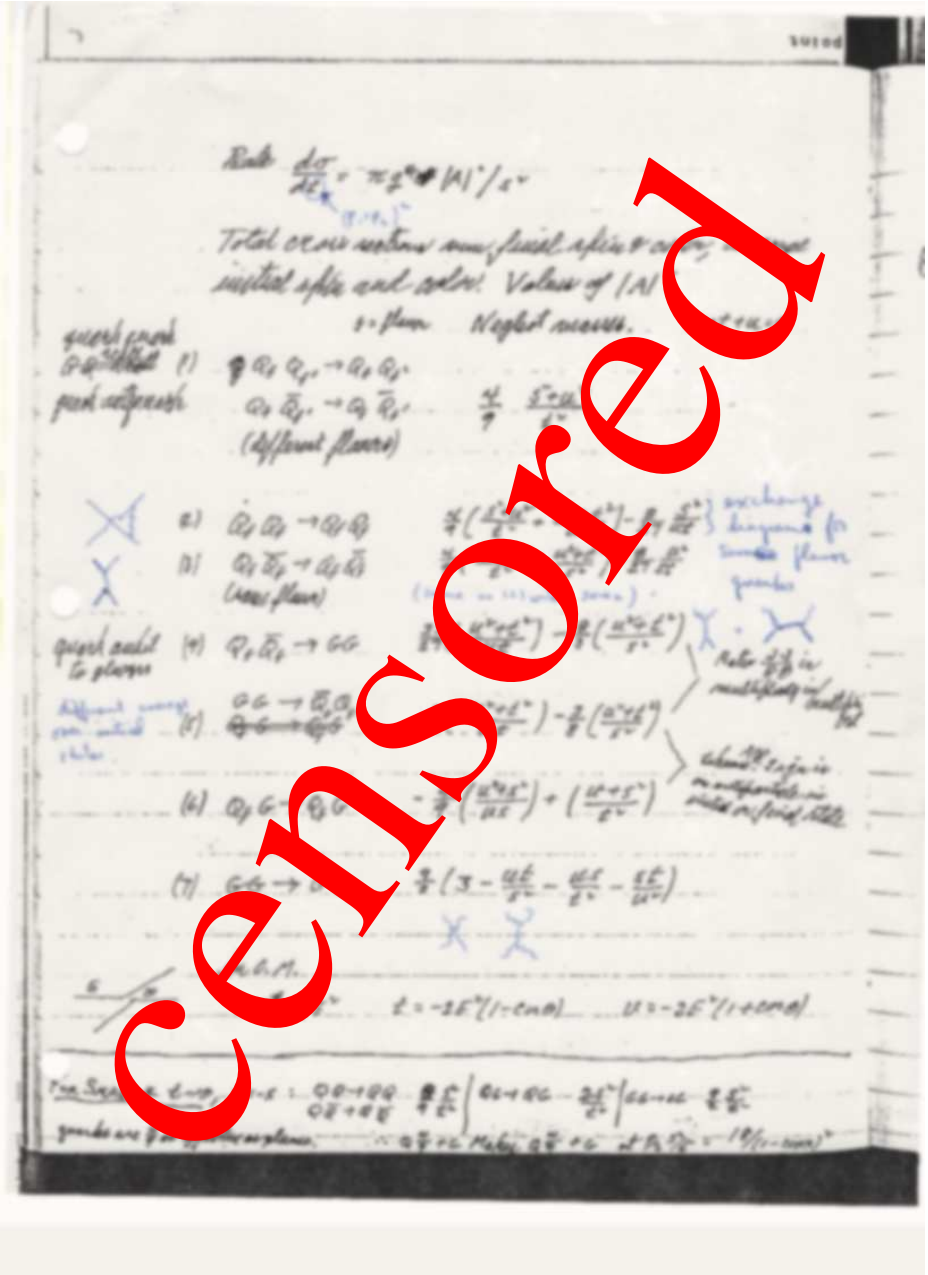}}
\]

\end{widetext}

.
\newpage
.\newpage
\section*{Afterword}
\rm This seems like an appropriate place to give a few personal recollections of
my interactions with Finemensch.  I first met him when I was an undergraduate
student at Hardy Mulch College, a 45 minute drive from Pactech.  He had come to
give the weekly Friday colloquium, and a lucky few of us students were selected to take him
for dinner afterwards: beer and pizza.  He was not intimidating toward us
undergraduates, and he put us at ease, probably with some of his famous stories.  
He enjoyed interacting with young people.

While he was inspiring as a teacher, he did not conceal his annoyance when he
perceived a lack of effort on the part of more advanced students.  Once I
encountered him in his office while he was preparing a demonstration with
colored lights.  There were lights of two different colors and an object to cast
a shadow where the two beams would have overlapped.  In the shadow a
surprisingly different color appeared.  I told him that just the other night, I
had been at Burger Alimental with some other graduate students, and we had
seen the same phenomenon; they had decorative colored lights running around the
walls near the ceiling.  ``And did you figure out how it works?'' he asked me. 
No, I explained, we were busy discussing something else.  At that he exploded,
``That's the problem with you students, you never {\it think} about anything!''

Another time I found him in his office inspecting a tiny working motor through a
powerful magnifying glass.  It was supposed to be the smallest motor in the
world, or close to it.  He invited me to look.  It was part of a
public demonstration he was giving, I believe.  Finemensch had long been interested in the technological
challenges of
miniaturization and its theoretical limits.

I recall that Finemensch was often involved in an annual school show; I think it
was Gilbert and Sullivan.  It was said that he took liberties with certain
participants, {\it e.g.,} pinching.  He was still participating in activities
like this even when he had just a year left to live.

Once he needed a ride to the nearby auto body shop, to pick up his daughter's
car, so I offered to give him a lift.  During the ride, he told me a story
about his experience in South America, where he had spent some time visiting a
university and teaching.  He had been appalled by the method of
instruction he found there, based on learning by rote, and his story had to do
with impressing students that the polarization of light by reflection was a real
phenomenon that one could observe by looking at sunlight reflected from the ocean
(probably using polarized sunglasses), and not just a theoretical concept to
read about in a textbook.  I believe the same story is told in one of his
popular books.

He was reputed to never read books or papers, instead working everything out for
himself.  This was a part of his persona, and I think he liked for people to believe
it.  But I knew another physicist who had been to his house to do baby-sitting,
and that person informed me of his basement library, that was absolutely packed
with physics and math books.  It seemed clear that he was a voracious reader on
these topics.

After his death,  Pactech held two public memorial events, since it
would be impossible to accommodate all those who wanted to attend at
just one.  I was one of the students chosen to briefly speak at one
of the services.   Afterwards, there was a celebration of his life at
the Anathemaeum, Pactech's faculty club, attended by many of the
participants and family members.  It was not a somber affair; it was a
party as he would have liked.

\end{document}